# Roadmap for Optical Tweezers

**Giovanni Volpe[1], Onofrio M. Maragò[2], Halina Rubinzstein-Dunlop[3]**, Giuseppe Pesce[4], Alexander B. Stilgoe[3], Giorgio Volpe[4], Georgiy Tkachenko[6], Viet Giang Truong[6], Síle Nic Chormaic[6], Fatemeh Kalantarifard[7], Parviz Elahi[8], Mikael Käll[9], Agnese Callegari[1], Manuel I. Marqués[10], Antonio A. R. Neves[11], Wendel L. Moreira[12], Adriana Fontes[13], Carlos L. Cesar[14, 15], Rosalba Saija[16], Abir Saidi[16], Paul Beck[17, 18], Jörg S. Eismann[17, 18, 19], Peter Banzer[17, 18, 19], Thales F.D. Fernandes[20], Francesco Pedaci[20], Warwick P Bowen[3], Rahul Vaippully[21], Muruga Lokesh[21], Basudev Roy[21], Gregor Thalhammer[22], Monika Ritsch-Marte[22], Laura Pérez García[1], Alejandro V. Arzola[23], Isaac Pérez Castillo[24], Aykut Argun[1], Till M. Muenker[25], Bart E. Vos[25], Timo Betz[25], Ilaria Cristiani[26], Paolo Minzioni[26], Peter J. Reece[27], Fan Wang[52], David McGloin[28], Justus C. Ndukaife[29], Romain Quidant[30], Reece P. Roberts[31], Cyril Laplane[31], Thomas Volz[31], Reuven Gordon[32], Dag Hanstorp[1], Javier Tello Marmolejo[1], Graham D. Bruce[33], Kishan Dholakia[33, 34, 35], Tongcang Li[36], Oto Brzobohatý[37], Stephen H. Simpson[37], Pavel Zemánek[37], Felix Ritort[38], Yael Roichman[39], Valeriia Bobkova[40], Raphael Wittkowski[41], Cornelia Denz[40], G.V. Pavan Kumar[42], Antonino Foti[2], Maria Grazia Donato[2], Pietro G. Gucciardi[2], L. Gardini[43, 44], G. Bianchi[43, 45], A. Kashchuk[43, 45], M. Capitanio[43, 45], Lynn Paterson[46], P. H. Jones[47], Kirstine Berg-Sørensen[7], Younes F. Barooji[48], Lene B. Oddershede[48], Pegah Pouladian[49], Daryl Preece[49], Caroline Beck Adiels[1], Anna Chiara De Luca[50], A. Magazzù[2], D. Bronte Ciriza[2], M. A. Iatì[2] and Grover A. Swartzlander, Jr.[51]

1 - Department of Physics, University of Gothenburg, 41296 Gothenburg, Sweden
2 - CNR-IPCF, Istituto per i Processi Chimico-Fisici, Messina
3 - ARC CoE for Engineered Quantum Systems, School of Mathematics and Physics, The University of Queensland, Australia
4 - Department of Chemistry, University College London, 20 Gordon Street, London WC1H 0AJ, UK
5 - Dipartimento di Fisica, Università degli studi di Napoli Federico II, Italy
6 - Okinawa Institute of Science and Technology Graduate University, Japan
7 - Department of Health Technology, Technical University of Denmark, Lyngby, Denmark
8 - Physics Department, Boğaziçi University, Istanbul, Turkey
9 - Chalmers University of Technology, Gothenburg, Sweden
10 - Departamento de Física de Materiales, IFIMAC & Instituto Nicolás Cabrera, Universidad Autónoma de Madrid, Spain
11 - Centro de Ciências Naturais e Humanas, Universidade Federal do ABC (UFABC), Santo André, São Paulo, Brazil
12 - Tratamento de dados geofísicos, Tecnologia e processamento de geologia e geofísica, Exploração, Petrobras - Petróleo Brasileiro S.A., Rio de Janeiro, Rio de Janeiro, Brazil
13 - Departamento de Biofísica e Radiobiologia, Universidade Federal de Pernambuco (UFPE), Recife, Pernambuco, Brazil.
14 – Departamento de Física, Universidade Federal do Ceará (UFC), Fortaleza, Ceará, Brazil.
15 - National Institute of Science and Technology on Photonics Applied to Cell Biology (INFABIC - IB and IFGW, UNICAMP), Campinas, São Paulo, Brazil
16 - Dipartimento di Scienze Matematiche e Informatiche, Scienze Fisiche e Scienze della Terra, Università di Messina, Italy
17 - Institute of Optics, Information and Photonics, University Erlangen-Nuremberg, Staudtstr. 7/B2, D-91058, Erlangen, Germany
18 - Max Planck Institute for the Science of Light, Staudtstr. 2, D-91058 Erlangen, Germany
19 - Institute of Physics, University of Graz, NAWI Graz, Universitätsplatz 5, 8010 Graz, Austria
20 - Centre de Biologie Structurale, CNRS, INSERM, Univ.Montpellier, France
21 - Indian Institute of Technology Madras, India
22 - Institute for Biomedical Physics, Medical University of Innsbruck, Austria
23 - Instituto de Física, Universidad Nacional Autónoma de México, C. P. 04510, Ciudad de México, México
24 - Departamento de Física, Universidad Autónoma Metropolitana-Iztapalapa, San Rafael Atlixco 186, Ciudad de México 09340, Mexico
25 - Third Institute of Physics - Biophysics, Georg August University Göttingen, Germany
26 - Department of Electrical, Computer and Biomedical Engineering, University of Pavia, Italy
27 - School of Physics, The University of New South Wales, Australia
28 - School of Electrical and Data Engineering, Faculty of Engineering and IT, University of Technology Sydney, Australia
29 - Vanderbilt Institute of Nanoscale Science and Engineering and Department of Electrical Engineering and Computer Science, Vanderbilt University, Nashville, Tennessee 37235, United States
30 - Nanophotonic Systems Laboratory, Department of Mechanical and Process Engineering, ETH Zürich, Zürich, Switzerland

31 - ARC CoE for Engineered Quantum Systems (EQUS), School of Mathematical and Physical Sciences, Macquarie University, 2109 NSW, Australia
32 - University of Victoria, Canada
33 - University of St Andrews, UK
34 - University of Adelaide, Australia
35 - Yonsei University, South Korea
36 - Department of Physics and Astronomy, and Elmore Family School of Electrical and Computer Engineering, Purdue University, Indiana, United States
37 - Czech Academy of Sciences, Institute of Scientific Instruments, Královopolská 147, 612 64 Brno, Czech Republic
38 - Small Biosystems Lab, Departament de Física de la Matèria Condensada, Facultat de Física, Universitat de Barcelona, Barcelona, Spain
39 - Tel-Aviv University, Israel
40 - Institute of Applied Physics, University of Muenster, Germany
41 - Institute of Theoretical Physics, Center for Soft Nanoscience, University of Muenster, Germany
42 - Department of Physics, Indian Institute of Science Education and Research, Pune, India
43 - LENS – European Laboratory for Non-linear Spectroscopy, University of Florence, Italy
44 - National Institute of Optics – National Research Council, Florence, Italy
45 - Department of Physics and Astronomy, University of Florence, Sesto Fiorentino, Italy
46 - Institute of Biological Chemistry, Biophysics and Bioengineering, School of Engineering and Physical Sciences, Heriot-Watt University, UK
47 - Department of Physics & Astronomy, University College London, UK
48 - Niels Bohr Institute, University of Copenhagen, Denmark
49 - Beckman Laser Institute, University of California Irvine, USA.
50 - IEOS-CNR, Institute of Experimental Oncology and Endocrinology "G. Salvatore", Napoli, Italy
51 - Center for Imaging Science, Rochester Institute of Technology, Rochester (NY), USA
52 - School of Physics, Beihang University, China

**Guest Editors:**
Giovanni Volpe: giovanni.volpe@physics.gu.se
Onofrio M. Maragò: onofrio.marago@cnr.it
Halina Rubinzstein-Dunlop: halina@physics.uq.edu.au

**Abstract**

Optical tweezers are tools made of light that enable contactless pushing, trapping, and manipulation of objects ranging from atoms to space light sails. Since the pioneering work by Arthur Ashkin in the 1970s, optical tweezers have evolved into sophisticated instruments and have been employed in a broad range of applications in life sciences, physics, and engineering. These include accurate force and torque measurement at the femtonewton level, microrheology of complex fluids, single micro- and nanoparticle spectroscopy, single-cell analysis, and statistical-physics experiments. This roadmap provides insights into current investigations involving optical forces and optical tweezers from their theoretical foundations to designs and setups. It also offers perspectives for applications to a wide range of research fields, from biophysics to space exploration.

**Introduction**

Optical forces emerge from momentum exchange between light and matter [1]. Despite their minuteness, they yield macroscopic consequences in extraterrestrial affairs. For example, already in the 17th century, Kepler recognized that the tails of comets are a consequence of the light pressure produced by the Sun light [2]. More recently, radiation pressure has been shown to be a key factor to ensure the stability of a star against gravitational collapse [3]. However, their role in laboratory experiments on Earth was elusive for many years [4,5]. The turning point came with the advent of the laser age that offered intense and coherent light sources [6]. These were used by Arthur Ashkin [7] and co-workers in a series of pioneering experiments to trap and manipulate particles [8,9], atoms [10-12], and biological matter [13,14].

In its simplest configuration, an optical tweezers is created by a single laser beam focused down to its diffraction limit, which creates the light intensity gradients needed to hold particles in three dimensions. Nowadays, optical tweezers have become sophisticated and versatile instruments that can hold, guide, push, stretch, poke, probe, sort particles, ranging from single atoms to microparticles, to cells and bacteria in a variety of environments, such as liquids, gases, or vacuum (**Figure 1**).

In this Roadmap, we collected a series of perspectives that give an overview of the different research areas where optical tweezers have found application and have advanced the science of light–matter interactions further. We also look at current and future challenges in the field and provide insights to what are possible future directions for the field. This Roadmap is organized as follows. We start by presenting information on experimental designs and setups, from the standard single-beam configuration to more complex designs, including recent developments employing intracavity feedback and metamaterials to generate optical forces. Then, we explore some fundamental aspects of theory and simulations that are crucial for a deeper understanding of experiments. Subsequently, we highlight some key applications for the accurate measurement of forces and torques. We show how the tweezers can be used for the characterization of the microrheological properties of fluids, the control and manipulation of nanoparticles, the trapping of particles in gas and vacuum, the exploration of statistical physics and nanothermodynamics, their combination with spectroscopy, their use for the study of biological matter from biomolecules to cells, and the use of optical forces in space applications. This Roadmap provides both an overview of the current state of the art in optical tweezers research and a guidance towards its future.

**Acknowledgements**

O.M.M. and G.V. acknowledge support from the MSCA-ITN-ETN project ActiveMatter sponsored by the European Commission (Horizon 2020, Project No. 812780). O.M.M. acknowledges support from the agreement ASI-INAF n.2018-16-HH.0, project "SPACE Tweezers". H.R-D acknowledges support under Australian Research Council's Discovery Projects funding scheme (project number DP180101002) and Australian Research Council Centre of Excellence for Engineered Quantum Systems (EQUS, CE170100009).

We wish to dedicate this Roadmap to the memory of Arthur Ashkin, Ferdinando Borghese, Michael Mishchenko, and Juan José Sáenz. These pages are part of their scientific legacy made of visionary and pioneering ideas that led to novel interdisciplinary research fields and applications.

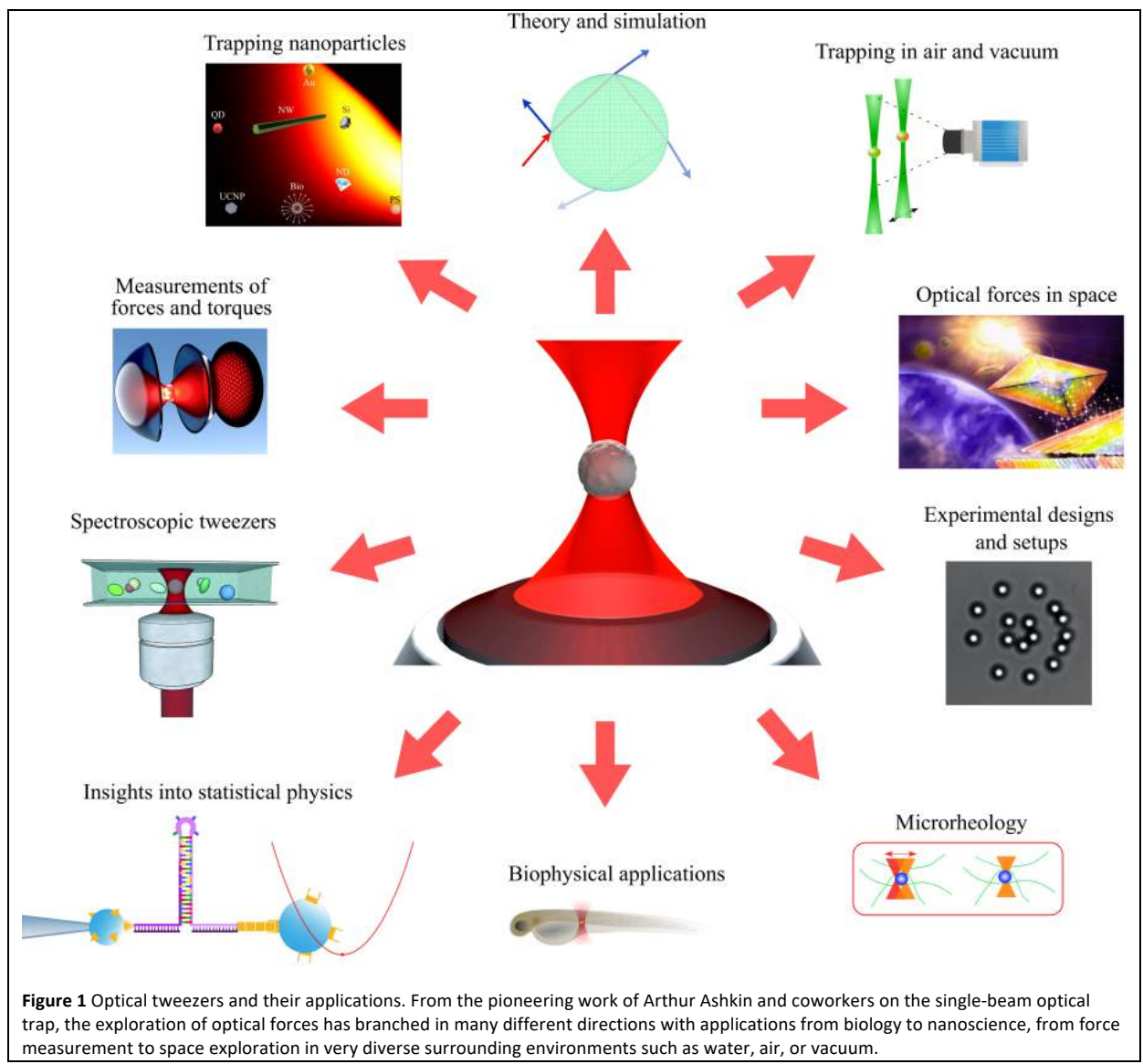

**Figure 1** Optical tweezers and their applications. From the pioneering work of Arthur Ashkin and coworkers on the single-beam optical trap, the exploration of optical forces has branched in many different directions with applications from biology to nanoscience, from force measurement to space exploration in very diverse surrounding environments such as water, air, or vacuum.

**EXPERIMENTAL DESIGNS AND SETUPS**

## 1 — Optical tweezers setups

*Giuseppe Pesce*

Dipartimento di Fisica, Università degli studi di Napoli Federico II

**Status**

Back in 1986, Arthur Ashkin and colleagues successfully trapped in 3D micrometer and nanometer size particles, dispersed in water, with a laser beam focused by a high numerical aperture objective lens [1]. This is how the amazing and powerful world of optical manipulation began. Since then, optical manipulation has been used in a myriad of experiments across different fields: biology, statistical mechanics, rheology, opto- and bio-fluidics, plasmonics are only few examples of research fields where optical tweezers (OT) played a key role. The ability to manipulate microscopic objects contactless was rapidly overcome by more quantitative experiments. OT are not just simple "tweezers": they are force transducers, as they can exert and measure forces, very tiny forces. The available range spans a very interesting scale: from few piconewtons (pN) down to few femtonewtons (fN). As a matter of fact, OT expanded this range, until then dominated but also limited by atomic force microscopy. This force range is extremely interesting since it comprises many biological, chemical and, more in general, condensed matter interactions. Moreover, forces around 1 pN correspond to energies of few $k_B T$, i.e., energies comparable to those of thermal baths. In other words, the dynamics of trapped particles can be used to understand thermodynamics properties of the medium in which the particles are dispersed. Thus, OT were immediately employed in statistical mechanics as well as in microrheology [2].

The need to unveil so many different phenomena stimulated researchers to develop new setup schemes and novel trapping techniques [3,4]. The single-laser-beam OT immediately turned into OT with two or more laser beams, first dividing the beam with beam-splitters then using acousto-optical deflectors (AOD) to obtain time shared traps. Later, very versatile devices like spatial light modulators (SLM) allowed researchers to explore other directions far away from simple trapping and manipulation. With these devices the light wavefront can be modified and engineered on demand: initially, multiple trapping, optical lattices and sieves demonstrated the power of this new technique, but it did not take long to see experiments where the angular momentum of light played a key role.

Despite its nearly forty-year history, optical trapping and manipulation is still an active research field. The recent advances in trapping of nanoparticles in vacuum have opened new challenges on testing fundamental physics, and quantum information science, many-body physics, metrology, quantum optics and ultracold chemistry.

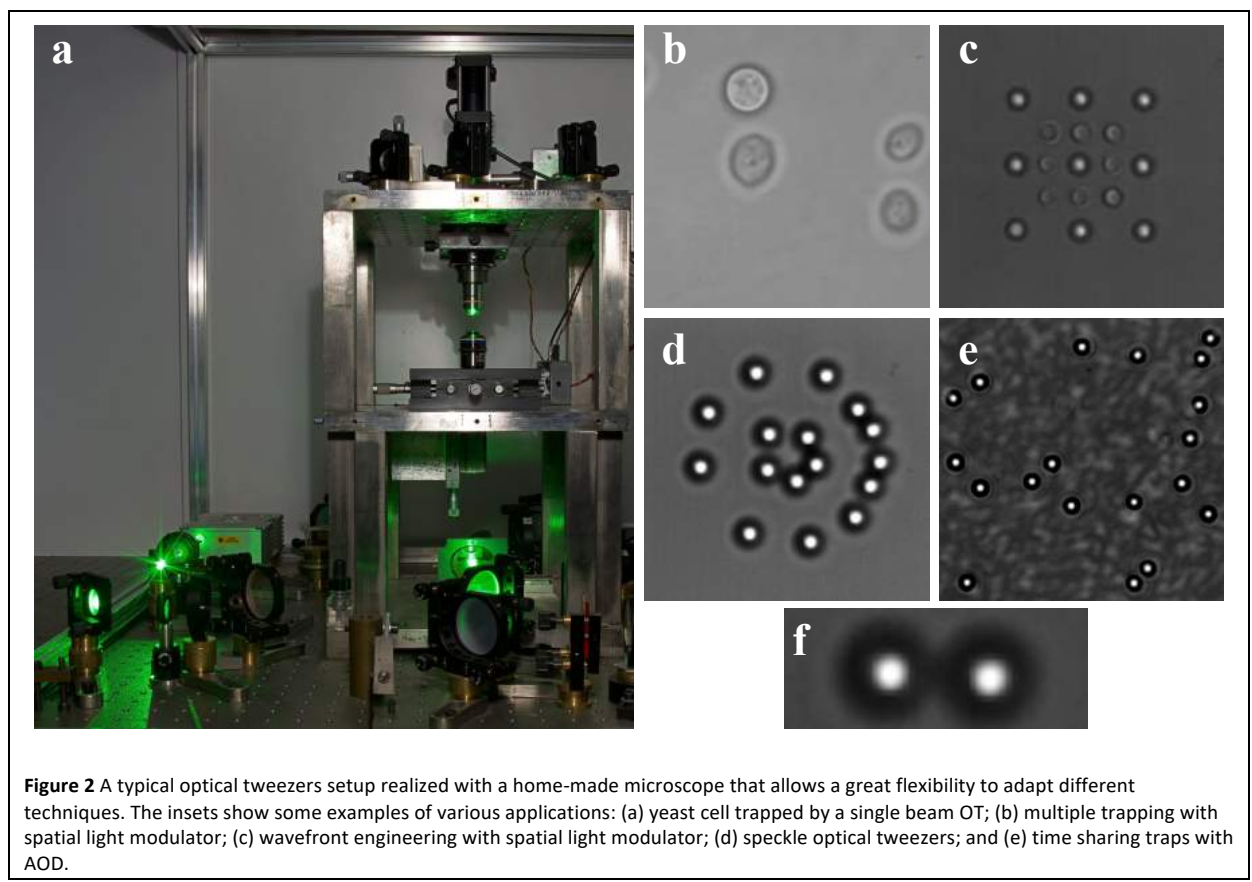

**Figure 2** A typical optical tweezers setup realized with a home-made microscope that allows a great flexibility to adapt different techniques. The insets show some examples of various applications: (a) yeast cell trapped by a single beam OT; (b) multiple trapping with spatial light modulator; (c) wavefront engineering with spatial light modulator; (d) speckle optical tweezers; and (e) time sharing traps with AOD.

**Current and Future Challenges**

In biological applications, OTs have contributed to reach exceptional results. Nevertheless, one of their main limitations still needs to be resolved. To exert reasonable forces on biological samples, OT use several milliwatts of light power that, even if at infrared wavelengths where the absorption is reduced, could damage the sample. Several attempts have been made to solve this issue, mainly using handles (beads or structures made with appropriate shape) that once trapped can be used to manipulate cells avoid heating due to incident light. Solving this issue is of a fundamental importance to boost the usefulness of OTs in biomedical applications especially for *in vivo* scenarios [5].

Another aspect that currently hampers OT applicability, limiting it prevalently to fundamental research only, is the size and cost of setups. Turn-key systems, provided by some companies, have quite high prices, because of the costs of optical components and those related to the development of stable systems. In fact, they are expected to provide easy and reliable operation.

OT require microscope objectives to obtain gradient forces high enough to reach the trapping condition, otherwise two counter propagating beams can be used to cancel the scattering force, this representing a great limit in setups development. Home-made systems can be less expensive, but often are too sophisticated and very large in size to be handled by unexperienced personnel and in routine applications, even if built with commercial microscopes. In any case, they comprise microscope objectives whose price is in the order of thousands of euros. Alternative solutions to microscope objectives are based on using optics fibers, which can be arranged in counterpropagating beams or creating a lens at their tip able to strongly focus the laser light and obtain a trapping condition. Microfluidics and fiber optics have greatly helped to move towards automatization and compactness, but a stand-alone trapping chip device is far to be realized.

Very recently, research in optical manipulation has shown a renewed interest in trapping in air and in vacuum. In vacuum the motion of an optically levitated dielectric nanosphere is well isolated from the environment and with appropriate techniques it is possible to cool down the motional energy of a trapped particle to demonstrate the quantum behaviour of a macroscopic object at room temperature. To this end, setups are becoming more and more sophisticated and new technical challenges must be faced [4].

**Advances in Science and Technology to Meet Challenges**

To overcome the issue of thermal damage to biological samples caused by laser power, several approaches can be pursued. A possibility is to develop new trapping schemes that use higher trapping efficiency. Very recently, a new idea based on intracavity trapping has been proposed and demonstrated [6]. Trapping occurs inside a laser cavity instead of using its output laser beam as usually done. The trapped object increases the power losses when it is at the beam waist centre so that the power is low: as it moves out, losses decrease and the power increases pulling the object towards the centre and the power decreases again.

Other possibilities rely on developing special handles so that the parts in contact with the cells are not held directly by the laser beam. This is quite simple to do, and it is already widely used, but is limited to manipulation only. If quantitative measurements are required, it is necessary to explore new ideas. Other physical effects can be used as well. Li and colleagues proposed an interesting solution where the laser beam cools down a small volume of appropriate material and samples are trapped by thermophoresis [7].

However, the main issue, especially when looking at the establishment of OT as routine technique, remains the cost and the size of the setup. OT are strictly bond to microscope objectives. Efforts should be made to develop new technologies able to produce smaller and less expensive lenses with the same focusing properties. Metamaterials have been successfully used to create a very tiny meta-lens using nanopillars on a glass coverslip [8,9]. This has still some limitation in terms of flexibility and applicability, since the lens is fabricated on the substrate, and its position and trapping depth are fixed or very limited; nevertheless, this may pave the way towards new and interesting directions.

**Concluding Remarks**

The ability to manipulate small objects with focused laser beams opened a broad spectrum of opportunities in fundamental and applied studies such as biosciences, sorting, guiding, and analysis of materials. These experiments require precise control over mechanical path and high stability. OT setups are cumbersome and are limited by a lack of flexibility and integrability. The cost of optical components is another key point that restricts the diffusion of trapping and manipulations setups. Since its beginning, about forty years ago, OT have shown a constant tendency to evolve by exploiting new research quests which are only limited by the advances in technology.

## 2 — Structured light for the tuning of light–matter interactions

*Alexander B. Stilgoe, Halina Rubinsztein-Dunlop*

ARC Centre for Engineered Quantum Systems, The University of Queensland, St. Lucia, Brisbane, Australia

School of Mathematics and Physics, The University of Queensland, St. Lucia, Brisbane, Australia

**Status**

The use of structured light in optical tweezers and its novel application built a deep understanding of several complex biological and physical systems [1]. Light can be structured/sculpted by modifying its intensity distribution, phase, and polarisation states with various spatial light modulator technologies. Two common spatial light modulator (SLM) technologies to create structured light are derived from liquid crystal SLM and digital micromirror devices (DMDs). These devices enable manipulation of the intensity and phase of the light as well as polarisation states. Structured light enables unprecedented control over light-matter interactions. The historical killer application of beam shaping using these devices has tended to be the multiple independent control of focused Gaussian spots or the introduction of angular momentum through azimuthal phase ramps in the Fourier plane [1,2]. The operational effect is the transfer of momentum to particles with refractive index contrast to the surrounding medium through refraction and interference. More recently, spatially varying phase, amplitude, and polarisation have been developed to enhance control of light–matter interactions such as the transfer of the exotic transverse angular momentum [3] (also **Figure 3**), or the multiple independent control of radial, linear, or azimuthally polarised laser beam traps [4, 5]. An opportunity for the development of a powerful new technique based on the theory of the optimal control of light–matter interaction to precisely control not merely linear, but also angular, transfers of momentum exist. Demonstrations of exceptional change in linear momentum have been demonstrated using carefully structured light [6, 7], and the modification of particle properties [8, 7].

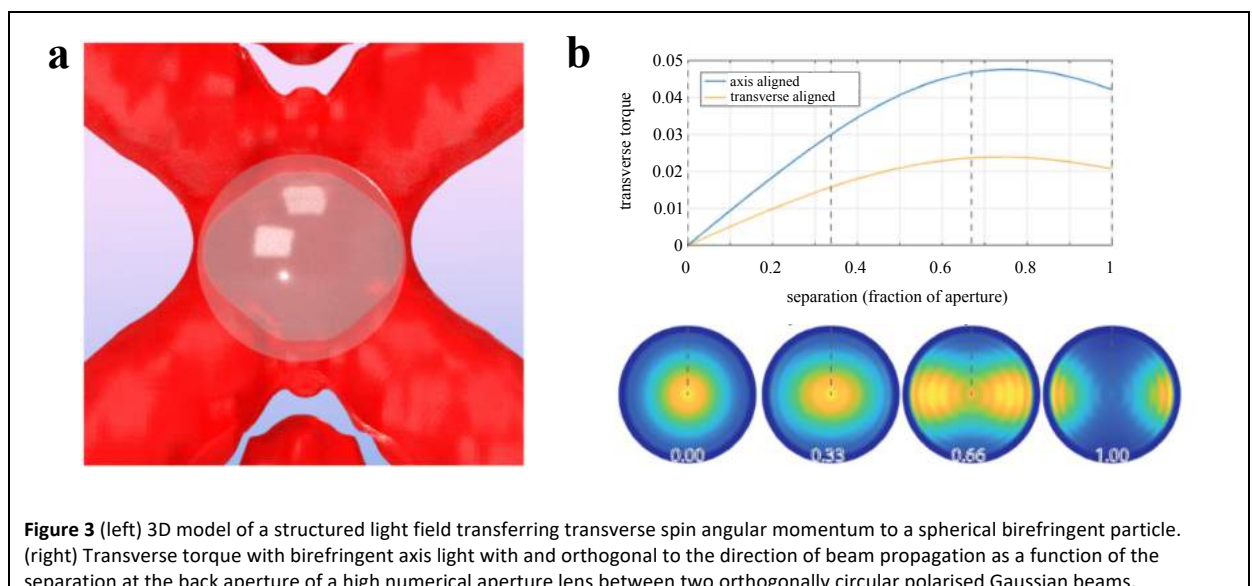

**Figure 3** (left) 3D model of a structured light field transferring transverse spin angular momentum to a spherical birefringent particle. (right) Transverse torque with birefringent axis light with and orthogonal to the direction of beam propagation as a function of the separation at the back aperture of a high numerical aperture lens between two orthogonally circular polarised Gaussian beams.

**Current and Future Challenges**

The idea of controlling the electromagnetic coupling between the incident and scattered components of light scattered by a particular particle is well established [6, 9] but it is limited by a set of challenges.

Ordinarily, optical tweezers act on spherical particles with simple structured laser beams. Simple and not necessarily complementary transfers of linear and angular momentum occur. For example, Laguerre–Gaussian type beam modes with large azimuthal number are notoriously poor at confining small particles to the focus of a beam in 3D. Simultaneous control of stable linear and angular momentum transfer to matter is elusive. Partly, this is due to the homogeneity of the plastic and glass particles commonly used in optical tweezers. Strongly anisotropic particles such as those with shape and material birefringence, trirefringence, or exceptionally high refractive index objects [8] have much richer interactions with the electromagnetic field. Exotic particles can have capabilities beyond that of ordinary materials [8,7] such as forces on the nano-newton scale and the transfer of spin or orbital angular momentum. In order to precisely control these exotic particles, we have to understand what the best arrangement of modes is to do so. Gaussian and higher order Laguerre– or Hermite–Gaussian modes are not good candidates to control these particles as their optimal interactions are influenced by interference and emergent dielectric resonances [9].

Optimised light fields have been experimentally realised for a range of different momentum operators in 2D [9]. The extension to 3D is possible. However, there are challenges with these approaches, namely, interactions are generally derived from the eigenspectrum of the effective measurement of momentum. Two problems arise that need to be addressed: (1) Not all fields are realisable, particularly in an optical tweezers where light moves through an aperture, and (2) optimisation of one parameter, such as spin angular momentum transfer or stiffness in one dimension, will not necessarily yield a field in which a particle can be either levitated or three dimensionally trapped (**Figure 4**). There are some strategies that can address this, ranging from modification of the theories to allow more stable configurations of less optimal transfers of momentum, to increasing the number of manipulated states, and to using a specially designed structure. For example, self-stabilizing levitation of a surface in low pressures [10].

The technical problems producing a stable configuration allowing for less optimal transfers of momentum runs deep as there are a large number of ways that a momentum transfer may be made sub-optimal but very few that will sufficiently improve stability. Subtle differences in the dimension and shape of particles are known to influence measurements in ordinary optical trapping experiments. In this circumstance it will also prevent the achievement of both optimal interactions and the correction for trapping stability.

To increase the number of states used in the interaction vectorial beam shaping techniques can be used. The challenge with implementation will be ensuring a good beam quality, stability (reliability), and alignment of the different polarisation channels remain an ongoing battle [5].

Specially designed surfaces are an interesting approach. The challenge here is to design materials/structure that also allows for further enhancement of the light–matter interaction with the structured light that may be needed for a particular type of momentum transfer.

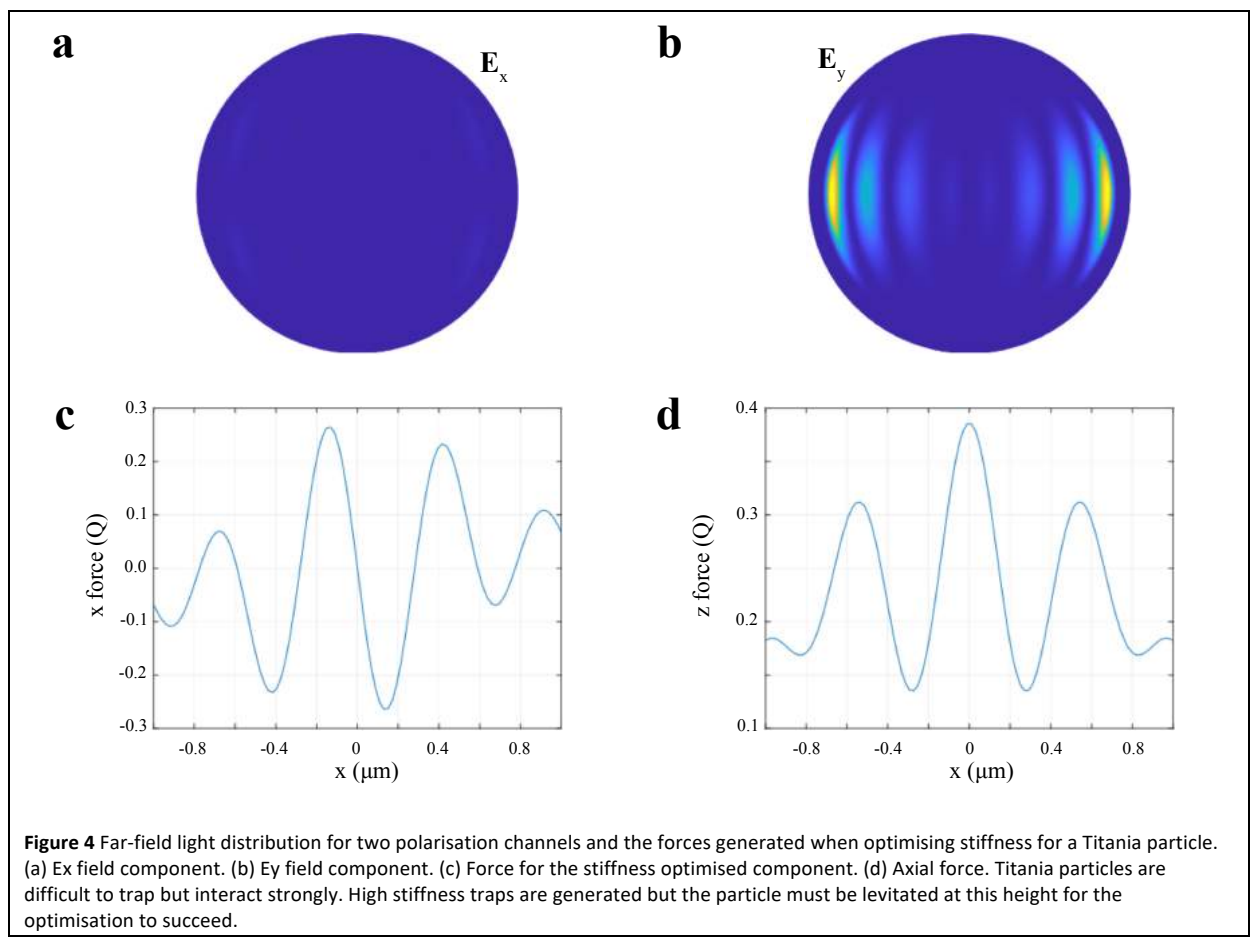

**Figure 4** Far-field light distribution for two polarisation channels and the forces generated when optimising stiffness for a Titania particle. (a) Ex field component. (b) Ey field component. (c) Force for the stiffness optimised component. (d) Axial force. Titania particles are difficult to trap but interact strongly. High stiffness traps are generated but the particle must be levitated at this height for the optimisation to succeed.

**Advances in Science and Technology to Meet Challenges**

Marrying these exotic particles with beam shaping could prove to be a powerful technique that has applications not only to trapping in water, but the examination of macroscopic levitated quantum states. There have been impressive and interesting developments in this area [11,12] that have developed subtle control and sensitive detection of particles that may be more generally usable for fluid environments.

The power of optimised structured light and materials lies in the ability to precisely tune the light–matter interaction, for example, to enable new high-precision measurements of position and momentum, or extend trapping to low or negative refractive index contrasts, or to self-contained lab-on-a-chip devices. Recent developments in the consistent and rapid measurement of viscosity at ballistic timescales [13] would be enhanced through the inclusion of shaped particles and optimised/shaped trapping fields.

Addressing the stable configuration challenge may be the hardest of those listed here. The solution requires a clearly laid out problem where an optimisation can be performed that also preserves desired properties of the optical trap, for example, a stable trapping position, or a consistent transverse of torque with rotation (see the left panel of **Figure 3** where two alignments of the optical axis have different levels of torque transfer). Perhaps moving away from the optimisation using an effective momentum measurement operator is a good idea as it cannot by its very construction have a good solution with these constraints.

The preferred solution for vectorial beam shaping would be the production of Huygens' style sources of on-demand beams with spatially varying polarisation. The use of polarised light shaping is growing and this may be a product that becomes available in the near future. Otherwise, better designs of the polarisation shifting interferometers will be needed to increase stability and allow optimal light–matter interactions.

As fabrication technology continually advances, new and structured materials will become available over time. There are two levels of the complexity of the design of these materials. The first, where simple beams and plane waves supply the momentum to control them. The situation where structured illumination is used is much less certain. It provides an opportunity to investigate various forms of iterative design where a light field and its target object can be subject to individual constraints such as uniform polarisation, pulse mode operation, feature size, and so on. It provides one of the most flexible approaches to the problem of enhancing optical trapping in multiple translational and rotational degrees of freedom.

**Concluding Remarks**

Unprecedented versatility and superb control of light -matter interaction on nano- and microscale has been enabled with optical beam shaping. The progress made till now is extremely encouraging although there are still outstanding challenges in applying these methods to wide variety of systems. There are remaining questions on how to tune light–matter interactions for versatile uses in an arbitrary system. The availability of well-defined and dynamic structured light fields could enable novel studies in complex biological systems where there can be limitations in knowledge about the scattering properties in the surrounding environment. New shaped fields could open up investigations and further exciting studies in the interface of classical and quantum mechanics by allowing for more precise tuning of the momentum transferred to and from a target system, allowing for better understanding of phenomena such as: the cooling to the ground state of systems over a large size range, studies of quantum behaviours at this scales, and out-of-equilibrium phenomena. The combination of optical tweezers using highly structured light fields with other imaging and spectroscopic techniques could enable insights into the burgeoning fields of quantum biotechnology and continue its contribution to the understanding of mechanobiology.

**Acknowledgements**

The authors acknowledge support from the Australian Research Council Discovery Project DP180101002 as well as from the Australian Research Council Centre of Excellence for Engineered Quantum Systems (EQUS,385 CE170100009).

### 3 — Speckle optical tweezers

*Giorgio Volpe*
Department of Chemistry, University College London, 20 Gordon Street, London WC1H 0AJ, UK

**Status**

Speckle optical tweezers are an optical manipulation technique that relies on disordered light fields to exert optical forces on micro- and nanoparticles [1] (**Figure 5**). They are named after optical speckle patterns, which are light fields with a random appearance, yet with universal statistical properties [2]. These optical patterns result from the interference of several optical waves with random phases. For a long time, the field of optical trapping and manipulation has considered these light patterns as an undesirable consequence of the propagation of coherent light through optically complex media, such as diffusers, multimode optical fibres, colloidal dispersions, turbid liquids, and biological tissue. Differently from most optical manipulation techniques where disorder is an issue to be minimized, speckle optical tweezers take advantage of the random nature of these light fields to address fundamental physical questions and for applications in microfluidics [1-3]. Historically, the atom cooling community was the first to adopt speckle light fields to trap small particles [4]. The last decade has experienced a revival of the technique to tackle questions in mesoscopic physics, where both static and time-varying speckle optical potentials have been used to study the emergence of anomalous diffusion (from subdiffusion to superdiffusion) in colloidal dispersions [2,5-8], to tune effective dispersion forces between small colloidal particles [9], to perform standard microfluidic operations, such as particle guiding and sorting [1], to assemble two-dimensional crystal-like and glassy colloidal materials [10], to control collective behaviours in active matter [11], and to reproduce first-passage statistics [12].

The future appeal of the technique (subject to further technical advances) is both fundamental and applied. On a more fundamental side, speckle optical tweezers can be further developed into a fully controllable proxy for many natural phenomena where particles move within potentials in the presence of disorder. Examples range from the emergence and propagation of defects in materials to the motility of molecules and organelles in cells, from the individual and collective motility of groups of animals to the diffusion of stars within a galaxy. The desirable feature in this model system would be that all variables and degrees of freedom of interest can in principle be controlled at the touch of a knob, i.e., light. On a more applied note, speckle optical tweezers can outperform standard optical manipulation techniques in instances (e.g., in microfluidic and biomedical applications), where simplicity of operation and robustness to environmental noise are key.

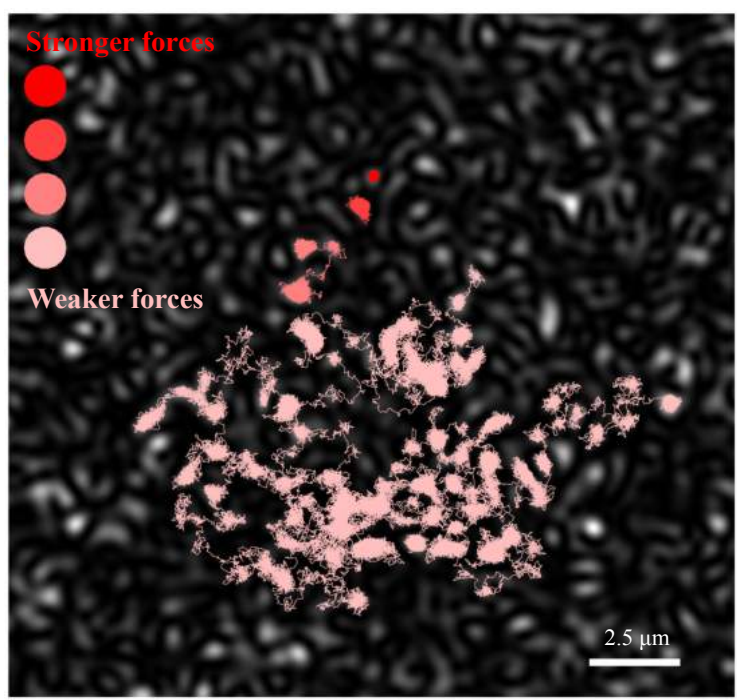

**Figure 5** Speckle optical tweezers at work. The trajectories of a Brownian microsphere become more and more confined the stronger the average force exerted by the bright spots of a speckle pattern (black and white background). The trajectories are color-coded (legend) for the average force experienced by the particle.

**Current and Future Challenges**

The main research challenge behind further advancements in speckle optical tweezers is intimately connected with the nature of optical disorder and its control. As the statistical properties of standard optical speckle patterns (i.e., their intensity distribution and correlation function) are universal [2], advancements in the field will go hand in hand with advancements in our capabilities to engineer disordered optical potentials in space and time:

- Structured speckle tweezers. Despite their random appearance, the universal properties of speckle fields have been harnessed to influence the translational diffusion of microscopic particles [2,5-7]. Customizing speckle properties by modulating the amplitude, phase, and polarization of the light field can enable the implementation of more complex speckle patterns with, e.g., tailored intensity distributions and correlations. Speckle optical tweezers could therefore be applied to perform complex optical manipulation tasks based on controlling both the translational and rotational diffusion of microparticles, such as by introducing short-, mid-, and long-range correlations in the particles' motion or by generating torques with advanced disordered non-conservative force fields.

- Three-dimensional speckle tweezers. A key feature of many optical manipulation techniques is the ability to trap and guide particles in three-dimensional space using single or multiple laser beams. A downside of most current speckle tweezers setups is that they can be operated primarily near a surface or at an interface. Indeed, the diffraction processes typically used to generate speckle patterns lead to out-of-plane divergence and longitudinal speckle grain elongation. Both aspects hamper non-trivial three-dimensional manipulation tasks with speckle optical potentials. Controlling the statistics of optical disorder over a volume (i.e., along the direction of light propagation as well as along the transverse plane) could enable the systematic extension of the technique to the third dimension [3].

- Near-field speckle tweezers. While trapping with random plasmonic islands has been demonstrated [13], using the localized fields generated by random metallic or dielectric nanostructures to manipulate small particles has an untapped potential for scaling speckle tweezers down towards the molecular level and for embedding them with molecular spectroscopy and sensing capabilities.
- *In vivo* speckle tweezers. Particularly alluring is finally the possibility of using speckle optical tweezers to directly manipulate matter *in vivo*, where the exploitation of other non-invasive optical manipulation techniques is compromised by the natural optical turbidity of biological tissue. Nonetheless, for this to happen, optical methods need to be able to penetrate deeper into tissue and guarantee that adequate levels of power and imaging capabilities are met simultaneously.

**Advances in Science and Technology to Meet Challenges**

As speckle optical tweezers operate exploiting complex optical potentials, developments in the field will benefit greatly from further scientific and technological advances in the fields of structured light and complex photonics. The possibility of structuring disordered optical potentials in three dimensions is intimately connected with our technical capabilities of controlling a large number of optical modes as well as all three components of the electric field simultaneously. Current wavefront shaping devices, such as spatial light modulators (SLMs) and digital micromirror devices (DMDs), have driven significant scientific progress so far in these fields. Further development of these technologies (in terms of, e.g., efficiency, number of pixels, spatial homogeneity, pixel crosstalks, spectral bandwidth and speed) along with the advent of novel devices capable of independently controlling all classical degrees of freedom of light at once will undoubtedly expand the range of applications also for speckle optical tweezers. Similarly, the realization and adoption of turnkey laser systems capable of supplying light beams with tailored structures would enable the straightforward generation of a great variety of high-power complex optical potentials for optical manipulation purposes. Alternatively, the fabrication of disordered plasmonic and dielectric surfaces and metasurfaces for the generation of structured light within optofluidic chips or at the distal end of an optical fiber can help the miniaturization of speckle optical tweezers, their extension to trapping applications on the nanoscale, and their use for *in-vivo* optical manipulation tasks. This possibility, while alluring, places high strains on current nanofabrication methods and modelling techniques. Moreover, non-invasive *in-vivo* applications of speckle optical tweezers require light shaping methods to guarantee higher efficiencies in terms of enhancing and stabilizing trapping potentials deep within tissue. Improvements in deep tissue imaging techniques (e.g., based on acoustic guiding), fast data handling and machine learning methods will also be a welcome development for the *in-vivo* implementation of speckle optical tweezers. Finally, from a more pragmatic point of view, simplicity of operation, robustness to environmental noise and low cost are major selling points for current speckle optical tweezers setups. Thus, any technical development that extends the range of operability of this technique will need to be affordable and easily available to the broader scientific and engineering community. To date, affordability remains a major practical challenge to the integration of novel technology in speckle optical tweezer setups.

**Concluding Remarks**

In conclusions, despite the random appearance of optical speckle fields, their statistical properties have been exploited to control and manipulate the diffusion properties of small particles. Although carefully engineered periodic potentials can perform better in specific optical manipulation applications, speckle optical tweezers offer additional advantages being intrinsically simple to operate, low cost, widefield and robust to aberrations from the optics and the surrounding environment. Beyond their applied interest for microfluidic, biomedical, and *in-vivo* applications, speckle optical tweezers are an ideal tool in the hands of researchers to tackle fundamental scientific questions in both equilibrium and non-equilibrium statistical physics and materials science. To expand the applicability of this tool further, the possibility of structuring optical disorder in two and three dimensions in space and over time is an attractive avenue to directly engineer new light-matter interactions in random optical potentials. Nonetheless, to maintain the broad appeal of speckle optical tweezers, it is key that the integration of novel technologies does not alter the fundamental character of the method, i.e., its accessibility to an extended scientific and technical audience.

**Acknowledgements**

I would like to acknowledge Sylvain Gigan and Giovanni Volpe for critical reading and fruitful discussions on the future of the technique.

## 4 — Optical trapping and manipulation mediated by optical nanofibres

*Georgiy Tkachenko, Viet Giang Truong, Síle Nic Chormaic*
Okinawa Institute of Science and Technology Graduate University, Japan

**Status**

Optical nanofibres (ONF) are waveguides produced by controlled drawing of a transparent material (most commonly glass softened by local heating), a technology largely developed over the last two decades [1]. The thinnest region or "waist" of an ONF has a diameter comparable to the working wavelength. Strong transverse confinement of the guided light at the waist leads to a significant evanescent field near the fibre surface and to radical restrictions of the allowed guided modes. In particular, if an ONF is sufficiently thin, it operates in the single-mode regime where the internal and external field profiles are the most predictable.

ONFs have many attractive features which explain their broad range of applications. Here, we focus on optomechanical ones, where material objects can be trapped or acquire forces and torques by interactions with ONF evanescent fields (see the overview in **Figure 6**). The large gradients and intensities of these fields are comparable to those of standard optical tweezers. Therefore, much like a tightly focussed laser beam, an ONF can stably trap dielectric particles in 2D by gradient optical forces. However, unlike a free-space beam, a single guided mode of an ONF has no restrictions in the axial direction, and the captured particle is optically propelled along the fibre [2]. When an ONF simultaneously captures multiple particles, they become optically bound to each other through the waveguide and the surrounding medium [3]. A recent study has demonstrated that ONFs show great potential for optical manipulation of composite metallo-dielectric particles [4] which are notoriously difficult to handle with free-space optical fields. Besides linear manipulation (trapping, propulsion, binding), ONFs can mediate light-induced rotation of material objects [5]. Free-space fields are often structured for spatial filtering and advanced optical manipulation and, similarly, the ONF-guided light can be made highly inhomogeneous in intensity or polarisation by using interfering fundamental modes or higher-order modes [6]. One of the unique features of ONF technology is the possibility to structure the fibre itself by nanomachining [7] in order to incorporate cavities or metastructures, an option largely unexplored in the context of optical trapping and manipulation. A major and arguably the most promising area of research involving ONFs concerns optical trapping and interfacing of laser-cooled atoms and other quantum emitters [8].

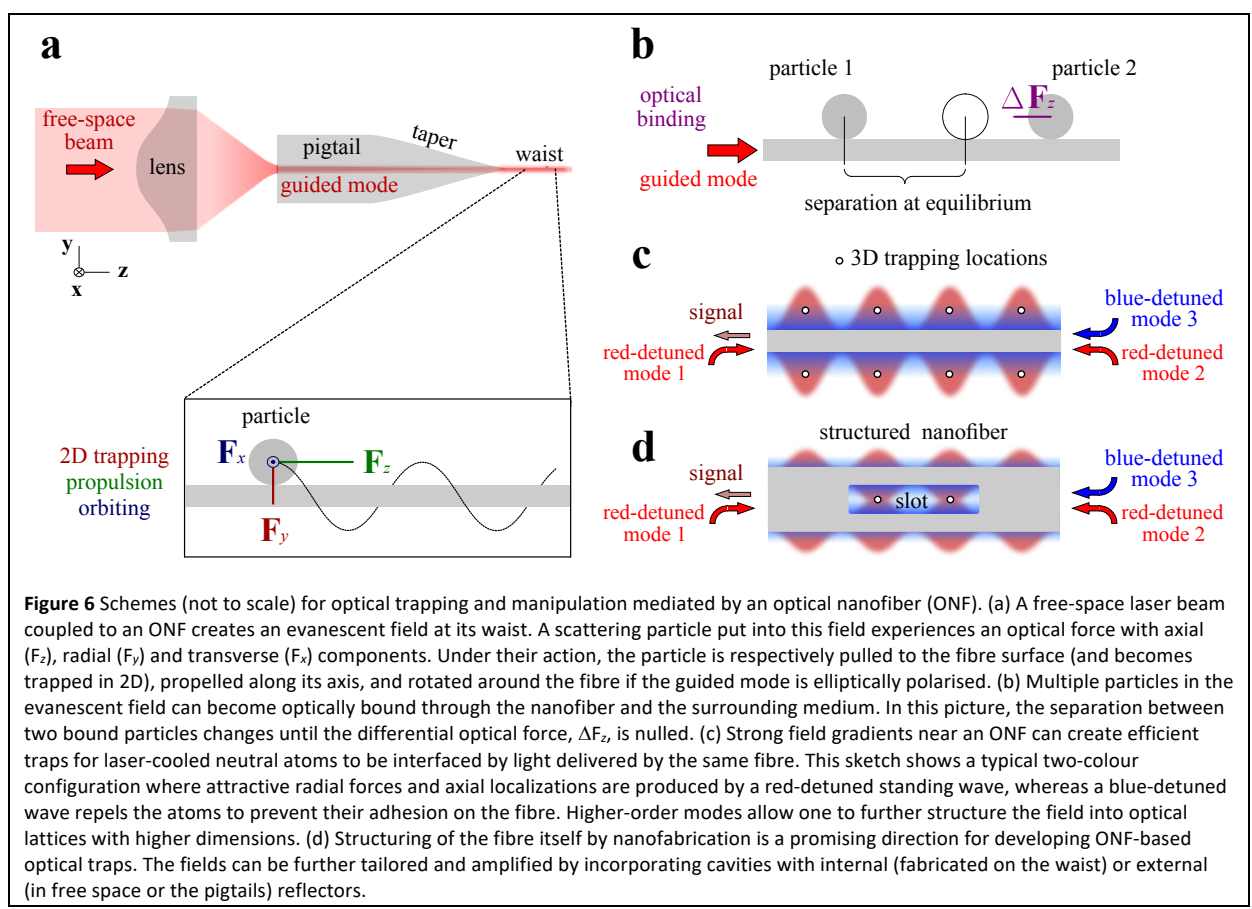

**Figure 6** Schemes (not to scale) for optical trapping and manipulation mediated by an optical nanofiber (ONF). (a) A free-space laser beam coupled to an ONF creates an evanescent field at its waist. A scattering particle put into this field experiences an optical force with axial ($F_z$), radial ($F_y$) and transverse ($F_x$) components. Under their action, the particle is respectively pulled to the fibre surface (and becomes trapped in 2D), propelled along its axis, and rotated around the fibre if the guided mode is elliptically polarised. (b) Multiple particles in the evanescent field can become optically bound through the nanofiber and the surrounding medium. In this picture, the separation between two bound particles changes until the differential optical force, $\Delta F_z$, is nulled. (c) Strong field gradients near an ONF can create efficient traps for laser-cooled neutral atoms to be interfaced by light delivered by the same fibre. This sketch shows a typical two-colour configuration where attractive radial forces and axial localizations are produced by a red-detuned standing wave, whereas a blue-detuned wave repels the atoms to prevent their adhesion on the fibre. Higher-order modes allow one to further structure the field into optical lattices with higher dimensions. (d) Structuring of the fibre itself by nanofabrication is a promising direction for developing ONF-based optical traps. The fields can be further tailored and amplified by incorporating cavities with internal (fabricated on the waist) or external (in free space or the pigtails) reflectors.

**Current and Future Challenges**

Overall, ONF-based photonics and optomechanics are in their early development stages, with many challenges remaining on the route to realisation of the full application potential of this technology. By the nature of their applications, ONFs must have low losses and highly predictable geometrical parameters of the waist region. Losses are especially important in experiments with quantum emitters where the signal-to-noise ratio is critical and ONFs are held in ultrahigh vacuum making heat dissipation very difficult. Obviously, heating increases the noise and severely lowers the optical damage threshold for the fibre. The existing fabrication techniques allow one to produce low-loss ONFs with the needed parameters, but currently there are no industry-standard implementations of an ONF-drawing rig. As a result, the fabrication quality varies greatly between custom-built systems, strongly depending on the operator's ability to calibrate and maintain the equipment, and even to handle the fibres. Moreover, the existing methods for precise characterisation require near-field scanning, which renders the probed nanofibre unusable for most applications. Therefore, improvement of the fabrication repeatability remains one of the major challenges for ONF technology.

In theory, unstructured ONFs are cylindrically symmetric and therefore do not alter the guided mode during propagation. However, experiments show that the symmetry is broken even in single-mode ONFs where uncontrolled stress birefringence transforms the polarisation of guided light. While for an adiabatic single-mode ONF these transformations can be undone using the Poincaré-sphere representation of the polarisation states at the waist [5], reliable control of higher-order modes in ONFs still evades experimentalists. This practical issue limits the use of ONFs to the single-mode regime and theoretical predictions involving vectorial evanescent fields [6] remain unrealised.

Another big challenge with ONF-based systems is scalability. To date, they are built around a single ONF waist which is used for optical manipulation or interfacing of colloidal particles or quantum emitters. While high-throughput lab-on-a-chip devices for handling colloidal particles with multiple ONFs are relatively easy to imagine, multiplexing for quantum ONF-based applications is much more challenging. One very promising solution involves Rydberg atoms coupled to an ONF, however it is not clear whether such atoms can be arranged in a chain and addressed independently to work as a quantum simulator [9]. It is feasible that multiple ONFs coupled to quantum emitters and linked to each other could form a functional quantum network, or the system could gain a higher dimensionality by the use of structured ONFs with a series of nanoscale cavities each representing a node in the network. These and silimar ideas still lack proof-of-concept experimental demonstrations, let alone implementations in actual photonic devices.

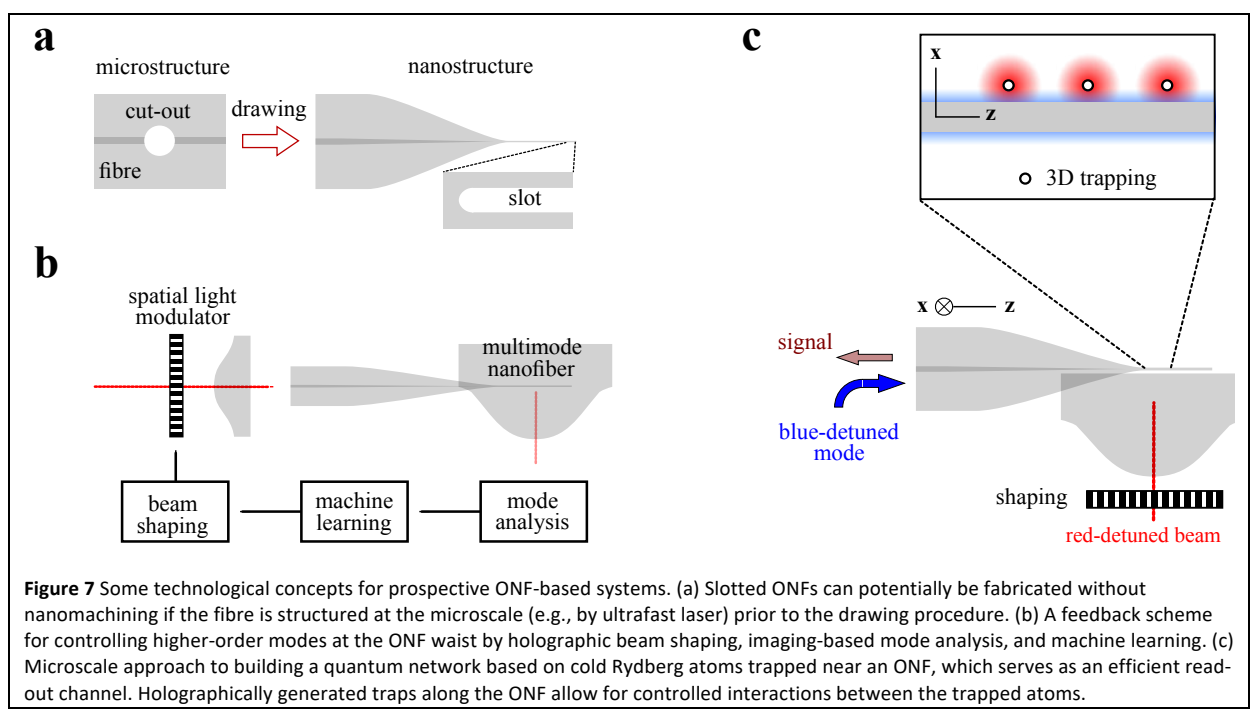

**Figure 7** Some technological concepts for prospective ONF-based systems. (a) Slotted ONFs can potentially be fabricated without nanomachining if the fibre is structured at the microscale (e.g., by ultrafast laser) prior to the drawing procedure. (b) A feedback scheme for controlling higher-order modes at the ONF waist by holographic beam shaping, imaging-based mode analysis, and machine learning. (c) Microscale approach to building a quantum network based on cold Rydberg atoms trapped near an ONF, which serves as an efficient read-out channel. Holographically generated traps along the ONF allow for controlled interactions between the trapped atoms.

**Advances in Science and Technology to Meet Challenges**

Repeatability of ONF fabrication should be improved by standardisation and automation of the drawing process. Commercial fusion splicers solve the analogous problem related to joining telecom optical fibres; similarly, one could imagine an automated machine allowing the operator to reliably produce ONFs with desired parameters verifiable by in situ probing. On a larger scale, major engineering advances are needed for consolidation of the ONF-drawing technology with integrated photonics based on lithographic printing. Perhaps nanowaveguides similar in quality to ONFs could be produced by reflowing of printed structures, as is often done with on-chip whispering-gallery microcavities. Thus, the need for precise mechanical manipulation associated with drawing could be eliminated.

State-of-the-art methods for structuring ONFs are based on nanomachining the fibre with a focussed beam of ions (gallium is more common, while helium offers contamination-free milling and a 10-fold higher resolution reaching 1 nm). These methods are expensive, low-throughput, and time-consuming, especially given the high chances of the nanofibres being damaged during the process. A promising alternative could be to use micromachining (such as femtosecond laser writing which has a wavelength-scale resolution) of the fibre and then to downsize the structured region in the radial direction by drawing (**Figure 7**a), as is common for fabrication of photonic crystal fibres.

To improve heat dissipation of vacuum-clad ONFs, it is worth researching atomic-scale coatings. Carbon, having a 1000-fold higher thermal conductivity compared to silica, is one excellent candidate. ONFs wrapped in graphene (e. g., via the Langmuir-Blodgett technique) would be a powerful tool for photonic and optoelectronic applications.

In order to control the guided modes in ONFs, one can structure the input beam, for which purpose the mode content at the waist region must be analysed. This task is very challenging but feasible, for instance by spatial spectroscopy of beating modes [10]. We expect reliable mode control for ONFs to be achieved by merging this "reading" capability with free-space holographic "writing" (spatial modulation) and a machine-learning algorithm to link the two, see **Figure 7**b. An alternative strategy would be to structure the fibre itself and filter the modes by dichroism or resonant excitation.

Besides the obvious approach of linking multiple ONFs, scalability of quantum systems based on trapped atoms can be pursued at the microscale, e. g. by combining an optical lattice (for localisation and selective interfacing of qubits) with an ONF (for efficient channelling of the information-carrying photons), see **Figure 7**c.

**Concluding Remarks**
Optical nanofibres offer a seamless connection between free-space beams and tightly confined evanescent fields which brings optical trapping, manipulation, and probing of material objects to a new level of performance. After nearly two decades of research, nanofibres and their application potential are well understood and have been clearly demonstrated in numerous experimental studies, the foremost being optical trapping and probing of cold atoms in the pursuit of a nanofibre-based quantum network. Despite its advantages, optical nanofibre technology is still at the research laboratory level, a long way from being implemented in commercial photonic instruments. In our opinion, the technology will only mature when the fabrication process and mode control become sufficiently accurate. Towards this goal, fabrication should incorporate in-situ probing, whereas the issue with the modes could be overcome by combining holographic beam shaping in free space, mode analysis at the nanofibre waist, and machine learning to complete the feedback loop. We also envisage advances in chemically applied coatings to tackle the heat dissipation problem for vacuum-clad nanofibres, and more active implementation of micro- and nanomachining for precise structuring of the nanofibre and the guided fields. Advances in the next decade should showcase the true potential of these miniature devices.

**Acknowledgements**
This work was supported by OIST Graduate University and the Japan Society for the Promotion of Science (JSPS) Grant-in-Aid for Scientific Research (C) Grant Number 19K05316, Grant-in-Aid for JSPS Fellows Grant Number 18F1836, and International Research Fellowship (Standard) P18367.

## 5 — Intracavity optical trapping

*Fatemeh Kalantarifard*
Department of Health Technology, Technical University of Denmark, Lyngby, Denmark

*Parviz Elahi*
Physics Department, Boğaziçi University, Istanbul, Turkey

**Status**

Optical trapping based on feedback has drawn much attention due to its unique and precise control on the position of the trapped particle, laser intensity, and the exerted force or torque. There are different reports where the authors took advantage of the feedback control on their optical trapping setup. For instance, the re-positioning of the trap is achieved by monitoring the particle's position and sending a fast electronic signal as feedback in force clamping [1], implementing feedback on the input polarization angle to maintain the torque [2], stabilizing the sample stage by a feedback loop [3], achieving power stability by monitoring the trap laser intensity with a photodetector and adjusting the laser power by a feedback loop [3].

All the feedback mechanisms mentioned above are extrinsic: They all require dedicated electronics and imaging as a part of the feedback loop. In 2019, F Kalantarifard et al. [4] presented a novel technique of intracavity optical trapping (ICOT) and introduced intrinsic nonlinear feedback force, i.e., a force that arises only because the optically trapped particle is placed inside a laser cavity (**Figure 8**). The main novelty of ICOT lies in the identification of a new mechanism for trapping microparticles using intracavity optical feedback, which is fundamentally different from the standard techniques already demonstrated in the past. This technique is based on intracavity optical forces that emerge from the influence of the optically trapped particle on the laser cavity.

According to the working principle shown in **Figure 8**, there is an optomechanical coupling between the laser cavity and the trapped particle such that the particle's positions, *r* and *z*, and laser signal power, *P*, are correlated. When the particle moves to the right, left, and down, the power increases, while by lifting the particle up, the power drops.

The ICOT scheme works using an aspheric lens with a low numerical aperture (NA about 0.12 or less) compared to the standard optical tweezers that employ an objective with typically NA>1. Therefore, as the most important benefit, this new kind of optical tweezers permits trapping at lower average laser intensity than alternative techniques. In general, any feedback modulation of the laser beam intensity can reduce the intensity on average. However, the internal feedback due to the laser cavity dynamics occurs on the timescale of nanoseconds, making it several orders of magnitude faster than any external feedback.

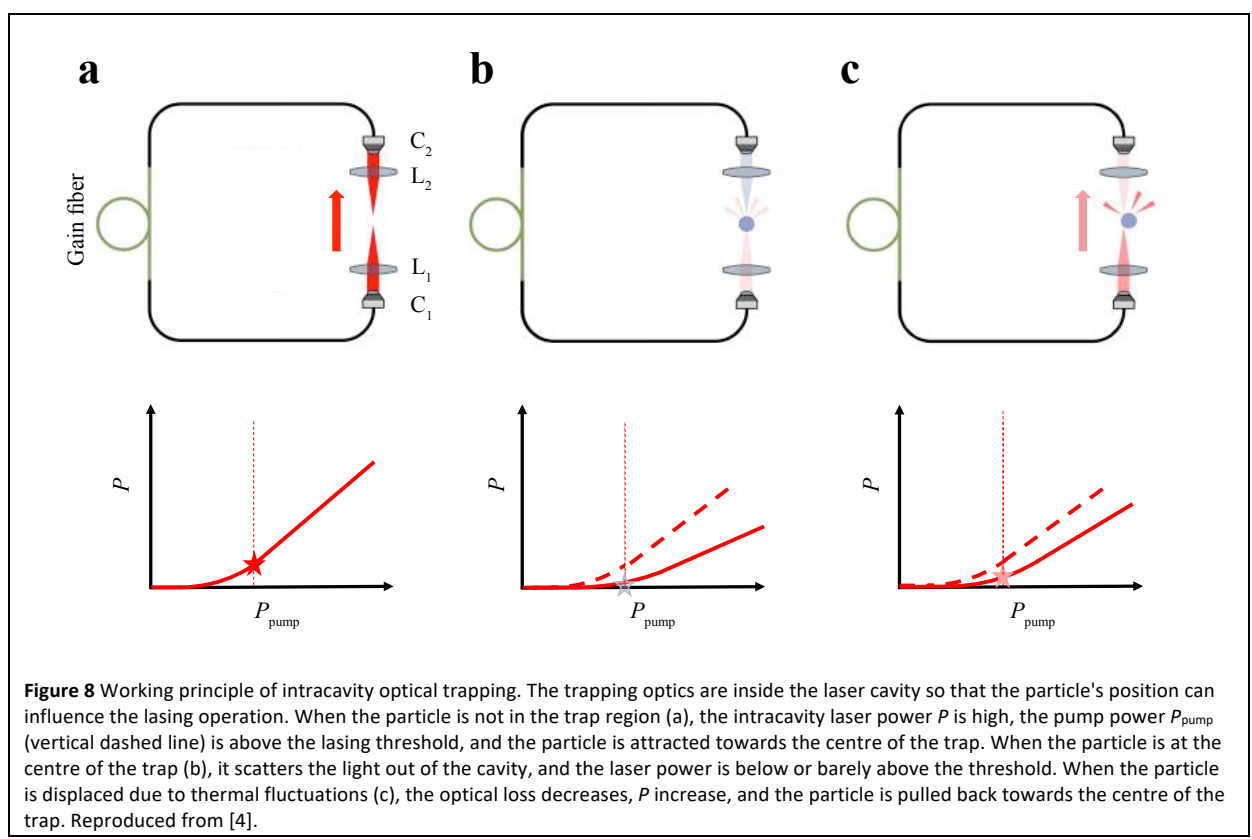

**Figure 8** Working principle of intracavity optical trapping. The trapping optics are inside the laser cavity so that the particle's position can influence the lasing operation. When the particle is not in the trap region (a), the intracavity laser power $P$ is high, the pump power $P_{pump}$ (vertical dashed line) is above the lasing threshold, and the particle is attracted towards the centre of the trap. When the particle is at the centre of the trap (b), it scatters the light out of the cavity, and the laser power is below or barely above the threshold. When the particle is displaced due to thermal fluctuations (c), the optical loss decreases, $P$ increase, and the particle is pulled back towards the centre of the trap. Reproduced from [4].

**Current and Future Challenges**

Employing a low-NA lens instead of a high-NA objective lens in an ICOT system is one of the main advantages. Although the low-NA lens brings many advantages compared to the more costly high-NA objective, the weaker gradient force requires excellent adjustment and is accompanied by some challenges. Scattering loss of the trapped particle plays a significant role in the ICOT systems. To take advantage of the feedback in the ICOT, the insertion loss of the laser cavity must be low, and the particle loss due to scattering should be significant. For small dielectric non-absorbing particles, where the particle size is much smaller than the wavelength ($a \ll \lambda$) and the dipole approximation is valid, the power scattered by the particle in the trap rapidly decreases and the scattering loss in terms of particle's displacement is much smaller than that of microparticles. Therefore, the nonlinear feedback that regulates intracavity trapping is negligible, and ICOT behaves as standard single-beam optical tweezers and cannot efficiently trap small particles at low intensity. This clearly shows that, while the intracavity feedback trapping is efficient at the microscale, it reduces to a standard single-beam optical trapping at the nanoscale.

Due to the use of a low numerical aperture lens in ICOT, the particle weight plays an important role in the stability of the trap in the axial direction for trapping the objects with a high refractive index. As a result, the setup design is limited to inverted microscopes for particles with very low density and high refractive index such as polystyrene. Depending on the particle's refractive index and density, the intracavity trapping needs to be designed accordingly. So, trapping microscopic beads in a vacuum or in a horizontal configuration could be challenging.

Position detection, especially in the axial direction, can also become challenging to some extent. As a standard tool for position detection, a quadrant photodetector requires a stable and low fluctuation in laser power. However, since in ICOT, the intracavity laser power, and hence the output power, varies by the particle's position, the high-resolution measurement of the position requires understanding the effect of power variation due to the position. Due to this complication, so far, the 3D tracking of the particle is mainly limited to 3D digital video microscopy, which is relatively slow, being limited by the camera frame rate.

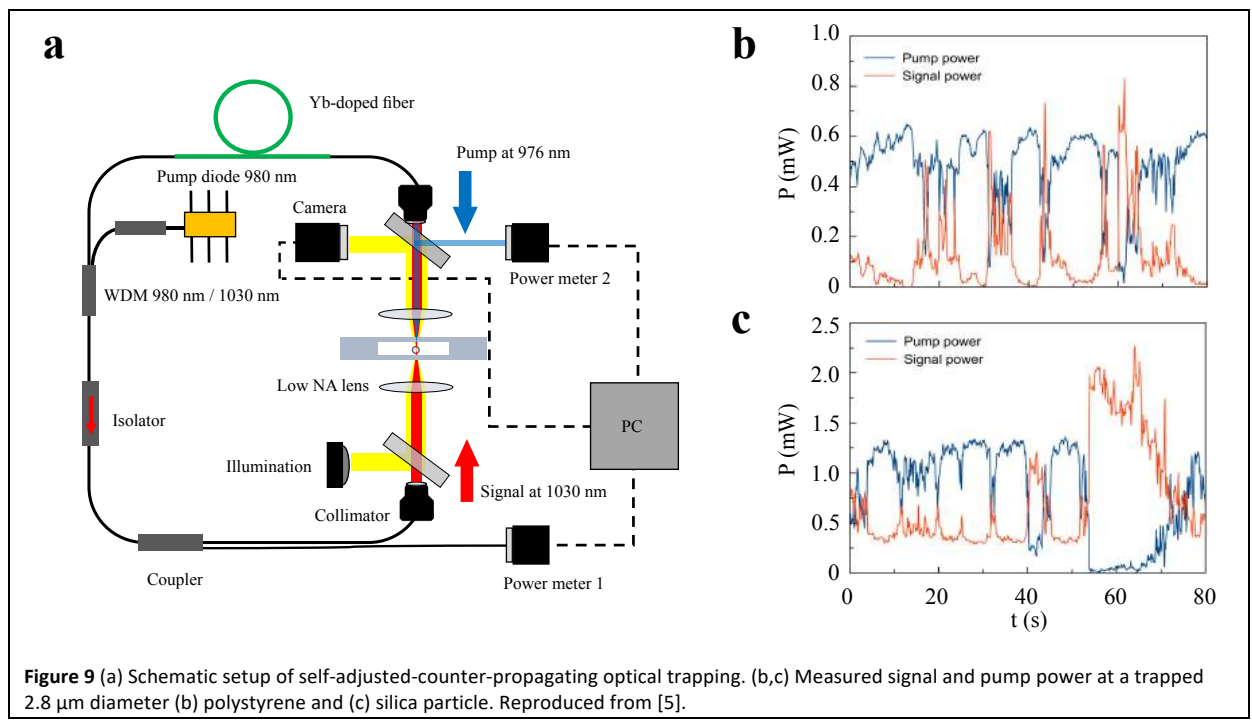

**Figure 9** (a) Schematic setup of self-adjusted-counter-propagating optical trapping. (b,c) Measured signal and pump power at a trapped 2.8 μm diameter (b) polystyrene and (c) silica particle. Reproduced from [5].

**Advances in Science and Technology to Meet Challenges**

The advantage of optical feedback compared to electronic feedback is its uniquely high bandwidth (≈10 MHz). The novelty of this approach opens many new possibilities to be explored in future works.

One of these possibilities is to study systems where the characteristic timescales associated with the laser and the motion of the optically trapped particle become comparable. In particular, one can expect its uniquely high bandwidth to be crucial when dealing with particles trapped in the air or a vacuum.

However, the intracavity optical trapping in liquid can also, in principle, be scaled down to particles significantly smaller than the wavelength by changing the experimental parameters. A small innovative change in fibre laser cavity [5] led to a new configuration called self-adjusted-counter-propagating optical trapping with a single Gaussian beam and using an even smaller actual NA (0.087) that is capable of trapping smaller polystyrene particles compared to ICOT. In this scheme, the power at the sample switches between the signal and pump (which is now propagating in the opposite direction of the signal) such that when the particle is in the centre of the trap, the laser signal power drops and pump power rises, and this occurs automatically by the displacement of the particle itself (**Figure 9**). Having two self-adjusted counter-propagating beams improves the axial direction trap for smaller particles and brings another advantage to the trapping system in the horizontal design.

Another change, such as increasing the numerical aperture of the lens, would allow to increase the losses due to the particle and therefore make ICOT more efficient. In another reconfiguration of the setup, the authors showed that by removing the isolator in the ring cavity, to make the light travels bidirectionally and then sending the light from both directions to the silica microparticle, the optical confinement per unit intensity is improved [6].

In terms of data acquisition and particle tracing, thanks to the advances in machine learning, particularly in image processing and deep learning, position detection and particle tracking in ICOT are no longer cumbersome, especially in the axial direction.

**Concluding Remarks**

ICOT is fundamentally different from other techniques as it is on the basis of nonlinear feedback forces arising because the particle is inside the laser cavity. Position clamping, for instance works based on (external) feedback on the trap power depending on the explicit measurement of the particle position. However, in the ICOT scheme, the feedback is intrinsic to the laser cavity, passive and extremely fast. In addition, due to the existance of an opto-mechanical coupling between the particle position and the laser cavity, the power is self-regulating, and it does not require recalibration each time that experimental details are modified.

The ability of trapping employing a very low numerical aperture lens in ICOT can yield advantages when dealing with biological samples that are sensitive to light intensity. In comparison with alternative trapping techniques, where a low-NA objective lens is employed [7-9], the ICOT achieves excellent results in capturing and trapping particles and yeast cells [4] by achieving a major reduction in optical intensity exposure at the sample.

Being the trap inside a laser cavity as a working principle of the ICOT has attracted attention [10] to a new research field that is curious how a trapped object can influence the lasing operation and what more exciting science and applications will be obtained.

**Acknowledgements**

The work is partially supported by Boğaziçi University Research Fund, Start-Up project 21B03SUP3 awarded to PE. FK acknowledges support from the Novo Nordisk Foundation (grant no NNF20OC0061673)

## 6 — Metasurfaces for optical manipulation

*Mikael Käll*

Chalmers University of Technology, Gothenburg, Sweden

**Status**

An optical metasurface is a dense 2D layer of scatterers with sub-wavelength dimensions and separations such that the layer appears locally homogeneous to an impinging electromagnetic field. The scatterers (or metaatoms) are typically designed to induce a specific phase and/or polarisation state of the scattered field by utilizing optical resonances, such as localized surface plasmons in the case of metallic metasurfaces (primarily Au) or geometrical resonances (Mie or waveguide modes) in the case of metasurfaces composed of high-index dielectric materials. By systematically varying the scatterer dimensions, shape, orientation and/or separation across the metasurface, it is then possible to shape the light that is transmitted through or reflected from the surface. This makes it possible to construct compact counterparts to a wide variety of classical refractive or reflective optical components [1], the prototypical example being a phase-gradient metasurface operating as a "flat" lens [2], but also multifunctional devices and miniature structures with tailored optical responses.

There are so far only a few experimental studies of metasurfaces in the context of optical manipulation, most of them very recent. Polystyrene beads have been trapped by using a reflective Si metasurfaces operating as a parabolic mirror [3] as well as by Si metalenses [4], in which case multifunctionality obtained by overlapping the phase-gradients corresponding to a lens and a q-plate to transfer orbital angular momentum was also demonstrated. Si metalenses with NA ≈ 0.9 have recently been used to optically levitate $SiO_2$ particles in vacuum, with trapping efficiency on par with conventional microscope objectives [5]. These results demonstrate that lithographically produced metastructures can exert optical forces without the use of traditional optical elements. Microscopic $SiO_2$ particles with embedded anisotropic Si metasurfaces able to deflect light at high angles to provide optical propulsion and torque were recently demonstrated [6]. These "metavehicles" are steered by altering the polarization of a single incident plane wave and were able to transport microscopic cargo across a surface in water, see **Figure 10**.

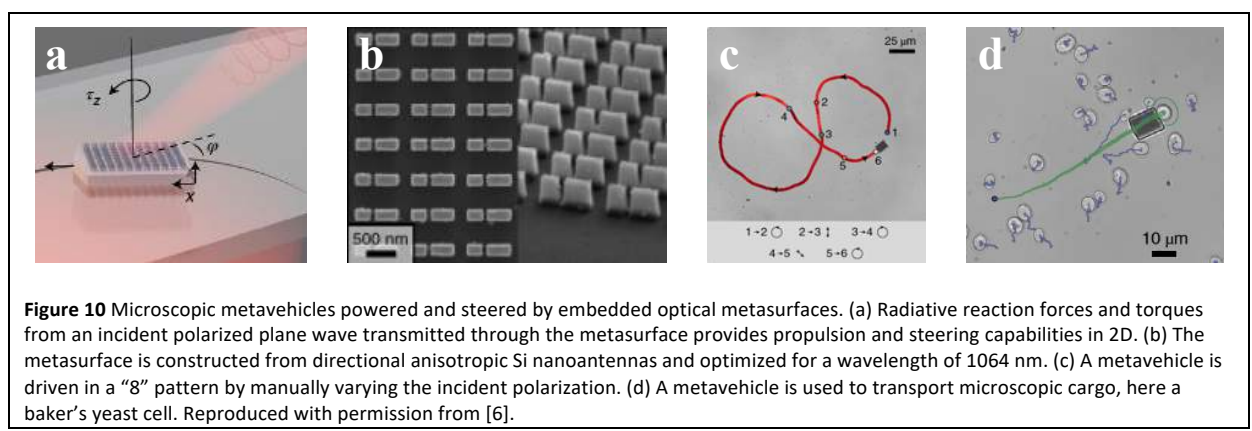

**Figure 10** Microscopic metavehicles powered and steered by embedded optical metasurfaces. (a) Radiative reaction forces and torques from an incident polarized plane wave transmitted through the metasurface provides propulsion and steering capabilities in 2D. (b) The metasurface is constructed from directional anisotropic Si nanoantennas and optimized for a wavelength of 1064 nm. (c) A metavehicle is driven in a "8" pattern by manually varying the incident polarization. (d) A metavehicle is used to transport microscopic cargo, here a baker's yeast cell. Reproduced with permission from [6].

**Current and Future Challenges**

Several challenges need to be tackled to make metaoptics a competitive alternative to traditional optical components used for optical manipulation. These include, but are not limited to:

- Poor efficiency and polychromaticity: Most current metasurfaces have comparatively low transmission or reflection efficiency compared to ordinary focusing optics, in particular outside the design wavelength.
- Poor focusing ability: Optical tweezing typically requires high NA optics, meaning that the corresponding metalens phase gradient needs to be very steep at the periphery of a lens.
- Costly and advanced manufacturing processes: Metasurfaces are typically fabricated using expensive electron beam lithography (EBL) and advanced layer deposition and etching protocols, which hampers their widespread application.
- Computational complexity: Current electromagnetic solvers cannot simulate an entire phase-gradient metasurface, which typically contain >$10^4$ optically interacting metaatoms, in a self-consistent manner. This limits the possibility to optimize devices.
- Tunability: Metaoptics devices are typically static in nature, which is a clear disadvantage compared to, for example, spatial light modulators.

The challenges listed above are general to the metaoptics field. All are active areas of research, though each one may be more-or-less critical to the field of optical manipulation. For example, efficient high-NA metalenses are needed for laser-tweezing in 3D, but they do not necessarily need to be polychromatic. Apart from the important but obvious advantage of shrinking the size and cost of components used for optical manipulation, the most impactful challenge in the long run will be to realize applications that are useful but practically unachievable using conventional optics and photonics devices. For example, ultracompact systems for optical trapping in lab-on-a-chip applications would probably need to incorporate both light source(s), metaoptics and detectors to bring real advantage; metaparticles could be made responsive to environmental cues to act as mobile sensors driven by light; and metasurfaces with simultaneously optimized phase, amplitude and polarization response could be used to create entirely novel optical force fields for diverse applications.

**Advances in Science and Technology to Meet Challenges**

Several technological advances, many already under development, can be expected to impact optical manipulation research and applications based on metasurfaces in the short to medium time frame, some of the most important being:

- More precise and/or more cost-effective fabrication methods, such as extreme ultraviolet lithography, nanoimprint lithography etc.
- More powerful optimization methods, including techniques based on artificial intelligence (e.g., deep learning) and, perhaps, even quantum computation.
- New materials with lower loss, higher refractive index and/or with externally controllable optical properties (e.g., phase change materials).

Advancements based on combining metasurface concepts and technology with those developed in neighbouring areas, such as diffractive optics (see, e.g., [7]), nanolasers, optical sensors and actuators, and nanoantennas are also very likely.

**Concluding Remarks**

The field of optical metasurfaces has developed into one of the most active research areas in fundamental and applied photonics. Building on foundations in plasmonics, metamaterials, and nanooptics, optical metasurfaces offer the prospect of extremely thin, compact, and multifunctional optical components and devices. These could be integrated into optical manipulation systems to drastically cut size, complexity, and cost, thereby accelerating their widespread use, and enabling new applications of such systems. Moreover, because of their potentially very small size and mass, metasurfaces could themselves be designed as targets for optical manipulation, thus realizing a new class of microscopic structures driven by optical forces. Given the rapid advances of the field and the enormous flexibility in terms of possible metasurface designs, there is certainly room for both novel creative applications and further refinement of concepts and devices for optical manipulation.

**Acknowledgements**
This work was funded by the Knut and Alice Wallenberg Foundation.

# THEORY AND SIMULATION OF OPTICAL FORCES

## 7 — Geometrical optics

*Agnese Callegari*

Department of Physics, University of Gothenburg, 41296 Gothenburg, Sweden

**Status**

Geometrical optics was already known to the ancient Greeks. The most ancient surviving treaty on optics is Euclid's *Catoptrics*, written around 280 BC. Euclid writes that light travels in straight lines (rays) in homogeneous media. The law of reflection and the phenomenon of refraction were also known, and Hero of Alexandria demonstrated that, if we assume that the light always travels along the shortest path, we can derive the law of reflection using the laws of geometry. The modern formulation of geometrical optics stems from the works of Snell, Descartes, and Fermat. Later, geometrical optics was rederived in the framework of the general electro-magnetic theory from Maxwell's equation [1].

Geometrical optics' fundamental concept is the ray, which has a power P, a propagation direction $\hat{r}$ and a polarization $\hat{p}$. The ray describes the light and its propagation though media. A homogeneous medium is characterized by its refractive index $\tilde{n}$, which is related to the medium dielectric constant $\varepsilon$ through the relation $\varepsilon = \tilde{n}^2$. The refractive index is in general a complex quantity: $\tilde{n} = n + i\,k$, where the real part defines the light speed in the medium $v = c/n$, and the imaginary part $k$ is non-zero for absorbing materials only and dictates how fast the optical power of the ray exponentially decays in its spatial propagation through the medium. The three fundamental laws of geometrical optics consist of (i) rectilinear propagation of a light ray in homogeneous media, (ii) the Snell's law of reflection, and (iii) refraction happening at the interface of two homogeneous media: $\theta_r = \theta_i$ and $n_t \cdot \sin\theta_t = n_i \cdot \sin\theta_i$. The intensity of the reflected and transmitted rays following a ray impinging on the surface between two media in Snell's law is based on the Fresnel's coefficients, which are defined for each component ($p$ parallel, $s$ perpendicular) of the ray polarization with respect to the interface. The Fresnel coefficients are derived from the electromagnetic theory by imposing the continuity condition of the electric and magnetic fields (ruled by the Maxwell's equations) at the interface between the two media [1]. Snell's law and the electromagnetic stress-energy tensor, which defines the energy and momentum carried by the light, are fundamental to explain the experimental evidence that light can apply forces to particles, which is the base of optical tweezers [2-4].

Geometrical optics has been successfully applied to model optical tweezers [2] and predict the optical force applied on a dielectric particle [2] (**Figure 11**). Geometrical optics is particularly useful in the case of microscopic particles, whose size $a$ (in the range of few hundreds of nm to several microns) is much larger than the wavelength $\lambda$ of the trapping laser beam, which for visible light is in the range of 380 to 780 nm in vacuum (and consequently reduced by a factor n in homogeneous media).

Geometrical optics has the limitation of not allowing to predict effects related to the phase and spin of light, which are peculiarity of the wave nature of light, because these features do not enter in the geometrical optics model. Despite of this limitation, geometrical optics has been successfully employed to describe, for example, optical forces on cells [5], the deformation of microbubbles in an optical field [6], the optical lift effect, the emergence of negative optical forces [7], the forces on non-spherical particles [8,9], Janus particles [10], absorbing particles [11], particles in a speckle light field [12], or trapped at water-oil interface [13].

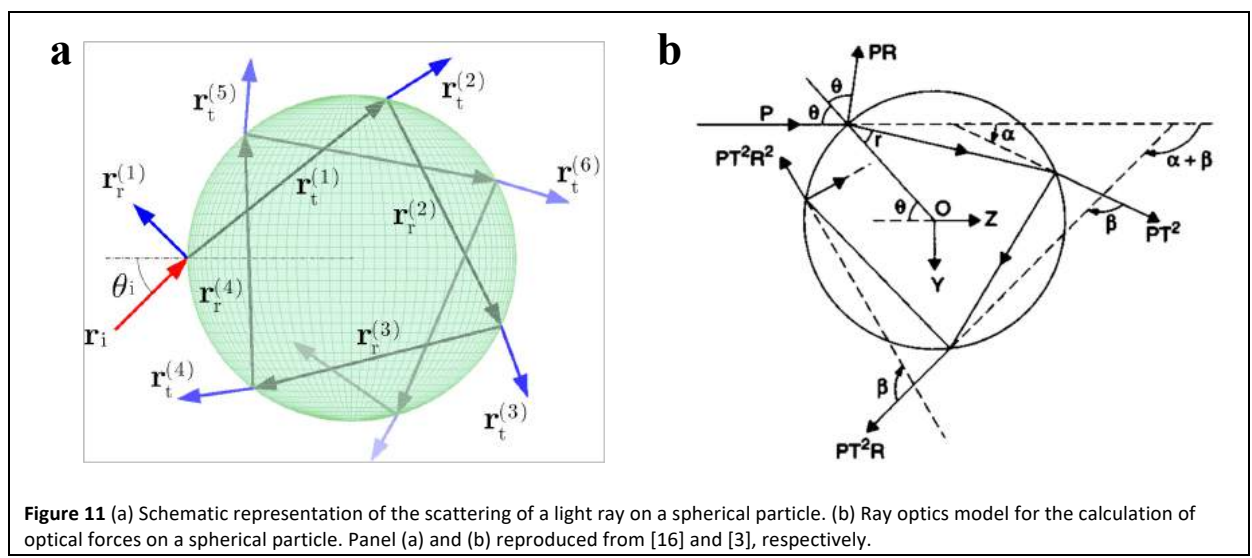

**Figure 11** (a) Schematic representation of the scattering of a light ray on a spherical particle. (b) Ray optics model for the calculation of optical forces on a spherical particle. Panel (a) and (b) reproduced from [16] and [3], respectively.

**Current and Future Challenges**

Geometrical optical provides an effective description of the interaction of light with matter and, in relation with optical tweezers, an intuitive and relatively simple explanation of the trapping mechanism. It is especially convenient in all the cases where the dimension of the particles is not small enough to allow full electromagnetic theory calculations. However, geometrical optics does not provide only a "fall-back" remedy, because it often captures the essence of the physical mechanism and provides an insightful picture without overwhelming non-essential details.

Geometrical optics is a well-established technique that works well in most of the standard cases, where there are analytic formulas provided by the theory [3,11] and for symmetric conditions, like a spherical particle, or a focused or uniform Gaussian beam, and counterpropagating beams. For the standard and non-standard cases, codes and toolbox are available [14-16], in different programming languages, with different possibilities for customisation.

At present, the main challenges for the calculation of optical forces based on geometrical optics are related to the number of rays that are required to accurately calculate force and torque (or in general the scattering) on a particle in a non-symmetric situation, which can be a rather time-consuming task, e.g., in the case of complex particle geometries (non-spherical and/or non-convex in shape, non-uniform in refractive index), or complex optical field dependence (non-uniform optical field). In general, an accurate calculation of the optical torque on a particle requires a larger number of rays than the calculation of the optical force. If, moreover, we are required to calculate the dynamics of a non-spherical, homogeneous dielectric particle, the force and the torque do not only depend on the particle position, but also on its orientation. In the case when the particle is non-homogeneous in refractive index, or it moves in a non-uniform optical field the situation is made even worse: there is often no available analytic expression for the force and the torques and a pre-calculation of forces and torques on a fine spatial grid for all the orientational degrees of freedom is a very daunting task and not a viable option. Even if pre-calculation were a possibility, in the case of a different value of the particle's radius or refractive index we would be forced to restart from the beginning. Also, depending on the circumstances, a grid of different sizes would be preferable, complicating further the pre-calculation task.

Another situation where the application of geometrical optics can be time consuming is the case of optical binding and, in general, the situation where interactions of many particles introduce non-linearity, and many-body calculations are required. A way to partially solve the speed problem in geometrical optics calculations could be providing the toolboxes the option of parallelization or exploit the presence of a GPU. However, adapting or rewriting the code is not optimal from the point of view of becoming dependent from a specific hardware that might become obsolete. On the other hand, a machine learning approach is starting to be used in combination with geometrical optics [17].

**Advances in Science and Technology to Meet Challenges**

The parallelization of the ray scattering calculations and the exploitation of GPUs with ad-hoc adapted codes for geometrical optics could result in an increased calculation speed, but do not solve entirely the problem of the complexity and of having to repeat the scattering calculation several time.

Currently, a promising approach is offered by the application of machine-learning techniques. The fundamental idea behind the machine learning approach to geometrical optics is that, once a network is trained for performing well for a certain set of parameters, then it can be used a set of parameters that are in the range of the training parameters without re-training, even if the parameters do not match perfectly. Because of the intrinsic "smoothness" of the transfer function characterising the building blocks of a neural network neural network, it is also possible to overcome the spurious discontinuities observed in geometrical optics calculations emerging in the case of grids with insufficient number of rays. Therefore, the computational effort seems to lay in the robust and smart training of the network. This would allow, in the future, the training of single neural networks which can predict accurately the optical forces and torques for a given general situation characterized by several parameter ranges, allowing a faster calculation for dynamics and statistical properties of optical systems.

**Concluding Remarks**

Geometrical optics is a well-known and established approach for effectively modelling optical tweezers and accurately calculating optical forces and torques on microscopic particles in the ray optics approximation. Several analytic and computational tools are available, which give accurate results within a fast or moderate computational time. More complex cases (non-spherically symmetric, non-homogeneous particles, non-uniform and/or complex optical fields, non-linear effects) require a longer and more complex computational approach and, in many such cases, the calculation speed is still a challenge. The introduction of the use of GPUs and parallelization could partially solve the challenges. The currently emerging machine-learning approaches seem promising in addressing consistently the challenge of computational speed, taking it away from the calculation and moving it upstream into the training process of the neural network.

## 8 — The dipole approximation

*Manuel I. Marqués*

Departamento de Física de Materiales, IFIMAC & Instituto Nicolás Cabrera, Universidad Autónoma de Madrid, Spain

**Status**

The use of light (with wavelengths of several hundred nanometres) to manipulate a nanoparticle (with size on the order of the nanometre) can be analysed using the single dipole approximation. In this approach, the optical interaction is directly obtained from the Lorentz force, and it is composed of two elements: the gradient force, proportional to the real part of particle's polarizability and the gradient of the field's intensity, and the scattering force or radiation pressure, proportional to the imaginary part of the polarizability and the gradient of the field's phase. In the case of inhomogeneous spinning fields, the last force component, usually considered to be proportional to the Poynting vector, depends also on the value of the curl of the spin density of the light field [1].

Slightly larger particles and high refractive index materials may also be modelled considering a dipole approach but, in this case, the magnetic response must be also contemplated. Dual equations are obtained for the force on these particles, together with a new term coming from the electric-magnetic dipoles interaction [2].

The knowledge of the mechanical response of these small particles is fundamental not just from the nanometric point of view. Force on larger particles or complex systems may be analysed studying the collective response of groups of electric dipoles, through the so-called discrete dipole approximation (DDA). In this case, the interaction among the dipoles is analysed through a multiple scattering process where the field on every particle is given by the incoming field and the field scattered from all other dipoles.

Thus, the knowledge of the dynamic behaviour of electric dipoles in complex environments and under structured light fields is important not just from a fundamental point of view, but also as a basic tool to analyse the optomechanical response of larger systems with huge applications on the fields of photonics and nanotechnology. Among these applications we may highlight; the nanometric control of the diffusivity, the improvement of radiation pressure propelling systems and the full control of binding at the nanoscale.

**Current and Future Challenges**

Although the calculation of the force on an electric dipole is an analytically manageable problem, it still presents fundamental challenges like, for instance, the measurement of the Belinfante's curl force, coming from the non-homogeneous spinning nature of the electromagnetic radiation. Another type of spin force, coming from the interaction between the electric and magnetic dipoles, has been recently measured [3] but the detection of the curl of the spin force acting merely on a single dipole, is still lacking.

Another challenge is the use of optical forces on electric dipoles to manipulate the diffusivity of nanoparticles using electromagnetic fields. These fields could be used to either increase the diffusivity or decrease it. Specifically, the challenge is to achieve active matter propelled by optical forces. Also, by attaching groups of nanoparticles to larger systems, a noticeable radiation pressure could be achieved, growing harmonically with the number of electric dipoles and lacking the mechanical stress induced by the torque expected at the anchor point [4].

Another future challenge consists of being able to manipulate the optical interaction among nanoparticles, with the final goal of either structuring larger and more complex structures or tunning the coagulation properties of colloids at the nanoscale.

Optical forces on electric dipoles have been also proposed as a method for sorting different types of nanoparticles based on their mechanical response. For example, a possible challenge is chirality sorting, based on the different optical forces felt by a particle with a non-scalar polarizability tensor when illuminated with circularly polarized light.

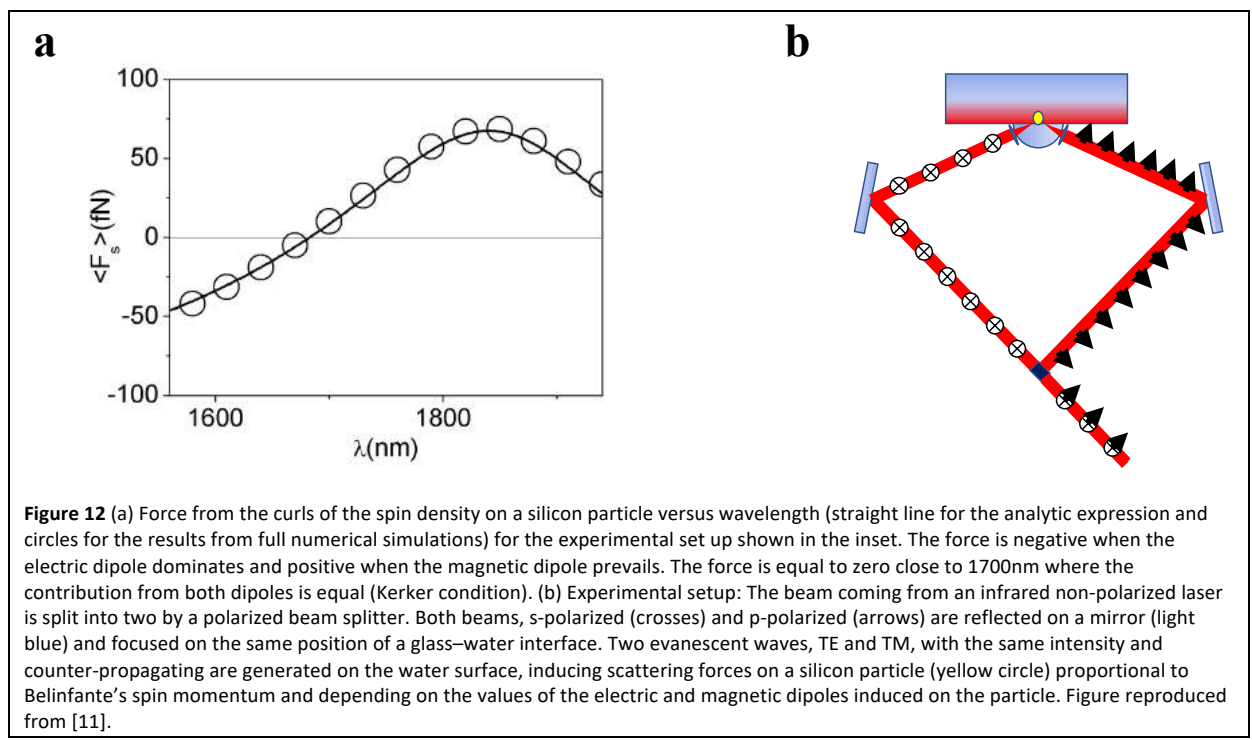

**Figure 12** (a) Force from the curls of the spin density on a silicon particle versus wavelength (straight line for the analytic expression and circles for the results from full numerical simulations) for the experimental set up shown in the inset. The force is negative when the electric dipole dominates and positive when the magnetic dipole prevails. The force is equal to zero close to 1700nm where the contribution from both dipoles is equal (Kerker condition). (b) Experimental setup: The beam coming from an infrared non-polarized laser is split into two by a polarized beam splitter. Both beams, s-polarized (crosses) and p-polarized (arrows) are reflected on a mirror (light blue) and focused on the same position of a glass–water interface. Two evanescent waves, TE and TM, with the same intensity and counter-propagating are generated on the water surface, inducing scattering forces on a silicon particle (yellow circle) proportional to Belinfante's spin momentum and depending on the values of the electric and magnetic dipoles induced on the particle. Figure reproduced from [11].

**Advances in Science and Technology to Meet Challenges**

The technical advances needed to achieve the above-mentioned challenges, using optical forces on dipoles, have been theoretically analyzed and, from those studies, several experimental set ups have been proposed.

Genuine forces coming from the spinning nature of the electromagnetic radiation and acting on a single electric dipole have been predicted, but never detected or gauged. Recently, an experimental setup to measure this force, proportional to the Belinfante momentum, has been proposed (**Figure 12**). The experiment is based on the Kerker condition, a requirement subject to the coincidence between the electric and magnetic response of a particle. The set up operates at the Kretschmann configuration, using two counterpropagating transversal waves (TE-TM) with the same intensity. Dipole optical forces on this system are only given by the curl of the spin density of the light field acting on each single dipole (electric and magnetic) independently, and not by the force coming from interaction between both dipoles.

On the other hand, to achieve a full control of the diffusivity of nanoparticles, interfering light fields have been proposed. For example, a field consisting of two perpendicular stationary waves with tunable phases induces an optical vortex lattice. This vortex lattice can increase the diffusivity of the nanoparticles and induce collective motion [5]. Also, gradient traps may be used to decrease the diffusion. For example, speckles are a good option to tackle this challenge [6] using the value of the intensity as a trigger to reduce diffusion.

The manipulation at will of the equilibrium distance between two dipoles at the nanoscale, using optical forces, is also a challenge from the technological point of view. A possible solution is based on the use of the near field binding (orders of magnitude larger than the usual far field binding), tunning the equilibrium distance by, either a periodic change of the polarization angle, or by applying an external magnetic field [7].

However, the technological proposals to achieve isotropic bonding are fully different and, in this case, fluctuating random electromagnetic fields must be generated. This technological problem could be tackled by using scattering of light on disordered systems [8]. Analytically, these fields have been proved to generate different types of isotropic interactions including long range forces [9]. As already mentioned, the control of these central forces, using isotropic fields, could be used to induce attractive or repulsive regimes promoting or avoiding coagulation at the nanoscale.

Finally, a new type of optical interactions has been predicted for dipoles with non-scalar polarizabilities, like, for instance, the ones found in nature for chiral particles. These type of polarizability tensorial-like responses could be artificially generated by a technology based on the magneto-optical response of some materials, like the n-doped polar semiconductor InSb. In these systems, the tensor properties of the polarizability are tuned by applying an external magnetic field. These "artificially generated" tensorial dipoles also experience optical forces and torques strongly dependent on the spin and helicity of the electromagnetic field [10].

**Concluding Remarks**

In conclusion, the optical forces on nanoparticles with size much smaller than the radiation wavelength may be studied considering the response of an electric dipole. Although being analytically treatable, results obtained from these small systems are far from being trivial and turn out to be very useful when analysing the response of more complex systems. Among others, future applications of the force on these systems are the nanometric control of diffusion and coagulation, the design of building up methods based on the control of the dipole's binding properties and the propulsion or sorting induced by electromagnetic radiation.

**Acknowledgements**

Present work has been supported by the Spanish Ministerio de Ciencia e Innovación (MELODIA PGC2018-095777-B-C-22), UAM-CAM project (SI1/PJI/2019-00052) and through the "María de Maeztu" Programme for Units of Excellence in R& D (CEX2018-000805-M).

## 9 — Generalized Lorenz-Mie theories for optical forces

*Antonio A. R. Neves*

Centro de Ciências Naturais e Humanas, Universidade Federal do ABC (UFABC),Santo André, São Paulo, Brazil.

*Wendel L. Moreira*

Tratamento de dados geofísicos, Tecnologia e processamento de geologia e geofísica,Exploração, Petrobras - Petróleo Brasileiro S.A., Rio de Janeiro, Rio de Janeiro,Brazil.

*Adriana Fontes*

Departamento de Biofísica e Radiobiologia, Universidade Federal de Pernambuco (UFPE), Recife, Pernambuco, Brazil.

*Carlos L. Cesar*

Departamento de Física, Universidade Federal do Ceará (UFC), Fortaleza, Ceará,Brazil.

Instituto de Física "Gleb Wataghin", Departamento de Eletrônica Quântica, Universidade Estadual de Campinas (UNICAMP), Campinas, São Paulo, Brazil.

National Institute of Science and Technology on Photonics Applied to Cell Biology (INFABIC - IB and IFGW, UNICAMP), Campinas, São Paulo, Brazil

**Status**

The Lorenz-Mie scattering theory was presented in the early 20$^{th}$ century for an infinite plane wave incident beam, providing correct analytical results for arbitrary particle size and refractive index. After the development of the laser, however, plane waves were no longer satisfactory, and a generalization for an arbitrary incident beam was needed, which became known as the generalized Lorenz–Mie theory (GLMT) [1]. GLMT is, therefore, an extension of the classical Lorenz–Mie theory, in which the solution is usually written in the form of an infinite series of partial wave terms for any type of incident beam.

From the vast applications of GLMT, here we will focus on its role in determining optical forces for arbitrary incident fields. Since the invention of optical tweezers, there has been a continuous and significant effort towards a complete description of optical trapping forces. Early models yielded a theoretical description for extreme ratios between particle radius (a) to the wavelength (λ) either in the Rayleigh (a/λ<<1) or geometrical optics regime (a/λ>>1). To close this gap, a full electromagnetic solution was required and, since GLMT used Maxwell's stress tensor to deal with mechanical effects of light, it became one of the formalism for rigorous optical force calculations. The advances in optical trapping experiments, in which force components can be precisely measured, allowed a comparison with theoretical forces, refining the models used. Also, most reports now present their results in terms of the forces instead of radiation pressure cross-sections that were more common in the past.

One of the earliest applications of GLMT to optical forces dates back to 1988, when the radiation pressure on a scatterer was determined from the reduced momentum of the Poynting vector [2]. The partial waves coefficients, also known as beam–shape coefficients (BSCs), were determined for a beam model introduced a decade before, developed to produce higher-order corrections of paraxial Gaussian beams [3]. This model showed a reasonable agreement with experimental trapping forces and efficiencies values when using a mild focused beam. High numerical aperture (NA) objectives, however, are required for stable 3D trapping of particles, and these high NA beams were not correctly described by the higher-order correction. Although the initial beams were a solution of Maxwell equations (i.e., Maxwellian beams), this was not true for the higher-order corrected beam model. A non-Maxwellian beam resulted in a radial coordinate-dependent BSCs. These strongly convergent beams needed to be adequately described by a proper beam model, a Maxwellian beam, and preferably with parameters that are accessible experimentally.

**Current and Future Challenges**

To validate GLMT for the description of optical forces, it is necessary to obtain the BSCs of arbitrary beams and show that they correctly describe the incident beam. Moreover, the optical force predictions have to be consistent with experimental results. Most of the early results determined a single force for each particle size, termed escape force, which was the maximum optical force that could be exerted on the trapped particle to balance the drag force, which showed good agreement with GLMT calculations for displacements within the particle radius [4]. Later methods were richer yielding the optical force profile on the particle as a function of its displacement from the focused beam [5].

The challenge to find a proper beam model for optical tweezers, and an analytical method to solve the beam shape coefficients, was found and demonstrated by a series of experiments, which also detected the diffraction effects present in the high numerical beams, morphology dependent resonances, and spherical aberration effects, described in a recent tutorial for optical forces [6]. As a result, the BSCs could be determined analytically from their definitions, and the same model and procedure has been successfully applied in other fields.

Today, GLMT is a mature framework to describe the scattered wave by a symmetric scatterer for any type of Maxwellian beam. It has been successfully applied to investigate: trapping aggregates of interacting plasmonic nanoparticles; the trapping of meta-materials such as negative refractive index particles; stability of the trapped particle in terms of the switching rate in time-division multiplexing of the trapping beam; optical binding using a resonant cavity evanescent wave trap; microrheology of gaseous media, photophoretic forces in aerosols; the effect of polarization on trapping of chiral spheres; and switching photonic nanojets within optical traps.

The application of GLMT has not been restricted to only highly focused Gaussian beams, but it has also been extended to other Maxwellian beams. An alternative and more general method to determine an analytical exact expression for the BSCs, using a Fourier transform of the angular momentum operator was demonstrated [7]. Different types of incident beams have been applied for optical trapping within the GLMT: zero–order Bessel beams; vortex Bessel beam; standing waves; Laguerre–Gaussian beam; Airy beams and arrays of Airy beams; optical stretcher; optical conveyors as long-ranged tractor beams; Frozen waves providing localized stationary longitudinal fields; pulsed laser beams considering linear and nonlinear effects; and guided modes within hollow-core photonic crystal fibers.

**Advances in Science and Technology to Meet Challenges**

Research is a never-ending story, and most models are always tweaked for the upcoming trapping application. A few issues are worth highlighting:

- There are presently no analytical solutions to guided modes in hollow-core photonic crystal fibers, but to model the force within GLMT an approximate mode model is used. We would expect a rise in experiments and simulations to validate guided mode models, possibly via optical forces.

- Optical forces calculations for particles inside optical cavities where the electromagnetic field is given by the electromagnetic interaction between the particle and the cavity walls are also expected.
- Near-field interactions is another growing area of spectroscopy such as tip-enhanced Raman scattering, which will probably be used to trap particles.
- Most optical force experiments that displace the beam with respect to the trapped bead are achieving this by tilting the beam at the back-objective aperture. It would be suitable to see a model for the focused beam in terms of this tilt, which would result in the trap displacement near the focus.
- GLMT has been applied to symmetric objects such as spheres, spheroids and infinite cylinders, but, if the correct boundary conditions are satisfied, or a superposition thereof, it would be possible to calculate the forces for other shaped objects yet to be determined.
- In structured metamaterials, optical cloaked objects have been theoretically demonstrated using plane waves incident on the object. Would it be possible to trap a cloaked object with a specially shaped beam?
- A field where GLMT was timidly employed was in optical binding to form optical matter. It is a field with several recent and important advancements, most of which could be well modelled with GLMT.

In summary future developments are hard to predict, but considering it as an extrapolation of the recent timeline, experiments would be more complex, aggregating other fields of research to optical trapping such as non-linear optics, non-equilibrium thermodynamics and/or machine learning. In any case, GLMT is a robust method to determine optical forces which is currently unmatched.

**Concluding Remarks**

Since its birth, GLMT has been used for a wide range of applications in optical trapping. Over the years, new beam models have been presented as well as procedures for determining the BSCs. With the recent developments in instrumentation for optical trapping, for particle tracking, laser sources and beam shaping, finer tolerance would be needed from the comparison between experimental results and optical force models, for which GLMT holds great promise. In the future, we will see which scientific questions GLMT will address next, and which new upgrades it may receive.

**Acknowledgements**

AF acknowledges the National Institute of Photonics (INFo) and CNPq. CLC acknowledges CNPq grant (312049/2014-5); "Física do Petróleo em Meios Porosos", PETROBRAS-UFC (2016/00328-4) and F020/WIPPS II—Simulação numérica de invasão de água em poços produtores", PETROGAL-UFC.

## 10 — T-matrix approach for optical forces

*Rosalba Saija, Abir Saidi*

Dipartimento di Scienze Matematiche e Informatiche, Scienze Fisiche e Scienze della Terra, Università di Messina (Italy)

**Status**

Since its formulation in the mid-60s, the Transition matrix (T-matrix) method, summarized by Peter C. Waterman in his milestone paper dated 1971 [1], has become one of the most powerful, versatile, and popular techniques for a theoretical description of electromagnetic, acoustic, and elastodynamic scattering by particles and surfaces. The most recent attempt to outline the vast realm of this technique and its practical applications dates back to 2020 with a comprehensive publication database, that includes more than 290 references, authored by Michael Mishchenko [2].

The T-matrix method is a complete wave-optical modelling of the particle-light interaction, recently extended to calculate the mechanical effects of light [3,4]. It represents an ideal bridge between the dipole and the geometrical optics approximation, extreme cases of the classical theory of light scattering. It can be successfully applied in the intermediate range, when the scattering particle size is comparable with the light wavelength ($x \simeq 1$). The T-matrix method has also proved to be a good approach when dealing with inhomogeneous particles and nonspherical particles, that can be modelled as aggregates of spherical monomers.

The T-matrix method, also known as null-field method and extended boundary condition method (EBCM), is based on the representation of the scattering function on the orthonormal basis of the spherical multipolar vectors that are solutions of the Maxwell equations. In this case, the scattering function, which is called transition matrix (T-Matrix), acting on the amplitudes of the incident field, gives as a result the amplitudes of the scattered field. Among the different methods used to calculate the optical properties of model particles, in near and far field, the ones based on T-matrix are currently among the most accurate and efficient, particularly when applied to non-spherical particles composed by spherical constituents, i.e., clusters or aggregates of spheres, spheres with spherical (eccentric) inclusions, and multi-layered spheres [5]. This technique takes into proper account the multiple scattering processes occurring among the spherical subunits composing the aggregate and the contribution coming from all the details of the model structure. Optical properties and mechanical effects of composite scatterers are analytically derived and can be exactly calculated without introducing any approximation except the truncation of the expansion of the fields, being able to check the convergence of the results at every step.

In recent years, the T-Matrix method has been extended to the study of the mechanical effects of light in optical tweezers where the incident radiation, modelled through the angular spectrum representation [6], consists of Gaussian beams [7], cylindrical vector beams [8], or Laguerre-Gauss beams [9].

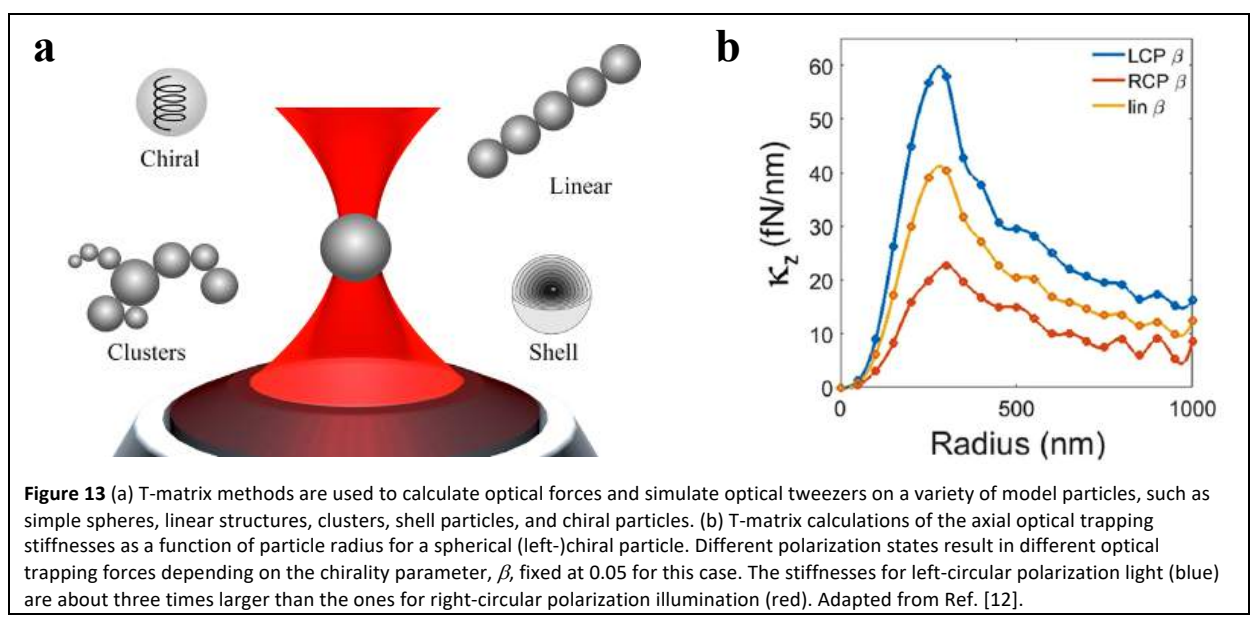

**Figure 13** (a) T-matrix methods are used to calculate optical forces and simulate optical tweezers on a variety of model particles, such as simple spheres, linear structures, clusters, shell particles, and chiral particles. (b) T-matrix calculations of the axial optical trapping stiffnesses as a function of particle radius for a spherical (left-)chiral particle. Different polarization states result in different optical trapping forces depending on the chirality parameter, $\beta$, fixed at 0.05 for this case. The stiffnesses for left-circular polarization light (blue) are about three times larger than the ones for right-circular polarization illumination (red). Adapted from Ref. [12].

**Current and Future Challenges**

As outlined above, the T-matrix is a consolidated method that manages to describe, analytically and numerically, the interaction between radiation and micro and nanostructured matter, possibly modelled as clusters or aggregates of spheres. Thanks to its high numerical accuracy, this approach is suitable for benchmark computations. However, some important issues of current interest may pose a challenge. First, advances in nanofabrication made available a novel class of materials and metamaterials with unprecedented optical properties [10]. Electromagnetic fields generated by such complex structures and metasurfaces can permit an optomechanical spatial control of submicrometric particles or molecules. T-matrix methods can accurately determine the resulting optical forces by accounting the full interaction between the surface and the interacting particles [11]. This is particularly interesting when considering optical forces applied to chiral particles [12]. In this context, properly engineered metasurfaces can represent a breakthrough in the selection and sorting of nano- and microparticles on the basis of their optical activity.

Related to the above considerations, a relevant improvement of the T-Matrix numerical tools concerns its applications for the inverse design of nanofabricated patterns or metasurfaces for optimal optical manipulation in the surface proximity. This improvement can be achieved by interfacing the force field data obtained with T-matrix simulations with automated procedures based on the use of neural networks [13].

Another future challenge concerns the development of some specific theoretical tools able to improve the calculations and modelling of optical forces on cosmic dust grains with complex shape and composition for applications of optical tweezers in space [14]. The relevance of these applications in space science lies not only in the possibility of an in-depth study of the chemical and physical properties of powders, but also in investigating the role played by cosmic powders on the composition of planetary atmospheres.

**Advances in Science and Technology to Meet Challenges**

In order to address the above challenges, using optical forces on micro- and nano-particles, a synergistic advance of the theoretical approach and of the experimental procedures is necessary.

As regards the study of the optical forces generated by metasurfaces, while the analytical approach based on the definition of the Maxwell stress tensor is complete and defined, providing in a closed form analytical expressions for transverse forces and radiation pressure, the numerical codes allowing their calculation must still be systematized. At the same time an improvement in the experimental methodologies capable of performing force measurements at the femtonewton order will be necessary. For this purpose, the force detection technique used in the optical tweezers has the required sensitivity and can be adapted for experiments.

The possibility of exploiting surface selection properties, based on the effect of optomechanical forces acting on nanoparticles with optical activity, pushes us to find a strategy based on artificial intelligence that can provide the best compromise in terms of sensitivity and ease of implementation. For this purpose, the neural network based on convolutive algorithms seems to be the most advantageous choice and in the near future research in this field will focus on the design of this type of numerical protocols.

Finally, the theoretical analysis for space science applications of optical trapping constitutes a test bed for the technologies that will be developed and can serve to address the strategies for the realization of experimental equipment.

**Concluding Remarks**

Since their first introduction light scattering techniques based on the T-Matrix approach have found applications in a wide area of science and technology, in particular as a fundamental tool for optical trapping. Over the years, the scientific community has found answers to many questions concerning the theory of optical trapping of particles at the micro and nanoscale, also in the presence of surfaces, both in far and near-field. With the recent advances in nanotechnologies, new challenges are arising related to new possibilities for the nanofabrication of complex surfaces and metamaterials that can open novel routes for optical manipulation. The expertise acquired in the field of optomechanics and nanophotonics will allow us to face these challenges through a multidisciplinary approach.

**Acknowledgements**
We acknowledge support by the agreement ASI-INAF n.2018-16-HH.0, project "SPACE Tweezers".

## 11 — Transverse angular momenta

*Paul Beck[1,2], Jörg S. Eismann[1,2,3], and Peter Banzer[1,2,3]*
[1] Institute of Optics, Information and Photonics, University Erlangen-Nuremberg, Staudtstr. 7/B2, D-91058, Erlangen, Germany
[2] Max Planck Institute for the Science of Light, Staudtstr. 2, D-91058 Erlangen, Germany
[3] Institute of Physics, University of Graz, NAWI Graz, Universitätsplatz 5, 8010 Graz, Austria

**Status**

The angular momentum (AM) of light plays a crucial role in the field of optics. A beam of light, described as an electromagnetic wave, can carry two different kinds of intrinsic AM. The first one is connected to a spinning electric or magnetic field vector and is denoted as spin AM (SAM). The most prominent example of the SAM of light is circular polarization [1]. Unlike SAM, orbital AM (OAM) is a consequence of the spatial structure of a beam or field. It usually emerges as an azimuthal phase gradient around singular points [2]. A plethora of applications reflects the significance of both OAM and SAM. Their utilization as information carriers in classical or quantum communications or for the manipulation of microscopic matter via optical AM transfer are just two examples. For a beam of light, the total AM density can be expressed as the sum of spin and orbital constituents, i.e., $j = s + l$, where $s$ corresponds to the SAM density and $l$ to the OAM density. The individual spin and orbital parts are defined via the electric and magnetic fields by

$$s = Im[\epsilon_0(E^* \times E) + \mu_0(H^* \times H)]/4\omega, \qquad (1)$$

$$l = r \times Im[\epsilon_0 E^* \cdot (\nabla)E + \mu_0 H^* \cdot (\nabla)H]/4\omega. \qquad (2)$$

Hence, each of them comprises contributions by the electric and the magnetic field. One can easily see that for a circularly polarized plane wave propagating along the z axis and described by its electric field vector $E_\pm = E_0(1, \pm i, 0)$, one obtains a position-independent non-zero longitudinal SAM density $s \propto E_0^2 \hat{e}_z$. In contrast, the OAM remains zero for an adequate choice of the coordinate system. Intrinsic optical OAM (independent of the coordinate system) arises in beams of light possessing a more intricate spatial structure. One prominent class of beams in this regard are so-called Laguerre-Gaussian modes, solutions to the paraxial wave equation, which carry discrete values of OAM. If a particle is interacting with a focused beam possessing AM, part of the AM might be transferred resulting in a mechanical action. In the case of conventional SAM or OAM, which is naturally aligned with the propagation or optical axis, either a spinning or orbiting motion of the particle about the optical axis has been observed experimentally [3]. However, it has been proposed that the orientation of AM is not restricted to be parallel or antiparallel to the optical axis, opening new possibilities for the manipulation of three-dimensional particle motion in an optical tweezers system.

**Current and Future Challenges**

Apart from providing a great platform for the study of fundamental light-matter-interactions, also the dynamics of the medium a particle is trapped in optically can be studied by optical momentum transfer [4]. The trapped bead then acts as a local sensor for, e.g., the local viscosity of the fluid. This is a huge advantage over other well-established, non-local methods. However, in terms of rotational motion, the employment of light carrying only longitudinal angular momentum may impose a severe restriction to such kinds of measurements. It would be thus highly desirable to transfer AM with an arbitrary orientation to trapped particles to enhance the potency of the aforementioned and many other techniques [5]. In order to achieve such full control over all rotational degrees of freedom of sub-micron-sized objects inside a simple optical trap, it is necessary to create a light field possessing not only longitudinal, but also transverse AM. Recently, it has been shown that upon tight focusing of light beams, a natural ingredient to optical tweezers, components of the SAM perpendicular to the direction of propagation arise [5, 6]. This effect occurs due to the superposition of out-of-phase transverse and longitudinal electric or magnetic field components [7]. Note that such a superposition does not take place for a collimated paraxial beam or plane wave which are assumed to feature only transverse field components. Schematically, the longitudinal electric and/or magnetic field components that are necessary for the formation of transverse SAM result from the process of tight focusing because of a tilt of the wave vectors of the incoming beam towards the focal spot. The distinction and emergence of longitudinal and transverse SAM are schematically shown in **Figure 14**. A spinning motion of the electric field vector about the direction of propagation results in longitudinal SAM (**Figure 14**a), whereas transverse SAM is a consequence of a rotation of this vector around an axis perpendicular to the optical axis (**Figure 14**b).

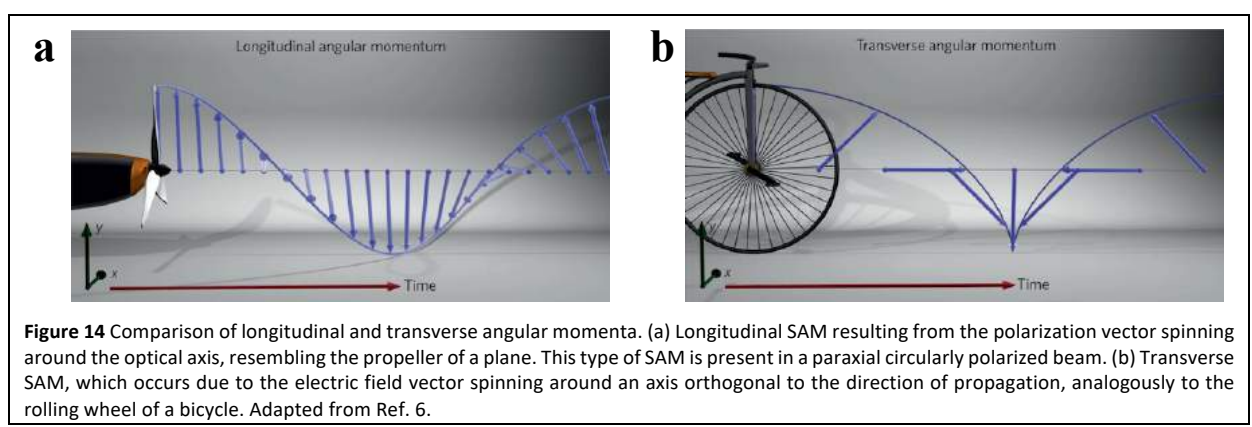

**Figure 14** Comparison of longitudinal and transverse angular momenta. (a) Longitudinal SAM resulting from the polarization vector spinning around the optical axis, resembling the propeller of a plane. This type of SAM is present in a paraxial circularly polarized beam. (b) Transverse SAM, which occurs due to the electric field vector spinning around an axis orthogonal to the direction of propagation, analogously to the rolling wheel of a bicycle. Adapted from Ref. 6.

**Advances in Science and Technology to Meet Challenges**

Recently, the utilization of transverse SAM in optical tweezers reached a new milestone. A trapped object was experimentally set in a spinning motion about an axis normal to the optical axis [8] following an earlier theoretical proposal [5]. The incoming field distribution consists in principle of two laterally displaced and circularly polarized Gaussian modes of opposite handedness (**Figure 15**a, left). Tight focusing of this beam results in a non-zero transverse SAM density distribution (**Figure 15**a, right) while the longitudinal SAM component vanishes [5]. The use of such an extraordinary field distribution is necessary here, as the utilized microspheres, measuring 3.5 μm in diameter, are only sensitive to the integrated SAM density, which is in fact zero for the majority of light beams that are used for trapping or in optics in general.

In contrast to the total transverse SAM, almost every tightly focused beam exhibits a locally non-zero transverse SAM density [6, 9, 10] that can be sensed by sub-wavelength sized particles. As an example, for a linearly *y*-polarized Gaussian beam (**Figure 15**b), we find a locally non-zero $s_x^e$ distribution that is antisymmetric with respect to the *x*-axis. As a result, if a nanoparticle is placed therein, the sign of the transferred torque will depend on its position relative to the *x*-axis. Generally speaking, different paraxial input fields will result in different distributions of the transverse SAM in the focal plane, providing an enormous variety of possible landscapes for particle manipulations. The whole situation becomes even more delicate when also considering the magnetic spin density, as in the non-paraxial regime, electric and magnetic transverse spin densities can feature utterly distinctive distributions [10] that can lead to material selective optical forces.

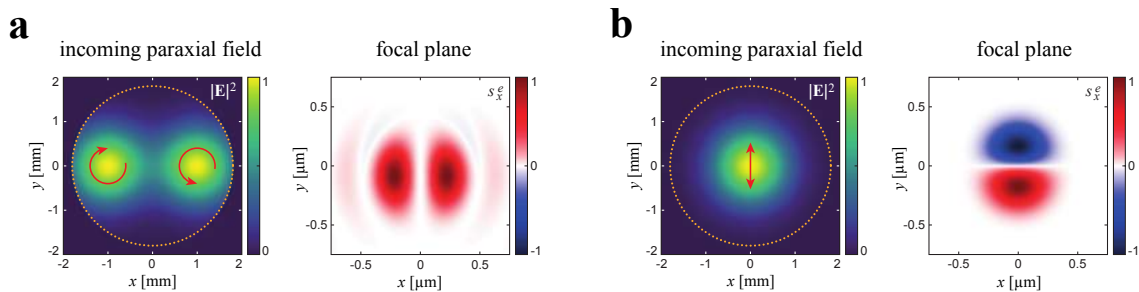

**Figure 15** Incoming field distribution and the resulting transverse spin density component $s_x^e$ in the focal plane for (a) two laterally displaced circularly polarized Gaussian beams of opposite handedness [5, 8] and (b) a linearly *y*-polarized Gaussian beam. Integration of the spin density distribution in (a) results in a non-zero value, whereas it is strictly zero for (b). The local polarization states of the input beams are indicated by red arrows, and the dotted circles indicate the entrance aperture of the focusing optics. Calculation parameters: wavelength = 532 nm, numerical aperture = 0.9, beam radius = 1.2 mm.

**Concluding Remarks**

Since the inception and first mention of transverse angular momenta of light in general and transverse spin in particular about a decade ago, the corresponding field has grown substantially, paving the way also for various applications. Especially in the context of optical forces, transverse components of the net spin or the spin density hold an immense potential to be explored in the future. Full control over rotational degrees of freedom in optical traps, the interaction with chiral matter and respective forces as well as the exploitation of three-dimensional spin density or spin vectors resulting from tailored electric and magnetic fields represent intriguing pathways. Moreover, the induction of position- and material-dependent forces and torques acting on trapped particles opens new pathways for applications, for instance, in the context of optically driven micromachines.

## MEASUREMENT OF FORCES AND TORQUES

### 12 — Photonic force microscopy

*Thales F.D. Fernandes, Francesco Pedaci*

Centre de Biologie Structurale, CNRS, INSERM, Univ.Montpellier, France

**Status**

Soon after their demonstration by Ashkin, the ability of Optical Tweezers (OT) to simultaneously manipulate and detect the position of microscopic particles at high spatio-temporal resolution was recognized as instrumental for the development of a novel near-field scanning imaging technique, termed Photonic Force Microscopy (PhFM) [1, 2, 3, 4]. In PhFM, an optically trapped particle is used to scan and probe the local environment surrounding the sample (**Figure 16**a). Similar to Atomic Force Microscopy (AFM), imaging in PhFM is achieved by moving the sample by a piezo stage while the trapped probe mechanically interacts with the surface of the sample. The probe displacement is then used to reconstruct the image of the surface topography. In PhFM, similar to AFM, the displacement of the trapped probe is used to define the signal in each pixel (or voxel), while a piezo stage moves the center of the optical trap over the sample. Crucially, with respect to AFM, the PhFM probe is held in position by its interaction with light in the laser focal region, creating a very soft optical spring, with a stiffness $\kappa$ of the order of 1 fN/nm, three orders of magnitude less stiff than standard AFM cantilevers. Such extreme softness is one of the main features of PhFM, allowing the characterization of materials of extremely low rigidity, which are outside the scope of AFM. In contrast, due to the stiffness of the cantilever, high-quality AFM signals can only be achieved with relatively rigid samples. Combined with the fact that it can be easily used in liquid, this opens the way for PhFM imaging of the topography of soft biological materials [5] such as the membrane of living cells under physiological conditions, which are notoriously difficult to image in AFM without chemical fixation (which artificially rigidifies and kills the cells). Another unique aspect of PhFM (and OT) is the absence of contact between the probe and the instrument, not achievable with other scanning techniques. This has been used to expand the reach of the technique from surface topography to volumetric scanning imaging [6]. Despite resolutions that currently remain below those of standard AFM, the potential of PhFM to image soft and biological surfaces or volumes in living systems remains unique, thus inviting further developments.

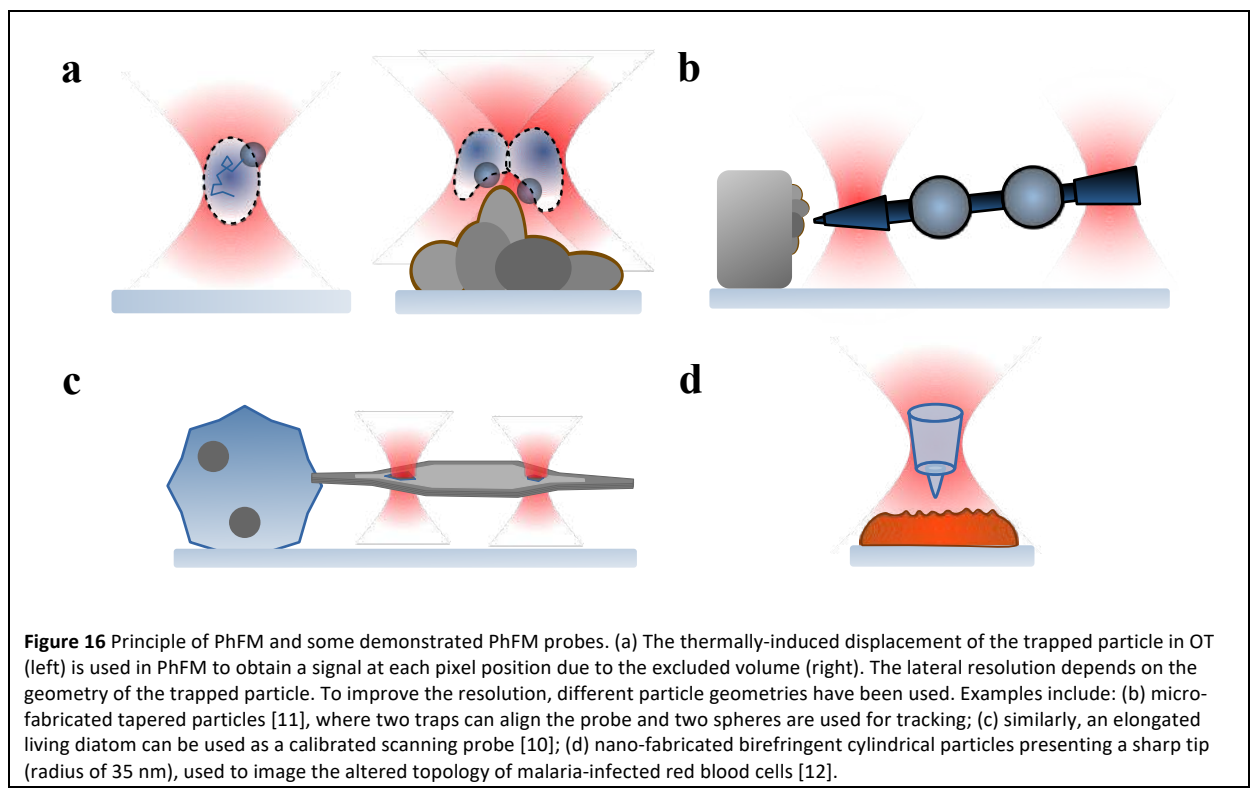

**Figure 16** Principle of PhFM and some demonstrated PhFM probes. (a) The thermally-induced displacement of the trapped particle in OT (left) is used in PhFM to obtain a signal at each pixel position due to the excluded volume (right). The lateral resolution depends on the geometry of the trapped particle. To improve the resolution, different particle geometries have been used. Examples include: (b) micro-fabricated tapered particles [11], where two traps can align the probe and two spheres are used for tracking; (c) similarly, an elongated living diatom can be used as a calibrated scanning probe [10]; (d) nano-fabricated birefringent cylindrical particles presenting a sharp tip (radius of 35 nm), used to image the altered topology of malaria-infected red blood cells [12].

**Current and future challenges**

The microscopic probe of the PhFM is, in liquid, an over-damped system where inertial effects can be neglected up to frequencies of several hundreds of kHz. When imaging with such a probe, its confined Brownian motion is relevant, setting the constrain of ergodicity: the probe requires a characteristic time $t_c = \gamma/\kappa$ (where $\gamma$ is the drag coefficient of the probe) to fully explore the volume of the laser trap. This is related to the precision of the measured signal chosen to characterize the pixels of the image (typically the average or median vertical displacement of the probe). As the error of the mean decreases with the square root of the number $N$ of independent measurements ($N = T/t_c$, where $T$ is the total measurement time at one position), it is interesting to note that there is no theoretical limit to the precision of the measured signal. The precision limit is set by practical constraints: the drift degrades the accuracy of the measurement at long times, and the total imaging time should remain reasonable for the system under study. Therefore, a good thermal, mechanical, and optical stability, that reduces the drift of the apparatus, together with a careful calibration procedure, is key to achieve high precision and accuracy. In order to minimize the total imaging time, it is necessary to reduce the characteristic time $t_c$. An increase in laser power leads to an increase of the trap stiffness $\kappa$, decreasing $t_c$. However, this is not always possible on delicate samples. A decrease in particle size decreases the drag coefficient $\gamma$ but also the stiffness $\kappa$, unless a (trappable) higher refractive index material is chosen.

Related to particle size is its geometry, crucial in any near-field scanning technique. In PhFM, the spherical dielectric particles typically used in OT have often been employed as scanning probes. While (sub-)microscopic spherical particles have geometrical limitations in achieving nanometer resolution in the lateral directions (the image is the convolution of the surface topology with the tip geometry), they are commercially available, the signals are straightforward to interpret, they allow nanometer resolution in the axial direction, and, when their diameter is reduced to a few hundreds of nanometers, they have been shown to allow imaging with sub-diffraction resolution [7]. To improve the lateral resolution, the potential of elongated probes has also been recognized and explored [8, 9, 10, 11, 12]. Elongated probes presenting a sharp feature also have the interesting capability to be manipulated by multiple independent traps (**Figure 16**b and **14**c), thereby probing unconventional samples where the topography of interest is perpendicular to the sample plane.

**Advances in Science and Technology to Meet Challenges**

For PhFM to reach full maturity, where it will be employed by a larger community outside of physics and developers, certain technical improvements will be required. The development of non-spherical probes will prove crucial to reaching lateral resolutions which approach AFM. Recent developments have enabled the massive nano-fabrication of microscopic probes presenting a sharp tip (a few tens of nanometer in radius, **Figure 16**d) which are easily able to be trapped [12]. An ideal scenario could be a trappable probe wherein: 1) its index of refraction is maximized to obtain higher trap stiffness, 2) its size is minimized to give low drag, and 3) a feature as sharp as an AFM tip (of few nanometers radius of curvature) is present to interact with the sample to increase the lateral resolution. This would allow high-resolution images of soft samples while minimizing the total imaging time. Additionally, more reliable surface passivation protocols are needed for the probe in different environments to avoid deleterious aspecific adherence of the probe to the sample during imaging, and novel solutions need to be developed to improve the ability to image thicker or non-transparent samples.

Being based on a versatile technique like OT, PhFM can conceivably be expanded in other directions that can be combined with measuring topography. Different functionalities that have been demonstrated on solid substrates can be transferred to PhFM, adding to them the capability of spatial scanning. Interestingly, adding plasmonic structures on the trapped probe has been used to produce a "nano-trap in a micro-trap" [13], where nanometer-sized objects can be trapped by the enhanced electromagnetic field created by a plasmonic structure on the trapped microscopic particle, effectively detaching the nano-tweezers from the surface and allowing 3D nano-manipulation. The use of fluorescence-based techniques such as super-resolution microscopy, Fluorescence Correlation Spectroscopy (FCS), and Fluorescence Resonance Energy Transfer (FRET), combined with an enhancement-optimized engineered structure built on top of a PhFM probe could be used to detect the presence, position, and conformational changes of single molecules, opening up the possibility of imaging heterogeneity. Finally, by employing birefringent particles, by controlling and measuring the polarization state of the trapping laser, PhFM can be expanded to control and measure the optical torque on the scanning probe with pN·nm accuracy [14]. This effectively scales a macroscopic rheometer to the (sub)micron scale [15], opening the way for simultaneous topography and micro-rheological imaging, for example on cell membranes and other soft interfaces.

**Concluding Remarks**

PhFM was recognized early on for its unique potential in topological and volumetric scanning of soft biological samples. However, due to key technical limitations, including a limited lateral resolution, the need for parallel fabrication of a high number of colloidal probes, and an involved detection scheme, its development lagged behind other scanning near-field techniques. Today, several of these limiting factors can be addressed. In particular, nanofabrication opens new possibilities in obtaining probes with sharp scanning features to achieve high lateral resolution. Moreover, coupling PhFM and its scanning capabilities with other optical modalities, from plasmonic manipulation and sensing to active micro-rheology, it is assured that future developments hold promise for this versatile technique.

## Acknowledgements

We thank Ashley L. Nord for useful comments. This work is supported by the ANR HiRes-PFM project grant ANR-20-CE42-0005-01 of the French Agence Nationale de la Recherche. The CBS is a member of the France-BioImaging (FBI) and the French Infrastructure for Integrated Structural Biology (FRISBI), two national infrastructures supported by the French National Research Agency (ANR-10-INBS-04-01 and ANR-10-INBS-05, respectively).

## 13 — Precision measurements with ballistic optical tweezers

*Warwick P. Bowen*
University of Queensland, Australia

**Status**

Optical tweezers have a truly broad range of applications, spanning biomedicine, biophysics, quantum physics and fundamental studies. Many of these applications rely on particle tracking and use measurements of the trajectory of the particle to learn about its environment. In the case of biophysics, one could for instance learn about the molecular forces applied to the particle, or about how energy is dissipated/stored in the medium within which the particle resides. This ability to extract information about the environment of the particle at nanoscale is truly powerful, for instance allowing discrete motor molecule steps and single molecule binding-events to be observed. However, a significant challenge has been to achieve measurements of the environment over timescales commensurate with the dynamics of the environment [1].

Typical optical tweezers operate in the Brownian regime, where the precision of the particle tracking measurement is sufficient to localise where the particle is in space, but insufficient to determine its instantaneous velocity. To determine the velocity, it is required to be able to resolve the position of the particle on a timescale fast compared to the momentum relaxation time [2], which for a microparticle in liquid is typically in the range of a microsecond. This is something that Einstein famously considered to be impossible [3]. As illustrated in **Figure 17**, the position of a particle in an optical trap equilibrises only slowly with its environment — at a timescale given roughly by the inverse of the trapping frequency. As a result, optical tweezers measurements tend to be relatively slow, limited to second to millisecond timescales.

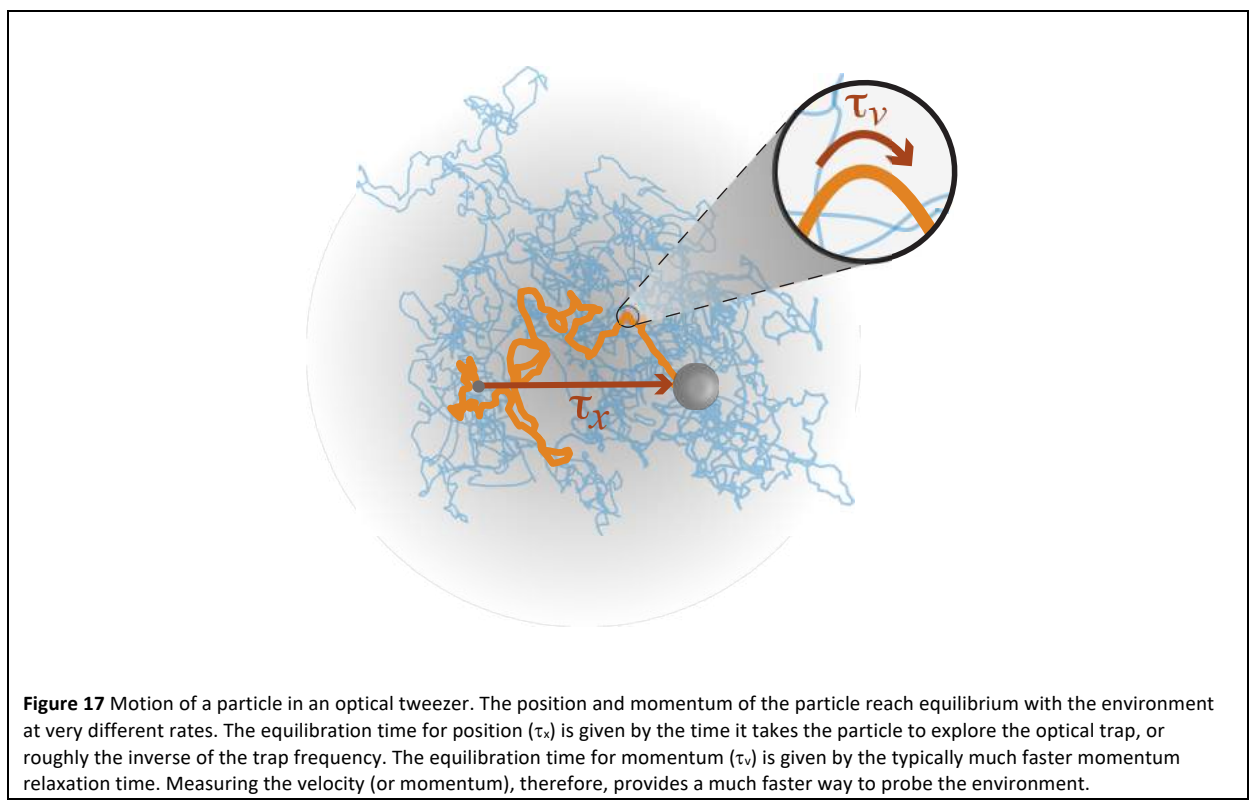

**Figure 17** Motion of a particle in an optical tweezer. The position and momentum of the particle reach equilibrium with the environment at very different rates. The equilibration time for position ($\tau_x$) is given by the time it takes the particle to explore the optical trap, or roughly the inverse of the trap frequency. The equilibration time for momentum ($\tau_v$) is given by the typically much faster momentum relaxation time. Measuring the velocity (or momentum), therefore, provides a much faster way to probe the environment.

**Current and Future Challenges**

The challenge then is whether it is possible to reach the ballistic regime of optical tweezers, where the instantaneous velocity of the trapped particle can be observed, and if so, whether this could enable faster optical tweezers measurements? The justification for expecting that faster measurements may be possible in the ballistic regime is that the speed at which a probe can measure its environment is directly connected to how fast the probe equilibrates with its environment. The position equilibrates slowly, as discussed above. However, the velocity equilibrates fast — so exactly the physics that makes it challenging to measure the instantaneous velocity also provides the prospect to probe the particles environment fast if one can measure it.

In principle, the precision of optical-tweezers-based particle tracking scales as the inverse-square-root of the detected optical power. Therefore, the solution to reach the ballistic regime appears obvious — just increase optical power. Unfortunately, the powers generally required to reach the ballistic regime for microscale particles are too high for commercial optical tweezers detection system to handle. This led the Raizen lab to develop alternative custom detection technologies for optical tweezers, including the fibre-bundle technique shown in **Figure 18** (bottom left) [4,5]. Using a custom photodetection system, they were able to reach the ballistic regime, proving Einstein wrong, and test the Maxwell-Boltzmann distribution for the first time for microparticles in a liquid [5].

Later in 2021, in a work by Madsen *et al.* [6], my laboratory demonstrated an alternative method to reach the ballistic regime, which we term *structured-light detection*. This method has the advantage of being compatible with commercial photodiodes. It works by first recognising that the vast majority of the optical trap photons that are transmitted through the optical tweezer carry no information about the position of the particle. By exploiting the different symmetry of the optical modes that carry and do not carry information, we were able to design a spatial filter to suppress the non-information carrying light by a factor of a thousand. This filter combines a phase plate with a single mode optical fibre, as illustrated in **Figure 18** (top left) and, when operating with around 200 mW of power allowed us to achieve record particle tracking sensitivity just above 1 femtometre-per-root-hertz (**Figure 18** (right)).

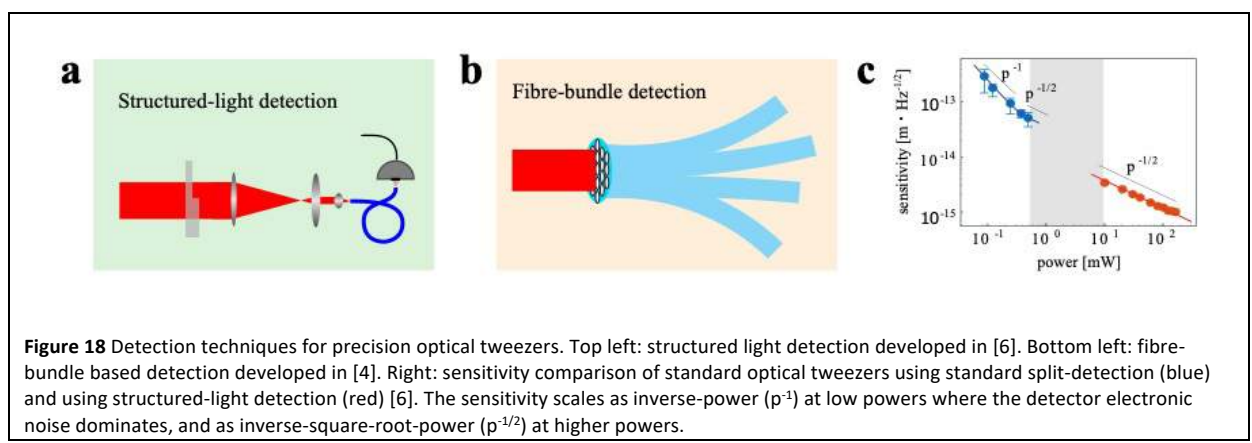

**Figure 18** Detection techniques for precision optical tweezers. Top left: structured light detection developed in [6]. Bottom left: fibre-bundle based detection developed in [4]. Right: sensitivity comparison of standard optical tweezers using standard split-detection (blue) and using structured-light detection (red) [6]. The sensitivity scales as inverse-power ($p^{-1}$) at low powers where the detector electronic noise dominates, and as inverse-square-root-power ($p^{-1/2}$) at higher powers.

**Advances in Science and Technology to Meet Challenges**

Applying our structured-light detection technique, we showed that it is possible to accurately characterise the momentum relaxation time at very fast speeds. Since the momentum relaxation time is directly connected to the viscosity of the medium in which the particle resides, this allowed us to make viscosity measurements that were around four orders-of-magnitude faster than has been possible before, at speeds as fast as 20 microseconds [6]. This now makes it possible to track the dynamics of viscosity, rather than simply treating it as a static material parameter. Some remaining open questions are: what viscosity dynamics are to be expected? in what sort of systems? and is it possible to apply these techniques to very small particles thereby suppressing force noise (which scales as particle area) and allowing more precise measurements of molecular forces?

To answer these questions requires advances in both science and technology. In science, a better understanding of out-of-equilibrium matter is required. We know that active matter can exhibit turbulence, vortex dynamics and a range of other rich physics [6]. Vortices are thought to form fast propagating bands in strain stiffening materials that carry along with them regions of high viscosity [7]. But whether such dynamics exists in other active materials, such as biological materials, and whether they might play a functional role in these materials, is not known. Models of biological active materials such as the cellular cytoplasm are needed to answer these questions.

Optical tweezers that can reach the ballistic regime within a living biological system are also required, and improvements in ballistic optical tweezers to enable operation with smaller particles. This will necessitate technology development. A key challenge is to reach the ballistic regime with optical powers that are sufficiently low that they do not intrude on the biological specimen. This can be expected to especially be a challenge when using small particles, since their lower optical scattering and faster momentum relaxation greatly increase the difficulty in reaching the ballistic regime. Solutions could involve employing resonant particles such as plasmonic particles, using structured light to increase the optical scattering [8], or even using quantum correlated light to improve the precision of the measurement at fixed optical power levels [9,10].

**Concluding Remarks**

It is now possible to reach the ballistic regime of particle tracking in precision optical tweezers, and to employ this regime to perform exceptionally fast measurements of the environment of the particle. This has been used to determine nanoscale viscosity at speeds in the range of tens of microseconds. Now that the technology has been proven, there are prospects for a wide range of applications from direct measurements of the energetics of molecular binding to experiments that shed light on the role of the dynamics of viscosity in biological function. Some of these applications are accessible with existing technology. Others will require new developments and innovations.

**Acknowledgements**

This material is in part based upon work supported by the Air Force Office of Scientific Research under award number FA9550-17-1-0397. It was also supported by the Australian Research Council Centre of Excellence for Engineered Quantum Systems (EQUS, CE170100009). W.P.B. acknowledges contributions to the figures from Dr Lars Madsen and Alex Terrason.

## 14 — Generation and measurement of torques

*Rahul Vaippully, Muruga Lokesh, Basudev Roy*
Indian Institute of Technology Madras, India

**Status**

Rotation of a non-absorbing birefringent particle in a single-beam optical tweezers was first shown by Friese et al. [1] using birefringent particles and circularly polarized light while the optical torque wrench was shown by La Porta and Wang [2]. The effect has since revolutionized optical tweezers where not only can one generate translational forces, but also apply torques. This facet has been used extensively in physics, where rotations have been used in topics ranging from fluid dynamics, rheology, exotic effects in vacuum and in biology where controlled torques were applied to cells and even single molecules [3], while in some cases new paradigms were opened with new insights into twisting of molecules like kinesin during stepping [4]. It is here that we have to realise that a rigid body can have three translational degrees of freedom while simultaneously having another three rotational degrees of freedom. The above-mentioned method only explores what can be called the in-plane (or yaw) motion. There are still the two out-of-plane modes, namely the pitch and roll motions. Pitch motion was partially realised when asymmetric cells were rotated using two trapping beams, but that was still far from generating controlled torques on spherical particles where the drag is minimised. Controlled roll torques have even today remained elusive. These out-of-plane modes become prominent particularly in proximity to surfaces when the rotational symmetry is broken. Moreover, there are plethora of biological problems where such a mode can provide wealth of additional information. The roll motion has so far only been explored with nanowires where nanoscale kinks were studied to ascertain the rotational motion.

**Current and Future Challenges**

Conventionally, in optical tweezers, the polarization of light was used to generate yaw rotational motion. However, it is not very useful when trying to turn in the pitch and roll sense because it is difficult to generate a single optical tweezers beam where the polarization rotates orthogonal to the Poynting vector enough. There have been some studies where small ellipticities were generated in the transverse sense with a single beam but that is hardly enough to spin a birefringent particle where the restoring torque has to be overcome. It is here that other means have to be explored. Nonetheless, the pitch and the roll mode dynamics differ from the yaw mode particularly in proximity to surfaces. The knowledge of the pitch mode is useful in problems like the study of membrane fluctuations in cells where one can envisage "slope fluctuations" [6] instead of the normal fluctuations alone, the tug of war between molecular motors, local rheology of cell cytoplasm [7], and the study of local diffusion of proteins on domains where pitch angle can provide additional information. It is believed that during the process of endocytosis in a cell, the particle turns. Moreover, interesting information can be obtained when tethers are pulled out of cells which then twist while contracting.

It is well known that birefringent particles turning under circularly polarized light in vacuum also exhibit pitch and yaw rotations which are all but uncontrolled. If ways can be designed to avoid this rotation, the cooling to quantum states might be more realisable. Moreover, it can be envisaged that out-of-plane rotations can have significant effects when forcing fluids into interfaces when spinning continuously at the micro-scale.

Molecular motors like Kinesin-1 and Myosin-5 are known to move in a hand-over-hand fashion while turning. It is believed that both the right-handed and left-handed motion is equally possible. One may ask whether that hypothesis is correct thereby providing valuable insights into functioning of life at the nano-scale. In an active bath, the correlation between the rotational mode and the translational mode of a spherical particle can provide a wealth of information.

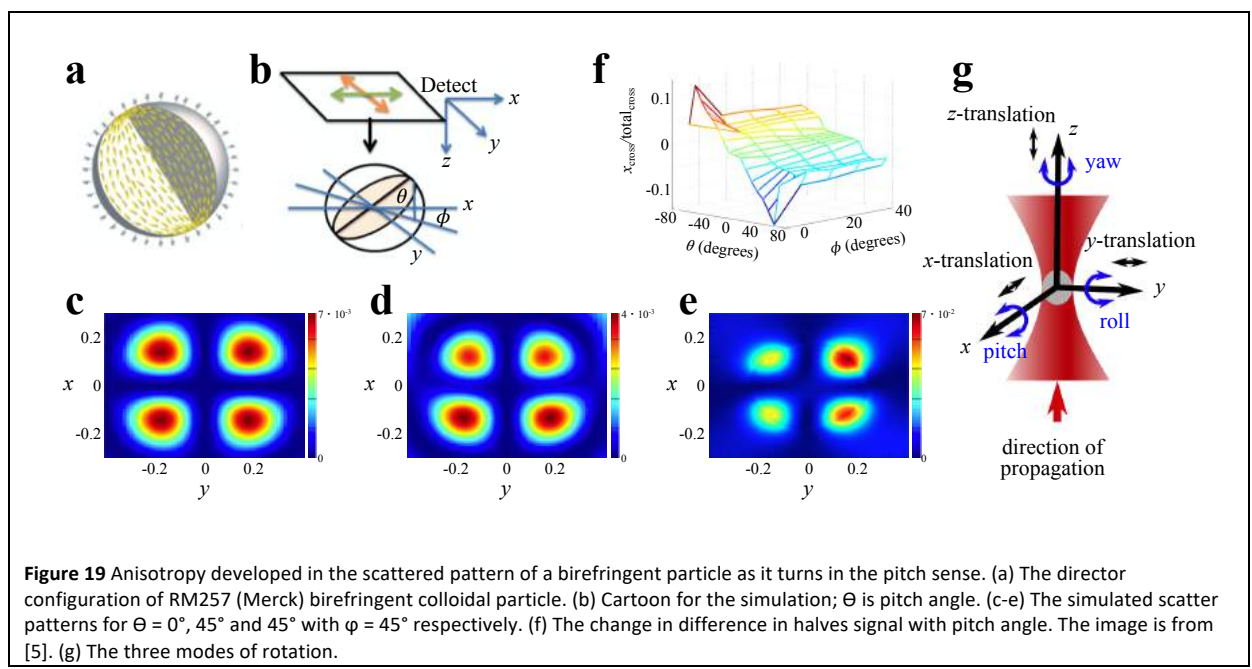

**Figure 19** Anisotropy developed in the scattered pattern of a birefringent particle as it turns in the pitch sense. (a) The director configuration of RM257 (Merck) birefringent colloidal particle. (b) Cartoon for the simulation; Θ is pitch angle. (c-e) The simulated scatter patterns for Θ = 0°, 45° and 45° with φ = 45° respectively. (f) The change in difference in halves signal with pitch angle. The image is from [5]. (g) The three modes of rotation.

**Advances in Science and Technology to Meet Challenges**

One of the first attempts at turning a particle in pitch sense uses two optical tweezers beams in holographic tweezers and moves one of the focal spots to reorient the particle. While this works for anisotropic particles very well and in some cases, even spherical particles, there is always a possibility that the particle would spin freely. It is for this reason that a scheme was suggested to use a birefringent particle in two trapping beams and then change the position of the focal spot of one beam. The problem with this scheme is that the particle has to be large enough to fit two independent beams. There are other schemes too, where other means can be used in addition to the confining single optical tweezers beam, like convection currents [8] or magnetic fields in the case of optically compatible magnetic materials which can even do roll motion [9]. Recently, a scheme was suggested which used two oppositely inclined tweezers beams to generate a pitch spinning polarization configuration [10]. In all these methods, there are two external stimuli used to turn in pitch sense. One can also put the optically trapped particle onto the stage gently and move the stage parallel to the surface to generate pitch motion [11]. This works for particles smaller than one beam waist of the trapping laser.

In order to detect the pitch rotational motion, a new scheme was suggested where the anisotropy in the lobes of the scatter pattern between two crossed polarizers for a birefringent particle was used, shown in **Figure 19**. This was used to determine cell membrane "slope" fluctuations [6], a new concept which derives from the gradient of the normal membrane fluctuations, as shown in **Figure 20**. Here the anisotropy in the scatter pattern is detected using difference in halves of a split photodiode. The technique of determining the pitch angle to high resolution can be used in combination with a two-beam rotation generation method on a birefringent particle to realise the optical pitch torque wrench [12]. Using this one can think of applying controlled pitch torques on surfaces like membranes which is relevant during situations like cell endocytosis.

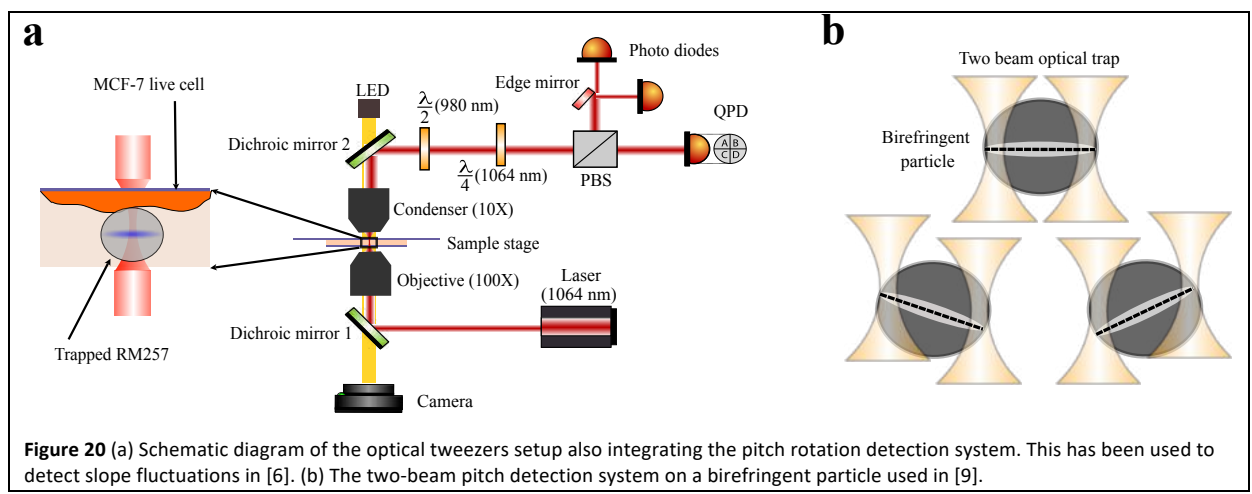

**Figure 20** (a) Schematic diagram of the optical tweezers setup also integrating the pitch rotation detection system. This has been used to detect slope fluctuations in [6]. (b) The two-beam pitch detection system on a birefringent particle used in [9].

**Concluding Remarks**

The generation and detection of yaw optical torques has been explored extensively over the past 20 years. However, the optical torques in other two degrees of freedom are only beginning to be explored. The pitch rotation was recently detected using the degree of anisotropy of the scatter pattern of a birefringent particle under crossed polarizers in the optical tweezers. This can be used in combination with two beam pitch rotation configuration to generate pitch torque wrenches. The roll mode is yet to be explored but a combination of optical and magnetic forces can be a start if the appropriate particles, which can simultaneously be optically confined without heating and bear magnetic properties, be used. Such modes of rotation can have many applications particularly in proximity to surfaces when the yaw dynamics varies from pitch and roll.

**Acknowledgements**

We thank the Indian Institute of Technology, Madras, India for their seed and initiation grants. This work was also supported by the DBT/Wellcome Trust India Alliance Fellowship IA/I/20/1/504900 awarded to Basudev Roy.

## 15 — Direct optical force and torque measurements

*Gregor Thalhammer, Monika Ritsch-Marte*

Institute for Biomedical Physics, Medical University of Innsbruck, Austria

**Status**

Force and torque measurements with optical tweezers [1] provide insight into mechanical effects on the micro-scale in a quantitative manner. Using an optically trapped particle as a probe, "photonic force microscopy" became a prime tool to study, e.g., forces acting on biomolecules such as DNA. The particular strength of optical trapping is its ability to measure forces in the pico-Newton range.

Originally, force measurements emerged from position measurements that used the deflection of the trapping beam by the particle, for a sensitive but robust detection of particle displacement induced by an external force. The force is then determined indirectly from the displacement [2]. Placing the detector to measure beam position in the back focal plane (BFP) of a lens collecting the transmitted trapping light makes this approach compatible with dynamic optical tweezers. In this configuration, the signal depends only on the relative position between particle and trapping spot. This method is straightforward to implement, employing a four-quadrant photo-diode (QPD) or a position sensitive detector (PSD). As it provides high sensitivity and high temporal resolution, it became the most widely used force detection method.

Inducing a rotational motion by applying optical torque [3, 4] has many applications, e.g., to measure viscosity in the constrained environment of a cell. More generally, torque measurements provide extra information that is crucial to obtain a complete notion of the mechanical effects of light, in particular for particles with asymmetric shape or non-isotropic materials. For such particles, torque measurements turn out to be more demanding due to the larger parameter space (orientation of the particles besides its position) that needs to be covered. Conventional force detection methods cannot be easily extended to torque measurements, because the sensor signals (from a QPD or PSD) are insufficient to uniquely determine all components of force and torque. For a continuously rotating particle one can instead resort to observing the periodicity of the motion as a measure of the applied optical torque.

An alternative way to force and torque measurements is provided by methods that directly measure the rate of linear or angular momentum transferred between the trapping light and the particle [5, 6, 7], which can be deduced from the change in the angular distribution, or equivalently, from the light distribution in the BFP. The principle is explained in **Figure 21**. Being based on the conservation of linear or angular momentum, respectively, this method is calibration free and suitable for particles of arbitrary shape and unknown optical properties, in complex optical trapping landscapes or in complex micro-environments.

**Current and Future Challenges**

As for any measurement method, the accuracy of the obtained results is a central issue. The indirect, position-based force detection methods require a calibration to relate the particle displacement to the force. To this end, a known force is applied. Often thermal forces or drag forces are used, as they can be calculated provided some parameters, such as the size of the particle and the viscosity of the buffer, are known. In practical applications, however, it is a non-trivial task to obtain these input parameters with good accuracy. Furthermore, statistical data analysis that takes into account all relevant effects requires careful operation [8]. As a consequence, experiments are largely restricted to simple settings, e.g., well-characterized microspheres in water, operated in the linear regime of the trapping potential. Force measurements with biological specimens of unknown or strongly varying properties are still challenging with indirect methods.

Direct force detection methods do not suffer from these limitations, but in order to bring to bear the advantage of this method being calibration-free, one has to ensure that sufficiently accurate results are obtained in a broad range of settings. One obstacle is the loss of some of the scattered light, as typically only the trapping light in the forward direction can be detected. For weakly scattering objects (such as cells) this is a minor issue, but for strong scatterers (such as polystyrene beads) one has to take the back-scattered light into account [7].

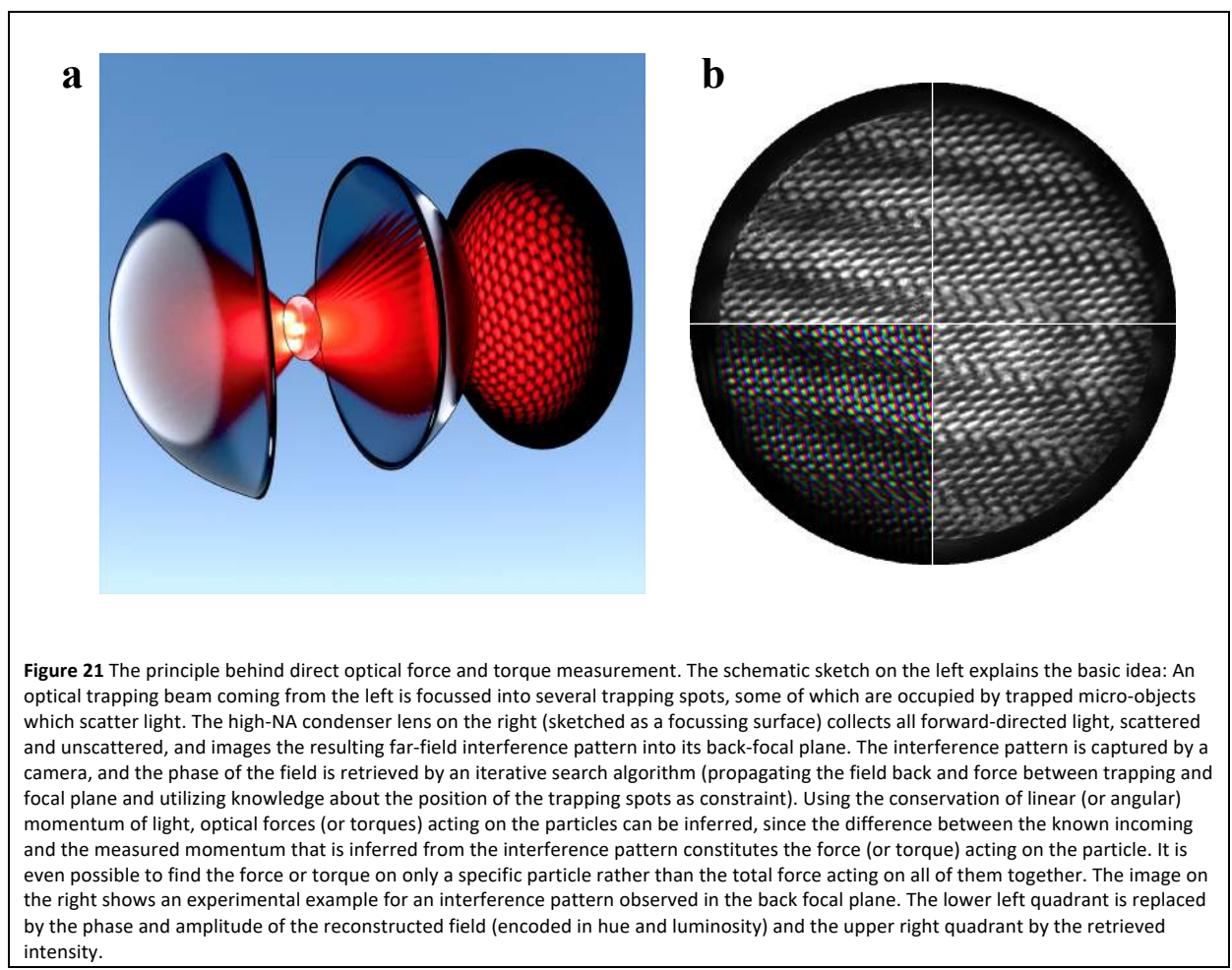

**Figure 21** The principle behind direct optical force and torque measurement. The schematic sketch on the left explains the basic idea: An optical trapping beam coming from the left is focussed into several trapping spots, some of which are occupied by trapped micro-objects which scatter light. The high-NA condenser lens on the right (sketched as a focussing surface) collects all forward-directed light, scattered and unscattered, and images the resulting far-field interference pattern into its back-focal plane. The interference pattern is captured by a camera, and the phase of the field is retrieved by an iterative search algorithm (propagating the field back and force between trapping and focal plane and utilizing knowledge about the position of the trapping spots as constraint). Using the conservation of linear (or angular) momentum of light, optical forces (or torques) acting on the particles can be inferred, since the difference between the known incoming and the measured momentum that is inferred from the interference pattern constitutes the force (or torque) acting on the particle. It is even possible to find the force or torque on only a specific particle rather than the total force acting on all of them together. The image on the right shows an experimental example for an interference pattern observed in the back focal plane. The lower left quadrant is replaced by the phase and amplitude of the reconstructed field (encoded in hue and luminosity) and the upper right quadrant by the retrieved intensity.

In many applications, in particular in the life sciences, the variability of the specimens and settings demand multiple repetitions of the measurements to obtain valid results. For this, high throughput by parallelizing data acquisition is desirable. However, all methods that analyze the light distribution in the BFP have the issue that in the far-field, contributions from simultaneously trapped particles overlap, which essentially limits force and torque measurements to the single particle case. With time-sharing techniques to create the optical traps, which require fast beam steering, this can be extended to sequential force measurements on several particles.

Holographic optical tweezers, which provide highly flexible trapping of many particles, are severely affected by this limitation of conventional force detection schemes. One possibility to overcome this is imaging-based position detection, which is, however, sensitive to drifts in the trapping position that may affect the accuracy. Alternatively, holographic force detection [9] measuring the complex field, i.e., phase and amplitude in the BFP, provides complete information about the field. From this one can calculate the field in the focal plane, in essence by a Fourier transform. There particles are separated and individual forces can be determined from the individual scattered fields. Simultaneous measurements for up to ten particles have been realized. While there is no fundamental limit to scale this up to even more traps, an experimental demonstration of highly parallel force measurements for hundreds of particles has yet to be demonstrated.

Torque measurements are still challenging. Current approaches usually rely on the idea of direct measurements: circularly polarized light, which exerts a torque on a birefringent or asymmetric particle, experiences a change of the polarization state, which reflects the torque strength. In a more general setting, also the orbital angular momentum of light needs to be taken into account. This, however, requires measurements that are sensitive to the phase of the light field, e.g., by recording interference patterns [10].

A general concern with trapping of small particles is the ubiquitous presence of noise due to thermal forces. While it masks other forces and thus limits the obtainable accuracy, thermal motion contains valuable information about, e.g., the trapping parameters.

**Advances in Science and Technology to Meet Challenges**

Force and torque measurements are essential to obtain complete information about mechanical conditions and, in consequence, to reach full control of optically trapped particles. This includes using feed-back control, for which high temporal bandwidth is sought.

High resolution detection of the scattered light with a camera preserves all information and thus delivers the highest flexibility. This, however, increases the burden for data processing. Eventually, the maximum frame rate of the sensor places restrictions to the temporal bandwidth, which ideally should reach above the cut-off frequency of the optical trap. Continuous progress in camera technology and computing power, especially of accelerators (GPU, FPGA), will push the limits further up. This needs to be matched by fast algorithms and optimized software implementation that fully employ the available computing resources.

Data analysis requires sophisticated statistical analysis to reveal the sought-after data, often hidden in noise due to thermal motion. Bayesian inference manages to do this in an efficient manner, i.e., requiring a minimal amount of data. There also appears to be a rising trend to utilize methods from machine learning [11], in particular deep neural networks, for data processing. This could also be helpful for the related task of determining the position and orientation of a particle of asymmetric shape, as often force and torque measurements are useless without corresponding position information. Unfortunately, for position there is no fundamental conservation law as for momentum.

**Concluding Remarks**

To date, powerful methods for force and torque measurements are available. Progress is expected to overcome the still existing limitations regarding accuracy and throughput and to broaden the range of applications, in particular for measurements with biological specimens of complex shape. Force and torque measurement methods should be studied together with methods for position and orientation measurement. Only together a complete picture about important physical parameters such as material stiffness or force/extension relationship is obtained. Furthermore, detailed information about forces together with knowledge of the structure of a specimen will provide insight into internal forces (stress) and induced deformations (strain). From this we expect a huge range of applications in biomedical research. To be successful in this area, retaining simplicity and robustness must be valued.

## Acknowledgements

The authors acknowledge support by the Austrian Science Fund (FWF projects P29936-N36 and F6806- N36).

## 16 — Using Bayesian inference for the calibration of optical tweezers

*Laura Pérez García*

Department of Physics, University of Gothenburg, 41296 Gothenburg, Sweden

*Alejandro V. Arzola*

Instituto de Física, Universidad Nacional Autónoma de México, C. P. 04510, Ciudad de México, México

*Isaac Pérez Castillo*

Departamento de Física, Universidad Autónoma Metropolitana-Iztapalapa, San Rafael Atlixco 186, Ciudad de México 09340, Mexico

**Status**

Optical tweezers enable sensing forces in the range from femtonewtons to piconewtons using probe beads to study, e.g., mechanical properties of macromolecules or to explore limits of small thermodynamic systems [1]. Generally, we infer the forces from the response of a trapped bead to the external stimulus following different mathematical and practical prescriptions that have been developed for more than 30 years since the invention of optical tweezers by Arthur Ashkin.

Calibration methods can be classified as either passive or active. Passive methods rely on the information provided by the Brownian trajectory of a particle trapped in static optical tweezers, while active methods use harmonic time-dependent external forces [2]. There are also direct force measurement techniques that aim to retrieve optical forces directly from the detection of the change in momentum that light suffers when it is scattered by the probe bead [3]. Here, we will focus our discussion on the passive approach, considering that the introduced concepts can be extended to other techniques. We will also consider spherical probe beads as the main subject. Extensions or generalizations can be found in the references. Also, here we primarily focus on discussing calibration in the overdamped regime; nevertheless, it is important to mention that most of the methods mentioned here can be extended to the underdamped regime (see [4,5] for instance).

At first approximation, only conservative forces are considered. Under this assumption, the Boltzmann distribution is widely used to retrieve arbitrary-shape potentials under any type of surrounding fluid, such as water, air, or viscoelastic fluids with only the need of a long-time statistically equilibrated trajectory of the trapped bead. This is a very general approach since it does not depend on the properties of the fluid. When the probe bead is tightly trapped by the optical trap, the force turns proportional to the distance from the equilibrium point of the trap, equal to the one of an ideal spring with stiffness $\kappa_x$. This regime is of broad interest in most applications where optical tweezers are used as force transducers. Several methods have been developed from this perspective, relying on analytical expressions that account for the main properties of the optical tweezers directly or through model fitting. The equipartition method is the simplest one, allowing us to estimate the stiffness directly from the variance of the trajectory of the bead at a known temperature. More sophisticated methods enable estimation of the diffusion, in addition to the stiffness. Typically, these include the autocorrelation function (ACF) [6], the mean square displacement (MSD) [6] and the power spectrum density (PSD) [7]. Based on the long-standing Fourier analysis, the latter comprises a sophisticated methodology that enables analysis in the frequency domain through fast computing routines, making it one of the most popular and reliable methods [1,7]. The accuracy, precision, and speed to estimate the force of optical tweezers depend on the selected method as well as on the intrinsic noise, frequency sampling, data length and bandpass filtering in data acquisition.

Nowadays, it is necessary to design methodologies enabling accurate estimates of the interesting parameters as well as their errors in experiments that are conducted in far from ideal conditions or in the limits of the accessible technology. For instance, there may be the interest to retrieve general force fields where the non-conservative forces are non-negligible, in systems out of equilibrium or where the potentials are far from the harmonic regime. For this purpose, the incorporation of all known information of the system, such as the nominal size of the bead and fluid viscosity at room temperature, together with their uncertainties, results relevant to improve estimations.

Recently the method FORMA (FOrce Reconstruction via MAximum-likelihood-estimator analysis) was introduced, tackling several of these drawbacks: FORMA does not require the potential to be conservative, it is less data-intensive, and its simplicity allows to perform fast and easy estimations. FORMA uses a maximum likelihood estimator or in the context of Bayesian Inference, a flat prior, i.e., we do not have a priori any knowledge about the parameters of interest (**Figure 22**). On the other hand, the method BEFORE (BayEsian FOrce REconstruction) uses a full Bayesian analysis which, in principle, can incorporate arbitrary prior information, that results into an enhancement of the inferred parameters [1].

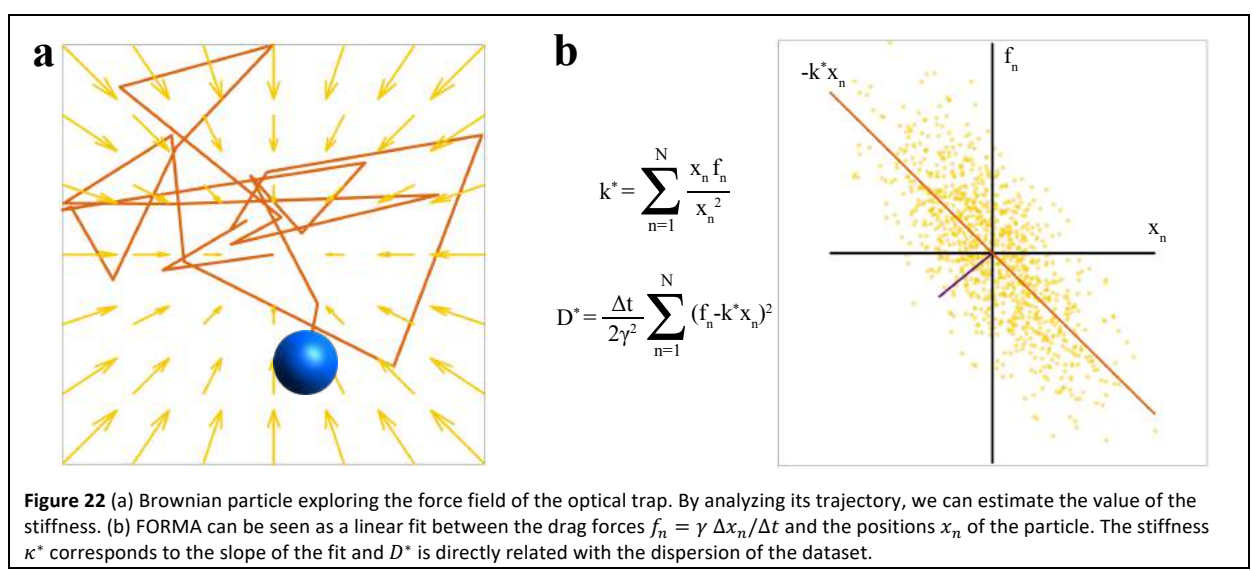

**Figure 22** (a) Brownian particle exploring the force field of the optical trap. By analyzing its trajectory, we can estimate the value of the stiffness. (b) FORMA can be seen as a linear fit between the drag forces $f_n = \gamma \Delta x_n/\Delta t$ and the positions $x_n$ of the particle. The stiffness $\kappa^*$ corresponds to the slope of the fit and $D^*$ is directly related with the dispersion of the dataset.

**Calibration by means of Bayesian inference**

We will focus on the case of a Brownian particle in a harmonic potential in the overdamped regime. The Langevin equation is then given by:

$$\gamma \frac{dx(t)}{dt} = -\kappa_x x(t) + \sqrt{2D} W_x(t),$$

where $\gamma$ is the friction coefficient, $\kappa_x$ is the trap stiffness, $D$ is the diffusion constant, and $W_x(t)$ is Dirac delta correlated white noise. Friction and diffusion are related by $D = \frac{k_B T}{\gamma}$. For this choice of model, we can derive the probability $P(\{x_n\}_{n \geq 1}|\theta)$ to observe a trajectory at a collection of measured positions $x_n \equiv x(t_n)$ given the model's parameters $\theta$.

From $P(\{x_n\}_{n \geq 1}|\theta)$, we can either obtain observables and contrast their formulas with experimental measurements or use Statistical (Bayesian) inference. Alternatively, we can think on the particle's trajectory as the signal carrying the information of the force field which is passed through a noisy channel (the thermal bath). In this new light, one can use information theory to reconstruct the optical forces [8]. Indeed, in the first case, the expectation of an observable $O(\{x_n\}_{n\geq 1})$ is

$$O(\theta) = \int \{dx_n\} \, P(\{x_n\}_{n\geq 1}|\theta) O(\{x_n\}_{n\geq 1}),$$

which can be used as a model fit to estimate the parameters of interest. This includes, for instance, the MSD and the ACF methods. On the other hand, since the statistical content of the force field is richer if we directly consider the particle's trajectory rather than a given observable, we can use Bayes' rule to write down the parameter's posterior distribution, that is:

$$P(\theta|\{x_n\}_{n\geq 1}) = \frac{P(\{x_n\}_{n\geq 1}|\theta)P(\theta)}{Z(\{x_n\}_{n\geq 1})},$$

where $P(\theta)$ contains information of the model's parameters prior to performing an experiment and $Z(\{x_n\}_{n\geq 1})$ is a normalization factor.

FORMA corresponds to looking for the optimal parameters $\theta_\star$ that maximizes the posterior distribution with a flat prior ($\theta_\star$ called maximum likelihood estimators) $\theta_\star = \mathrm{argmax}_\theta \, P(\{x_n\}_{n\geq 1}|\theta)$. BEFORE (BayEsian Force Reconstruction) considers instead a conjugate prior so that the posterior is easily computable and no maximization is needed, retaining all statistical information of the inferred parameters. One can show that using a Gauss-inverse Gamma prior distribution, the posterior distribution for the pair of parameters $(\kappa_x/\gamma, D)$ becomes

$$P_{(\kappa_x/\gamma, D)}(x, y|\{x_n\}_{n=1}^N) = \mathcal{N}(x|K_N, \gamma_N y) \, \mathrm{InvGamm}(y|\alpha_N, \beta_N),$$

where $\mathrm{InvGamm}(x)$ is the inverse Gamma distribution, $\mathcal{N}(x)$ denotes Gaussian distribution, and the parameters $\alpha_N, \beta_N, \gamma_N$, and $K_N$ depend on the particle's positions $\{x_n\}_{n=1}^N$.
In case we have no data, that is when $\{x_n\}_{n=1}^N = \emptyset$, then the posterior distribution is simply the prior that contains the information we have on the system *before* performing the experiment:

$$P_{(\kappa_x/\gamma, D)}(x, y|\emptyset) = \mathcal{N}(x|K_0, \gamma_0 y) \, \mathrm{InvGamm}(y|\alpha_0, \beta_0),$$

where $\alpha_0, \beta_0, \gamma_0$, and $K_0$ are so-called hyperparameters: they must be chosen to reflect the information a priori that we have about the physical system. The marginal posterior distributions have a series of advantages, as for instance, all their moments can be obtained explicitly, as for instance, the estimates for $\kappa_x/\gamma$ and $D$. The explicit formulas can be found in Ref. [1].

To illustrate the power of this method, in **Figure 23**a we show the results of Monte Carlo simulations. For them we have taken the values of $\kappa_x = 1 \, pN/\mu m$, $\gamma = 0.0188 \, pN \, s/\mu m$ (thus $\kappa_x/\gamma = 53.05 \, s^{-1}$), and $D = 0.215 \, \mu m^2/s$, and discretise the Langevin equation. The sampled position was taken every $\Delta t = 10^{-4} \, s$. We have then generated a single trajectory and see how the posterior marginals and estimates evolve in parameter space as a function of the number of samples.

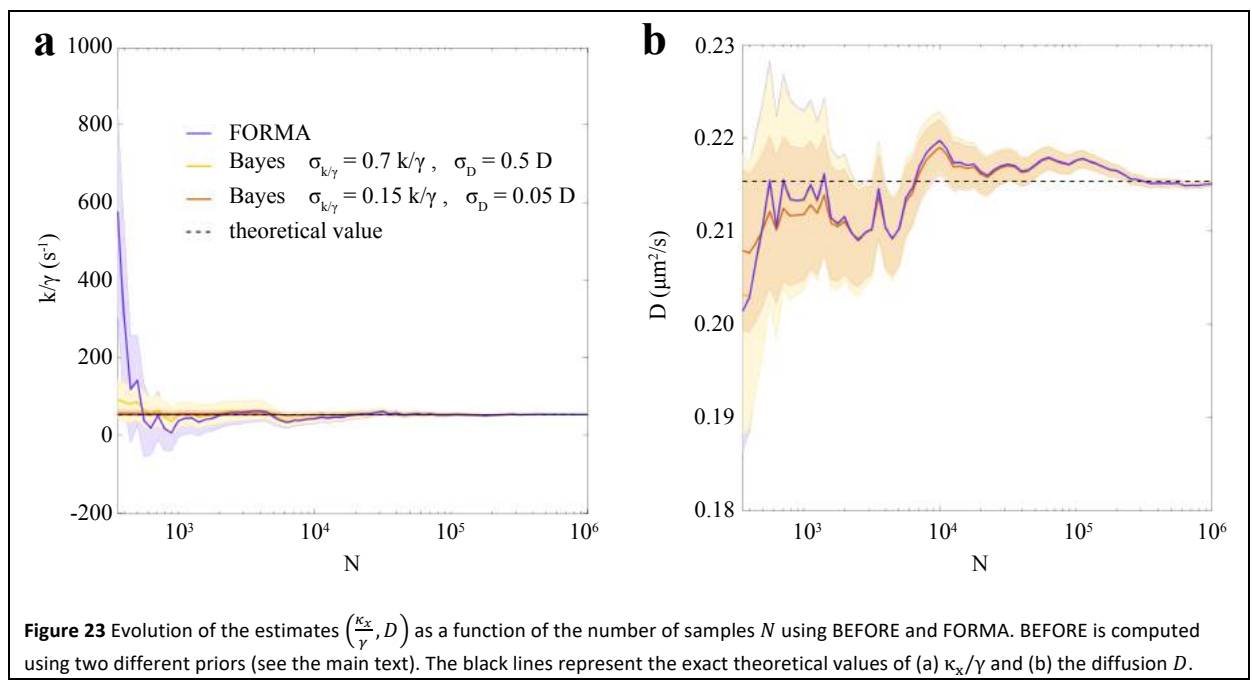

**Figure 23** Evolution of the estimates $\left(\frac{\kappa_x}{\gamma}, D\right)$ as a function of the number of samples $N$ using BEFORE and FORMA. BEFORE is computed using two different priors (see the main text). The black lines represent the exact theoretical values of (a) $\kappa_x/\gamma$ and (b) the diffusion $D$.

**Figure 23**b shows the evolution of $\kappa_x/\gamma$ and D, using FORMA and BEFORE as a function of the number of sampled points $N$. To understand the advantages of incorporating a priori information of the parameters, we performed two different estimations of BEFORE using two different priors with different levels of accuracy. Hypothetically, we assume a priori information to be taken by normal distributed random variables $\mathcal{N}(53.05\ s^{-1}, \sigma_{\kappa_x/\gamma})$ and $\mathcal{N}(0.215\ \mu m^2\ s^{-1}, \sigma_D)$. The error bars in FORMA correspond to the asymptotic value of the error when $N$ is large [1]. As the number of points in the trajectory increases, providing the method with more data, the inferred parameters improve accordingly. From this figure, as better information is provided, faster and more accurate estimations are retrieved from BEFORE. This contrasts sharply with FORMA and any other existing method, for which neither a priori information can be incorporated nor the whole posterior distribution of the parameters is obtained.

**Current and Future Challenges**

As the uses of optical tweezers evolve and the variety of phenomena explored increases in complexity, the accuracy of experiments become more challenging. Moreover, some attention has been brought to time-dependent processes, where the stiffness of the trap might change over time and the trajectory of the particles is shortened. In some experiments, optical potentials might not be harmonic, resulting in the system not reaching its equilibrium state. The study of living matter with optical tweezers has also brought additional challenges as the conditions of the surrounding fluid might not be ideal and the imaging of beads may be affected by the inhomogeneity of the living tissue, compromising, in turn, the quality of the data set.

**Advances in Science and Technology to Meet Challenges**

As we mentioned, the implementation of better statistical algorithms will enable the extraction of more information from a given data set. One important aspect here will be to assess the uncertainty coupled to a measurement given certain conditions and availability of information. This will not only give us a more reliable way to quantify the precision and accuracy of an estimate but also could bring light to core aspects of the behaviour of the particles under the influence of optical tweezers, such as the production of entropy and the irreversibility of stochastic processes.

**Concluding Remarks**

More general methodology that account for subtle forces in far-from-ideal conditions using all the information at hand is provided by the Bayes' rule (BEFORE). Here, we gave a brief introduction of BEFORE that illustrates the applicability and its conceptual advantages in the calibration of optical tweezers. We showed how the whole information of the parameter's distribution is retrieved, and how valuable a priori information can be incorporated to have faster and more precise estimations.

**Acknowledgements**

A.V.A. acknowledges financial support from UNAM-DGAPA-PAPIIT-IN111919.

## 17 — Optical tweezers calibration with deep learning

*Aykut Argun*

Department of Physics, University of Gothenburg, 41296 Gothenburg, Sweden

**Status**

Analysing microscopic force fields is crucial for a wide range of experiments including optical tweezers, DNA stretching and non-equilibrium physics [1]. This analysis is usually made by modelling the force field by some parameters that are estimated using particle trajectories. This information then can be used to understand the underlying dynamics, to predict a future state of the system, or to calibrate the experimental setup. The calibration of such systems sometimes needs to be done in real-time [2]. The most well-studied example is the force field of an optical tweezers, where the force has the form $F = -kx$ (where $k$ is the stiffness of the harmonic trap and $x$ is the distance from equilibrium). For this case, there are many standard calibration methods that already exist, such as the variance method, autocorrelation method and power spectrum methods [1]. The variance method directly determines the stiffness from the variance of the particle position in the trap ($k = k_\mathrm{B}T/\langle x^2 \rangle$ where $k_\mathrm{B}T$ represents the thermal energy). The autocorrelation method determines $k$ by fitting the decorrelation curve of the particle position in the trap. The power spectrum method is also a powerful method especially at high frequencies, which fits the power spectrum of the Brownian particle to a Lorentzian and infers the stiffness. All these standard algorithms work well when the measurements contain sufficient data points and are error free [1]. More efficient standard algorithms include linear maximum likelihood methods [3] that applies a linear regression, which makes a first order approximation to the force around the equilibrium point. This allows more accurate and faster calibrations of force fields compared to standard algorithms [3] as well as the ability to calibrate non-conservative force fields, such as a rotational force field.

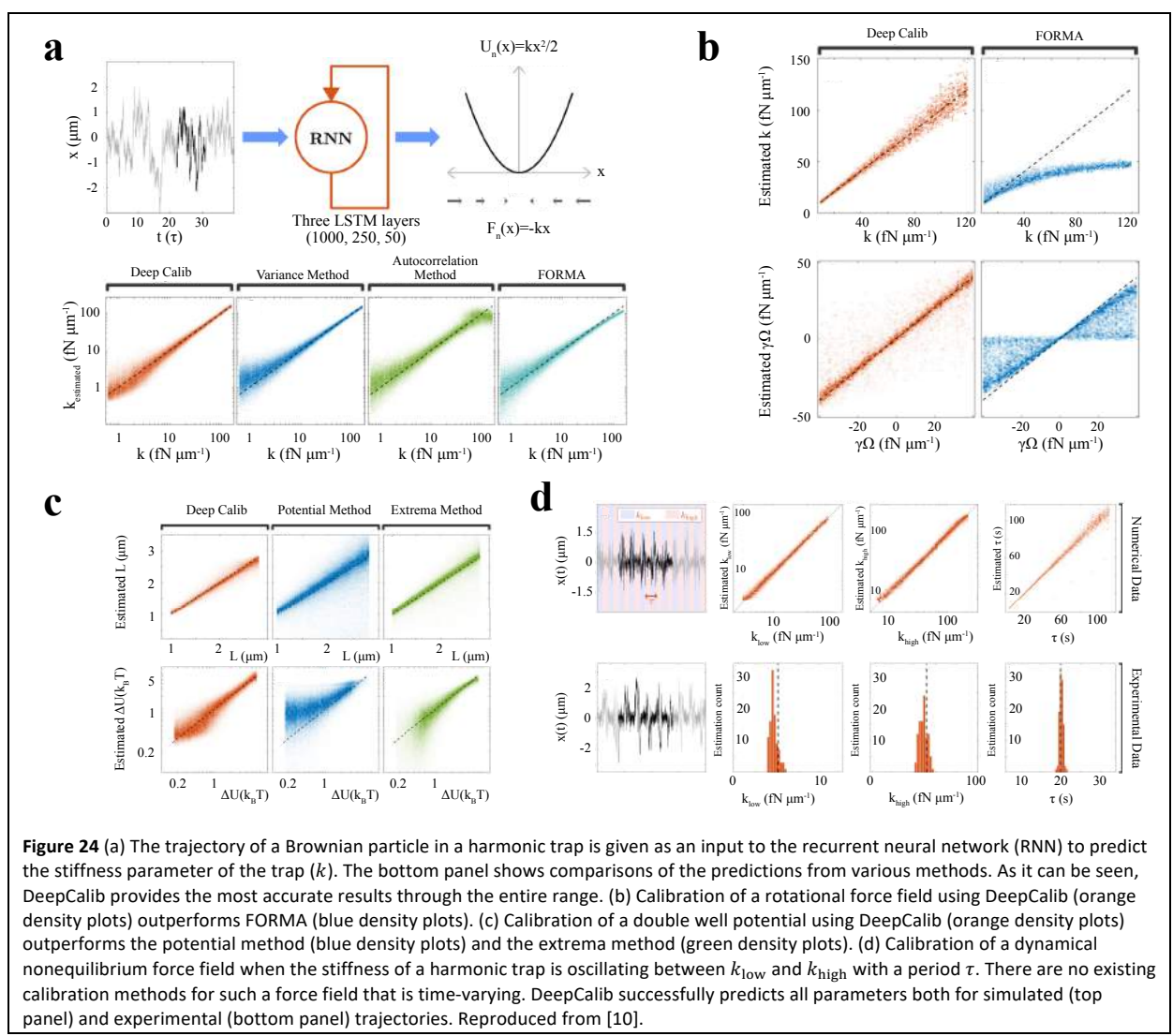

**Figure 24** (a) The trajectory of a Brownian particle in a harmonic trap is given as an input to the recurrent neural network (RNN) to predict the stiffness parameter of the trap ($k$). The bottom panel shows comparisons of the predictions from various methods. As it can be seen, DeepCalib provides the most accurate results through the entire range. (b) Calibration of a rotational force field using DeepCalib (orange density plots) outperforms FORMA (blue density plots). (c) Calibration of a double well potential using DeepCalib (orange density plots) outperforms the potential method (blue density plots) and the extrema method (green density plots). (d) Calibration of a dynamical nonequilibrium force field when the stiffness of a harmonic trap is oscillating between $k_{low}$ and $k_{high}$ with a period $\tau$. There are no existing calibration methods for such a force field that is time-varying. DeepCalib successfully predicts all parameters both for simulated (top panel) and experimental (bottom panel) trajectories. Reproduced from [10].

**Current and Future Challenges**

As microscopic environments are stochastic, the calibration methods require more data for higher accuracy, due to the need for averaging. If there is enough statistical information in the data, averaging of displacements provides accurate force measurements [4]. However, this is often not possible to have in biophysics. Moreover, the available calibration techniques for non-standard force fields are very limited. Examples include non-harmonic potentials, rotational force fields and dynamic non-equilibrium force fields. As the number of parameters that is to be estimated from a trajectory increases, the amount of data required for an accurate measurement grows exponentially for complex potential landscapes. Although the maximum likelihood algorithms such as FORMA can do better in these potential landscapes, they only work if the data acquisition has high frequency and the force field is not time-varying.

There are also experimental setups which need real time calibration if the system is subject to change or degradation over time. Examples to such systems are forces generated by electrodes inside liquid environments, chemical forces and bacterial forces.

There are challenges in the field also in data acquisition. The more advanced techniques allow faster and more accurate particle localization. Specifically, the advancements in the photodetectors have recently improved single particle tracking and enabled trajectory measurements at very high frequencies [5]. In addition, the advancements in data-driven image analysis methods facilitated the possibility of enhanced image analysis [6,7].

**Advances in Science and Technology to Meet Challenges**

While the need for more advanced calibration methods for experimental setups increase, data-driven analysis methods, such as recurrent neural networks, have become known as a very powerful tool to extract information from time series data. This has led to the development of DeepCalib [8], a free software package that uses deep neural networks to calibrate force fields from trajectories. Specifically, DeepCalib uses LSTM layers to extract information from trajectories and uses simulated data to train the neural network. Particularly for shorter trajectories or smaller force values, when the calibration is challenging, DeepCalib is proven to work better than other algorithms for harmonic potential, as shown in **Figure 24**a. In addition, DeepCalib can accurately calibrate non-conservative rotational force fields better than the existing methods. In this case, the force field has two parameters, the central force stiffness $k$ and the rotational parameter $\Omega$. Although being accurate for lower values of the $k$, FORMA struggles at higher force values because of the data points becoming uncorrelated. DeepCalib is able to estimate the parameters of the force fields better than FORMA, as shown in **Figure 24**b. The difference of DeepCalib becomes more easily visible as we look at more non-standard cases, such as a double-well potential. Here, the force field is parametrised by the equilibrium distance $L$ and the energy barrier height $E_\mathrm{B}$. As shown in **Figure 24**c, DeepCalib provides more accurate results when predicting the parameters of a double well potential. Finally, and most importantly, DeepCalib can calibrate any force fields, such as time varying force fields that yield a dynamical non-equilibrium system. In this case, a harmonic potential with switching stiffness from $k_\mathrm{low}$ to $k_\mathrm{high}$ with period $\tau$ is considered. It is shown in **Figure 24**d that DeepCalib can very accurately calibrate all the parameters of a short trajectory for such a challenging case, both for numerical data and for experimental data. It is also shown that the DeepCalib is more robust than the standard algorithms against measurement noise, diffusion gradients and reality gap [8].

**Concluding Remarks**

With the introduction of DeepCalib, it becomes evident that data-driven, neural network approach for the calibration of microscopic force fields outperforms the standard methods in challenging conditions and limited available data. More importantly, DeepCalib can be applied to non-conservative or time varying force fields that no standard calibration methods exist. With these advantages and the great robustness to measurement noise, inhomogeneous environments and reality gap between experiments and theory, DeepCalib is a very powerful and flexible tool for analysing trajectories to extract the force fields. It allows for just a minor change in the code to immediately adapt for a new force field, while standard techniques totally change if a different force field is considered. Therefore, DeepCalib is ideal to calibrate complex and non-standard force fields from short trajectories and it is readily available as a free Python software package [9]. It also proves that the recurrent neural networks are a very powerful tool to analyse Brownian trajectories. Similar techniques have also been applied to, for example, to characterize anomalous diffusion trajectories and have also been shown to be very powerful [10,11].

**Acknowledgements**

I acknowledge support from H2020 European Research Council (ERC) starting grant complex swimmers (Grant no. 677511).

## MICRORHEOLOGY

## 18 — Microrheology for quantitative non-invasive mechanical measurements

*Till M. Muenker, Bart E. Vos, Timo Betz*

Third Institute of Physics - Biophysics, Georg August University Göttingen, Germany

**Status**

Optical-tweezers-based microrheology has emerged as a powerful tool to measure the local mechanical properties of soft viscoelastic materials in a minimal-invasive way [1,2]. Unlike other rheological methods such as atomic force microscopy (AFM) or bulk rheometry, measurements are not just limited to the surface of a material and do not require large volumes. Therefore, applications range from the measurement of bio-polymers [1] to the determination of local mechanical properties inside living systems like cells [2] and even whole organisms.

In general, there are two distinct forms of microrheology as illustrated in **Figure 25**. Passive microrheology is most common due to its experimental simplicity [3]. The position of a probe particle, embedded inside the material, is monitored by means of video microscopy or laser interferometry. In thermodynamic equilibrium, particle fluctuation are purely due to thermal motion and can be linked to the complex shear modulus G* using the generalized Stokes-Einstein relation [3]. Due to relatively small displacements, this method is prone to experimental drift and general noise-related errors. Additionally, it is fundamentally limited to passive systems that remain in thermodynamic equilibrium during the full measurement period. Recently, great scientific efforts have been devoted to active systems such as living cells, which introduce besides the thermal fluctuations an additional, active motion on the probe particles [2,4]. This active component is interesting as it allows to understand general active systems, yet it also renders passive microrheology as performed today unusable because the activity breaks the thermodynamic equilibrium. Active microrheology, in contrast, circumvents these limitations [1,2]. Here, optical tweezers are used to apply well-defined, typically oscillating, forces to the particle. At the same time, the particle position is observed. This can be done using the same laser, an additional detection laser or video microscopy. The ratio of particle displacement and force yields access to the response function of the material from which in turn the complex modulus G* can be deduced. Being less susceptible to experimental noise and able to quantify even active and soft non-linear systems, this method has become the standard way to investigate the local mechanics of living systems [2]. Furthermore, comparing the resulting mechanical properties to the spontaneous fluctuations of the free system allows a direct quantitative access to the active forces acting on the probe particle [5].

Understanding active, but also non-linear microscopic systems is among the most pressing problems hindering our advancement in the generation of complex smart materials, but also in the quantitative description of complex and fast changing systems like biological and artificial cells.

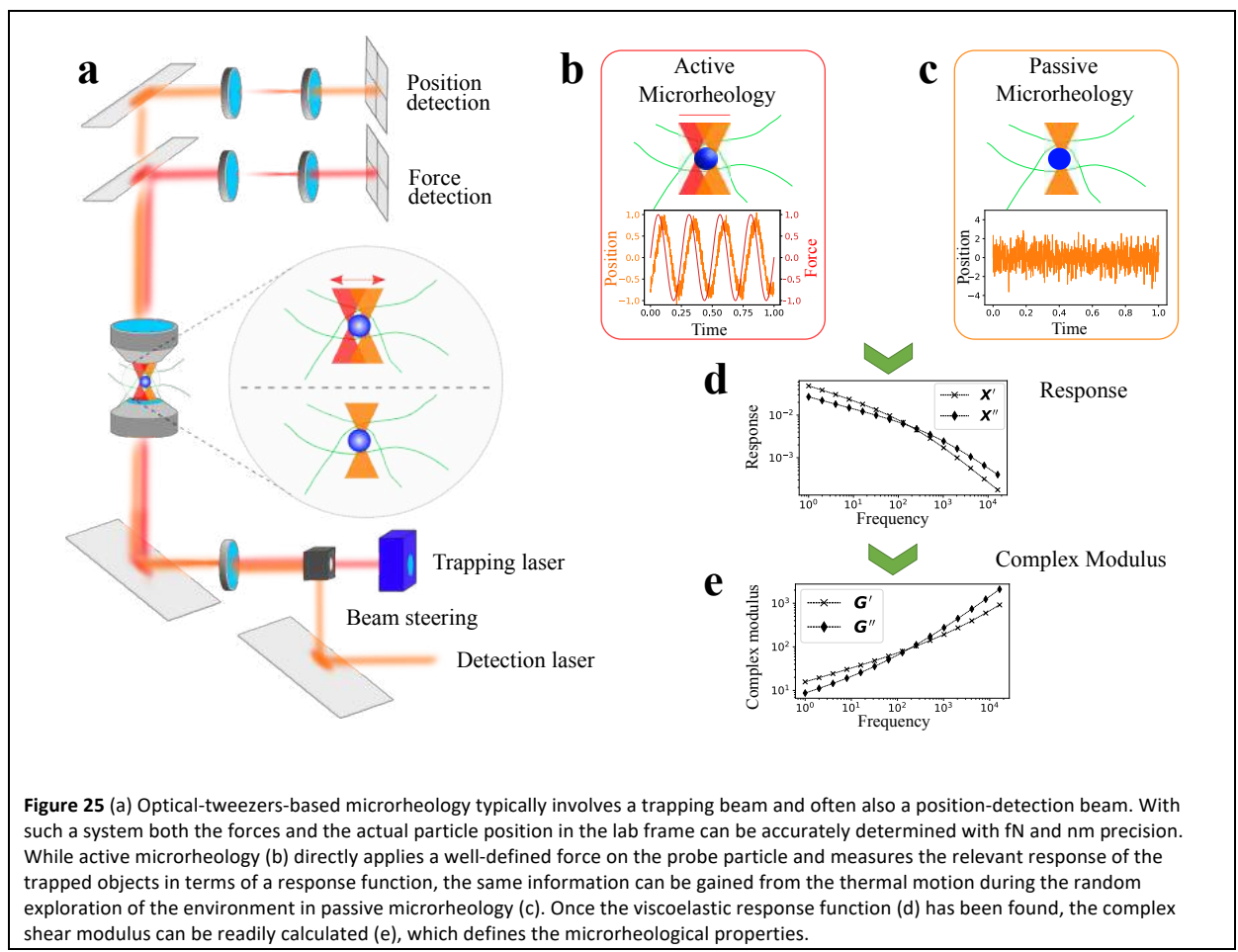

**Figure 25** (a) Optical-tweezers-based microrheology typically involves a trapping beam and often also a position-detection beam. With such a system both the forces and the actual particle position in the lab frame can be accurately determined with fN and nm precision. While active microrheology (b) directly applies a well-defined force on the probe particle and measures the relevant response of the trapped objects in terms of a response function, the same information can be gained from the thermal motion during the random exploration of the environment in passive microrheology (c). Once the viscoelastic response function (d) has been found, the complex shear modulus can be readily calculated (e), which defines the microrheological properties.

**Current and Future Challenges**

One major challenge in microrheology measurements is the instrument's complexity. The limited availability of commercial systems and the specialist skills and knowledge required to run optical tweezers based microrheology measurements puts it in a scientific niche. Additionally, since particle deflections in the nanometer regime are measured, these setups are extremely sensitive to noise. Mechanical vibrations (e.g., from camera cooling fans), stage drift, electronic noise and temperature fluctuations all disturb these highly sensitive measurements. As the overall optical forces are limited, only soft materials of up to approximately 1 kPa can be measured. Furthermore, microrheology experiments have a limited throughput as only single particles are trapped at a time. Also, since the optical pathway of each measurement position is unique, the optical trap needs to be calibrated at each measurement position.

A second challenge lies in the measurement of materials with a non-linear stress-strain relationship. In the past decades, there is an increasing appreciation of the importance of non-linear mechanical responses, which are ubiquitous in nature. However, with few exceptions [6], all reported optical tweezers microrheology experiments are performed in the linear regime, i.e., a mechanical regime where the measured stiffness is independent from the applied strain. The reason for measuring the linear regime are two-fold. Firstly, the trap stiffness, limited by laser power and sample heating [7], does not allow large deformations, and secondly, the theory for analysing data is only valid for small deflections.

A third challenge is a fundamental limitation of measuring living matter, namely its ability to adapt to external stimuli at sufficiently long timescales. Microrheology measurements that induce slow deformations thus do not measure a mechanical response function, but rather cellular adaptation. Infinitesimal small deformations would not stimulate the cell, but would also not provide a signal in inherently noisy systems. In addition to deformations, local heating induced by the optical trap also affects cellular behaviour.

Hence, the main challenges are related to experimental complexity, measurement precision and force generation.

**Advances in Science and Technology to Meet Challenges**
The roadmap to improve or even overcome these current limitations can be drawn by a mixture of technical improvements, which are mostly already within reach. The force limitation can be improved by two approaches. Firstly, using 800 nm lasers reduces the heating in an aqueous environment by almost one order of magnitude if compared to the typically used 1064 nm lasers [7]. Unfortunately, in the 800nm regime, only expensive Titan Sapphire lasers are available at the required power of more than 1 Watt. The introduction of more powerful near-infrared laser at a wavelength of about 800 nm with high quality beam profiles would immediately increase the reachable optical forces in the non-heating regime by an order of magnitude. Secondly, this can be further enhanced by protocols that allow generation of probe particles with higher refractive indices. Unfortunately, such particles would inherently lead to an increase of reflection, which can only be overcome by anti-reflection coatings that are achieved by refractive index gradients [8].

To gain access towards the non-linear regime, relevant for many contemporary questions, again, the force application capacities of tweezers are the main hurdle. Firstly, the nonlinear stiffening often leads to mechanical properties that are not testable anymore with the common 100 – 200 pN forces, and require several nN. Hence, so far active optical tweezers based nonlinear microrheology has only been done in soft DNA gels [6]. Besides deforming the material to the nonlinear regime using the tweezers itself, the systems can also be prestressed.

A further improvement will be a systematic multiplexing of the lasers, thus allowing multiple measurements in parallel to overcome the drawback in throughput. Using high speed beam steering, this is already possible [9]. Multiplexing enables also two-particle microrheology [10] which allows to determine force propagation in complex and structured materials.

Another approach to overcome the limited measurement efficiency would be fully automatized approaches, where the system will rapidly and automatically switch from one probe bead to the next one. This would not only allow to increase the measurement statistics, but also to observe mechanical changes over a long time, for example in cells during migration or differentiation. Such automation approaches are currently under development, but might require the usage of deep learning approaches to correctly identify suitable probe particles.

A final way to boost microrheology would be to extend current theoretical analysis methods of passive fluctuation observation, which mostly rely on equilibrium thermodynamics. Recent approaches like the observation of broken detailed balance have opened new directions to extract the active forces from simple passive measurements. More refined and general approaches here would allow to perform high speed video tracking to directly access the viscoelastic material properties and active forces in driven non-equilibrium systems.

**Concluding Remarks**
The potential of optical-tweezers-based microrheology has not yet been unleashed, but to gain a substantial advancement of the method, it needs to become more accessible to a wide number of potential users. An outstanding potential for future experimental breakthroughs lies in an automated user-friendly system, which could ideally be even commercially available and exploits 800 nm, high mode-quality lasers. Such systems would pave the way for a high throughput, local characterization not only of simple passive materials, but also of currently engineered smart materials, as well as living biological and active biomimetic systems. Especially, characterizing the viscoelastic transitions within such materials with high temporal and spatial resolution is only possible using optical tweezers based microrheology, making it the relevant tool for local characterization. Hence, if the field manages to follow the here-suggested roadmap, optical tweezers based microrheology has the potential to become a main method for quantitative cell biology as well as the key characterization tool for the currently emerging fields of active and smart materials.

**Acknowledgements**

This work was supported by the European Research Council ERC-Consolidator grant PolarizeMe (771201) and by the DFG under Germany's Excellence Strategy (EXC 2067/1- 390729940).

## 19 — Active microrheology

*Ilaria Cristiani, Paolo Minzioni*

Department of Electrical, Computer and Biomedical Engineering, University of Pavia, Italy

**Status**

The evaluation of the response of a material to an external mechanical stress is of utmost importance, both in fundamental research and in industrial product development [1]. Indeed, the study of rheological properties, namely viscosity and elasticity, allows getting fundamental information about both the material structure at the microscopic level and the characteristics of the involved molecular bonds. In recent years, a large attention has been devoted to the development of micro-rheometers, that means devices able to evaluate the response of small volumes of materials (in the range of nano- or pico-liters) This fact is particularly relevant for biomaterials, as an example, or when dealing with samples available in small volumes because of their cost or their low reproducibility. In addition, micro-rheometers allow investigating the material response with high spatial resolution, helping to understand the supramolecular organization and mesh-size of network-structured materials.

Active micro-rheology techniques are based on the insertion of micro-probes in the material under test. Such probes are moved under the effect of an external force that induces a local stress σ, i.e., a force per unit area. Rheological properties can be inferred by recording the relative material response in term of strain, that is the amount of deformation, and strain rate.

Magnetic or optical forces [2] are ideally suited for this application because they can be applied on the probe without physical contact (thus not modifying the medium under test) and their intensity and spatial distribution can be controlled with high precision. In particular, the exploitation of optical forces was pushed by the development of micro-opto-fluidic systems which have been largely investigated and developed in the last decade [3]. Opto-fluidic chips, based on the integration of microfluidic channels and optical waveguides on the same substrate, offer the chance to develop miniaturized micro-rheometers in which the optical force is applied by counter-propagating (and diverging) beams. We note that optical systems for active micro-rheology can rely either on focused-single-beam optical tweezers or on counter-propagating (and freely diverging) beams [4]. In the first case the beads are usually moved by exploiting the gradient component of the force, whereas in dual-beam tweezers the motion is determined by the scattering component of the optical force. The latter configuration, as will be shown in the following, can provide several benefits for rheology applications, making this method highly versatile and well suited for a wide range of materials [5].

In **Figure 26**, we show the basic scheme of a micro-rheometer, fabricated in fused silica by femtosecond laser writing [6]. The laser beams emitted by the two opposite waveguides are used to trap one micro-probe immersed in the fluid under test (beam power ≈ 1 mW). By unbalancing the power levels of the two beams a net force is exerted on the bead: its motion along the channel is dependent on the optical force spatial profile and on the fluid viscoelastic properties. Two different kinds of measurements can be performed:

(i) Using an external modulator it is possible to periodically unbalance the optical power coupled to the waveguide. With this approach a time-varying optical force and the consequent microprobe oscillation is induced thus mimicking the behaviour of oscillatory rheometers [7]. By recording and analysing the bead motion both the elastic and viscous response of the fluid can be recovered.

(ii) Trapping the microprobe close to one side of the microchannel, and then abruptly decreasing the power emitted by the waveguide on the opposite side, it is possible to "shoot" the microprobe across the microchannel [5]. The measurement of the probe motion provides information on the viscous response of the fluid.

The integrated micro-rheometer has been initially validated by characterising well-known fluids like water-glycerol mixtures providing very reliable results. The potential of the system has then been confirmed by accurately measuring the regimes of the mechanical response of Non-Newtonian fluids like DNA hydrogels and yield stress materials [8,9].

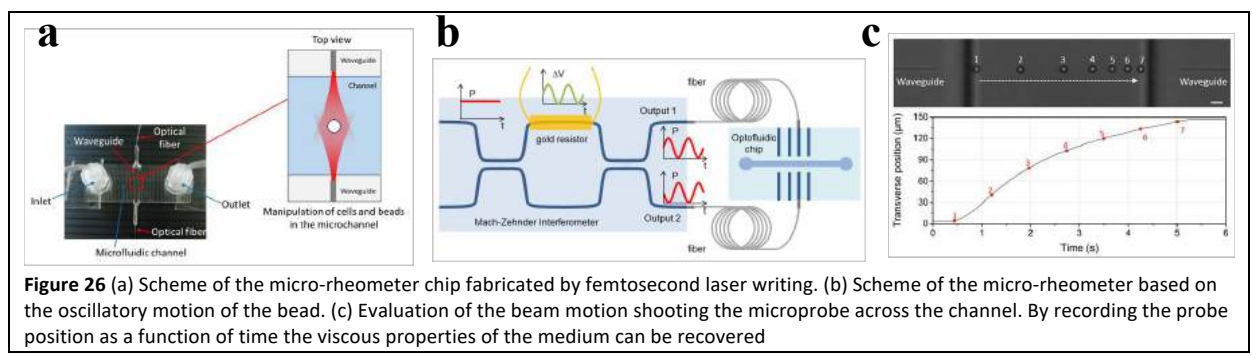

**Figure 26** (a) Scheme of the micro-rheometer chip fabricated by femtosecond laser writing. (b) Scheme of the micro-rheometer based on the oscillatory motion of the bead. (c) Evaluation of the beam motion shooting the microprobe across the channel. By recording the probe position as a function of time the viscous properties of the medium can be recovered

**Current and Future Challenges**

The exploitation of the scattering component of the force ($F_S$), instead of the gradient component ($F_G$), offers several advantages for rheological measurements. Indeed, in the large particle regime optical force can be obtained by decomposing the optical beam into rays, and then by calculating the scattering and gradient force components ($F_S$ and $F_G$ respectively) for each ray [10-11]. The forces exerted on a spherical particle of index $n_P$, immersed in a medium of refractive index $n_M$, by each optical ray are given by the following equations:

$$F_S = \frac{n_M P}{c} \left\{ [1 + R\cos(2\theta)] - T^2 \frac{[\cos(2\theta - 2\gamma) + R\cos(2\theta)]}{[1 + R^2 + 2R\cos(2\gamma)]} \right\}$$

$$F_G = \frac{n_M P}{c} \left\{ [R\sin(2\theta)] - T^2 \frac{[\sin(2\theta - 2\gamma) + R\sin(2\theta)]}{[1 + R^2 + 2R\cos(2\gamma)]} \right\}$$

where $P$ is the power carried by each ray, $c$ is the light velocity in vacuum, $R$ and $T$ are, respectively, the reflection and transmission Fresnel coefficients considering an optical ray forming an angle of incidence $\theta$, and having a refraction angle $\gamma$ inside the particle Summing up the forces due to each ray the overall $F_S$ and $F_G$ for the whole beam is calculated. It is important to point out that by increasing $\Delta n = n_P - n_M$ the scattering force grows monotonically, while the gradient force displays a more complex behaviour and can even be reversed in the case of high refractive index particles.

Moreover, the graph reported in the top panel of **Figure 27** shows the dependence of the optical scattering force exerted by a freely propagating gaussian beam (power 10mW, waist 3.5μm/5μm) on particles with a radius varying from 1 to 5 μm and positioned immediately at the waveguide output. It can be easily noticed that choosing a radius comparable with the bead size provides the condition of maximum applied force, maintaining at the same time a good trap stability in the transversal direction. On the other hand the bottom panel of **Figure 27**, reporting the pressure (stress) applied by the particle on the external medium, shows that reducing the probe size allows a significant increase of the applied stress, although this is paid by a reduction of trap stability.

In a fully integrated device, thanks to the excellent spatial and temporal control over force application an optimal trade-off can be obtained between these constraints. Force intensity up to 1 nN, with good trapping stability, have been demonstrated, making possible the reliable investigation of viscoelastic properties of a vast class of materials also in a nonlinear viscosity regime. As an example, [9] reports the results obtained on an aqueous suspension of packed, swollen microgels, a well-studied example of simple yield stress fluids.

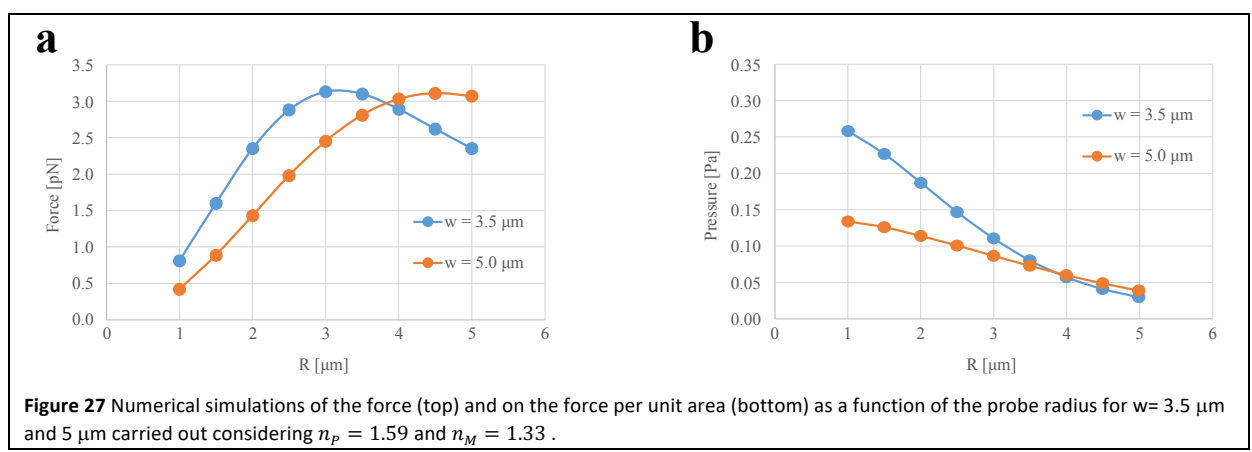

**Figure 27** Numerical simulations of the force (top) and on the force per unit area (bottom) as a function of the probe radius for w= 3.5 μm and 5 μm carried out considering $n_P = 1.59$ and $n_M = 1.33$.

**Advances in Science and Technology to Meet Challenges**

Different scientific and technological advancements could largely contribute to opening new scenarios in optical microrheology.

1. The versatility of the micro-rheometer is strongly dependent on the range of stress that can be applied to the fluid. Higher stress calls for high optical power coupled to the microfluidic channel (up to 2 W were used for yield stress materials, producing a stress of ≈ 1 Pa). Therefore, a great limitation is now given by propagation losses of the on-chip waveguides that are reported in dB/cm range. Moreover, a major source of power loss in the currently available systems is due to the coupling losses between optical fibers, used to carry the laser radiation from the source to the chip, and the integrated waveguides. This can be ascribed to optical mode mismatch between the two guiding structures, other than coupling misalignments

2. The integration of tunable lenses on-chip could enable to focus the optical beams down to the desired size and position, enabling better spatial control in the measurement and optimizing trap's strength and stability. Although the integration of tunable lenses is quite challenging, it must be highlighted that different approaches for realizing microfluidic lenses are currently being tested and considered in the scientific literature, even if for different applications.

3. Exploitation of high-index microbeads could sensibly increase the applied stress for a given optical power, thus also decreasing the limitations associated to thermal effects that can in principle hinder the measurement. At the state of the art, the availability of high-refractive index (n>2) microbeads transparent in the 980-1070 nm range is quite limited, but it must be noted that the described configuration theoretically allows using probes with $n_P$ up to 5 (a value much higher than that usable with focused-beams optical tweezers) without losing the trapping effect, while largely increasing the applied scattering force.

**Concluding Remarks**

Integrated micro-rheometers based on a dual beam laser trap constitute a new class of devices able to provide low sample consumption while ensuring an excellent calibration and measurement repeatability, as no alignment of optical components is required, differently from microscope-based optical tweezers. Additionally, thanks to the possibility of using high-index beads, and to the lack of beam focalization, the constraints imposed by localized sample heating are extremely relaxed with respect to standard configurations.

By combining these aspects, the reported micro-rheometers ensure an unprecedented versatility, making it possible to carry out stress-controlled active measurements, both in linear and nonlinear regime, on a wide range of materials, from aqueous polymer solutions and soft gels to stiffer materials, thus opening the way to the development of low-cost, miniaturized and largely automated micro-rheometers.

**Acknowledgements**

Tha authors would like to acknowledge G. Zanchetta, V. Vitali, G. Nava, T. Bellini, F. Bragheri, R. Osellame, P.Paié and A. Crespi who collaborated with the authors for the achievement of the results described in this section.

**TRAPPING NANOPARTICLES**

## 20 — Optical trapping of nanostructures

*Peter J. Reece*
School of Physics, The University of New South Wales, Australia

*Fan Wang*
School of Physics, Beihang University, China

*David McGloin*
School of Electrical and Data Engineering, Faculty of Engineering and IT, University of Technology Sydney, Australia

**Status**

The trapping of nanostructures within the field of optical trapping is long established. The original 1986 optical tweezers paper devotes much of its attention to the limitations of the new 3D trapping technique, and demonstrates trapping of a 26 nm diameter silica sphere, with clear predictions that smaller particles should be feasibly trapped as well. These ideas were quickly taken forward by Ashkin and Dziedzic to explore biological material in the form of tobacco mosaic viruses, with a diameter of around 20 nm, but a length of about 3 μm, hence trapping nanorods for the first time and setting up a discussion around the role of shape and size of trapped particles, as well as initiating the use of optical tweezers in the biological sciences.

This early work quickly progressed from trapping of dielectric particles, which had been considered optimal due to their low scattering cross sections. Svoboda and Block [1] demonstrated that in the nano-regime (here around 35 nm) metallic and dielectric particles had comparable scattering properties and, as the metallic particles had much higher polarizabilities, trapping was stronger.

Work building on these early ideas has mushroomed in the subsequent decades, and there is a plethora of target particles in a range of fields, spanning nanotubes and nanowires, made from polymers, glass, semiconductors and graphene among others, through to more conventional particles made from more exotic materials, such as diamond, quantum dots and up-converting material [2]. Trapping has been demonstrated in liquid media, in air and in vacuum, and has been shown to enable the creation of nanoparticle clusters. Near-field approaches, such as plasmonic nanotweezers, moving away from the traditional free space optical tweezers, have also made significant demonstrations in this area.

Applications of nanoscale trapping have become increasingly viable with the development of these approaches, making use, for example of naturally occurring nanoparticles. These have included exploring sub-diffusion dynamics inside cells, trapping inside living organisms, assembly of nanowire devices using holographic tweezers and probing quantum limits using vacuum trapped nanoparticles.

The push in recent years has been in research aiming at functional nanoscale optical tweezers, aiming to probe truly nanoscale phenomena, within cells, for example, and to make use of the properties offered by the nanoparticles themselves to seek new opportunities in imaging, rheology, magnetic field sensing, temperature sensing and as a means to explore quantum effects in bulk materials, such as optical cooling. Hybrid approaches, utilising optical field to induce other effects, such as thermal forces have also received attention.

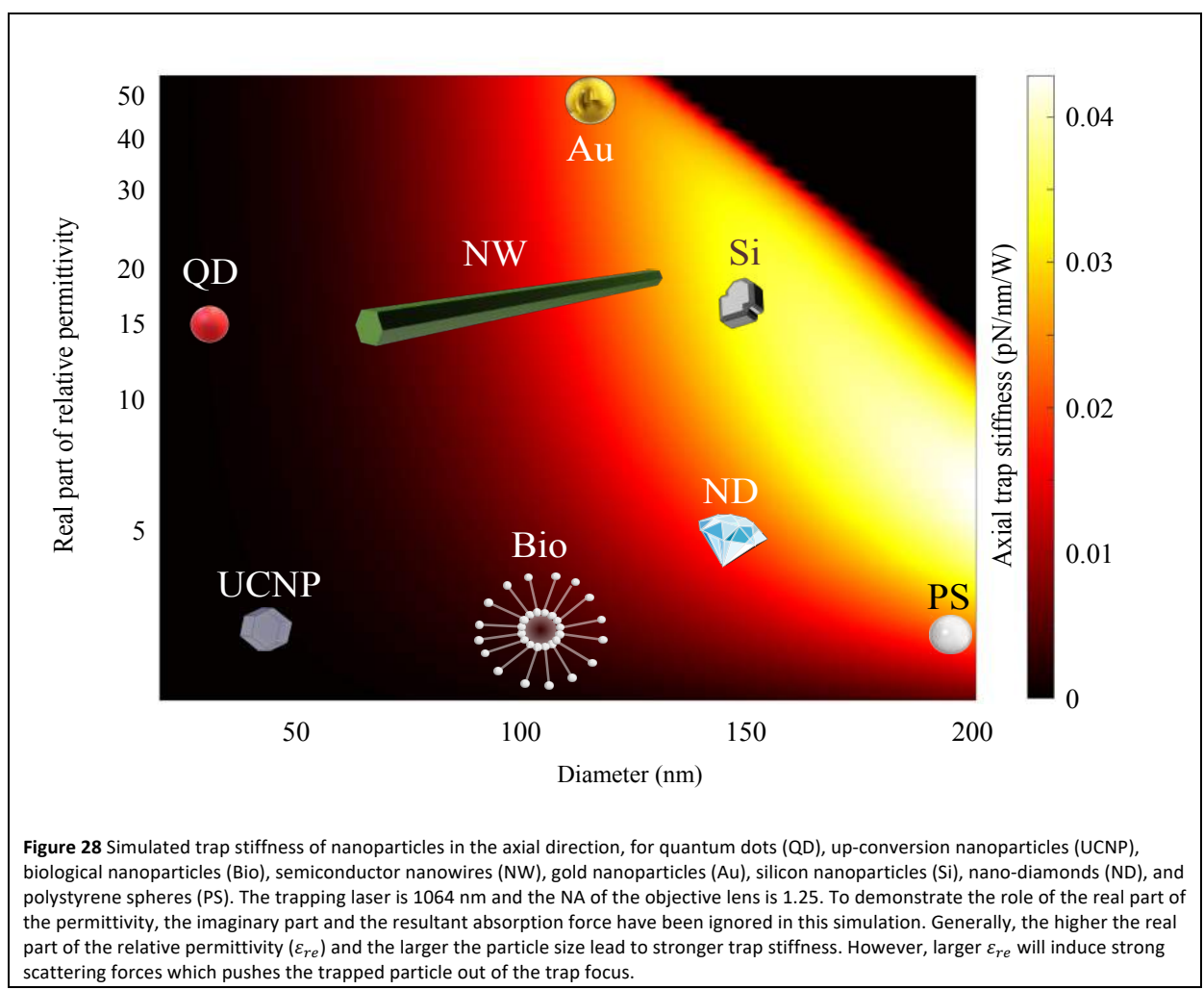

**Figure 28** Simulated trap stiffness of nanoparticles in the axial direction, for quantum dots (QD), up-conversion nanoparticles (UCNP), biological nanoparticles (Bio), semiconductor nanowires (NW), gold nanoparticles (Au), silicon nanoparticles (Si), nano-diamonds (ND), and polystyrene spheres (PS). The trapping laser is 1064 nm and the NA of the objective lens is 1.25. To demonstrate the role of the real part of the permittivity, the imaginary part and the resultant absorption force have been ignored in this simulation. Generally, the higher the real part of the relative permittivity ($\varepsilon_{re}$) and the larger the particle size lead to stronger trap stiffness. However, larger $\varepsilon_{re}$ will induce strong scattering forces which pushes the trapped particle out of the trap focus.

**Current and future challenges**

These advanced functional nanomaterials expand the sensing ability of optical tweezers, yet there are still important issues concerning trapping efficiency and manipulation capabilities, which currently hinders the progress in this field.

Generally, higher permittivity and larger size lead to stronger trap stiffness (**Figure 28**). Nanostructures mostly come at the price of small polarizability (∝ volume), limiting their trap stiffness. Optimal performance can be achieved using metallic nanoparticles (≈50nm), but typically, to obtain stable trapping, one might use ≈100 mW power in the near-infrared and the heat generated prevents sensitive biological measurements. Other types of nanoparticles, including biological nanoparticles, have lower trapping efficiencies due to their smaller refractive index (≈1.4-1.5). Near-field optical trapping, including evanescent field trapping, plasmonic trapping and photonics hot spot trapping, have a demonstrated potential for trapping nanostructures, especially for low index materials; however, the trapped probe is usually located in a fixed position or within 2D, losing the 3D manipulation capability.

The low polarisability of nanostructures leads to a low scattering cross-section, which prevents the use of traditional scattering detection to measure the trap stiffness, limiting the capability of systems for real-time position tracking, particle localisation within the trap volume, 3D force detection and particle metrology. It also poses a challenge for detecting single particle trapping events as the particle sizes are well below the diffraction limit.

Intracellular trapping nanostructures for sensing and studying the interaction between nanoparticles with organelles has potential to lead to new insights in cell biology. However, the intracellular trapping environment can make direct application of tweezers technologies difficult. In particular, when nanostructures enter a cell via endocytosis, they will be encapsulated by a vesicle, leading to aggregation, changing the trapping properties and destroying the functionalized particles' surface. Another challenge of intracellular trapping is that the cytoplasm has a higher refractive index and viscosity than water, decreasing the efficiency of trapping probes. The high laser power required to overcome the lower trapping efficiency may heat and damage the organelles.

The manipulation capability of nanostructures, for example, to control the trapping orientation of nanoparticles remains a challenge. Multi-spot holographic optical tweezers have been demonstrated to orient micrometre long nanowires, however orientation control of asymmetry nanostructures with nanometre-size dimension is currently limited to polarisation effects, and predominantly 2D trapping. Finally, due to the diffraction limit, the laser trapping spot is usually about ten times larger than nanoparticles, limiting the trapping efficiency and the trapping confinement area.

**Advances in science and technology to meet challenges**

Enhanced polarisability of materials is one of the key tools for increased optical forces and extending the trapping limits at the nanoscale. Highly polarisable materials, including semiconductors (e.g., Si, GaAs) are suitable candidates that can avoid issues of heating. High dielectrics can also exhibit geometric resonances (e.g., Mie resonances) that can further improve trapping efficiency [3]. Beyond this, new mechanisms that enhance polarisability, for example, Zeta potential [4], super-radiation [5], ion resonance effect [6] and plasmonic shells [7], may be employed to enhance the effective gradient force. Hybrid trapping schemes that incorporate fluidic or thermal forces can also be highly effective for nanoscale trapping [8].

Engineering the physical geometry of particles has been profoundly successful in micrometre scale trapping and micromanipulation. With, for example, two photon polymerization or other lithographic processes, shape optimization has been used for enhancing forces and torques, and enhancing force detection. Nanoscale lithography can similarly be useful for structuring objects of particular sizes and shapes, that can be used for control orientation. Nanolithography will also be advantageous in bringing uniformity to objects that may otherwise have substantial size dispersion. Other types of nanostructures have natural shapes that will maintain a fixed orientation in a trap can also be exploited in certain applications. The alternative to particle engineering is optical beam shaping, to optimise the optical potential, an approach which has to date shown more promise for larger particles over nanoparticles [9]; advances in approaches using point spread engineering are likely needed to enable flexible nano-trapping approaches in the future

High fidelity detection of forces and positions are essential for force sensing applications. The challenge for extending back-focal plane interferometry is the need to detect a diminishingly small signal in a strong background. Calibration of position detection from individual objects is also essential, as the variability in size can be substantial (> 10%) even in the most well controlled fabrication processes. With the smaller particles also comes faster Brownian motion, meaning that both sensitivity and bandwidth need to be maintained. Fluorescence based methods are a viable alternative, and in certain circumstances, also solve the individual sizing issues. Bespoke detection schemes that enhance the measurable signal also hold much promise. Detection of orientation for anistropic particles will also be important. Total force sensing approaches avoid the problem of force calibration, but for strongly scattering objects both forward and backscattered light needs to be recorded for the accurate determination.

Finally, localization remains a key challenge. Trapping volumes in gradient force tweezers generated by diffraction limited optics are necessarily large compared to the nanoparticle dimensions. So, whilst confined, the nanoparticles are free to move within this space. One possibility for reducing the nanoparticle motion is to use laser-based cooling, which suppresses the Brownian motion [10]. Whilst there are opportunities for creating smaller trapping volumes, using, for example, super-oscillations, it is near-field geometries, such as plasmonic nanostructures and photonic crystal cavities, that can deliver truly nanoscale trapping volumes with reduced power requirements. The progress of near-field trapping towards advanced applications requires the development of a number of auxiliary functions, such as reconfigurability for translation, 3D position detection, and integration with the intended application; enabling technologies, such as meta-optics [11], will help advance across all these areas.

**Concluding Remarks**

In summary, the field of nanoscale trapping is very much still an open and active area of research. Delivering new technologies and approaches will be crucial in unlocking the full potential of trapping of nanostructures and their intended applications. These innovations will come from the application of new materials, nanoscale engineering techniques, novel trapping techniques, and bespoke detection schemes and beam shaping.

## 21 — Optical nanotweezers

*Justus C. Ndukaife*

Vanderbilt Institute of Nanoscale Science and Engineering and Department of Electrical Engineering and Computer Science, Vanderbilt University, Nashville, Tennessee 37235, United States

*Romain Quidant*

Nanophotonic Systems Laboratory, Department of Mechanical and Process Engineering, ETH Zürich, Zürich, Switzerland

**Status**

Optical trapping [1], recently recognized with one-half of the 2018 Physics Nobel Prize, has emerged as an important tool in biology and life science by providing the means to manipulate biological objects and measure forces associated with biomolecular interactions. In standard optical tweezers, a tightly focused laser beam is used to exert strong gradient forces to stably trap microscopic specimens near the laser focus. However, the diffraction limit precludes propagating light to be focused to subwavelength volumes, thus preventing stable trapping of nanosized objects (<100 nm). Evanescent electromagnetic fields in the vicinity of optical nanoresonators have the capability to scale optical traps down to the nanometer scale. This is achieved by spatially confining the light field to deeply subwavelength spots, while enhancing the local light intensity. Furthermore, the high sensitivity of nanoresonators to minute perturbations of their close surrounding offers a convenient way to track the trapped specimen dynamics by monitoring small changes of the resonant light. Different approaches, based on plasmonic resonances [2] in noble metals, Mie resonances [3] in high refractive index low-loss dielectric or semiconductor materials, and dielectric photonic crystal cavities [4] have enabled to extend optical trapping beyond what conventional optical tweezers can do, ultimately reaching the single molecule level (**Figure 29**)[5].

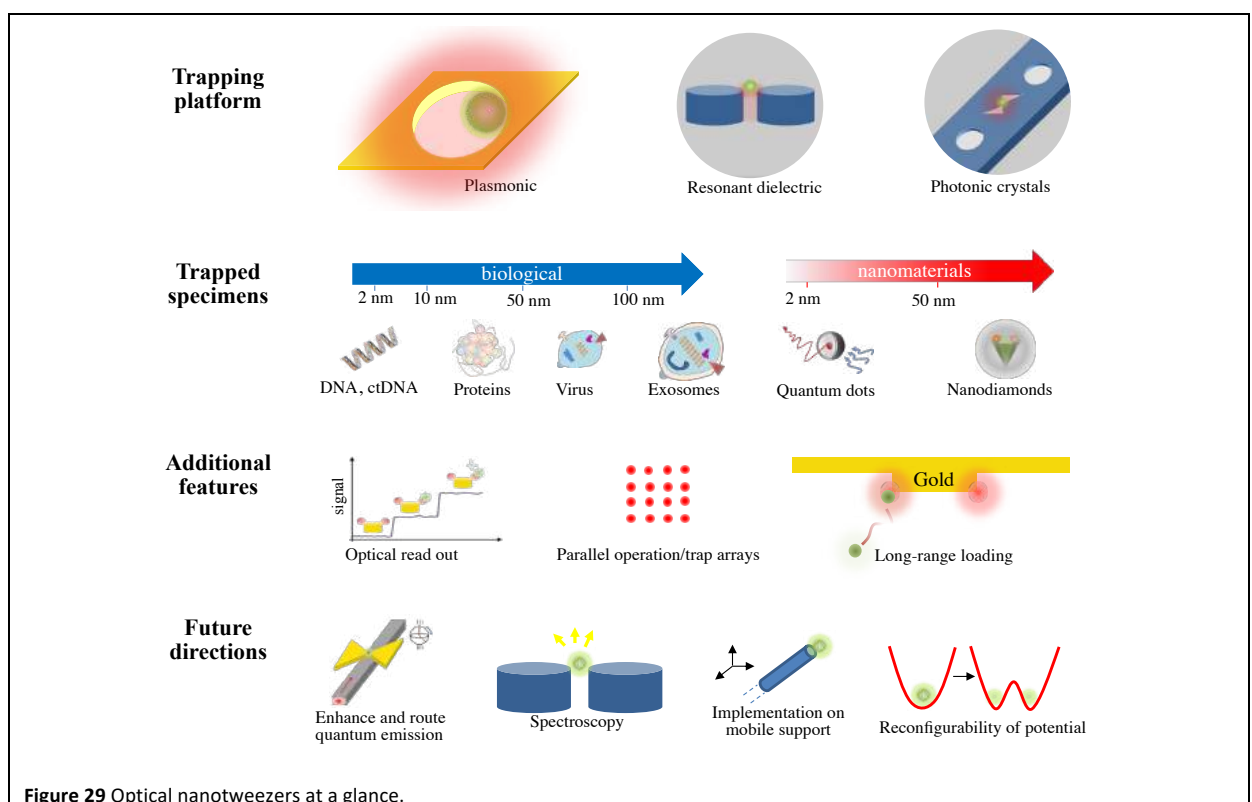

**Figure 29** Optical nanotweezers at a glance.

**Current and future challenges**

Despite their clear potential, optical nanotweezers have not yet become a transversal enabling tool used by researchers from other communities to address scientific questions. This is in part due to several remaining challenges related to trap loading, three-dimensional manipulation of the trap object, and photothermal heating.

(i) *Trap loading* – While nearfield optical potential enables extending trapping to nanometric specimens, their subwavelength confinement raises the practical challenge of efficient trap loading. One default approach relies on Brownian diffusion but it suffers from being slow and non-deterministic. The applicability of nanotraps is conditioned controlled methods to deliver the specimen to the trapping site. A high loading throughput is desired for systematic experiments and reaching statistically relevant data.

(ii) *Material losses* – Depending on the operating wavelength and the choice of material platform, optical losses can become a limitation. First, they lead to a local temperature increase which, if excessive, can alter the integrity of sensitive specimens like biological systems. Additionally, temperature gradients also introduce other dynamics such as buoyancy-driven fluidic convection, increased Brownian motion and thermophoresis. Thermophoretic motion when repulsive will reduce the overall trapping stability. Another consequence of optical losses is the broadening of the resonance bandwidth which can limit the efficiency of feedback-based trapping mechanisms like the SIBA effect [6] as well as of the readout of the specimen dynamics.

(iii) *Control over the trapped specimens* – While most effort so far have focused on new nanophotonics platforms capable to push trapping to smaller sizes down to the single protein level, the field would greatly benefit from further control over the trapped specimen. In particular, accurate 3D manipulation would enable a controlled interaction with other systems, relevant for instance to assembly of nanomaterials or molecular interaction. Furthermore, reconfigurability of the trapping potential would open new opportunity in the study of complex nanoscale dynamics or collective effects in trap arrays.

**Recent advances and future prospects**

The past decade has seen substantial research activities on new optical nanotweezer approaches to address the aforementioned challenges in the field. Capitalizing on recent advances in nanophotonics, all-dielectric optical nanotweezers can overcome the photoheating effect present in their plasmonic counterpart. While Mie-resonances provide lower field enhancements, the reduced temperature rise drastically reduces any influence of fluid convection, Brownian motion, and thermophoresis and thus relaxes the optical trapping potential requirements for stable trapping. For example, silicon nanoantennas achieved stable trapping of 20 nm dielectric spheres, while ensuring a negligible temperature rise of 0.04 K [3]. One exciting future direction to further relax the trapping laser power requirement with negligible heat generation and narrow bandwidth resonances involves harnessing non-scattering anapole modes or bound states in the continuum (BIC) in all-dielectric resonators [7].

With respect to dynamic particle loading and dynamic manipulation, hybrid approaches that combine optical and RF electric fields offer a solution to mitigate relying on slow Brownian diffusion. For example, a recent paper reported opto-thermo-electrohydrodynamic tweezers (OTET) [8] that harnesses the balance between optically-induced electrothermoplasmonic flow and AC electro-osmotic flow to initiate long-range transport and trapping of single protein molecules away from the high intensity laser focus to mitigate photothermal damage.

The use of optical nanotweezers to simultaneously perform trapping and sensing of small molecules and biomolecular interaction has also generated enormous interests. For example, aperture optical nanotweezers have been shown to be capable of measuring antibodies binding to proteins, proteins binding to DNA and small molecules (such as drugs) binding to proteins [9], which may find applications in for understanding the pathogenesis of diseases and for antibody screening. Also, they have lately proved practical to monitor fast conformation dynamics of single proteins and tackle biological questions, which could not be addressed even using very sophisticated characterization techniques.

Another important future direction is to combine nanoscale optical trapping with spectroscopic techniques such as Raman, fluorescence or infrared absorption to probe the biochemical properties of trapped objects at the single particle level. Such a capability is particularly attractive for single extracellular vesicles (EV) analysis towards understanding the heterogeneity of single nanosized vesicles such as exosomes and exomeres.

Last but not least, when integrated at the extremity of a mobile support, optical nanotweezers are also promising for deterministic positioning of single quantum emitters to nanophotonic cavities for generation and manipulation of non-classical light on-chip [10]. The use of optical nanotweezers combined with rapid electrothermoplasmonic loading may help overcome the challenges of limited scalability and throughput of AFM-based manipulation.

**Concluding remarks**

Optical nanotweezers are crucial towards the isolation, manipulation and interrogation of nanoscale objects. Future directions and capabilities will benefit from latest advances in nanophotonics enabled by all-dielectric nanostructures, combination with other non-optical forces as well as from an extensive engineering effort to turn them into a more accessible technology, readily available to other scientific communities.

We envision that optical nanotweezers will significantly impact both biological sciences and nanoassembly of heterostructures for instance, in the context of quantum photonics. In biology, they offer new opportunities to get further understanding on the interaction between biological molecules or to monitor single molecule response to external controls. In the actively investigated field of extracellular vesicles, optical nanotweezers could enable to trap and analyze individual nanosized EV population to further understand the impact of heterogeneity on EV molecular cargo content and function. In the rapidly growing field of quantum technology, we envision that optical nanotweezers could play a key role in the assembly of integrated hybrid quantum photonic circuitry as well as in the development of compact levitation systems for optomechanics and ultrasensitive force sensing [10].

## 22 — Opto-mechanical manipulation of nanodiamonds

*Reece P. Roberts, Cyril Laplane, Thomas Volz*

School of Mathematical and Physical Sciences, Macquarie University, NSW 2109, Australia
ARC Centre of Excellence for Engineered Quantum Systems (EQUS), Macquarie University, NSW 2109, Australia

**Status**

Diamond is a hard, ultrastable and non-toxic material that is chemically inert. These properties along with the possibility of efficient surface functionalisation make nanodiamonds (NDs) an ideal bio-compatible agent for conducting advanced biological studies [1]. NDs also exhibit strong optical activity, both in Raman scattering and through embedded fluorescent molecular complexes. They can host a multitude of optically active colour centres, such as nitrogen-vacancy (NV) and silicon-vacancy (SiV) centres, which provide direct sensing modalities to the surrounding thermal, magnetic and electric fields [2]. Finally, diamond has a high refractive index of about 2.4 and a low absorption in the visible and infrared spectrum, allowing for strong optical confinement and manipulation, while minimizing heating or damage to the surrounding environment. Ultimately, the combination of all these properties makes NDs containing optical defects a powerful tool for applications as fluorescence markers in biology as well as nanoscale sensors in liquid environments [1].

The NV centre in diamond holds a prominent position in quantum technology, as it is a robust room-temperature spin system with optically addressable and readable spins [1,2,3]. Many proposals investigate their potential in the context of hybrid quantum systems, suggesting the coupling of NV spin degrees of freedom to the motional degrees of freedom of a high-Q mechanical oscillator, realized for example through a levitated ND in vacuum [2,4]. Such hybrid systems could be employed for foundational tests, such as for probing the boundaries between classical and quantum physics and exploring the realm of quantum gravity [5].

Ensembles of optically active colour centres in NDs offer enhanced optical forces thanks to their atomic-like polarizability [2]. Analogous to dipole traps in atomic physics, the associated forces with colour-centre resonances are wavelength and state-dependent, allowing the toolbox of atom trapping to be applied to the manipulation of a mesoscopic object, as illustrated in **Figure 30**. Uniquely, the colour centres are concentrated on a length scale much smaller than the wavelength of light, enabling cooperative effects to play a significant role [6]. This goes well beyond the typical atom trapping context. Addressing the internal degrees of freedom might also enable optical cooling methods from atomic physics to be applied to a mesoscopic levitated object. Yet, at this point in time, a number of important challenges related to the interplay of trapping laser, colour centres and the diamond material itself remain unsolved.

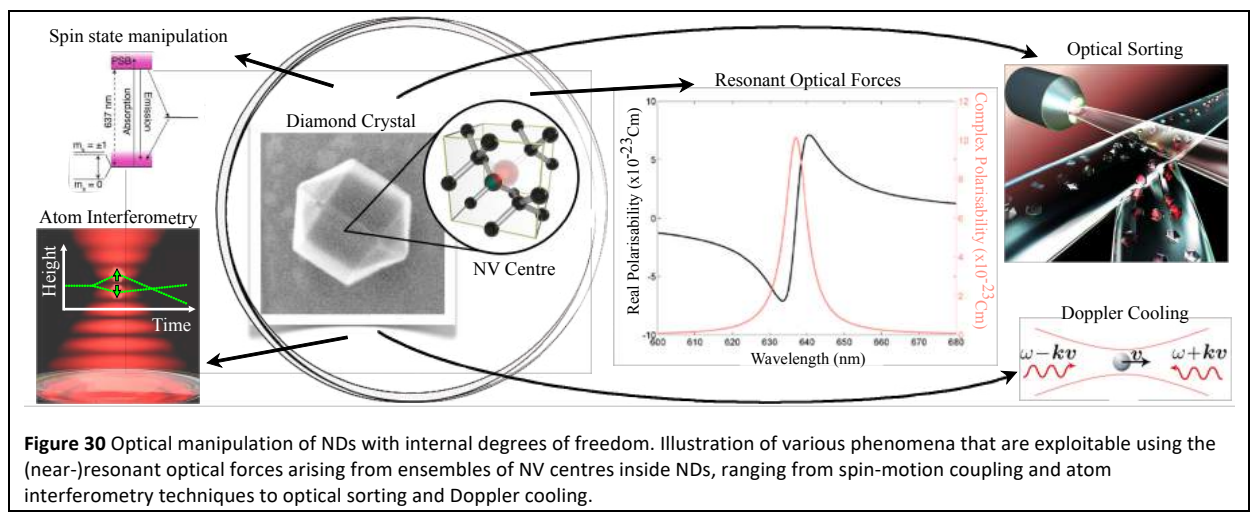

**Figure 30** Optical manipulation of NDs with internal degrees of freedom. Illustration of various phenomena that are exploitable using the (near-)resonant optical forces arising from ensembles of NV centres inside NDs, ranging from spin-motion coupling and atom interferometry techniques to optical sorting and Doppler cooling.

**Current and Future Challenges**

Two key challenges for unlocking the full potential of optically trapped NDs lie in the understanding of the photo-physics of the different colour centres within a strong optical trapping field and determining the interplay of material impurities with the laser light [2].

Controlling material purity is paramount in optical trapping as embedded defects can induce strong trap-laser absorption causing heating of the particle [4,7]. As mentioned previously, a key requirement for many of the aforementioned levitation proposals is that the NDs must be trapped in ultra-high vacuum in order to ensure the highest possible Q factors, and to minimize decoherence arising from background gas collisions. The removal of gas collisions, however, eliminates conductive cooling leaving radiative cooling as the only mechanism for heat loss. Hence, whilst diamond has a high thermal conductivity, it is of no use in this regime. Further, being optically transparent in the near-to-far-infrared, diamond has low emissivity and therefore exhibits minimal radiative cooling. Without a mechanism to lower the internal temperature of a levitated nanoparticle, NDs have been reported to heat up and eventually burn or graphitise below ≈20 mBar due to the absorption of the trapping light by defects and impurities on the surface and inside of commercially available NDs [7].

Adding onto the challenge of material purity, understanding the interplay between the unwanted ND impurities and the desired colour centres is essential for optimising the optical and spin properties of the ND system [2,8]. Traditionally, the NV centre, or more specifically the NV- centre, is considered as a stable, optically addressable and readable spin system. However, when involving an additional trapping laser, the optical and spin dynamics of the NV centre as a whole are impacted heavily by the surrounding NV centres, the surrounding impurities, as well as all of the bound surface charges. These interactions lead to many undesirable effects such as blinking, heating, fluorescence quenching, ionisation, spin depolarisation and spin decoherence which all diminish or fully suppress the desirable properties of the NV centre [9].

Finally, unlike silica spheres which can be made to a high degree of purity and uniformity, NDs typically come with large variations in size, shape and composition/colour centre concentration. This is eminent since observations of different NV centre samples have exhibited contradictory effects [9], to the point where dramatic differences in optical properties within a single sample have been observed [6]. Size variations lead to strong variations in the trapping strength and asymmetric/non-spherical shapes can strongly enhance the noise in experiments and lead to unwanted complex in-trap dynamics, such as rotation and libration. Furthermore, many of the aforementioned detrimental effects are exacerbated by the lack of ND particle consistency in terms of colour centre concentration, which limits the ability for systematic study and analysis of each contributing effect.

**Advances in Science and Technology to Meet Challenges**
Due to the potential opportunities and the anticipated impact of exploring trapped ND systems with optical defects, significant effort has been devoted to overcoming the aforementioned challenges. The different approaches can be summarised into four distinct groups.

The most obvious approach and ultimately the most ideal avenue in the long run is material engineering, i.e., to comprehensively study, analyse and grow ND material under better-controlled conditions [8]. Significant work has already been undertaken to analyse the charge donor and acceptor effects arising from surrounding impurities on the spin dynamics of NV centres in NDs [2].

One promising approach for addressing the strong variability in ND material properties is the sorting of ND samples according to specific desirable criteria [10]. Passive size sorting based on mechanical methods (routinely used for nanoparticles) can be combined with sorting according to colour centre concentration. The latter sorting modality again relies on the exploration of optical scattering forces arising from the (near-)resonant complex polarisability of NV centres (**Figure 30**). Ultimately, it might even be possible to sort NDs according to their radiation (quantum) properties such as the degree of cooperativity [6].

In the meantime, a number of research groups are putting up with the existing ND material but instead are applying in-situ experimental techniques to mitigate some of the negative effects of the trapping laser on the spin-optical properties of the NV centres. Fast gating of the trapping laser has been demonstrated, which enables the ND-NV spin sensor to be manipulated and read-out during the dark intervals of the gated trapping light [9]. Trapping with 1550 nm light reduces demonstrably the detrimental effects of the trapping laser, although this does not entirely eliminate the problem [4]. Typically, these approaches only solve a single issue and are more a band aid than a complete elimination of the root cause.

Lastly, there is also the possibility of using alternative optical defects for various proposals [2]. Whilst NV centres are the most prominent defect in diamond, there are more than 500 optically active diamond defects, each with their own individual optical, magnetic and spin properties. The most prominent ones are SiV centres and germanium-vacancy (GeV) centres [8]. Yet, many of the more intrinsic material-related issues persist for those defects too. Finally, other optically active nanoparticles such as nanorubies or rare-earth-ion-doped nanocrystals might offer a viable alternative for realizing advanced levitation proposals.

**Concluding Remarks**
Overall, combining NDs with embedded optically active defects and optical trapping has become a robust tool in biological applications, now routinely used in the community. Material improvements, such as better uniformity in terms of ND shape and colour centre concentration, might enable a completely new regime of experimentation where optically trapped NDs could enable quantitative fluorescence measurements. NDs with engineered colour centre content could enable the precise and externally controlled mapping of nanoscale magnetic and thermal fields, e.g., in living cells.

Understanding the complex interplay of the diamond matrix and the internal charge dynamics is key to realizing better ND oscillators, and as a consequence, successfully implement a number of optical levitation proposals that would enable experimentalists to build hybrid quantum systems and advanced sensors capable of probing the boundaries between classical and quantum physics and quantum-gravitational effects.

## Acknowledgements

This work was funded in part by the Australian Research Council (ARC) Centre of Excellence for Engineered Quantum Systems (CE110001013) and through an ARC Discovery Project Grant (DP170103010). C.L. is a Sydney Quantum Academy (SQA) Fellow and acknowledges support through the SQA.

## 23 — A label-free approach to seeing individual biomolecules and their interactions

*Reuven Gordon*
University of Victoria, Canada

**Status**

In 2012, the optical trapping of a single protein was demonstrated using a double-nanohole (DNH) in a metal film [1], as shown in **Figure 31**. The protein, bovine serum albumin, showed natural conformational switching between open and closed states in the trap, as confirmed by changing pH and forcing the protein into the open state. Compared to other works that looked at the folding dynamics of proteins [2], the DNH approach was tether-free and label-free. The optical tweezer held the protein in place, so no tether was required. The label-free attribute, however, really makes this technique transformative.

Labels change the nature of the biomolecule. They are expensive, finicky, and they give an indirect signal. They limit the observation time to the lifetime of the label and limit the speed to the rate at which photons can be counted from the label [2]. While they have advanced our understanding profoundly, many are looking to new label-free approaches like iScat [3] and NEOtrap [4], to analyze biomolecules in their native state. The DNH (or more generally shaped-aperture) approach may become a standard tool in single molecule biophysics, drug discovery and antibody screening.

Using the DNH tweezer, it has been possible to measure antibodies binding specifically to proteins, proteins binding to DNA, small molecules (such as drugs) binding to proteins and many other protein interactions [5]. It is possible to measure the affinity of these interactions, which is important for drug discovery, antibody screening etc. These affinity studies are possible at equilibrium and on a single protein – very little sample is required. We have worked on unprocessed samples, which shows potential for speeding up analysis, removing expensive purification steps and looking at active processes in real time: e.g., looking at the secretion products from individual hybridoma cells in real-time antibody screening.

The shaped-aperture approach has been combined with nanopores, which directs the protein/biomolecule to the aperture [6], for studies of both weak and strong interactions with immunotherapy targets [7,8]. The tweezer also solves the overly fast translocation problem of nanopores. It is simpler, however, to use the tweezer without the nanopore (and vice versa), and simplicity often trumps performance. The shaped-aperture allows for literally watching proteins work. There are no other ways of doing this, allowing for unprecedented observations of all the states involved. It possible to watch a single protein for many hours, but with a time resolution that is limited only by the detector speed (i.e., the application is not photon-starved).

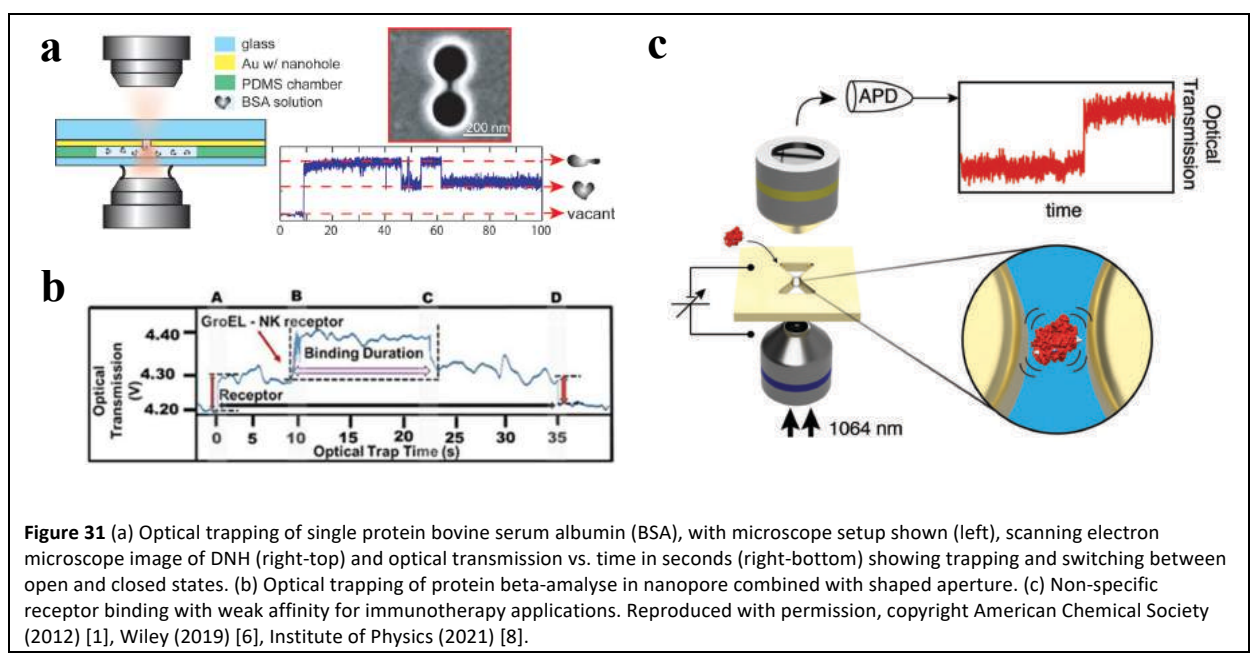

**Figure 31** (a) Optical trapping of single protein bovine serum albumin (BSA), with microscope setup shown (left), scanning electron microscope image of DNH (right-top) and optical transmission vs. time in seconds (right-bottom) showing trapping and switching between open and closed states. (b) Optical trapping of protein beta-amalyse in nanopore combined with shaped aperture. (c) Non-specific receptor binding with weak affinity for immunotherapy applications. Reproduced with permission, copyright American Chemical Society (2012) [1], Wiley (2019) [6], Institute of Physics (2021) [8].

**Current and Future Challenges**

Significant effort has been spent on developing the protein trapping technique since 2012 [1]. A common challenge with any new technique is that there is a healthy level of scepticism for these approaches. Since 2019, however, other groups have demonstrated the application to biomolecules in the single digit nanometer range, so this challenge is now resolving [6-8].

Considering the promise for applications in drug discovery, antibody screening, and single molecule biophysics in general, the main remaining challenge is ease-of-use. The approach combines three fields: nanotechnology, laser tweezers and biochemistry. It is complicated. Widescale adoption of this approach will likely require a simplification and a combination of the laser-tweezer and the nanotechnology parts. In this way, biochemists in academia and industry can treat those parts as a single "black-box" and focus on probing biomolecules and their interactions.

A second challenge is heating. Typically, biomolecules become more thermophobic as temperature is increased. Proteins are thermophilic at lower temperatures but become thermophobic just above room temperature. Nanostructured metals are prone to extreme heating due to plasmonic resonances. This means that the laser tweezer could repel biomolecules of interest, making it harder to trap them. It is still possible to trap thermophobic particles, but the time-to-trap increases, and this makes the trapping inefficient.

On the topic of not repelling biomolecules, surface interactions should also be controlled. Certain surfaces, like glass, pick up a large surface charge in aqueous environments. This serves to repel molecules of interest and it should be avoided for the same reason as with thermophoretic effects. We have noticed that switching to a non-aqueous environment, such as hexane, allows for much more rapid trapping of nanoparticles. Based on the concentration of nanoparticles in solution, we expect that the time-to-trap is milliseconds, which is true for hexane but for aqueous samples it is longer.

Another challenge is extending to different types of proteins and environments. So far, most studies have focussed on globular or other free-solution proteins, whereas membrane proteins are also of great interest. Looking at proteins within a cell would also be an outstanding achievement, although this may prove too challenging with aperture-based trapping.

A final challenge is in throughput and scalability. Looking at biomolecules one-by-one is likely too time-consuming to have the level of impact required for antibody screening and drug discovery applications. Multiple traps are desired that can rapidly analyze biomolecules in parallel, each quickly gating between trapping events to obtain the biomolecule of interest.

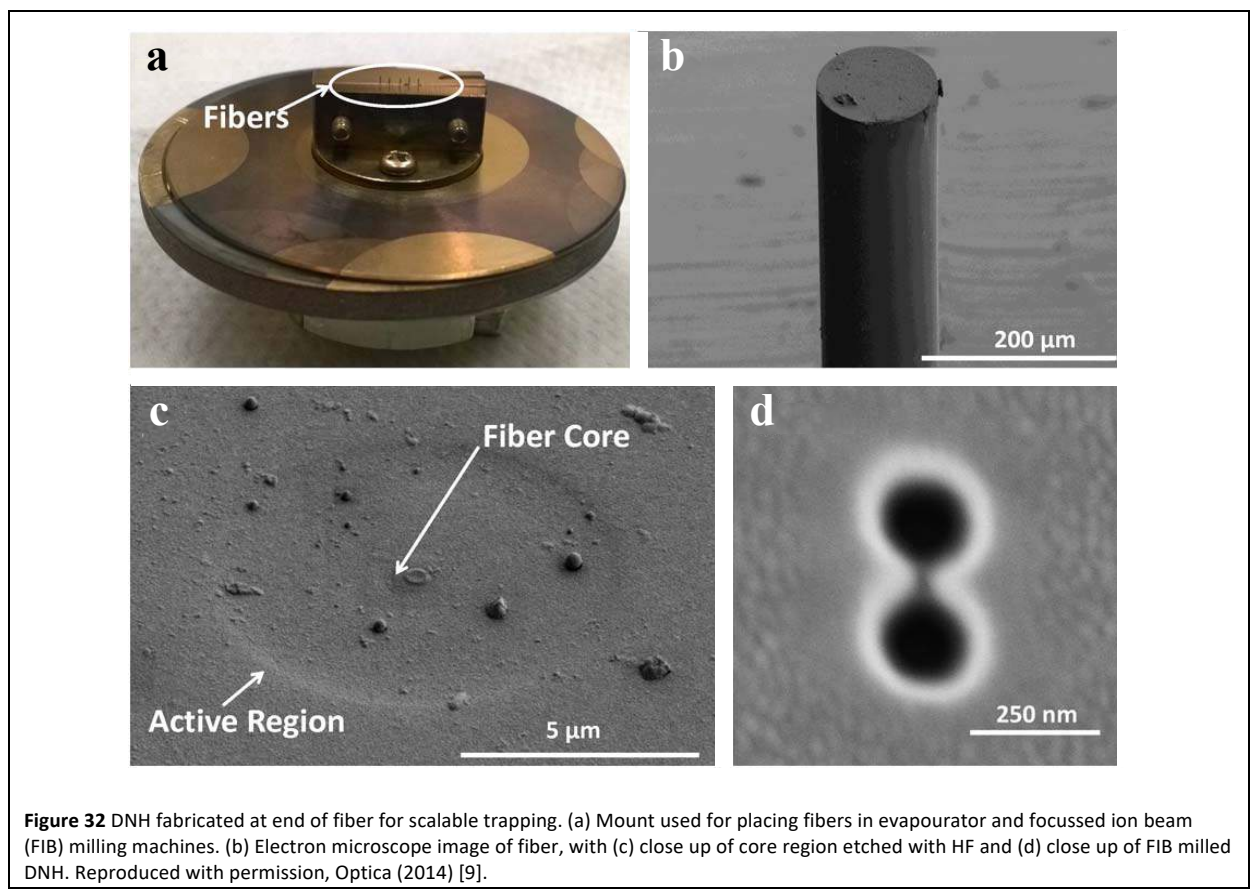

**Figure 32** DNH fabricated at end of fiber for scalable trapping. (a) Mount used for placing fibers in evapourator and focussed ion beam (FIB) milling machines. (b) Electron microscope image of fiber, with (c) close up of core region etched with HF and (d) close up of FIB milled DNH. Reproduced with permission, Optica (2014) [9].

**Advances in Science and Technology to Meet Challenges**

A simpler way to do shaped aperture trapping is to integrate the platform on the end of a standard optical fiber [9], as shown in **Figure 32**. Ideally this approach will be easy to fabricate and use standard parts that are robust and low-cost. Making the nanofabrication low cost will likely make use of colloidal nanofabrication that has high throughput, uses inexpensive materials and does not require esoteric nanofabrication equipment. With the fiber-end approach, the optical tweezer can be used to probe standard microwells, which is the playground of many biochemists.

To avoid heating, the likely solution is to work away from plasmonic resonances. It is for this reason that I do not call these plasmonic tweezers, because I believe you specifically want to stay away from plasmonic resonances to avoid heating. It is possible to have field enhancements with nanostructured metals without invoking plasmonic resonances. In fact, early theory on "nanometric optical tweezers" recognized that "non-resonant" illumination was desirable to avoid thermal damage and strong fields can be achieved with sharp features [10] and/or gaps without invoking plasmonic resonances.

To avoid surface effects, it is necessary to use surfaces that do not charge as much, or to treat the surfaces. We have done surface coating to minimize surface interactions [1], and we have promising results from plastic substrates that charge less than glass. Several groups have worked at creating phospholipid bilayers on surfaces in plasmonics and nanopore studies. For metals, this usually requires some additional thin film coating. It is possible that this can be added to the shaped aperture platform as well to allow for trapping of membrane proteins.

The fiber-based approach above also solves the scalability challenge. With successful incorporation at the end of an optical fiber, it is easy to see how this can be scaled up and parallelized. For example, 8 fiber probes can be attached to a single probe station and simultaneously address 8 different microwells. Stepping through these microwells, one may envision 96 well plate scanning. Having 96 fibers probing the entire well plate at once is another possibility. The fiber-based approaches benefit from this modularity that has already enabled tremendous advances in fiber optic communication. An alternative approach is to use a holographic optical tweezer approach to address multiple traps at once.

**Concluding Remarks**

Biotechnology is advancing rapidly. New technologies have given unprecedented control of our genetic material and the proteins they produce. Yet our understanding of how these proteins function is still lacking. Single molecule methods, such as shaped-aperture optical trapping, give a new window into the biophysics of proteins and other biomolecules. To expedite real impact, these tools will have to remove complexity barriers associated with nanofabrication and optical tweezers, so that they become standard tools for biochemists and other scientists. If massively scaled up, they present disruptive potential to the fields of drug discovery, antibody screening and other related industries. While only a handful of groups around the world have demonstrated shaped-aperture trapping of proteins so far, their exciting demonstrations show potential for vast impact.

**Acknowledgements**

This work is supported by an NSERC (Canada) Discovery Grant RGPIN-2017-03830.

## TRAPPING IN GAS AND IN VACUUM

## 24 — Optical levitation in gaseous media

*Dag Hanstorp, Javier Tello Marmolejo*

Department of Physics, University of Gothenburg, 41296 Gothenburg, Sweden

**Status**

Optical manipulation was introduced in 1970 when Arthur Ashkin trapped a single particle in water using two counter-propagating beams to balance the scattering forces. A year later he managed to trap a particle with a single, vertical beam in air where the particle's weight counteracted the scattering force [1]. This optical levitation technique has since then been used to levitate particles in gaseous media. The field of optical trapping, however, really took off when Ashkin demonstrated the optical tweezers in 1986. Here the same microscope objective is used both to trap and observe an object, and due to the particles' buoyancy, very low trapping powers are needed. These huge advantages propelled optical tweezers to a standard, built-in tool in many commercial microscopes. They can be combined with all kinds of optical detection techniques and are now frequently applied in fields outside optics.

The design of optical traps has always been fundamentally intertwined with the surrounding media, which could be a liquid, gas, or vacuum. When trapping in a liquid, there are strong mechanical and chemical interactions between the trapped object and the media. Here solid particles, liquids enclosed by a membrane (vesicles or cells), or insoluble liquids can be trapped. In vacuum, on the other hand, the particles can be non-intrusively trapped, but only solid particles or liquids with very low vapor pressure can be investigated. As a consequence, the intermediate case of trapping in gaseous media constitutes the most general method for experimental work on isolated samples of microscopic liquids.

Although trapping in air was performed at the very beginning of the field of optical manipulation, it is only in recent years that the activities in the field have emerged [2]. In optical levitation, particles are trapped with a long focal length lens producing a trap with low stiffness. This allows a trapped particle to be used as an extremely sensitive force probe [3], and environmental properties like air friction can be easily controlled and have led to unique visualizations of fundamental physical properties [4]. A very important and emerging application of optical levitation is in the field of micro-chemistry, where samples with a very high surface to volume ratio can be investigated. Here the long working distances from the trapping lens allows access for various excitation and detection schemes. Optical levitation has also been used to investigate optical phenomena such as whispering gallery modes (WGM) [5] and optically induced rotation [6].

JPhys Photonics (2021) ####

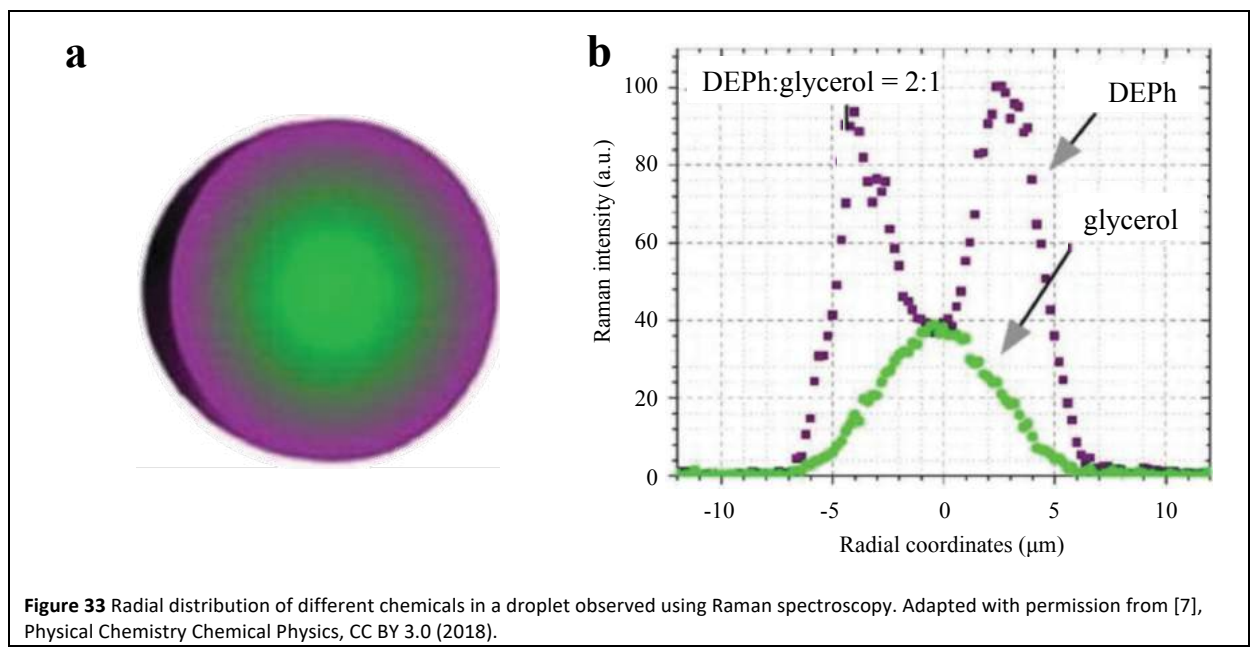

**Figure 33** Radial distribution of different chemicals in a droplet observed using Raman spectroscopy. Adapted with permission from [7], Physical Chemistry Chemical Physics, CC BY 3.0 (2018).

**Current and Future Challenges**

Optical levitation is the ideal system for the development of methods to investigate the chemistry of microscopic isolated samples of liquids. It is then the liquid itself, due to the surface tension, that creates the interface between the trapped liquid and the surrounding gas. This is very different from other types of experiments on microscopic amounts of fluids where the liquid is confined by the wall of, e.g., a micro-pipette or a micro-fluidic system. Many interesting experiments have been conducted, but the field is still in its infancy.

A main challenge in the field is to develop the optical levitator to a complete system for microchemistry, where three classes of experiments can be identified. First, optical levitation can be used to study the intrinsic properties of the trapped liquid. Trapping is then conducted in an inert gas which only has the function to create a viscous force assisting in the trapping process. The feasibility of this has been shown by Kalume et al. [7], who used Raman spectroscopy to observe the formation of a shell structure in a droplet consisting of two different solvents (**Figure 33**).

Second, methods for studies of the interaction with the surrounding media can be developed. This is of high relevance in aerosol science [8] since the absorption and emission of molecules and also larger particles have implications in a wide range of fields, such as environmental science, cloud formation, or the spreading of viruses.

Third, the interaction between two different droplets can be studied by a dual optical levitator where the spatial position of the two trapped droplets can be independently controlled. **Figure 34** shows a proof of principle experiment where the interaction between two different liquids is studied with high spatial as well as temporal resolution. These experiments show the potential of using optical levitation in microchemistry research. However, a lot of development is needed before the full potential of the experimental scheme is obtained.

Another important application of optical levitation is to study optical processes, and in particular scattering. A single scatterer with a fixed location where the size can be tuned gives opportunities to reveal information that normally is hidden when light scattered from multiple sources are integrated. Mie scattering, for example, can be used to measure the size and refractive index of the particles [9]. Optical levitation here gives unique experimental possibilities since the size of the trapped particle easily can be varied by evaporation or condensation.

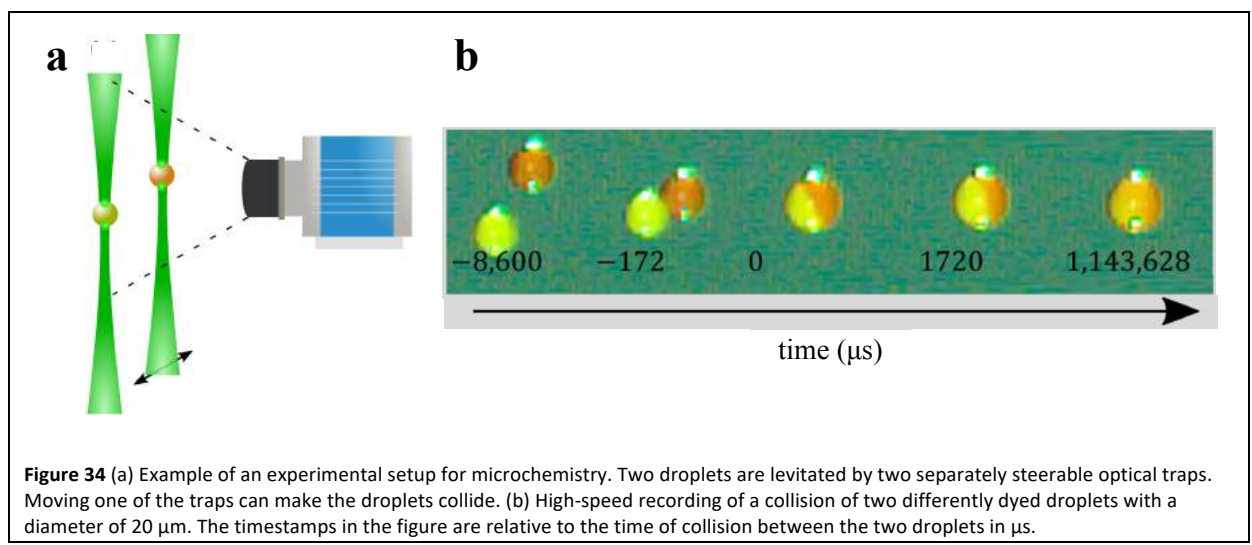

**Figure 34** (a) Example of an experimental setup for microchemistry. Two droplets are levitated by two separately steerable optical traps. Moving one of the traps can make the droplets collide. (b) High-speed recording of a collision of two differently dyed droplets with a diameter of 20 µm. The timestamps in the figure are relative to the time of collision between the two droplets in µs.

**Advances in science and technology to meet challenges**

There are currently many interesting developments in the field of trapping in gaseous media, at the same time as there are many remaining challenges. The main technological advancement required to make optical levitation a widespread technique would be to construct an experimental system where particles can be readily dispensed and trapped, just in the same way as there now are plug-in units for optical tweezers that can be incorporated in commercial microscopes. The most important needed advances in science and technology are:

- *Development of more efficient trapping to avoid the very powerful lasers needed today.* This is of particular importance for experiments using multiple traps. One possible solution is the use of acoustic levitation to complement the optical trap. In an acousto-optical trap, the acoustic field could do the main part of the work to counteract gravity, while the optical trap would provide the stability and control. This would reduce the laser power necessary to trap, and also expand the range of sizes that can be trapped to six orders of magnitude ranging from nm up to mm.
- *Development of a reliable dispensing mechanism.* Optical tweezers can simply pick up particles dispensed in a liquid sample, whereas van der Waals forces makes it difficult to release particles into the gas phase. Many different techniques exist, including laser-induced acoustic desorption, micro-droplet dispensers, piezoelectric transducers, and ultrasonic nebulizers, but there is still no reliable dispense-on-demand system available.
- *Incorporation of an environmental control in the optical levitator.* Today various experiments have been conducted under different environmental conditions such as the composition of the surrounding gas, pressure, and temperature. However, there is not yet a system available where these parameters can be freely varied, at the same time as the liquid droplets can be dispensed into the system on-demand.
- *Development experiments and theory of directional Mie resonance spectra.* Mie spectra have shown to be made up of a series of Fano resonances where their evolution is directly dependent on the material properties, i.e., refractive index and optical absorption. Therefore, the evolution of material properties on the surface of droplets could be measured in situ during processes like evaporation or chemical absorption.

- *Development of the use of cavity ringdown spectroscopy (CRS) together with Mie resonances.* Levitated particles have been placed inside mirror cavities to investigate their properties through CRS [10], but, at a Mie resonance, particles themselves act as cavities. The observation of a decay in the scattering of a particle during a resonance would mean that CRS could be performed without the need for an external cavity and allow instant measurement of trace contaminants in a droplet.

**Concluding Remarks**

Optical levitation is an eye-catching technique with a broad range of applications from demonstrations of physical principles to fundamental research. Many experiments using optical levitation of liquid droplets in gaseous media have been initiated. The main experimental challenge is to combine the different technologies to dispense and trap droplets, control the environment, induce reactions and finally detect the processes using advanced detection schemes. When an experimental system that combines these technologies has been developed it is expected that optical levitation becomes a universal tool that will be applied for experimental studies of physical, chemical as well as optical processes of importance in many different applications such as combustion, printing, devices for inhalation, and atmospheric sciences, just to mention a few.

**Acknowledgements**

Financial support from the Swedish Research Council (2019-02376) is acknowledged.

## 25 — Rotational optomechanics of spinning particles

*Graham D. Bruce*
University of St Andrews, UK

*Kishan Dholakia*
University of St Andrews, UK
University of Adelaide, Australia
Yonsei University, South Korea

**Status**

Levitated optomechanics is the trapping and isolation of mesoscopic particles at low pressure. As the pressure is reduced, the translational motion of the particle changes from overdamped to underdamped oscillations, i.e., the oscillations become increasingly coherent. This decoupling from the thermal environment also allows the motion to be cooled with feedback schemes. Overall, this leads to very high-quality factor systems with very long decoherence times, ideal for a suite of sensing and measurement studies. Ashkin first levitated dielectric particles under vacuum conditions in 1977, but the field of levitated optomechanics has only taken shape in the last decade and subsequently made rapid progress including, in 2020, the first demonstration of cooling a levitated nanosphere to the quantum ground state [1].

Going beyond translational motion, introducing a rotational degree of freedom opens exciting prospects. An anisotropic particle (either due to its shape or to birefringence) experiences a torque in the presence of circularly polarized trapping light which causes continuous rotational motion around its centre of mass, at an angular velocity restricted by viscous drag. In low pressure environments, such spinning particles can reach rotational frequencies in the GHz regime (**Figure 35**) [2]. The rotation can add an inherent stability to the particle motion [3] and also couple in intriguing ways to the translational motion of the particle. Translational stabilization is beneficial in studies of the interactions between particles and surfaces (e.g., quantum friction as detailed below) and multi-particle interactions such as optical binding (covered elsewhere in this roadmap). Moreover, rotation can be used to average-out the effects of inhomogeneities such as fixed electric dipole moments in the particles.

More broadly, harnessing the rotational motion of optically trapped particles offers enticing prospects in both applied and fundamental science. Pressure sensing with levitated rotating particles has been demonstrated as an alternative to mm-scale magnetically levitated spinning-rotor pressure gauges, with advantages of higher spatial resolution and insensitivity to external magnetic fields [4]. External forces and torques influence the rotation rate: a frequency-locked rotating nanorod has been demonstrated to offer a non-resonant force sensor [4], while external torques as small as $10^{-28}$ Nm can be measured with rotating particles [5], surpassing torque sensitivities in cryogenically cooled tethered systems by orders of magnitude. Ultimately, in ultrahigh vacuum where the drag on the particle is negligible, material properties of the rotor such as the tensile strength can be probed [2].

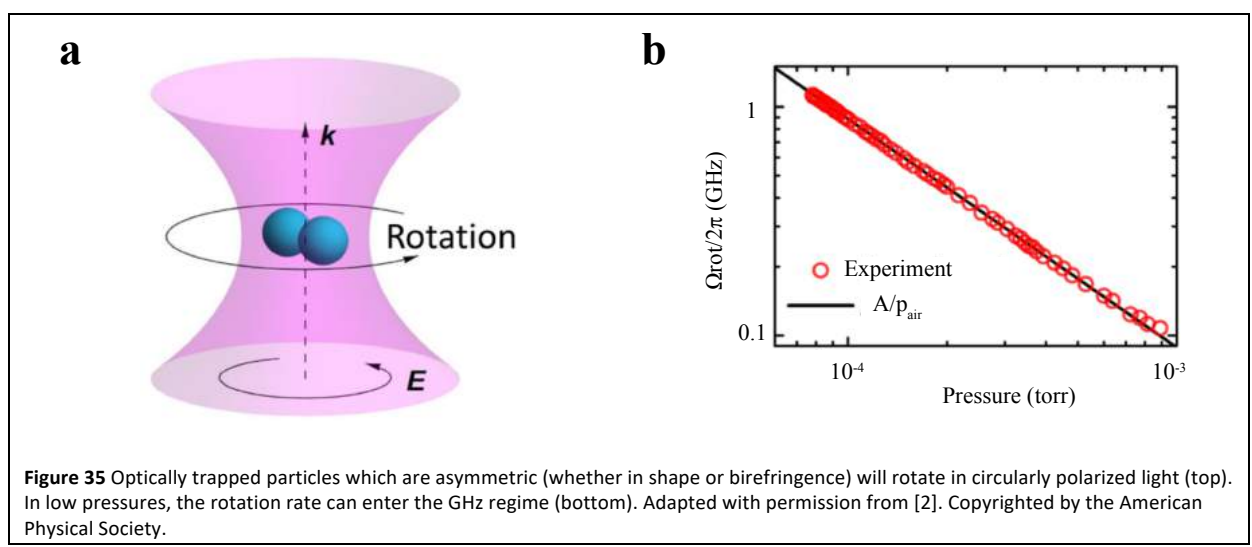

**Figure 35** Optically trapped particles which are asymmetric (whether in shape or birefringence) will rotate in circularly polarized light (top). In low pressures, the rotation rate can enter the GHz regime (bottom). Adapted with permission from [2]. Copyrighted by the American Physical Society.

**Current and Future Challenges**

One major challenge in levitated optomechanics is the demonstration of mesoscopic quantum phenomena, taking advantage of the excellent isolation from environmental noise afforded by levitation. Cooling all rotational and translational degrees of freedom to the quantum ground state is an open challenge. However, while the observation of quantum phenomena with translational degrees of freedom requires significant cooling to the motional ground state, quantum phenomena may be seen in rotational dynamics without this cooling requirement [6], with potential applications in quantum enhanced torque sensing at the $10^{-30}$ Nm level and tests of wavefunction collapse models.

Rotating levitated particles provide a promising platform in the search for so-called quantum friction. For two surfaces undergoing parallel relative motion at constant speed, the vacuum electromagnetic fluctuations are predicted to cause a force counteracting the motion. The existence of quantum friction is itself under debate [7], and experiments have yet to unambiguously measure its effects, due primarily to the fact that the force is incredibly weak. However, for a room temperature, 150 nm diameter silica sphere, rotating at 1 GHz, 200 nm from a silica surface, the quantum frictional torque is predicted to be $10^{-28}$ Nm, comparable to recently achieved torque measurements with levitated particles rotating far from any surface [5]. An unambiguous measurement of the quantum frictional torque in a regime of negligible air damping will require a background pressure below $10^{-9}$ torr (**Figure 36**a). Therefore, the principal current challenge is to combine state-of-the-art achievements in rotation rate, background pressure and stable confinement close to a substrate, in a single experiment. A future challenge will be the detection of the orders-of-magnitude weaker vacuum friction: the drag force exerted on a rotating particle in free space.

An intriguing prospect is the possibility of observing another elusive fundamental phenomenon: Zel'dovich amplification of electromagnetic waves by a spinning object [8]. In this phenomenon, rotational energy can be transferred from matter to the light if the object's rotation rate is greater than the angular frequency of the light divided by its topological charge. State-of-the-art experiments can generate beams of light with topological charges as high as $10^4$, and optically-trapped particles can achieve rotation rates of $10^9$ Hz, suggesting that Zel'dovich amplification of infrared light might be observable. The challenge will be in sufficiently focussing such long wavelength light with such high values of topological charge onto nanoscale particles, although future increases in angular velocity would reduce the required wavelength of light and ease this challenge.

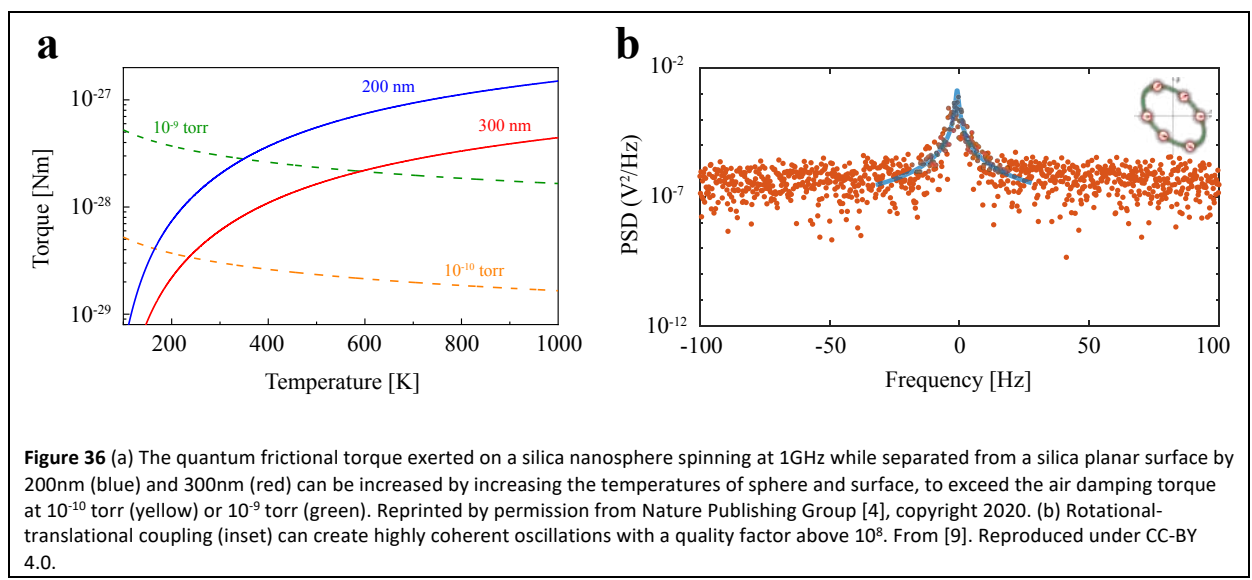

**Figure 36** (a) The quantum frictional torque exerted on a silica nanosphere spinning at 1GHz while separated from a silica planar surface by 200nm (blue) and 300nm (red) can be increased by increasing the temperatures of sphere and surface, to exceed the air damping torque at $10^{-10}$ torr (yellow) or $10^{-9}$ torr (green). Reprinted by permission from Nature Publishing Group [4], copyright 2020. (b) Rotational-translational coupling (inset) can create highly coherent oscillations with a quality factor above $10^8$. From [9]. Reproduced under CC-BY 4.0.

**Advances in Science and Technology to Meet Challenges**

Advances in the material properties of particles are likely to have wide-ranging impacts on rotational levitated optomechanics. An early material used in rotational studies was vaterite, a positive uniaxial birefringent crystal of calcium carbonate: micron-sized spheres of vaterite were rotated at MHz rates [3]. The birefringence of the vaterite particle allows for unforeseen new understanding of light-matter interaction. From a particle perspective, this results in effects such as the nonsymmetric coupling between rotational and translational degrees of freedom, leading to forces that are linearly nonconservative. A repeated sequence of small translations and rotations, which return the particle to its initial configuration, transfer energy between the optical field and the particle. Some trajectories increase in energy, accumulate momentum, and show coherent characteristics, leading to self-sustaining motion arising even from stochastic thermal forces. This generates oscillations with a quality factor in excess of $10^8$ (**Figure 36**b), which is highly attractive for sensing [9]. Separately, from the perspective of light, inhomogeneous circularly polarised light has components of optical momentum that circulate around the beam axis, resulting in additional forces which may be revealed by the birefringent nature of the trapped object. However, synthesis of nanoscale vaterite is challenging, and rotational levitated optomechanics has so far required the development of alternative rotors, such as rods or dumbbells of silica. The introduction of more exotic materials to the levitated toolbox are likely to herald further new advances.

The choice of particle material also plays an important role in the challenges outlined above. The quantum frictional torque is resonantly enhanced if the surface and sphere have common phonon polariton modes (e.g., if they are constructed from the same material). A further judicious choice of the material of the particle and surface may increase the value of this torque by orders of magnitude, e.g., by exploiting a resonance between the surface polariton frequency of the material and the rotational frequency of the sphere, to promote photon tunnelling between sphere and medium [10]. Predictions suggest this will allow experiments to detect quantum friction at much higher pressures (around $10^{-4}$ torr). Meanwhile, the plausibility of observation of Zel'dovich amplification increases with the rotation rate of the particle: new materials may need to be investigated which have sufficient tensile strength to withstand rotation rates in the tens or hundreds of GHz.

**Concluding Remarks**

The effects of rotation has, to varying degrees, surprised and encouraged research in the area of levitated optomechanics. This is a degree of freedom that is adding immense value to the field and opening up new studies in both the classical and quantum domains. As we progress in cooling, torque fluctuations due to the quantisation of the field may dominate the motion. Entering this domain means control of rotation using concepts such as squeezing of the light field may be used [11]. The next decade is likely to see exciting twists in the light-matter interaction story.

**Acknowledgements**

The authors acknowledge funding from the UK Engineering and Physical Sciences Research Council (EP/P030017/1).

## 26 — Dynamics and applications of an optically levitated nanodumbbell

*Tongcang Li*

Department of Physics and Astronomy, and Elmore Family School of Electrical and Computer Engineering, Purdue University, Indiana, United States

**Status**

Due to their great isolation from the thermal environment, optically levitated nanoparticles in vacuum have great potentials in precision measurements and studying fundamental physics. While most experimental works in levitated optomechanics have focused on spherical particles, there are growing interests in using nonspherical nanoparticles [1,2]. Compared to spherical particles, nonspherical particles offer the opportunity to study the rotation and libration (torsional vibration) besides the center-of-mass motion. Among nonspherical particles, silica nanodumbbells stand out as they have been experimentally levitated in a high vacuum (**Figure 37**) [3]. A silica nanodumbbell can be created by chemical synthesis or by merging two levitated silica nanospheres together in optical tweezers. The bonding between the two nanospheres is remarkably strong [3]. Because of its anisotropic shape, its optical polarizability along its long axis is larger than those along the two short axes. Circularly polarized optical tweezers can transfer the spin angular momentum of photons to the levitated nanodumbbell and drive it to rotate (**Figure 37**a) [3, 4]. Record high rotation speeds beyond 5 GHz have been reported with a levitated nanodumbbell in a high vacuum (**Figure 37**b) [5]. In linearly polarized optical tweezers, the long axis of the nanodumbbell will tend to align with the electric field of the optical tweezers. It will be a novel analogy of the famous Cavendish torsion balance with an unprecedented torque detection sensitivity on the order of $10^{-28}$ Nm/$\sqrt{\text{Hz}}$ (**Figure 37**d) [3]. Recently, a levitated nanoparticle has detected a torque as small as $4.7 \times 10^{-28}$ Nm in just 100 seconds [5]. Such an ultrasensitive nanoscale torque detector will be able to study Casimir torque [6] and quantum vacuum friction near a surface [5]. Besides torque-sensing, a levitated nanodumbbell can also be used to create macroscopic rotational quantum superposition states to test the limit of quantum mechanics. Cooling will be required for most quantum applications. The three center-of-mass motion modes and two librational modes of a nanodumbbell have been cooled simultaneously with active feedback (**Figure 37**) [7]. Recently, parametric feedback cooling was used to cool the libration of a levitated nanodumbbell to sub-kelvin temperatures [8]. Depending on the laser power, the heating due to photon shot noise can dominate when the air pressure is below $10^{-7}$ torr (**Figure 37**d) [3, 8].

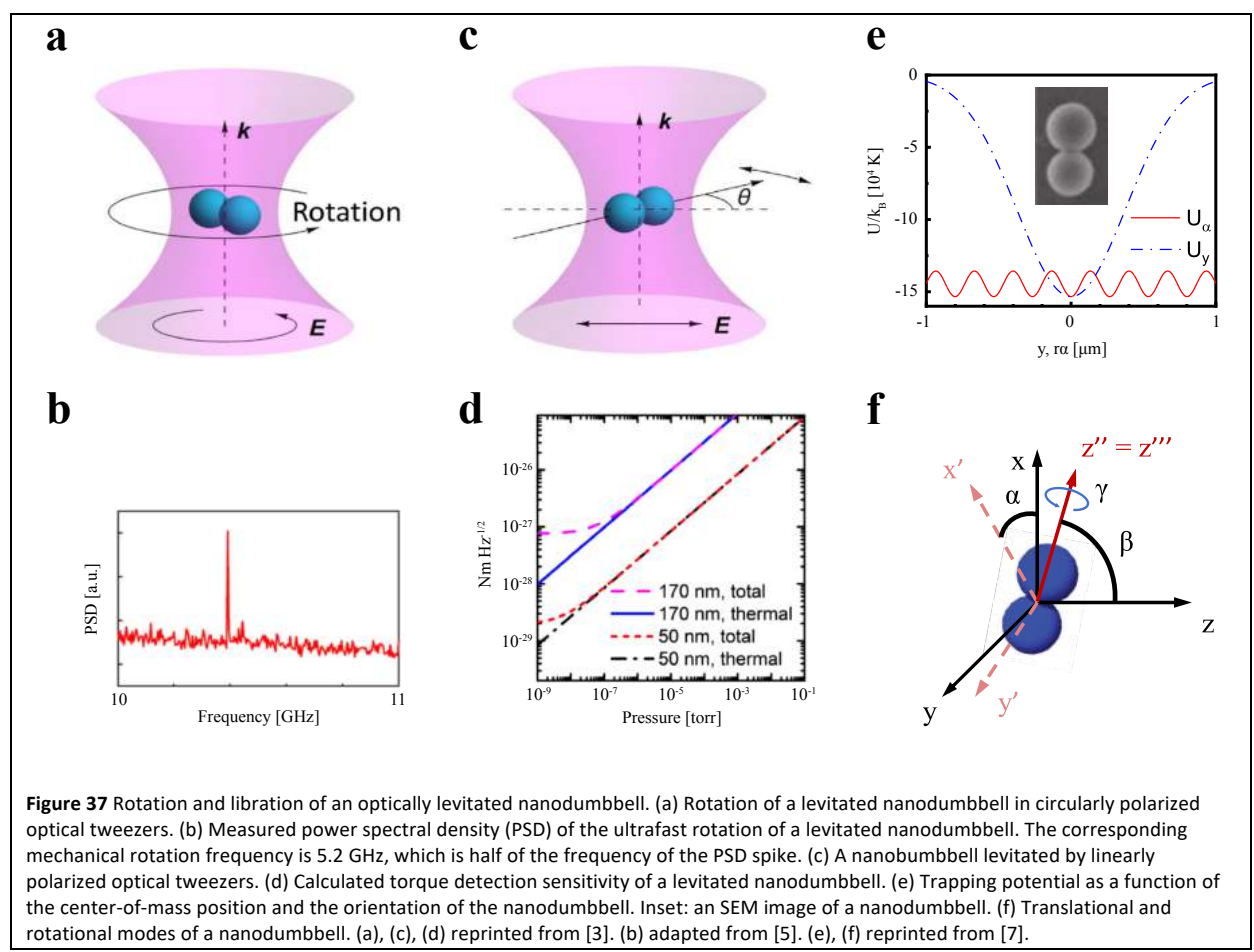

**Figure 37** Rotation and libration of an optically levitated nanodumbbell. (a) Rotation of a levitated nanodumbbell in circularly polarized optical tweezers. (b) Measured power spectral density (PSD) of the ultrafast rotation of a levitated nanodumbbell. The corresponding mechanical rotation frequency is 5.2 GHz, which is half of the frequency of the PSD spike. (c) A nanobumbbell levitated by linearly polarized optical tweezers. (d) Calculated torque detection sensitivity of a levitated nanodumbbell. (e) Trapping potential as a function of the center-of-mass position and the orientation of the nanodumbbell. Inset: an SEM image of a nanodumbbell. (f) Translational and rotational modes of a nanodumbbell. (a), (c), (d) reprinted from [3]. (b) adapted from [5]. (e), (f) reprinted from [7].

**Current and Future Challenges**

The maximum rotation speed of a levitated nanodumbbell is limited by its breakdown due to the centrifugal force [3]. At a rotation speed of 5 GHz, the maximum stress in a nanodumbbell with a diameter of 150 nm is expected to be more than 100 GPa. To detect the vacuum friction, a nanoparticle will need to be driven to rotate at high speed near a surface with a sub-micrometer separation in a high vacuum. To make sure the vacuum friction is larger than the traditional friction due to air damping, the pressure of the residual air will need to be below $10^{-9}$ torr [5], which will be challenging to achieve.

Ground state cooling will be required for many quantum applications. But it has not been achieved with a levitated nanodumbbell. The feedback cooling of a levitated nanodumbbell was hindered by the coupling between different modes due to the uncooled rotational motion around its long axis [7]. Because of the symmetry of a nanodumbbell, its free rotation around its long axis does not couple to the optical tweezers directly. Hence, this rotation mode cannot be cooled by a laser beam directly. This uncooled mode can cause heating of other modes.

Because a nanodumbbell (about 150 nm) used in optical levitation experiments so far is much smaller than the wavelength of the trapping laser (typically 1550 nm), the nanodumbbell appears as a simple spot on the optical image of the scattered light [3]. The shape and orientation of the nanodumbbell is not directly observable from the optical image. It is thus often difficult to understand the system and interpret experimental results. It is also time-consuming to pump a vacuum chamber to a high vacuum after a nanodumbbell is captured in air at 1 atm. Only silica nanoparticles have been trapped in a high vacuum successfully, which limits the application of the system. Other nanoparticles such as diamond have not been optically trapped in a high vacuum due to laser heating. For certain applications, the torque signal to be detected is static. Thus lock-in amplifier cannot be used to improve the sensitivity of the system.

**Advances in Science and Technology to Meet Challenges**

Currently, the geometry of a levitated nanodumbbell is confirmed indirectly by the asymmetry of its air damping rates. Direct visualization of the shape and orientation of a levitated nanoparticle will provide more information about the system, especially when the nanoparticle has an irregular shape. Many super-resolution optical imaging techniques beyond the diffraction limit have been developed for biological imaging. Some of these techniques, such as the structured light illumination, will also be suitable for imaging the shape and orientation of an optically levitated nanoparticle. Currently, a nanodumbbell is first captured in the air at 1 atm, and then pumped to vacuum. This process is slow, especially if a pressure below $10^{-7}$ torr is required. Bake out is often required to achieve a pressure below $10^{-7}$ torr, which is time-consuming and may cause the loss of a levitated nanoparticle. Capturing a nanodumbbell in a vacuum directly will speed up such a process. A cryogenic environment can also reduce the air pressure and heating effects due to residual air molecules. Only silica nanodumbbells have been trapped in a high vacuum so far. To trap other functionalized nanoparticles, internal temperature cooling or a combination of optical tweezers and an ion trap will help to trap other nonspherical particles in a high vacuum.

Quantum ground state cooling is required for many applications. To avoid the cooling limitation due to the free rotation of a nanodumbbell around its long axis, we can use more asymmetric nanoparticles, such as a nanodumbbell with a small nanoparticle on its side. This can break the rotation symmetry of a perfect nanodumbbell and couple all 6 degrees of freedom to the optical tweezers. To achieve 6D ground state cooling of a levitated nanodumbbell, we can use cavity coherent scattering cooling with elliptically polarized and shaped optical tweezers [9]. Coherent scattering strongly enhances the interaction by using the high-intensity trapping laser for cooling as well. We will need to trap a nanoparticle near a surface for studying surface interactions. The reflected light and the incoming light will interfere and form a standing well near the surface, which has a potential minimum about ¼ wavelength away from a high-refractive-index surface. It will automatically determine the position of the nanoparticle and the surface. We may also enhance the nanodumbbell-photon interaction by trapping a nanoparticle using a nanoscale photonic crystal cavity [10].

**Concluding Remarks**

Silica nanodumbbells have been optically levitated in a high vacuum [3,7,8]. The dynamics of their motion are highly nonlinear. Circularly polarized optical tweezers have driven them to rotate at GHz frequencies. A nanodumbbell levitated in linearly polarized optical tweezers will be an analogy of the famous Cavendish torsion balance. Optically levitated nanodumbbells will be excellent torque detectors for studying Casimir torque and quantum vacuum frictions near a surface. Once cooled to the quantum regime, they can be used to investigate rotational quantum mechanics in unprecedented regimes. Rotational quantum superposition states of nanodumbbells can be used to study continuous spontaneous localization. They may also be used to search for dark matter candidates.

**Acknowledgements**

T. L. is supported by the Office of Naval Research under Grant No. N00014-18-1-2371, and NSF under Grant No. PHY-2110591.

## 27 — Optical binding

*Oto Brzobohatý, Stephen H. Simpson, Pavel Zemánek*

Czech Academy of Sciences, Institute of Scientific Instruments, Královopolská 147, 612 64 Brno, Czech Republic

**Status**

Multiple scattering of light induces structured interactions, or optical binding forces, between collections of small particles (**Figure 38**). These forces depend both on the morphology of the optical field and on the shape, composition and distribution of the particles comprising the ensemble. When wide, loosely focused beams are used, many particles can be encompassed and, for modest laser power, the binding forces are sufficient to promote various modes of self-organisation within moderately sized ensembles. For symmetric configurations of dipolar particles, binding interactions appear to be conservative with systems satisfying the Boltzmann distribution. For lower symmetry systems, the underlying non-conservative characteristics of optical forces start to express themselves, giving rise to various non-equilibrium effects.

Optical binding of a pair of dielectric microspheres was first observed at an interface, then in an isotropic fluid [1]. Although the form of the interaction has been exhaustively analysed for simple, highly symmetric cases, larger and more complex systems are less tractable. Nevertheless, a range of structures have been synthesised that can be dynamically reconfigured in one or more [2] dimensions by holographically modifying the incident field. Due to the non-conservative nature of the binding interaction, optically self-organised structures can execute a range of complex, often periodic motions. Examples include sorting and separation of microparticles [3], the generation of self-sustained oscillations in optically bound chains [4] and complex, coordinated rotations of collections of spherical or elongated [5] particles in beams carrying angular momentum.

Shifting from overdamped, aqueous environments to vacuum opens multiple new frontiers. While the optical control of single particles in vacuum (i.e., optomechanics) continues to break new ground in the fields of weak force sensing and quantum technologies, collective optomechanics, involving optically interacting particles, remains relatively unexplored. The most common experimental systems rely on complex fabrication methods (e.g., coupled membranes, photonic nanobeams, toroidal microresonantors). In contrast, systems based on optical levitation of discrete particles constitute an entirely new type of light-matter interface, which removes many of the constraints of the established approaches and offers easy tunability of all system parameters. To date, only a few studies of optical binding in vacuum have been reported. Bykov et al. [6] reported long-range optical binding mediated by intermodal scattering within an evacuated hollow-core photonic crystal fibre, Arita et al. [7] showed binding between rotating birefringent microspheres, Svak et al. [8] exhibited tuneable longitudinal binding between multiple particles in cross-polarized counter-propagating laser beams and characterised their interaction, modal behaviour and nonlinearity, and Riesel at al. [9] studied the tunability of mutual optical coupling between two nanoparticles levitated in separated optical traps.

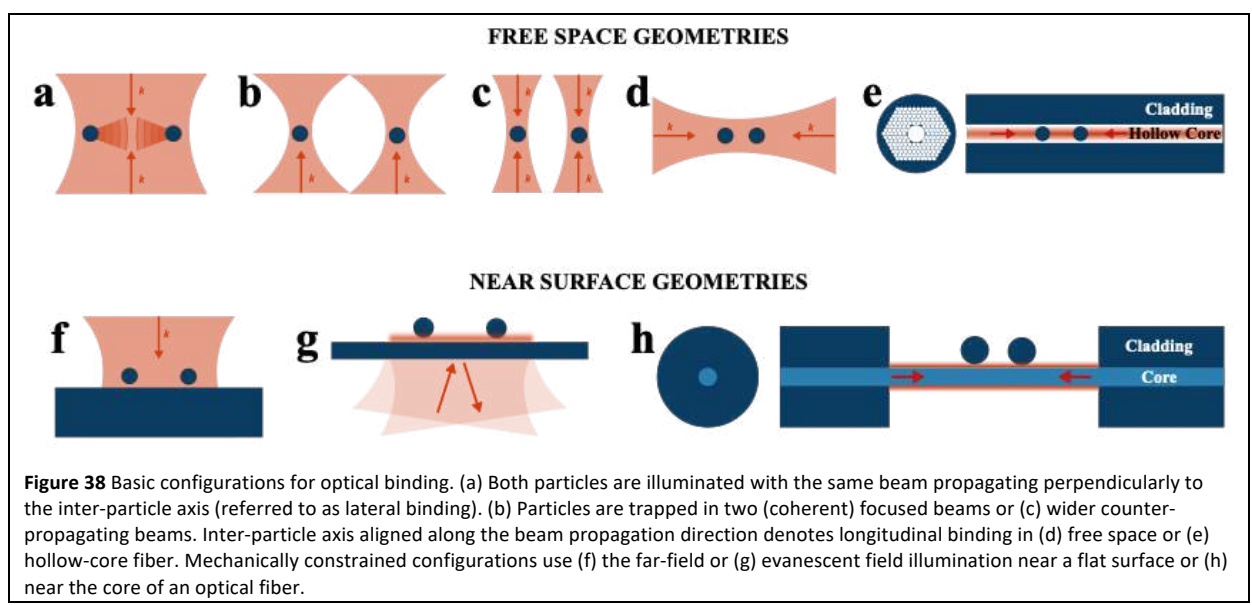

**Figure 38** Basic configurations for optical binding. (a) Both particles are illuminated with the same beam propagating perpendicularly to the inter-particle axis (referred to as lateral binding). (b) Particles are trapped in two (coherent) focused beams or (c) wider counter-propagating beams. Inter-particle axis aligned along the beam propagation direction denotes longitudinal binding in (d) free space or (e) hollow-core fiber. Mechanically constrained configurations use (f) the far-field or (g) evanescent field illumination near a flat surface or (h) near the core of an optical fiber.

**Current and Future Challenges**

Optically bound systems present an intriguing combination of scientific challenges in non-equilibrium statistical mechanics, stochastic thermodynamics, macroscopic quantum mechanics and, additionally, in control and information theory. Applications range from the synthesis of novel metamaterials or optical robotics to sensing and quantum technology.

A major challenge for optical self-organization in over-damped systems concerns the connection between the system parameters (e.g., the incident optical field, the shape and composition of the particles) and the probable result of the self-organization process. This relationship is highly non-trivial since, in general, the forces acting on the system are non-conservative so that the Boltzmann distribution will probably not apply. Indeed, the resulting ensemble may not be at equilibrium at all. Although a general theory may never be available, insight may be derived through deeper understanding of optical interaction forces, basic properties of non-equilibrium steady states (especially in terms of entropy production), and transitions between such states [10]. Such concerns are shared with the rapidly growing field of active colloids, and a more detailed understanding could lead to the engineering of novel metamaterials or even complex mobile assemblies performing useful work at the nanoscale.

The same challenges are present in under-damped systems, although the underlying dynamics are further complicated by the presence of inertia, permitting the formation of more general dynamical attractors, including limit cycles or even autonomous chaos (**Figure 39**). Ever present thermal fluctuations provide an opportunity to study the statistical mechanics of numerous exotic dynamical systems from the synchronization of noisy, under-damped oscillators (which could be periodic or chaotic), to the redistribution of energy in the famed Fermi-Pasta-Ulam-Tsingou problem. In addition to fundamental interest, the ability to prepare and control such systems could have applications in sensing.

Taking collective levitational optomechanics into the quantum regime presents the most substantial challenge. The combination of feedback and cavity cooling recently resulted in ground state cooling of a single nanoparticle [11,12]. An analogous approach, applied for example to the separation of two or more particles in appropriately non-linear potential landscapes, could lead to the macroscopic entanglement of multiple degrees of freedom. Finally, we note that collective levitational optomechanics contains the required ingredients to, in principle, allow the preparation of macroscopic classical or quantum time crystals, perhaps functioning as quantum memories or interfacing with other quantum technology.

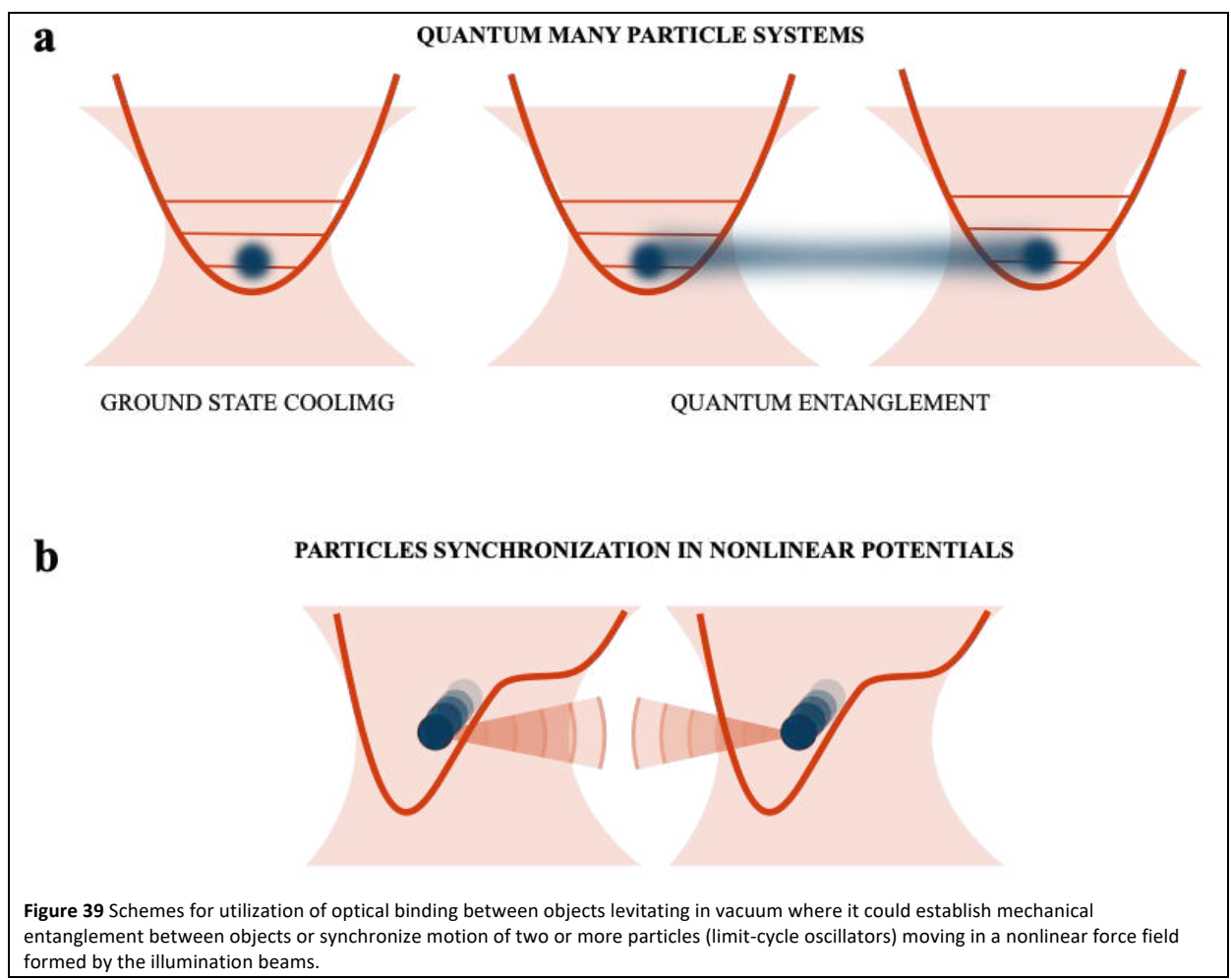

**Figure 39** Schemes for utilization of optical binding between objects levitating in vacuum where it could establish mechanical entanglement between objects or synchronize motion of two or more particles (limit-cycle oscillators) moving in a nonlinear force field formed by the illumination beams.

**Advances in Science and Technology to Meet Challenges**

The current progress in technology of beam shaping via spatial light modulators, amplitude and frequency beam modulation, and overall computer control of the experiments paves the way to the demanding experiments with a single levitating nanoparticle in high vacuum where its mechanical energy is decreased via cavity or feedback cooling schemes to reach the lowers levels of quantum harmonic oscillator. The first experiments reaching the ground state of the quantum harmonic oscillator [11,12] demonstrated the feasibility of this approach. However, synchronized cooling of several levitating nanoparticles is still in its infancy and advanced feedback protocols must be developed for cooling systems with many coupled degrees of freedom, manufacturing, and automatic launching of high quality spherical and non-spherical particles with controlled optical properties to avoid e.g., heating effects, advanced 3D tracking methods for systems with multiple particles and a wide range of time scales.

**Concluding Remarks**

Optical binding forces are an extremely complex and general form of interaction, dependent on the morphology of an external optical field and on the form of the particles comprising the ensemble. In highly symmetric cases, the interaction can appear conservative — allowing the formation of crystalline arrangements. More generally, optical binding is a non-linear, non-conservative, many-body interaction. These characteristics make optically bound systems an ideal platform for studying driven self-organisation of fluctuating structures that may be static or, alternatively, may execute a range of complex motions. In under-damped systems, the dynamical properties are modified by inertia, allowing systems to accumulate momentum, resulting in far richer behaviours which may include limit cycles, quasi-periodic or chaotic motion. In combination, these characteristics make the collective optomechanics of levitating systems an ideal test bed for studying the stochastic thermodynamics of collections of self-sustained periodic or even chaotic oscillators. By exploiting sophisticated feedback methods, developed for single particle optomechanics, these complex, coupled systems may be taken into the quantum regime permitting the synthesis of entangled systems of particles or even macroscopic time crystals.

**Acknowledgements**

This work has been partially supported by the Czech Academy of Sciences Praemium Academiae and Czech Science Foundation (21-19245K)

**INSIGHTS INTO STATISTICAL PHYSICS**

## 28 — Nanothermodynamics and statistical physics

*Felix Ritort*

Small Biosystems Lab, Departament de Física de la Matèria Condensada, Facultat de Física, Universitat de Barcelona, Barcelona, Spain

**Status**

Single-molecule tools, and optical tweezers in particular, have revolutionized the way we interrogate biological matter. By measuring energies on the scale of thermal fluctuations, optical tweezers can monitor individual molecular trajectories and reactions, opening new venues in thermodynamics and kinetics at the nanoscale [1]. With optical tweezers we can now follow the folding of individual nucleic acids and proteins with sub-millisecond time resolution, identify kinetic intermediates and misfolded structures previously hidden to bulk assays, reaching the bar of the 1kcal/mol energy accuracy in molecular thermodynamics (**Figure 40**). Besides investigating intramolecular reactions, optical tweezers have been used to characterize intermolecular interactions for biophysical and biomedical purposes, from protein-DNA and ribonucleotide complexes, to ligand binding (e.g., metal ions, polycations, peptides, dendrimers), and protein–protein interactions. Besides their many applications in biophysics, optical tweezers have found relevant applications to the thermodynamics at the nanoscale or nanothermodynamics. Known as the thermodynamics of small systems [2], the prospective of exerting and measuring tiny forces (in the piconewton scale) on small objects (colloidal particles, molecular systems, electrical devices) over nanometer distances, has spurred an intense research activity in stochastic thermodynamics [3]. Work in the 1990s in the field of nonequilibrium statistical mechanics led to the transient (Evans-Searles) and stationary (Gallavotti-Cohen) fluctuation theorems. These results demonstrated the existence of a symmetry relation for the entropy production (i.e., the heat dissipated to the environment) during a time interval t, $S_t$, in a nonequilibrium process, $P(S_t)/P(-S_t)=\exp(S_t/k_B)$, with $k_B$ Boltzmann's constant. Consequential to these results are the nonequilibrium work and fluctuation relations by Jarzynski and Crooks for isothermal processes. Jarzynski equality permits us to derive free-energy differences ($\Delta G$) from irreversible work (W) measurements by exponentially averaging the work over many repetitions of the experiment, $\Delta G = -k_B T \log \langle\exp(-W/k_B T)\rangle$, with T the temperature of the environment. The importance of these results is grounded on the fact that many molecular transformations can only be performed irreversibly: unfolding and folding melting curves often exhibit hysteresis. In mechanical unzipping experiments (**Figure 40**) the ends of a single DNA, RNA or a protein are mechanically pulled apart by moving an optical trap relative to an immobilized bead or surface (single-trap setup) or a dumbbell (dual-trap setup). By repeatedly unfolding and refolding a molecule, it is possible to extract the folding energy $\Delta G$ by using the Jarzynski and Crooks nonequilibrium work relations. Since their first experimental tests, many applications and extensions have been developed [4].

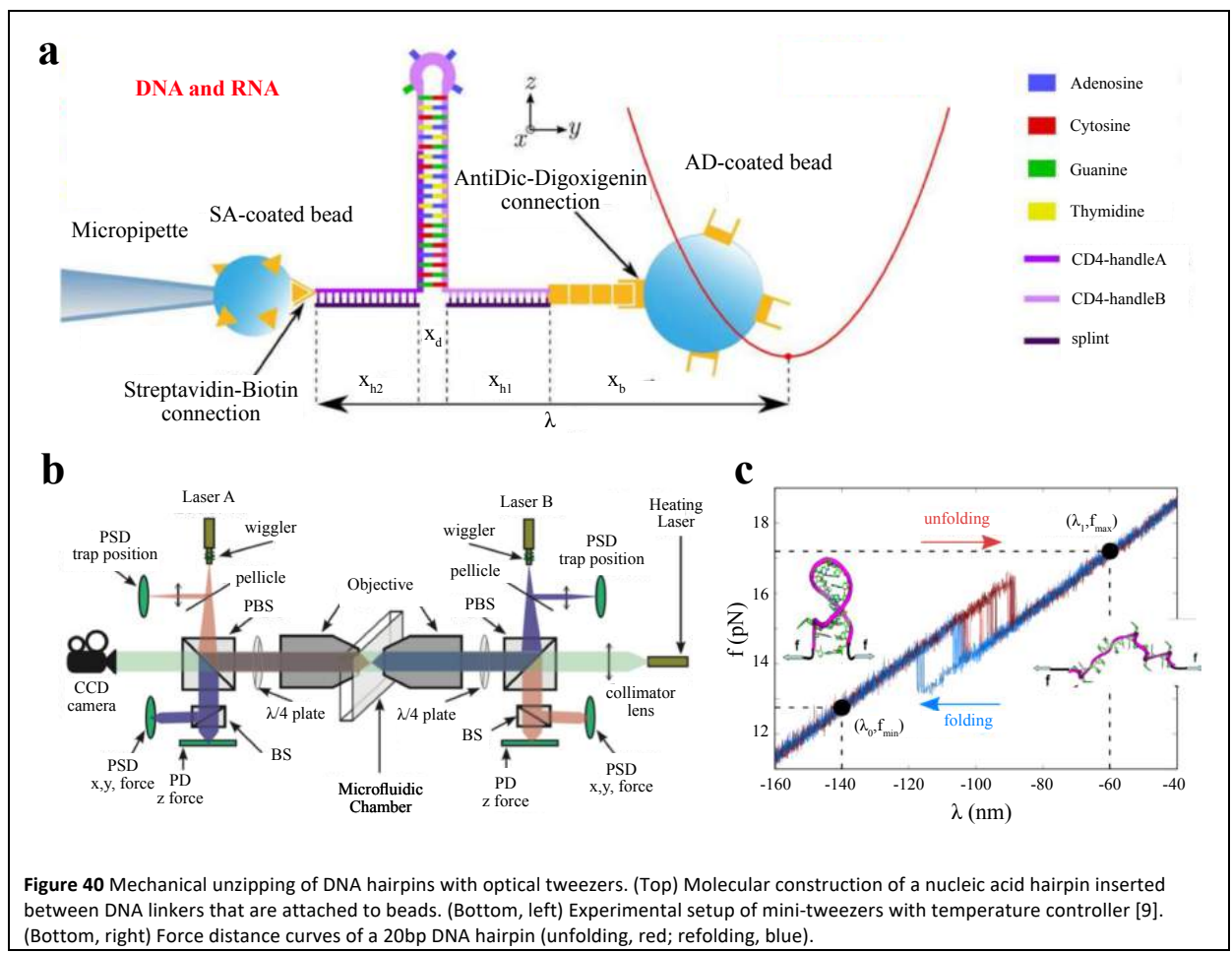

**Figure 40** Mechanical unzipping of DNA hairpins with optical tweezers. (Top) Molecular construction of a nucleic acid hairpin inserted between DNA linkers that are attached to beads. (Bottom, left) Experimental setup of mini-tweezers with temperature controller [9]. (Bottom, right) Force distance curves of a 20bp DNA hairpin (unfolding, red; refolding, blue).

**Current and future challenges**

Thermodynamics, a discipline inherited from the industrial revolution, applies to all scales. Elementary particles, atoms, molecules, living cells, superconductors, human beings, black holes up to the entire universe, all comply with the laws of thermodynamics. As Einstein put it "Classical thermodynamics [...] is the only physical theory of universal content which I am convinced [...] will never be overthrown". Thermodynamics is a conceptual framework encompassing many extraordinary events, from the self-assembly of viral particles to the birth of a cell from another cell. These systems obey thermodynamics, and despite we regularly use them in the lab, they escape our understanding. The second law inequality (i.e., the entropy of the universe keeps increasing) is probably the most striking law in nature, qualitatively distinct from conservation laws (such as energy and momentum). The second law has always amazed scientists, in 1887 Maxwell proposed his famous paradox of a small, smart, and effortless being, that beats the second law by meddling with the molecular motion. Only much later, in the 1960s, Landauer and Bennett understood that information was key for the logical operations of the demon (storage and erasure of bits) giving rise to the thermodynamics of data processing. Together with stochastic thermodynamics, the new field of thermodynamics of information has emerged, a new discipline that puts information, entropy, and work on common grounds [5]. The Landauer limit sets an equivalence between energy an information, 1 bit=$k_B T$ log 2, similar to Joule's mechanical equivalence between work and heat. This makes it possible to experimentally measure information and test feedback fluctuation theorems. The thermodynamic uncertainty relations [6], recently derived by Seifert and colleagues, set a lower bound for the entropy production in nonequilibrium states. The bound is given by the ratio between the (squared) average of a current (e.g., the speed of a molecular motor) and its variance. The key question of experimentally measuring the entropy production, stands out as a primary goal in the field. Besides the average entropy production, the spectral density of entropy fluctuations are key structural features related to the efficiency of molecular machines. Future challenges are the study of motor enzymes (e.g., helicases and polymerases) where chemical reactions are involved and operate out of equilibrium. Although optical tweezers permit us to monitor enzyme's movement and the work exerted by the optical trap, we cannot simultaneously monitor the ATP-fuelling hydrolysis reaction. In this case, the experimental variables needed to determine the entropy production are inaccessible and the accessible information is only partial.

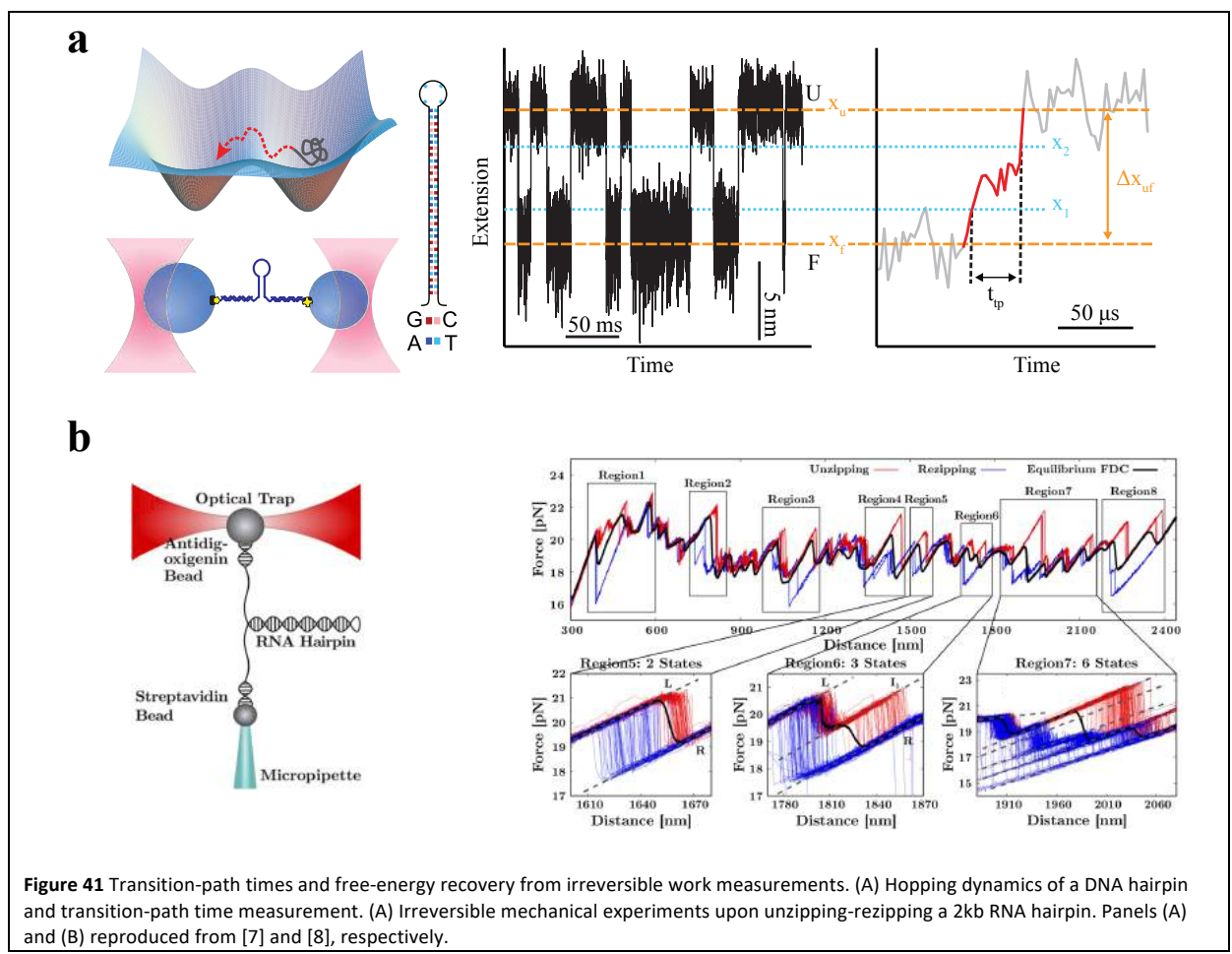

**Figure 41** Transition-path times and free-energy recovery from irreversible work measurements. (A) Hopping dynamics of a DNA hairpin and transition-path time measurement. (A) Irreversible mechanical experiments upon unzipping-rezipping a 2kb RNA hairpin. Panels (A) and (B) reproduced from [7] and [8], respectively.

**Advances in science and technology to meet challenges**

Optical tweezers permit us to measure kinetics [7] and thermodynamics [8] of molecular transformations with unprecedented resolution (**Figure 41**). From the energy of native structures of DNAs, RNAs and proteins to the binding of ligands (e.g. restriction enzymes, peptides, etc..) and the experimental verification of the law of mass action. Nanothermodynamics also requires measuring enthalpies ($\Delta H$) and entropies ($\Delta S$), which are strongly compensated for weak biomolecular interactions. The sole knowledge of $\Delta G$ does not tell us anything about the $\Delta H$ and $\Delta S$ contributions, which are typically much larger. Upon DNA hybridisation, the two strands of the DNA helix form specific hydrogen bonds between complementary bases (A-T and G-C). While the average $\Delta G$ is ≈1-2 kcal/mol per basepair, enthalpies and entropies are roughly ten times larger ($\Delta H$≈8 kcal/mol, $T\Delta S$≈6 kcal/mol). Key to nanothermodynamics is the heat capacity change ($\Delta C_p$) in a biomolecular reaction. $\Delta C_p$ tells us about changes in configurational entropy, a relevant quantity in protein folding and self-assembly reactions. The recent development of temperature-controlled optical tweezers has represented an important step in this direction which has permitted to experimentally demonstrate that protein barnase folds in a funnel-likeenergy landscape where the transition state forms a dry molten-globule [9].

Instrumental developments in optical tweezers should narrow the gap between the complexity of molecular systems and the limited number of measured variables. Future technological developments aim to combine optical tweezers with other techniques such as fluorescence (fleezers) and light polarization measurements (angular optical traps) [10] or even electrical measurements (e.g., nanopore translocation [11]). These developments aim to bring light onto additional reaction coordinates beyond molecular extension that will greatly contribute to expand our knowledge of energy and information. The possibility to simultaneously track several experimental variables permit us to monitor the evolution of multiple reaction coordinates in an enlarged multi-dimensional energy landscape, a crucial step to measure entropy production in nonequilibrium systems. Finally, one might also ask about a bottom-up approach for thermodynamics, i.e., whether developments in stochastic thermodynamics may have impact on macroscopic systems. An exciting venue of research might extend thermodynamics of information to particle assemblies and macroscopic objects [12], where trajectories of subsystem variables lead to entropy fluctuations, for example in bristle bots and kilobots (small but macroscopic robots) and even in animal behaviour (e.g. flocks of birds). These and other exciting questions wait for the future.

**Concluding remarks**

Nanothermodynamics investigates energy and information processes in nanoscale systems where energies involved are comparable to the thermal noise. Biomolecular systems are of particular interest because their thermodynamic behaviour is determined by evolutionary biological processes operating at multiple scales: from the molecular scale (i.e., the genotype) to the cellular scale and higher (i.e., the phenotype). Understanding those energy processes operating at the molecular scale in living matter paves the way to unifying nonequilibrium physics and biology within the framework of information theory, evolution, and life. By measuring energy and information, optical tweezers combined with other techniques, permit us accurate and better measurements of entropy production and information, maybe in many particle and macroscopic systems too. A key feature of biomolecular systems is the contrast between their small nanometric dimensions and their complexity. As put by François Jacob long ago, evolutionary processes are akin to bricolage (do-it-yourself) where nature uses everything that is available to improve what does already exist. It is for this reason that evolution operates at all scales and times, giving biological structures that make up living beings their unique and marvelous complexity.

**Acknowledgements**

I wish to acknowledge Icrea Academia prizes 2018 (Catalan Government) and the Spanish Research Council Grant No PID2019-111148GB-I00 for financial support.

## 29 — Experimental nonequilibrium statistical physics with optical tweezers

*Yael Roichman*
Tel-Aviv University, Israel

**Status**

Finding the fundamental rules governing the state and dynamics of systems far from thermal equilibrium remains an open challenge in physics. The leading approach to address this challenge, in a universal way, is to find generalizations of the theoretical framework of thermodynamics and statistical mechanics that apply to such systems. These include attempts to extend the definition of state functions, e.g., temperature and pressure, defining thermodynamic quantities as averages on a single phase-space trajectory level, and looking at large deviations of a system from its average state. All of these approaches rely on our understanding of the fluctuation spectrum of the system, which is generally unknown far from thermal equilibrium.

Onsager was the first to realize that very close to thermal equilibrium, and in contact with an ideal heat reservoir, temperature relates the system's response to a small perturbation to its thermal relaxation from such a fluctuation. This result is known as the fluctuation-dissipation theorem. Intuitively, this means that the same physical processes govern both the random fluctuations of the system and its dissipation. Focusing on the fluctuations in driven systems, a set of fluctuation theorems connecting the probability of thermodynamic favourable paths in phase-space to that of unfavourable ones was recently discovered and unified [1]. In the past three decades, optically driven colloidal particles were used to validate such theoretical predictions for systems out of equilibrium. These also include the generalized forms of the fundamental fluctuation-dissipation theorem [2] and a range of fluctuation theorems [3-5].

To study the spectrum of fluctuations experimentally requires mesoscopic small systems in which the probability of large fluctuations is finite. It is also beneficial that their dynamics be accessible on the single-particle level. These necessities led to the use of optically driven colloidal particles as a paradigmatic model system for the field [1]. The characteristic timescale for thermal relaxation of a suspended colloidal particle is several orders of magnitude faster than the typical driving timescale of an optically driven particle. This timescale separation implies that one can decouple the effect of the slow driving degrees of freedom and the fluctuation spectrum of the bath, a common feature of many experiments done on dilute colloidal suspension (see, for example, **Figure 42** [6], and [2]).

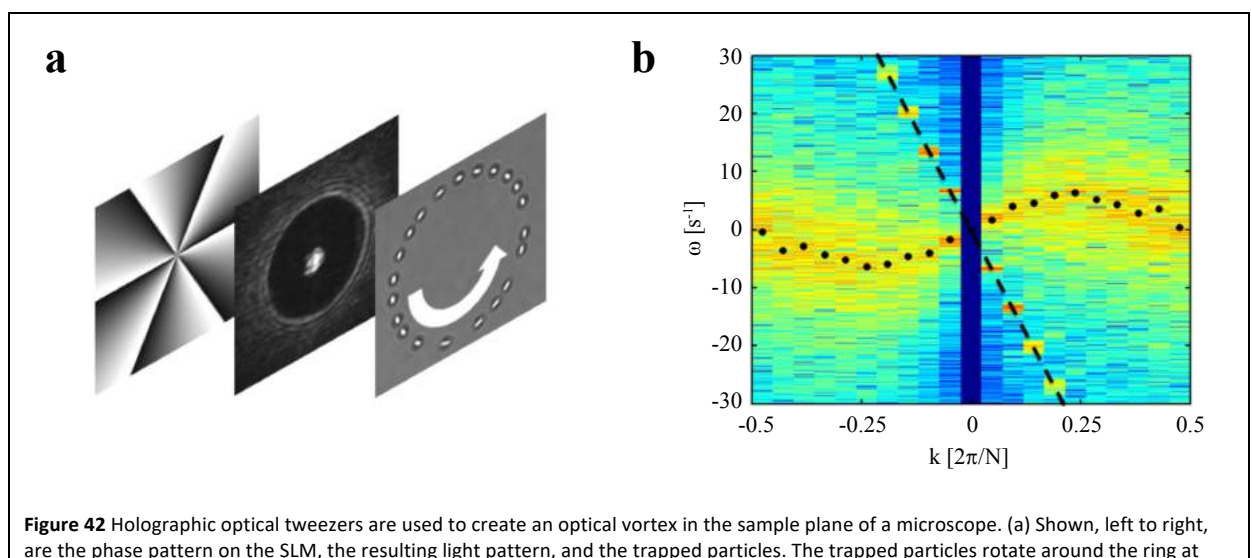

**Figure 42** Holographic optical tweezers are used to create an optical vortex in the sample plane of a microscope. (a) Shown, left to right, are the phase pattern on the SLM, the resulting light pattern, and the trapped particles. The trapped particles rotate around the ring at constant angular velocity while exhibiting phonon-like breathing modes. (b) The measured dispersion relation of these vibrations is given by deterministic equations of motion and is uncorrelated to thermal noise. The thermal noise determines the width of each dispersion mode.

**Current and Future Challenges**

The motivation to study nonequilibrium statistical physics is to describe better our world, which is generally out of thermal equilibrium. Natural systems, and especially living matter, are usually subjected to non-conservative forces and currents, they include many interacting entities, and they change with time. Even at a driven dissipative steady state, their spectrum of fluctuations is rarely purely thermal. Therefore, the next challenge in experimental studies of nonequilibrium statistical physics using optical tweezers is to construct experiments that address some of these differences. Optical tweezers and, specifically, computer-controlled optical tweezers based on time-multiplexing beams or holographic technology have been instrumental in offering a versatile range of possible driving protocols. These are essential also for future progress.

The first challenge is to go beyond timescale separation and decipher the effect of coupling between driving and fluctuation spectrum on the state of a system far from thermal equilibrium. Several ways to couple thermal fluctuations and driving in colloidal suspensions have been implemented. One approach to couple driving and fluctuations is to do so by hand. Specifically, to use feedback-controlled driving protocols, in which driving is dictated by the instantaneous state of the system at the time of measurement. This fruitful approach culminated in experimental studies of information machines: processes that convert measured information about a system to extractable work [7] and of stochastic resetting processes [8] (**Figure 43**). Irrespective of which approach is used, the relation between dynamics and fluctuations remains beyond the scope of present understanding.

The effect of interactions on the state of a many-body system is not trivial to calculate, more so far from equilibrium. Using dense colloidal suspensions and computer-controlled optical tweezers affords a direct way to study many-body model systems out of equilibrium. A range of interactions between colloidal particles beyond the ever-present hard-core repulsion and hydrodynamic interactions is available, e.g., by use of magnetic and electric fields. The challenge, however, in conducting experiments with driven many-body systems is double. The first is conceptual; a theoretical description of the system is either unavailable or hard to obtain due to the long-range hydrodynamic interactions between colloidal particles. Second, single-particle tracking of dense regions in the colloidal suspension is challenging and may give ambiguous results.

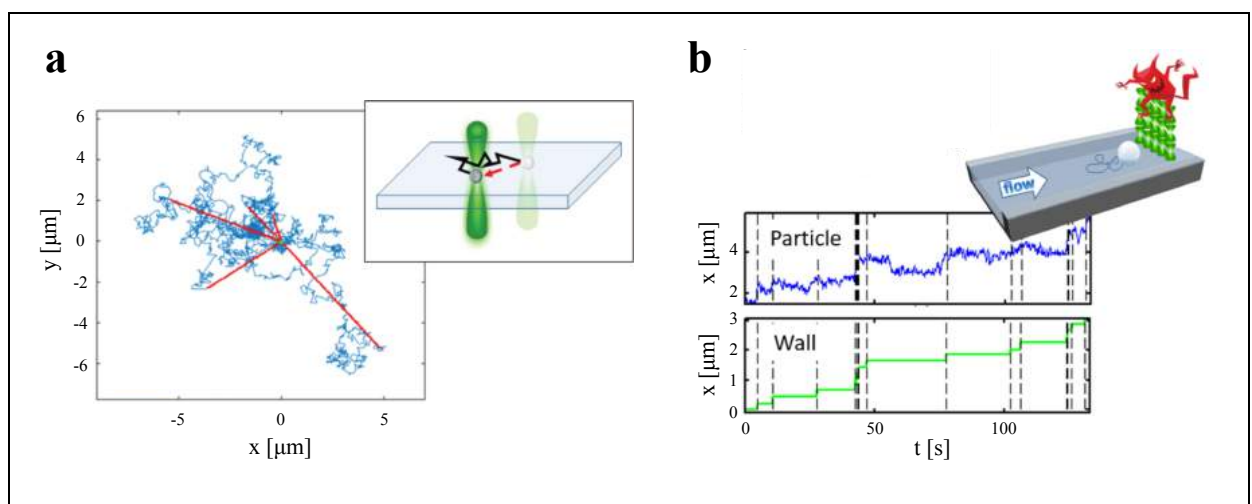

**Figure 43** Feedback and control driving protocols where driving depends on the instantaneous state of the system. (a) Stochastic resetting: Thermal fluctuations determine the trajectory of a diffusing particle in a quasi-2D colloidal suspension. The bead is trapped and returned

to the origin, resetting the system, after some random time. (b) Information engine: A particle diffuses in a quasi-1D channel near an optically repelling wall. Whenever the particle diffuses far enough from the wall, against the flow, the wall is brought slightly closer to the particle.

**Advances in Science and Technology to Meet Challenges**

The technical challenge of accurately tracking many close active and driven particles is a significant obstacle in the future studies of microscopic systems far from thermal equilibrium. One difficulty that arises in this context is distinguishing between very close objects. This is compounded by the fact that images of close particles are distorted in many cases, leading to inaccurate localization. This challenge can be addressed by modifying the colloidal particles, the imaging techniques, or the analysis method. The development of core-shell colloidal particles with dark or low refractive index shells and a visible core is an essential step towards obtaining distinguishable particles and is currently pursued [9]. Other imaging techniques such as holographic imaging enhance distinguishability and improve localization significantly, but require further development. Finally, machine learning for image analysis might provide a way to circumvent the need to localize each particle individually.

The second hurdle in this respect is the ability to track each particle. Specifically, if particles move between consecutive images a larger distance than half the distance between them, it becomes hard to determine which is which. Technically, this can be solved for many-particle systems using fast cameras, decreasing the time the particles move in between images. However, high-speed imaging comes at a price of limited measurement duration. The development of fast cameras that store images continuously for long durations is necessary.

**Concluding Remarks**

The extension of the framework of statistical mechanics to describe nonequilibrium conditions is limited by the lack of experimental observations in controlled and straightforward systems. Optical tweezers have emerged as a leading and ideal setup for such experiments. Taking these experiments closer to reality by coupling driving to fluctuations and looking at many-particle systems will provide crucial observations beyond current theory. These could pave the way toward establishing an extensive theoretical framework for statistical mechanics far from thermal equilibrium.

**Acknowledgements**

This work is partly supported by the Israeli Science Foundation grants No. 998/17 and 385/21, and partly by European Union's Horizon 2020 research and innovation programme (grant agreement No. [101002392]).

## 30 — Light-assisted organization of active colloidal matter

*Valeriia Bobkova*
Institute of Applied Physics, University of Muenster, Germany

*Raphael Wittkowski*
Institute of Theoretical Physics, Center for Soft Nanoscience, University of Muenster, Germany

*Cornelia Denz*
Institute of Applied Physics, University of Muenster, Germany

**Status**

The phenomenon of spontaneous pattern formation in nonlinear matter links together various fields of theoretical and experimental physics, highlighting similarities in externally driven nonlinear self-organization in different scientific branches from physics, chemistry or biology over sociology and economy to astrophysics. The investigation of periodic or quasi-ordered spatial structures arising out of local interactions under the presence of an external driving field above a certain threshold is of high importance for a deeper understanding of many natural and artificial dissipative systems far from equilibrium. Optics plays an important role in this investigation, since light is capable of creating (quasi-)periodic, disordered, or chaotic patterns in configurations similar to active resonators since the feedback inherent to these systems naturally creates a dissipative nature of the interaction with matter, and diffraction supplies spatial coupling. A simple optical single feedback model is described by Firth [1], where a functional nonlinear optical material is coupled to a feedback mirror, thereby creating a wealth of spatio-temporal patterns.

While most of the early experimental implementations of pattern formation have used bulk nonlinear optical materials, cold and hot atomic vapors were also shown to exhibit self-assembly, being driven by the collective interaction of single entities. This feedback driven collective behavior together with the previously shown self-assembly of colloids in an optical tweezer [2] are the solid reasons to expect also colloidal soft matter to exhibit self-organized pattern formation in a single feedback configuration.

In order to realize dissipative self-assembly of colloidal matter far from equilibrium, a colloidal suspension is placed in an optical feedback system. The driving mechanism [3] is schematically shown in **Figure 44**a: a homogeneous Gaussian wave propagates through a thin layer of a colloidal suspension and obtains therefore a random phase distribution, induced by the initially inhomogeneous density of colloids. The wave being reflected by a feedback mirror propagates back to the colloidal layer, such that phase fluctuations are transformed into intensity fluctuations due to diffraction. Each colloidal particle — in the particular experiment these are polystyrene spheres suspended in water — experiences gradient forces, caused by the inhomogeneity of the light field which pulls them into the intensity maxima while scattering forces are cancelled by the counterpropagating geometry. In this way, the optical feedback drives spontaneous pattern formation in a colloidal suspension. Thereby, the simple single feedback configuration enables colloidal pattern formation while paving the way for active control over the self-organization process in colloidal matter.

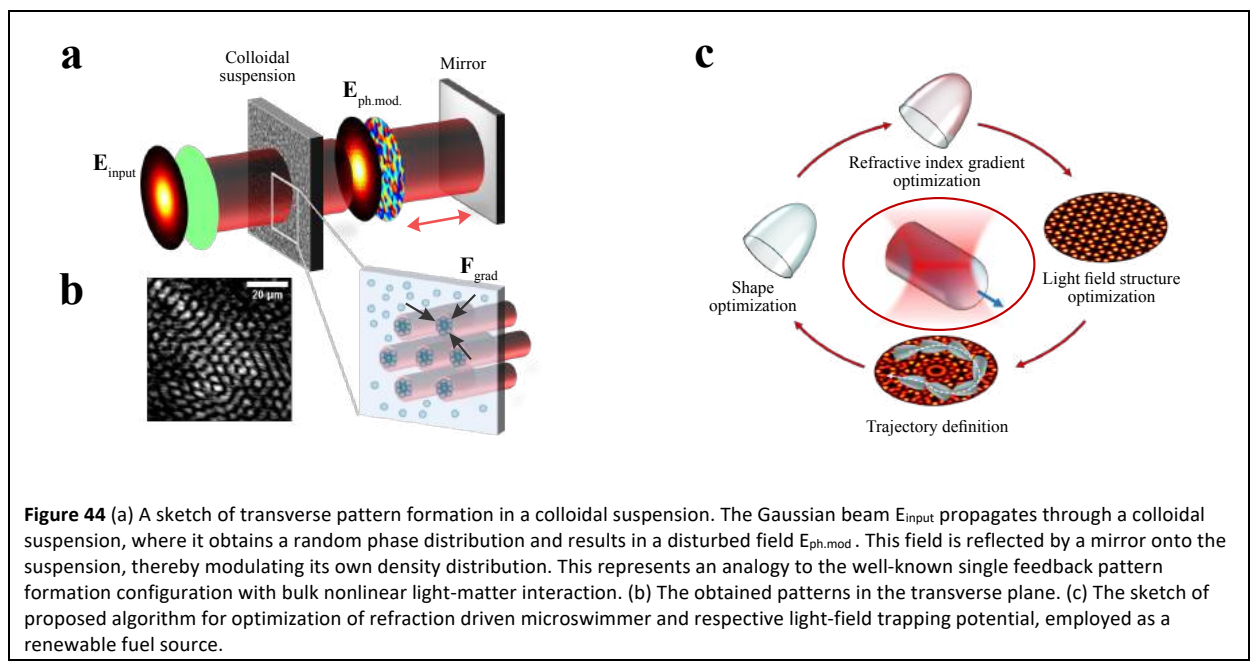

**Figure 44** (a) A sketch of transverse pattern formation in a colloidal suspension. The Gaussian beam $E_{input}$ propagates through a colloidal suspension, where it obtains a random phase distribution and results in a disturbed field $E_{ph.mod}$. This field is reflected by a mirror onto the suspension, thereby modulating its own density distribution. This represents an analogy to the well-known single feedback pattern formation configuration with bulk nonlinear light-matter interaction. (b) The obtained patterns in the transverse plane. (c) The sketch of proposed algorithm for optimization of refraction driven microswimmer and respective light-field trapping potential, employed as a renewable fuel source.

**Current and Future Challenges**

The single-feedback configuration represents a great model system for the investigation of spatial self-assembly of dielectric nanospheres, demonstrating complex collective behavior of soft matter under homogeneous illumination in various geometries. However, for many applications, such as targeted drug delivery or environmental remediation, desired is a defined directional and controllable motion of the colloidal particles. A promising, but challenging approach allowing to fulfil these requirements is the implementation of functional and active soft matter capable of converting external energy into a directional movement. The impressing progress in the field of active microrobots [4,5] in recent years exemplarily demonstrates the wide variation of mechanisms that can be employed to induce self-propulsion of colloidal particles thus creating an "artificial fuel" for the system. The most widely spread propulsion mechanism is based on chemical reactions, which may be catalyzed by light. However, the reaction products are often toxic for a living environment and, moreover, the induced propulsion direction can hardly be controlled. An approach often employed to control the particles' trajectories is their manipulation by a magnetic field [6]. Though a couple of successful experimental implementations of active control over spherical or more complexly shaped microswimmers with a magnetic field was demonstrated, sculpting a free-shaped three-dimensional trajectory of a propelled soft particle is a demanding challenge. Here, again, light comes into play. Employing light as a renewable fuel source for the soft colloidal particle propulsion is on the one hand of particular advantage with respect to biomedical and environmental applications. On the other hand, structuring light to obtain more complex artificial fuel landscapes is already a resolved question [7]. Despite these clear advantages, up to now, light structuring is only moderately employed for the control over the trajectory and speed of single swimming soft particles. In addition, the majority of up to now proposed light-induced micromotors are driven by ultraviolet (UV) or infrared (IR) light sources and thus cannot be tunable in a broad spectral range, limiting the potential application fields to specific light sources.

Microfabrication of such a versatile light-driven particle, which is not wavelength selective, biocompatible and able to demonstrate defined directional motion under illumination is one of the current challenges in the field. Here, employment of optical forces well-known from the optical tweezers approach can be a smart solution for next generation of soft colloidal microswimmers or even microrobots.

**Advances in Science and Technology to Meet Challenges**

As outlined above, a promising approach to tackle the challenge of externally controlled directional soft matter motion is employing light as a driving fuel. Thus, the motion of single entities strongly depends on their shape and refractive index distribution being defined by light scattering and refraction. Recently, light-driven spatial self-organization of symmetric and asymmetric nonspherical colloidal particles was demonstrated experimentally under a Gaussian beam illumination [8]. Varying the shapes and materials of the microswimmers together with spatially structuring illumination landscapes pave the way for observation of novel collective spatio-temporal dynamics in initially stochastic soft matter.

One promising procedure towards arbitrary sculpturing the three-dimensional shape of a colloidal microswimmer is its fabrication by soft laser lithography. At the same time, a two-photon polymerization process allows a continuous change of the refractive index distribution for certain materials enabling laser-writing particles with gradient refractive index distributions. Thus, incident light on that particle acts as a direction-sensitive fuel and controls the transfer of the resulting momentum difference into mechanical motion.

Hence, the motion of a single particle is completely defined by its refraction property and incident light field potential. The progress of the past decades in the field of spatial light modulation provides the control tools in order to tune both properties: refractive index and the shape of a colloidal microswimmer by the means of soft lithography approach, on one hand, and spatially structured light illumination, on the other hand. Moreover, a promising approach for computer-based definition of these control parameters is employment of machine learning algorithms (schematically shown in **Figure 44**c), which already find numerous applications in soft matter research in the recent years [9].

Colloidal microswimmers fabricated by the means of soft lithography can be driven by light with a broad spectral range, as required, for instance, by environmental applications due to increasing conversion efficiency of the broad sunlight spectrum into the particles' motion. Moreover, the employment of refraction driven microrobotics for biomedical applications, such as cell inspecting or drug delivery, takes advantage of the light source tunability and the driving wavelength can be chosen according to the absorption spectrum of the investigated object. Additionally, the substrate used to fabricate a microswimmer employed in the soft lithography process can even be biocompatible as, for instance, natural hydrogels. Thus, the proposed microfabrication approach has a great potential, since it covers a broad range of requirements for emerging applications of artificial soft micro robots easily, and does not request additional chemical functionalization of any type.

**Concluding Remarks**

Self-organization of soft matter driven by light is a broad rapidly developing field with versatile possible applications in medicine, biology, environmental studies and others. Moreover, the fundamental principles of self-assembly demonstrate complex nonlinear dynamics and are of particular theoretical interest. The evident potential of artificial soft-matter systems to mimic natural swarming dynamics drives research on functionalization and control over single microswimmers, making them transfer light energy into the directional movement. Both theoretical and numerical investigations in this field are strongly supported by machine learning approaches and, we believe, within the next decade will contribute to the development of highly efficient bioinspired light responsive materials and discover novel application potentials.

**Acknowledgements**

The authors acknowledge Matthias Rueschenbaum for his help in the pictures' preparation and Thorsten Ackemann for the valuable discussions during the project realization. This work was funded by European Union Horizon 2020 program (ColOpt ITN 721465); German Research Foundation (DFG) — Project-ID 433682494 — SFB 1459; University of Muenster.

## COMBINATION WITH SPECTROSCOPY

### 31 — Raman scattering in (thermo)plasmonic tweezers

*G.V. Pavan Kumar*

Department of Physics, Indian Institute of Science Education and Research, Pune, India

**Status**

Optical spectroscopic interrogation of small systems such as molecules, nanoparticles and microstructures in a solution is relevant to various aspects of science and (bio)analytical technology. Given the small size of such entities, they undergo Brownian motion in a fluid at room temperature. In order to effectively interrogate them over a period of time, there is a necessity to confine these Brownian objects in space and time. Over the past few decades, optical trapping and tweezing methods have been extensively utilized to trap and probe micron-scale object in a fluid with considerable success. Such tweezing platforms have also been integrated with optical spectroscopy methods including florescence, Raman scattering and some non-linear optical methods.

Particularly, Raman scattering, in the form of Raman Optical Tweezer has been utilized to study a variety of trapped systems including biological cells and micro-structures. Although Raman scattering is a powerful spectroscopy tool utilized to capture the chemical fingerprint of molecules and materials, it still suffers from low optical cross-sections compared to Rayleigh scattering and fluorescence. Surface enhanced Raman scattering (SERS) has emerged as a powerful tool to detect single-molecules in a solution [1,2,3] wherein molecules interact with metallic nanostructures. The enhanced Raman signature is predominantly facilitated by the electromagnetic enhancement mechanism, in which plasmonic fields around the metallic nanostructures play a critical role. This makes SERS a chemically-specific, single molecule spectroscopy tool at sub-wavelength scale.

There is an imperative to combine single-molecule Raman scattering with optical trapping, tweezing and pulling [4] methods especially at micro- and nanoscale. Such integration not only expands the toolbox of solution-phase single molecule optical spectroscopy methods, but also can be harnessed for micro- and nano-fluidic applications where temporary confinement, aggregation and release of molecules and nanostructures can alleviate the problem of clogging of fluidic channels. This has implications in clinical diagnostics and *in-vivo* probing of molecules and nanostructures, where spatial, temporal and spectral evolution of these entities can be probed concomitantly.

This has motivated researchers to combine nano-optical tweezer methods with single-molecule Raman microscopy methods, and to this end emerging platforms based on plasmonic and dielectric nanophotonic elements have been explored to enhance and control force fields, along with electric and optothermal fields.

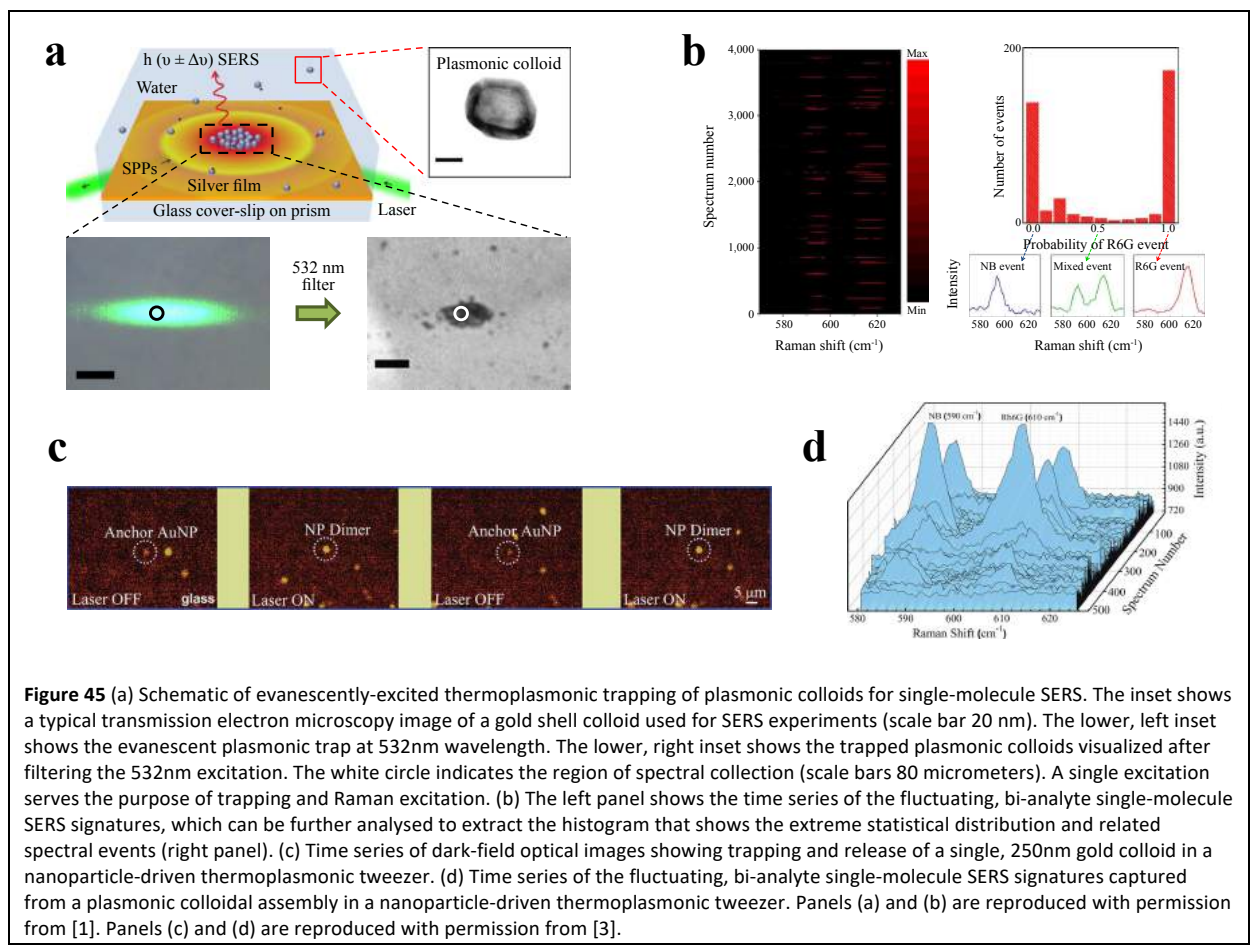

**Figure 45** (a) Schematic of evanescently-excited thermoplasmonic trapping of plasmonic colloids for single-molecule SERS. The inset shows a typical transmission electron microscopy image of a gold shell colloid used for SERS experiments (scale bar 20 nm). The lower, left inset shows the evanescent plasmonic trap at 532nm wavelength. The lower, right inset shows the trapped plasmonic colloids visualized after filtering the 532nm excitation. The white circle indicates the region of spectral collection (scale bars 80 micrometers). A single excitation serves the purpose of trapping and Raman excitation. (b) The left panel shows the time series of the fluctuating, bi-analyte single-molecule SERS signatures, which can be further analysed to extract the histogram that shows the extreme statistical distribution and related spectral events (right panel). (c) Time series of dark-field optical images showing trapping and release of a single, 250nm gold colloid in a nanoparticle-driven thermoplasmonic tweezer. (d) Time series of the fluctuating, bi-analyte single-molecule SERS signatures captured from a plasmonic colloidal assembly in a nanoparticle-driven thermoplasmonic tweezer. Panels (a) and (b) are reproduced with permission from [1]. Panels (c) and (d) are reproduced with permission from [3].

## Current and Future Challenges

Integrating nanophotonic tweezer platforms with Raman microscopy methods brings in new challenges and opportunities. Below we note specific question and aspects from various viewpoints:

*Optical trapping viewpoint:* What are the optimized tweezer configurations for Raman scattering from individual nanostructures and single-molecules? Can we utilize a single laser beam to concomitantly trap and perform Raman spectroscopy? If yes, what are the optimized illumination configurations? What are the advantages of using evanescent-wave Raman tweezers? Can they be extrapolated to optical fibre platforms for single-molecule detection? These questions are yet to addressed in detail, and combining optical tweezer toolbox with Raman spectroscopy can bring in new perspective and challenges.

*Molecular dynamics viewpoint:* What information can be obtained by studying Raman spectral evolution of macromolecules such as proteins and DNA in a nano-optical tweezers? Can we effectively probe the confirmation changes of individual biomolecules under nano-optical Raman tweezer confinement? Recently, plasmonic nano-pore based single molecule spectroscopy has gained prominence. Specifically, the extraordinary acoustic spectroscopy method of Gordon and co-workers [5] has opened a new direction to probe acoustic vibration (0.1 to 10 cm$^{-1}$) of macromolecules in a trap. Understanding spectral evolution of complex biological systems and their interactions in such a platform will add new insights.

*Raman spectroscopy viewpoint:* Traditionally, Raman spectroscopy in tweezer platforms probe spontaneous vibrational energy transitions of molecules and materials. Apart from this, non-linear optical methods such as coherent Stokes, and anti-Stokes Raman scattering and stimulated Raman scattering can add great value to interrogate molecules and nanostructures. Integrating these nonlinear methods with nano-optical tweezer is yet to explored in considerable detail.

*Plasmonics/Nanophotonics viewpoint:* Plasmonic and dielectric meta-surfaces and metamaterials have opened new avenues to control light at sub-wavelength scales. Adapting these materials for nano-optical Raman tweezing brings new opportunities where parameters such as optical phase, amplitude, polarization and spectral signatures can be tuned to control optical forces and Raman spectral enhancements. In such a scenario, design and development of meta-surfaces requires dual optimization of force fields and spectral enhancement features, and this will be an interesting direction for research in computation electromagnetism, where machine learning methods can be implemented.

*Optothermal interaction viewpoint:* Excitation of molecules and materials, even in a solution phase, leads to optical heating. This heating effect is pronounced when a wavelength-resonant nanostructure (such as plasmonic nanoparticle) is used for co-trapping and field enhancement. Such optothermal fields, if optimized, can be used as an advantage in nano-optical tweezers. An example of such a platform is thermoplasmonic tweezers [1,3], in which optothermal fields around a plasmonic nanostructure is harnessed to trap and manipulate nano-colloids and spectrally interrogate molecules in the trap. Attention is needed to address a variety of questions such as: How do (bio)molecules interact with such an optothermal field in a trap? What are the new opportunities of opto-thermophoretic interaction with molecules and materials in such a trap?

We wish to emphasize that the above viewpoints are not exhaustive, and there are many interesting questions and challenges that we foresee as this field evolves.

**Advances in Science and Technology to Meet Challenges**
Below we discuss some relevant developments in nano-optical tweezers that has been or can be adapted to perform single molecule Raman micro-spectroscopy.

*Single-molecule SERS tweezers:* Evanescent-wave thermoplasmonic tweezer with single-molecule detection sensitivity has been realized [1] for large scale plasmonic colloidal assembly (**Figure 45**a). In this study, a plasmonic metal film is excited in evanescent-wave configuration to trap millimetre sized assembly of plasmonic colloids. These assemblies are reversible, and can facilitate multiple plasmonic hot-spots for single-molecule bi-analyte SERS detection and analysis (**Figure 45**b). A relevant aspect is that such an assembly can also be used for remote trapping of plasmonic colloids.

Recently, single-nanoparticle driven thermoplasmonic Raman tweezers with single molecule detection sensitivity has been realized [3]. The motivation of such a platform is not only to trap and probe single plasmonic colloid (**Figure 45**c), but also to perform single molecule SERS (**Figure 45**d) in a small colloidal assembly driven by a single gold nanoparticle. This technique can be harnessed to trap and study the dynamic spectra of individual entities such as biomolecules and nanoparticles.

Examples of Optothermal tweezers which can be harnessed as single-molecule Raman tweezers:

- Käll and coworkers [6] were one of the first to show microfluidic optical aggregation of plasmonic nanoparticles, and resulting SERS from those structures in a trap. Their study showed the ability of SERS traps to be adapted to microfluidic platforms.
- Gordon and co-workers [5] devised a double nanohole based plasmonic tweezer platform that can probe vibrational dynamics of nano-objects in acoustic phonon regime. The

- significance of this work stems from the fact that Raman Stokes shifts as small as 0.1 cm-1 can be probed using this method.
- Cichos and co-workers [7] have realized thermophoretic tweezers that can confine single amyloid fibrils for a period of more than 60 minutes. They also track the evolution of individual fibrils and measure its rotation diffusion constants.
- Zhang and co-workers [8] have extensively studied opto-thermoelectric effects to capture and probe nanoscale entities and their adaptation of antenna-based structures hold promise for resonance-enhanced SERS trapping studies.
- Wenger and co-workers [9] have utilized so called "antenna-in-a-box" plasmonic tweezer platform to trap and study single quantum objects. This innovation can potentially be adapted to Raman trapping studies of individual biomolecules.
- Ndukaife and co-workers [10] have showcased the ability to trap and probe individual biomolecules using opto-thermo-electrohydrodynamic tweezers. This has opened up new abilities to manipulate molecular systems and can be useful for Raman trapping studies.
- Gucciardi and co-workers [11] have utilized Raman tweezers to trap and interrogate micro- and nano-plastics in sea water. This is one of the most important applications of Raman nano-tweezers, and has direct impact in water pollution monitoring and control.

The above list is no way comprehensive, but gives an overview of what has been achieved and what can be possibly adapted to study single-molecules in a Raman nano-optical tweezer.

**Concluding Remarks**

Combining Raman spectroscopy with optical tweezing methods based on nanophotonic platforms provides multiple opportunities to probe molecules and nanostructures down to a single copy limit. Such an endeavour has impact not only in probing fundamental questions in chemical and optical physics, but also in applications such as *in-situ*, *in-vitro* and *in-vivo* bio-analysis and pollution monitoring. With the development of scanning probe methods such as tip enhanced Raman scattering, optical nano-Raman tweezing methods and their limits can be pushed to study intra-molecular vibrations in a trap. Such innovations will indeed open up uncharted research directions for the future.

**Acknowledgements**

GVPK acknowledges Swarnajayanti fellowship grant (DST/SJF/PSA02/2017-18) from DST, India.

## 32 — Spectroscopic tweezers for micro- and nanoplastics detection

*Antonino Foti, Maria Grazia Donato, Pietro G. Gucciardi*
Consiglio Nazionale delle Ricerche, Messina, Italy

**Status**

Optical tweezers (OT) enable trapping and manipulation of micro and nanoparticles dispersed in fluids, by exploiting the tiny forces exerted on matter by tightly focussed laser beams [1]. The first demonstration of OT dates back to 1970 by A. Ashkin, Nobel prize in 2018. Raman analysis in an optical trap (Raman Tweezers, RT) was first demonstrated Ajito et al. in 2002 [2], who showed spectra of 40nm PS spheres. Hybrid instruments combining OT with fluorescence and Raman spectroscopy are becoming increasingly popular, as they enable measurements of size and composition in a contact-less fashion. A variety of spectroscopic optical tweezers (SOT) were implemented, integrating both single-trap and dual-trap configurations with confocal microscopy, total internal reflection, stimulated emission depletion microscopy, coherent anti-Stokes and Surface Enhanced Raman Spectroscopy (SERS) [1,3].

RTs are gaining attention as a unique analytical tool for environmental sciences and food analysis, capable to detect and chemically identify micro- and nano-plastics (MNPs) [4-7]. Plastic items subject to atmospheric weathering do fragment into small debris called microplastics (size < 5 mm), and then further down to small microplastics (< 20 µm) and nanoplastics (< 1 µm). Such particles are nowadays almost ubiquitous in water ecosystems, soil and air, from urban to remote areas [8]. Their spread into the alimentary chains is observed in food items (seafood, fruits, and beverages), and represent a potential threat for human health [9]. Occurrence data are, however, limited (for microplastics) or lacking (for nanoplastics), due to technological gaps in the detection and identification of sub-10 µm particles [10]. RTs, with their ability to measure the size and chemical composition of individual particles, in complex fluids and down to the nanoscale, feature peculiar advantages with respect to standard SEM/EDX, micro-FT-IR and micro-Raman.

Future technological advancements of SOT are expected to provide affordable and practical solutions to some of the major knowledge gaps for understanding the fate of MNPs in the environment and food. These include chemical identification and quantification of size, mass and number of particles; detection in complex media, including liquid digested samples; more sensitive light-based methods for fast nanoplastics identification down to 10 nm level; portable equipment for automated and intelligent analysis of MNPs in batch volumes; integrated strategies for the concentration of small particles and detection surface contaminants; methods to monitor the fragmentation of individual MNPs under weathering and digestion conditions.

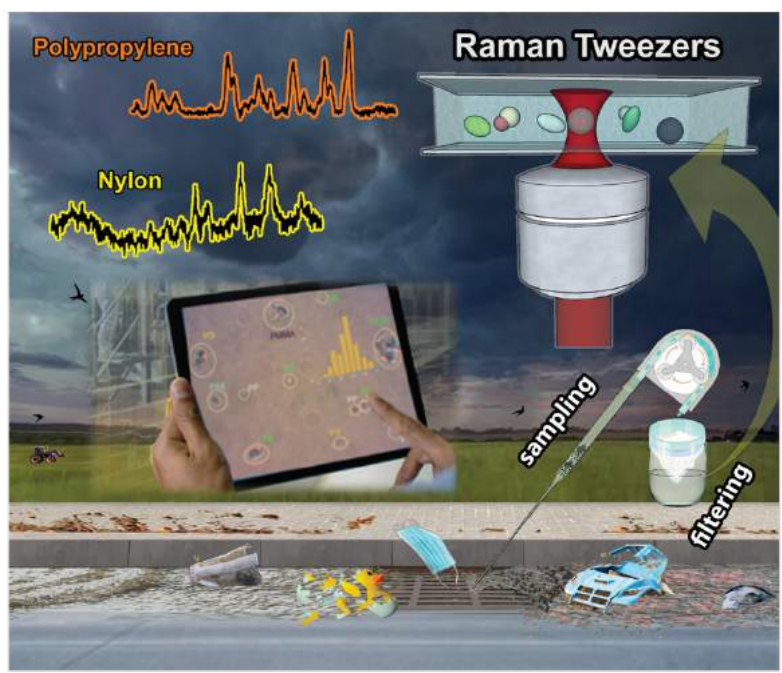

**Figure 46** Field-deployable spectroscopic optical tweezers will be designed to be compact and integrated with fluidic circuits capable to extract samples from the environment, filtrate and concentrate particles, prior to multi-spectroscopic analysis for plastic particles classification by size and materials.

**Current and Future Challenges**

In order to meet the methodological demands in environmental & food analysis (EFA), next generation SOT need to be compact, sensitive, fast, versatile and deployable on the field. Challenges will change and become harder as we move towards the nanoscale, and include:

- *Quantification of particles* by size, mass and material poses difficulties at different levels. Brownian diffusion of MNPs requires procedures to avoid re-counting. Ideally, particles should be driven in a confined stream to SOT, be identified and measured one-by-one, and then be (optically) disposed, or better, sorted according to their nature. MNPs weight from sub-femtograms (for 100 nm particles) to nanograms (10 µm diameter). Mass determination by optical means is a formidable challenge that should be afforded in liquid environment.
- *Multispectral analysis*. Detection and sizing of sub-200 nm particles is demanding because of the low visibility of dielectric particles and diffraction-limited resolution. Fluorescence-Tweezers, combined with far-field nanoscopy [11], is a promising solution to measure dye-stained nanoplastics [12] optically trapped in transparent media or in digested samples crowded of organic residues with sub-diffraction resolution. Faster, label-free classification tool could be obtained implementing SOT with deep-UV laser fluorescence, coherent anti-Stokes Raman, stimulated Raman or near-infrared absorption (780 – 2500 nm) spectroscopy [13].
- *Enhanced sensitivity* will be needed to tailor MNPs detection towards the 10 nm scale. Plasmonic optical tweezers (POT), by exploiting the enhanced, spatially confined near-fields at the surface of plasmonic nanostructures [14], can be used for trapping very small particles, for thermophoretic concentration and SERS. Integrating POT with SERS is a promising solution towards high sensitivity detection of surface contaminants.

- *Field-deployable setups* will require thoughtful consideration of system design trade-off, integrating fluidic circuitry for samples extraction (e.g., from ponds, rivers), multistage filtration (removal of debris and living organisms > 20 µm) and pre-concentration (real samples can be highly diluted). A major challenge is the design and coupling of microfluidic circuitries to concentrate particles and drive the stream towards the trapping and sensing area. Dielectrophoretic and acoustophoretic techniques integrated in microfluidic chips could do the job. Compact optomechanical components will be needed, such as lasers, spectrometers, light detectors, motorized and piezo stages for positioning. Accurate spectral assignment and large volumes of data (thousands of spectra) will pose challenges requiring artificial intelligence tools.
- *Particle's fragmentation*: Studying the fragmentation of individual (trapped) particles offers unique opportunities to understand the degradation paths of MNPs. Challenges involve the design of OT with sample compartments that emulate photo-oxidation, thermal/mechanical stress and digestive erosion, integrated with sensitive particle's size determination tools.

**Advances in Science and Technology to Meet Challenges**

As SOT are based on photonic, (micro)fluidic, nano-engineered components and materials, developments in the field will benefit from technological advances in the design of photonic compact components, a smart integration among optomechanical and fluidic parts, addition of efficient nanoscale elements.

To afford the challenges, SOT will likely exploit multiple laser beams for trapping and analysis, and be designed around complex lab-on-chip architectures. These will host input channels (for sample loading but also to add chemicals and/or enzymes and simulate digestion processes), followed by compartments for trapping and spectroscopic analysis, and output channels for selective disposal/sorting (e.g., by material and shape). Advances in microfluidics materials and design will be needed to fabricate compartments resistant to strongly acidic/basic environments, thin enough to provide access to high NA microscope objectives, from both the top and the bottom, that permit imaging, trapping, position tracking and spectroscopy. Compartments need to be properly coupled to fluidic channels large enough to grant laminar flows (Re < 2000) and process volumes at speeds of 1 L/h. Compact, multi-axis piezoelectric stages will be required for precise MNPs positioning under the OT. Integration of ultrasonic transducers would enable particle concentration by acoustophoresis.

Advances in lab-on-chip technology will embed nanostructured plasmonic and metamaterials surfaces in fluidic circuits, enabling POT platforms for nanocontrol of small dielectric particles, thermoplasmonic concentration and plasmon enhanced spectroscopy. New methodologies will be needed for SERS detection of MNPs and chemicals absorbed at their surface in a POT. Optical approaches to measure particles' mass in air and vacuum are emerging, based on force modulation and levitodynamics techniques [15], but new methods will be needed to tailor the sensitivity below the picogram regime and for measurements in liquids. Integrating wave front shaping with spectroscopic detection is expected to speed up MNPs classification, by the generation of multiple traps.

The development of dyes capable to selectively tag different plastics would permit fast materials discrimination, by implementing SOT with multi-wavelength lasers. Advances in deep-UV CW lasers technology ($\lambda_L$ < 250 nm) will enable label-free fluorescence detection of MNPs ($\lambda_{FL}$ > 300 nm) and fluorescence-free Raman analysis ($\lambda_R$ < 270 nm). Single particle photo-oxidation and fragmentation studies will benefit from variable wavelength UV-lasers in the UVB and UVA range (280 – 315 and 315 – 400 nm). White light lasers can enable SOT in the NIR. Objectives corrected for chromatic aberrations up to 2500 nm will be needed for precise light focussing and collection.

Portable, wi-fi controlled Raman spectrometers, even handheld, are on the market. A clever coupling of these instruments with SOT is a starting point towards compact, field-deployable setups. All parameters will be controlled remotely and results visualized on rugged notebooks or tablet screen, empowered by Artificial Intelligence for particles classification and big data analysis.

**Concluding Remarks**

In conclusion, SOT are moving the first steps in the field of environmental sciences and food analysis. First applications to real-world samples, limited to Raman analysis, have been shown. Either used to trap and analyse single particles or coupled to field-flow fractionation techniques, SOT have the potential to fill the technological gap in MNP detection. In the next years research and development will be oriented towards the exploration of the full potential and limits of SOT in terms of sensitivity, size limitations, multispectral analysis and chemical information that can be extracted from contaminated MNPs and in complex environments. Portable SOT configurations will enable on-site data acquisition (e.g., on a boat, dock, along a production line, or even in an underwater vehicle), allowing scientists for a broad spatial coverage (rivers, lakes, ponds, seas) and to collect time series data.

**BIOPHYSICAL APPLICATIONS FROM BIOMOLECULES TO CELLS**

## 33 — Mechanosensitivity of molecular bonds and enzymatic reactions

*L. Gardini[1,2], G. Bianchi[1,3], A. Kashchuk[1,3], M. Capitanio[1,3]*

[1] LENS – European Laboratory for Non-linear Spectroscopy, University of Florence, Italy.
[2] National Institute of Optics – National Research Council, Florence, Italy.
[3] Department of Physics and Astronomy, University of Florence, Sesto Fiorentino, Italy.

**Status**

Mechanical energy is one of the driving forces of living systems. Animals, tissues, and cells respond to mechanical stimuli and adapt to them. This capacity is inherently coupled to the susceptibility of enzymatic reactions and molecular bonds to mechanical force, which shape the energy profiles of biomolecule structural states (**Figure 47**a). Therefore, mechanochemistry and mechanoenzymology, i.e., how mechanical forces affect chemical and enzymatic reactions (**Figure 47**b), is at the base of mechanotransduction processes, which translate mechanical signals into cascades of biochemical reactions that shape gene expression profiles and live cell function.

Optical tweezers have been applied to investigate force dependence of enzymatic reactions since the mid 1990's. Studies on molecular motors such as myosin and kinesin have pioneered the field and allowed the measurement of nanometre-sized motor steps, piconewton forces, and kinetics on a time scale of few milliseconds. As the spatial resolution of the technique improved, it became possible to directly observe the movements of enzymes such as RNA polymerase, which reads DNA in single base pair steps (≈0.35 nm) to synthesize mRNA [1], and the ribosome, which moves mRNA in single codon steps (≈1 nm) during protein synthesis [2]. The achievement of optical tweezers with angstrom resolution was, thus, a major breakthrough in the field, revealing for the first time how force influence transcription and translation processes.

On the other hand, improvements in temporal resolution allowed researchers to dissect the chemomechanical cycles of motor proteins with increasing detail, unveiling multiple conformational changes during force production in myosin and kinesin motors. These discoveries pushed optical tweezers towards few microsecond temporal resolution, which represents current state-of-the art [3,4].

The achievement of such detailed measurements was made possible by the development of a plethora of different experimental configurations. Among the most common, a single-bead assay is usually applied to processive motors such as conventional kinesin and myosin V (**Figure 47**c), a suspended two-bead assay has been employed to investigate DNA and RNA processing enzymes with angstrom resolution (**Figure 47**d), while a three-bead assay is used for non-processive motors such as muscle myosin (**Figure 47**e) [5,6].

Another important class of experiments enabled by optical tweezers investigates the kinetics of molecular bonds and their force dependence. Such studies have allowed the understanding of important biomolecular interactions, such as receptor–ligand interactions, which underlie communication of cells with the external environment, or interactions of proteins with the cytoskeleton, which are implicated in cell shape, mechanics, and motility. Mainly two approaches have been used to study the load dependence of molecular bonds: dynamic force spectroscopy and force-clamp. In dynamic force spectroscopy, force on the bond is increased at different loading rates and the rupture forces are measured. In force-clamp, the bond lifetime is measured under different constant forces. While the latter method allows direct measurement of the load dependence of bond lifetimes (**Figure 47**b), the two approaches have their own pros and cons, and theoretical frameworks to switch from one measurement to the other have been developed [7].

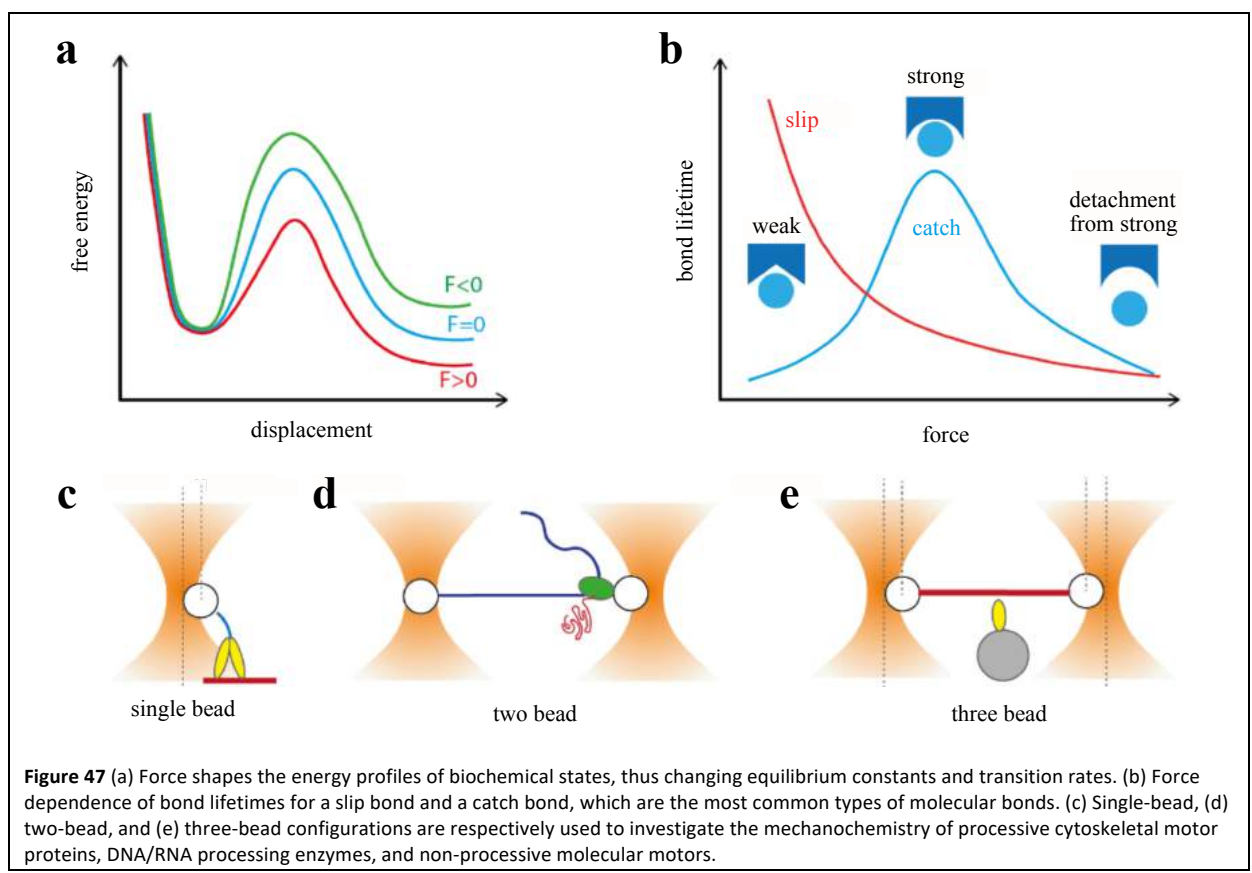

**Figure 47** (a) Force shapes the energy profiles of biochemical states, thus changing equilibrium constants and transition rates. (b) Force dependence of bond lifetimes for a slip bond and a catch bond, which are the most common types of molecular bonds. (c) Single-bead, (d) two-bead, and (e) three-bead configurations are respectively used to investigate the mechanochemistry of processive cytoskeletal motor proteins, DNA/RNA processing enzymes, and non-processive molecular motors.

**Current and Future Challenges**

The last decades have seen impressive advancements in optical tweezers resolution. Spatial resolution has reached the limit dictated by thermal forces, which are inherently present in single molecule studies in aqueous solution at room temperature. Temporal resolution has reached the limit imposed by the mechanical relaxation time of the bead probe and the connected biological molecules, which is also a constitutive element of the experimental method. Current resolution limits will be difficult to overcome, but the spatial and temporal resolution limits that have been reached, respectively on the angstrom and microsecond scale, will be sufficient to explore fine details of most enzymatic reactions and biomolecular interactions. The application of state-of-the-art techniques to a plethora of motor proteins and molecular bonds will allow us to explore fine details of their function and mechanical regulation.

On the other hand, the research field is now facing the big challenge of increasing the complexity of in vitro reconstituted systems to bring them closer to real-world biological systems. Typical optical tweezers experiments involve the study of the interaction between two single molecules, whereas multi-molecular interactions and complex regulatory mechanisms are most common in cells.

A prototypical example are the molecular complexes that form at junctions between cells (adherens and tight junctions), between cells and the extracellular matrix (focal adhesions), or between the nucleus and the cytoskeleton (LINC complex), which comprise a large number of adaptor proteins that link the cellular or nuclear membrane to the cell cytoskeleton (**Figure 48**a). Although these molecular complexes together with mechanosensitive ion channels are supposed to play a pivotal role in mechanotransduction processes, their mechanoregulatory properties are largely unexplored, also because of the lack of adequate techniques to study multi-molecular interactions with sufficient detail.

Similarly, the mechanistic properties of motor protein function have been mostly explored through the interaction between a single motor protein (or functional subdomains) and a cytoskeletal filament, whereas molecular motors in cells are regulated by a large number of binding partners, cytoskeletal filaments are arranged in complex networks through an array of actin and microtubule binding proteins, and the different motor proteins work in teams that are connected through lipid vesicles (**Figure 48**b).

Moreover, optical tweezers in vitro experiments are usually performed under non-physiological conditions. For example, ATP concentration in motor protein experiments is usually in the micromolar range to slow down motor speed and better distinguish steps, whereas ATP in cells is on the millimolar range. Analogously, experiments are performed at room temperature (22°C) instead of 37 °C, and salt concentration is usually lower than in the cell cytosol to increase the number of molecular interactions and statistics. However, extrapolation of these data to physiological conditions is not trivial and the emerging features might differ significantly from what occurs in cells.

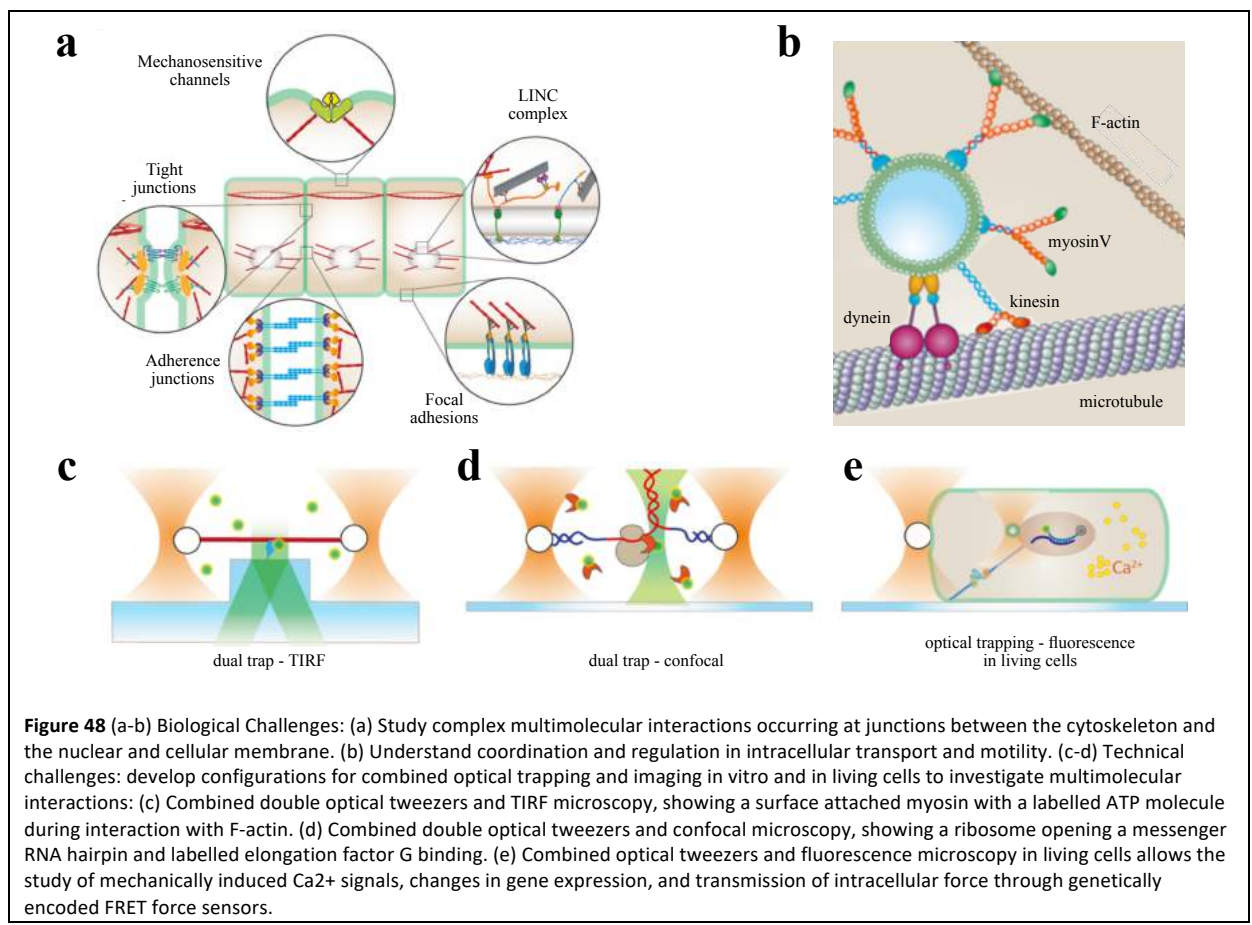

**Figure 48** (a-b) Biological Challenges: (a) Study complex multimolecular interactions occurring at junctions between the cytoskeleton and the nuclear and cellular membrane. (b) Understand coordination and regulation in intracellular transport and motility. (c-d) Technical challenges: develop configurations for combined optical trapping and imaging in vitro and in living cells to investigate multimolecular interactions: (c) Combined double optical tweezers and TIRF microscopy, showing a surface attached myosin with a labelled ATP molecule during interaction with F-actin. (d) Combined double optical tweezers and confocal microscopy, showing a ribosome opening a messenger RNA hairpin and labelled elongation factor G binding. (e) Combined optical tweezers and fluorescence microscopy in living cells allows the study of mechanically induced Ca2+ signals, changes in gene expression, and transmission of intracellular force through genetically encoded FRET force sensors.

**Advances in Science and Technology to Meet Challenges**

Future advances in the field will critically rely on the development of novel methodologies and techniques for the study of the mechanochemistry and mechanoenzymology of biological systems under experimental conditions that closely mimic the cellular environment. Ideally, future technologies should enable researchers to investigate single molecules in real cellular or multicellular systems.

The need to increase the complexity of nanomechanical measurements in vitro is pushing the field toward combining optical tweezers setups with techniques that allow the simultaneous detection of multiple signals. Optical tweezers have been frequently combined with fluorescence microscopy, owing to its specificity and single molecule sensitivity. Pioneering efforts in this regard were driven by Yanagida's lab, where a combination of three-bead assay with total internal reflection fluorescence (TIRF) microscopy was developed to detect the binding of fluorescently labeled nucleotides to muscle myosin to observe the chemomechanical coupling between force production and ATP hydrolysis [8] (**Figure 48**c). More recently, a suspended dumbbell assay was combined with single-molecule fluorescence microscopy and microfluidics to investigate homologous recombination by RAD51 proteins on single-stranded DNA [9]. Later, Comstock et al. combined Ångström resolution optical tweezers and confocal microscopy with single molecule sensitivity to detect changes in DNA extension induced by protein binding [10]. A similar configuration was later used to investigate how ribosome's internal conformational changes are coupled with the activity of translocation factors to unwind mRNA secondary structures [11] (**Figure 48**d). Heller et al. combined optical tweezers with STED super-resolution microscopy to investigate binding of transcription factors on DNA and their diffusive search of target sequences with high spatial resolution [12]. Such combinations of optical manipulation and fluorescence microscopy (also known as fleezers) will open the way to studies on the multi-molecular interactions occurring at cellular junctions between adaptor proteins and cytoskeletal filaments, a fundamental step toward a better understanding of mechanotransduction mechanisms.

Significant efforts have been done also in the field of motor proteins to study how teams of motors coordinate their stepping. Different assays have been developed to study motor cooperativity in teams of kinesin motors, multiple processive myosin V, or arrays of muscle myosin motors interacting with unregulated [13] and regulated [14] actin filaments. An impressive effort allowed Lombardo et al. to reconstitute in vitro three-dimensional actin networks onto which fluid-like liposomes were transported by Myosin Va teams [15]. The authors were able to define each actin filament's 3D spatial position using super-resolution STORM microscopy and the 3D trajectories of the myosin-transported liposomes moving within the actin network by single particle tracking. We foresee that the application of optical tweezers within this or similar assays will allow a better understanding of how motor proteins experience force in living cells.

**Concluding Remarks**

Methodologies that allow probing single enzymatic reactions and molecular bonds in living cells are the ultimate step toward understanding the complex multimolecular interactions that occur *in vivo*. Among the many applications, optical tweezers have been applied in living cells to measure the strength of single molecular bonds and single steps of motor proteins while, in combination with fluorescence microscopy, to probe mechanically induced ionic or genetic signals [16] (**Figure 48**e). A promising tool for the measurement of force in living cells with molecular detail is given by force sensors based on Forster resonance energy transfer (FRET) [17], which can be genetically encoded and combined with optical tweezers, opening the way to investigate how mechanical signals propagates in living cells [18]. Still, the complexity of the cellular environment poses notable limitations in accessing quantitative measurements with single molecule detail and future developments of novel and complementary techniques will play a key role in advancing the field.

## 34 — Biological applications

*Lynn Paterson*
Institute of Biological Chemistry, Biophysics and Bioengineering, School of Engineering and Physical Sciences, Heriot-Watt University, UK.

*P. H. Jones*
Department of Physics & Astronomy, University College London, UK

**Status**

Applications of optical tweezers to biological systems have been, and remain, one of the outstanding successes of the technique. Trapping of viruses and bacteria using visible laser light was demonstrated by Arthur Ashkin [1] very shortly after the discovery of the optical gradient force and the invention of optical tweezers, and was soon followed by the use of near infrared laser light which minimises optical damage to the trapped cell [2]. In this early work both whole cells (bacteria, protozoa, yeast, red blood cells), and sub-cellular structures (chloroplasts within green algae, organelles within protozoa) were trapped and manipulated without damage. Optical trapping of living cells and sub-cellular structures continues to have important impact in fields ranging from environmental biotechnology to molecular genetics and cancer.

Optical tweezers have been used to measure global-cell properties such as the elasticity of whole red blood cells, finding a link between the mechanical property of cell deformability and medical conditions such as diabetes [3]. Alternative trapping geometries such as the fibre-based optical stretcher [4] have also been used to induce whole-cell deformation for mechanical phenotyping of, e.g., cancer cells. Alternatively, mechanical properties of localised area of the cell, for instance a region of the cell membrane, can be measured by using a microbead handle to pull a membrane tether (**Figure 49**a). From the applied force (inferred from the displacement of the microbead in the calibrated trap) and the extent of the tether, the membrane stiffness can be determined. De Belly et al. [5] used this method to measure membrane stiffness of pluripotent stem cells, elucidating the signalling mechanism that regulated the decrease in membrane tension required for subsequent shape change at cell differentiation.

Bolognesi et al. [6] pulled tethers from a pair of synthetic vesicles by drawing them apart after adhesion, thus forming interconnected compartments which remained chemically distinct, as shown by fluorescence imaging (**Figure 49**b). Vesicles were also stimulated to fuse and merge using laser light to heat gold nanoparticles attached to the vesicle surfaces (**Figure 49**c). Here, three vesicles contained separate components required for protein synthesis, leading to expression of GFP in the fused vesicle, also referred to as the cell mimetic microreactor.

Insight into the biochemical response of a cell to mechanical stimuli was previously obtained by Wang et al. [7], who applied a force to the transmembrane integrins by means of an optically trapped microbead, and observed the subsequent wave of activation of Src across the cell using FRET (**Figure 49**d).

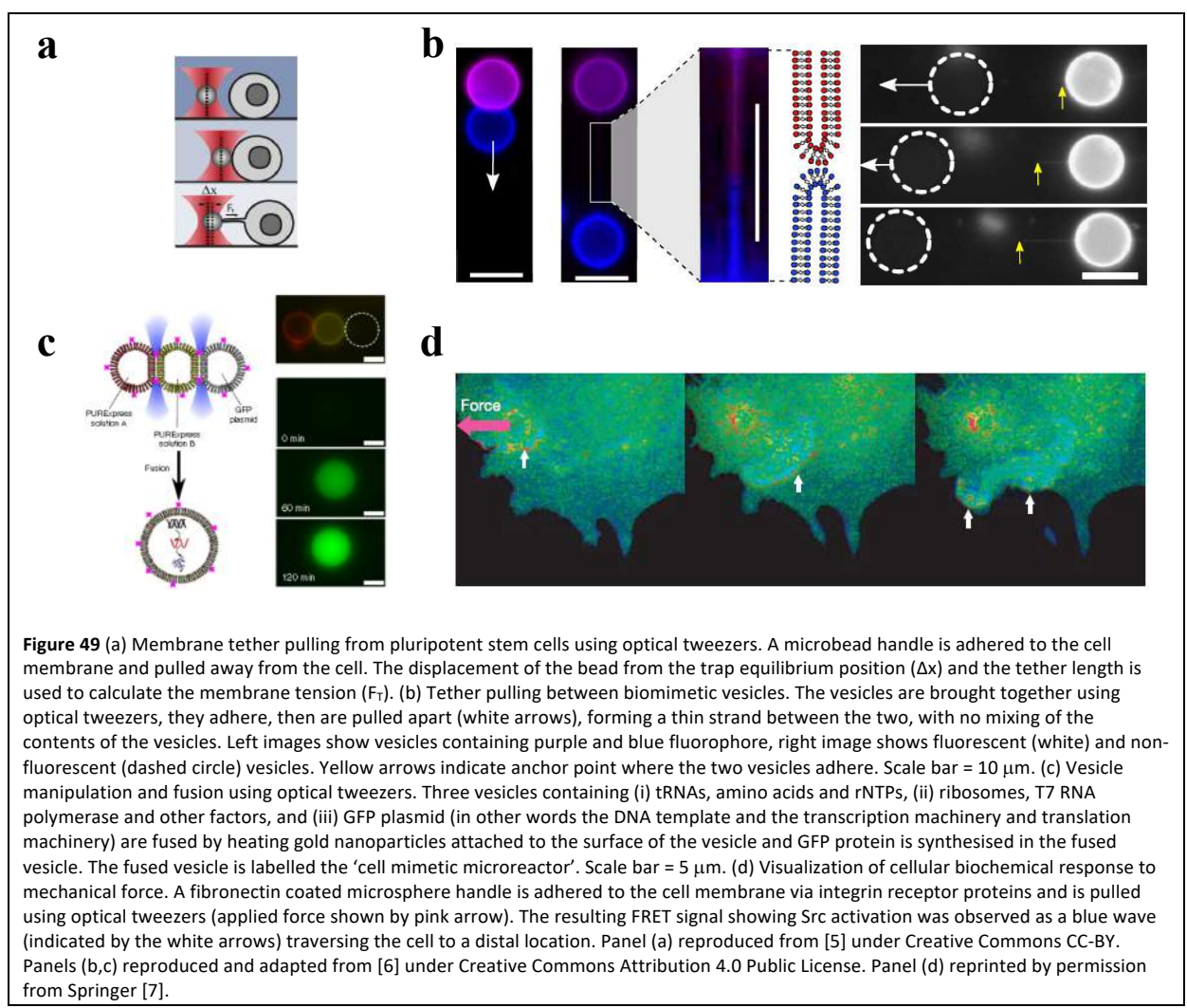

**Figure 49** (a) Membrane tether pulling from pluripotent stem cells using optical tweezers. A microbead handle is adhered to the cell membrane and pulled away from the cell. The displacement of the bead from the trap equilibrium position (Δx) and the tether length is used to calculate the membrane tension ($F_T$). (b) Tether pulling between biomimetic vesicles. The vesicles are brought together using optical tweezers, they adhere, then are pulled apart (white arrows), forming a thin strand between the two, with no mixing of the contents of the vesicles. Left images show vesicles containing purple and blue fluorophore, right image shows fluorescent (white) and non-fluorescent (dashed circle) vesicles. Yellow arrows indicate anchor point where the two vesicles adhere. Scale bar = 10 μm. (c) Vesicle manipulation and fusion using optical tweezers. Three vesicles containing (i) tRNAs, amino acids and rNTPs, (ii) ribosomes, T7 RNA polymerase and other factors, and (iii) GFP plasmid (in other words the DNA template and the transcription machinery and translation machinery) are fused by heating gold nanoparticles attached to the surface of the vesicle and GFP protein is synthesised in the fused vesicle. The fused vesicle is labelled the 'cell mimetic microreactor'. Scale bar = 5 μm. (d) Visualization of cellular biochemical response to mechanical force. A fibronectin coated microsphere handle is adhered to the cell membrane via integrin receptor proteins and is pulled using optical tweezers (applied force shown by pink arrow). The resulting FRET signal showing Src activation was observed as a blue wave (indicated by the white arrows) traversing the cell to a distal location. Panel (a) reproduced from [5] under Creative Commons CC-BY. Panels (b,c) reproduced and adapted from [6] under Creative Commons Attribution 4.0 Public License. Panel (d) reprinted by permission from Springer [7].

**Current and Future Challenges**

Although optical tweezers have enabled remarkable progress and discoveries in the life sciences, there remain many challenges, both physical and biological, that will require significant scientific and technological advances.

One such challenge is optical trapping *in vivo*. While considerable insight has been gained from *in vitro* experiments on single molecules and cells, even these experiments are not straightforward. Frequently the molecules or cells under study are purified or freed from their 'normal' surrounding complex environments of the intracellular space or the extracellular matrix, respectively. Furthermore, since cells, from bacteria to human cells, form part of larger 3D systems, it is clearly desirable to scale up experiments to multi-cellular systems, tissues or even whole organisms, particularly when considering signalling between cells, or physiological function.

One field where cell–cell interactions are key is immunology. Interaction between cells is at the core of the immune response, and influence virtually all aspects of immunity, however it is difficult to obtain precise control over experimental parameters such as interaction times or interaction strengths. Optical tweezers certainly have a role to play in immunological studies as they have the facility for precise control over cell–cell interactions (both temporally and spatially) which aids the development of quantitative immunological models.

Further challenges include the extraction of quantitative data from trapping in complex environments, which often requires precise calibration of the optical trap derived from observing the Brownian motion of a trapped particle. While this is straightforward and routinely performed for spherical particles in water, to calibrate the optical trap *in vivo* or *in cyto* is challenging for a number of reasons: the trapping beam is distorted as it passes through the cell membrane, the cytoskeletal network and membrane-bound organelles of the cytosol, plus the trapped object itself may be irregular in shape and in a crowded and complex environment with spatially and/or temporally varying refractive index and viscoelasticity throughout the volume. Additionally, active processes within cells, e.g., polymerisation and depolymerisation of the cytoskeleton, continually exert forces on a trapped particle. As a result of this complexity, standard calibration methods that are based on predictable thermal fluctuations alone can no longer be used. For optical tweezers to be used at their maximum potential as a force sensor or force actuator for living systems, alternative strategies are urgently required.

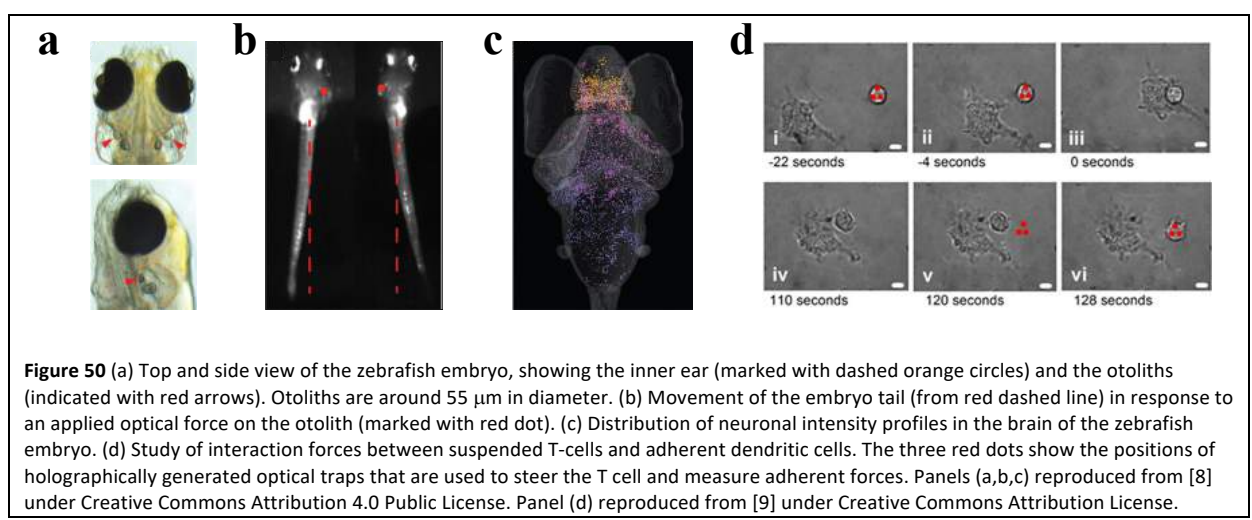

**Figure 50** (a) Top and side view of the zebrafish embryo, showing the inner ear (marked with dashed orange circles) and the otoliths (indicated with red arrows). Otoliths are around 55 μm in diameter. (b) Movement of the embryo tail (from red dashed line) in response to an applied optical force on the otolith (marked with red dot). (c) Distribution of neuronal intensity profiles in the brain of the zebrafish embryo. (d) Study of interaction forces between suspended T-cells and adherent dendritic cells. The three red dots show the positions of holographically generated optical traps that are used to steer the T cell and measure adherent forces. Panels (a,b,c) reproduced from [8] under Creative Commons Attribution 4.0 Public License. Panel (d) reproduced from [9] under Creative Commons Attribution License.

**Advances in Science and Technology to Meet Challenges**

Some progress towards meeting these challenges can be made through physical techniques. For example, imaging through highly scattering biological tissue can be enabled by using adaptive optics technology [8]. In adaptive optics the wavefront aberrations introduced by propagation through a turbid medium are corrected by means of phase control of the incoming laser beam. This can be achieved using a spatial light modulator (SLM) to pre-compensate the phase shifts produced by scattering. Furthermore, advances in technologies have enabled dynamic corrections with speeds of several kHz.

Alternatively, a judicious choice of specimen can aid greatly *in vivo* optical trapping experiments. The transparency of the zebrafish embryo has led to its role as model system for studying the mechanisms of development and disease. Otoliths (calcium carbonate crystals or 'ear stones') in the embryo inner ear (**Figure 50**a) are responsible for acceleration sensing. Optical tweezers have been used to apply a force to otoliths *in vivo* in order to mimic motion of the organism [8]. The embryo's tail was observed to move in response to the force exerted on the otolith, thereby demonstrating a behavioural response of an entire organism that was stimulated by optical tweezers (**Figure 50**b). Additionally, the use of optical tweezers on these structures meant that the organism could be held stationary and therefore imaged with high resolution using selective plane imaging microscopy. Whole brain calcium imaging with cellular resolution was achieved by using genetically encoded calcium indicators, generating a brain-wide map of the cells of the vestibular system (**Figure 50**c).

Holographic beam shaping using SLMs also enable dynamic, shaped traps, for example a triple beam trap used for the study (and control) of cell-cell interactions, such as in the initial stages of immune response (**Figure 50**d). Here, the force between T-cells and dendritic cells was measured, and a significantly greater force was required to separate T-cells from their specific antigen-presenting dendritic cells compared to those without the antigen [9]. Significantly, change in measured adhesion force due to the intervention of pharmacological therapies demonstrates the emerging role of optical tweezers in drug discovery for inflammatory disease and immunotherapy research.

Most work described so far measures or exerts forces in the image plane (2D), but cells and multi-cellular systems are 3D. Accurate and high spatial resolution 3D tracking of trapped microspheres is possible using a novel technique that images multiple focal planes simultaneously [10]. This method permits accurate 3D trap calibrations necessary to make force measurements and perform microrheology in 3D biological systems.

**Concluding Remarks**

Despite the complexities inherent in optical tweezers experiments on biological systems, the progress that has been achieved to date is remarkable. There remain, though, significant challenges in applying the method to arbitrary systems that are not chosen for favourable properties, e.g., embryonic stem cells that are not sensitive to external mechanical cues [5], or organisms that are optically transparent [8]. Recent technological development has gone some way to meeting these, and in the process revealed yet more fascinating biology. Further advances may be envisaged through the combination of optical tweezers with complementary techniques, such as super-resolution microscopy, selective plane imaging microscopy, multi-plane imaging, spectroscopy, high throughput (microfluidic) technology or acoustic trapping, as well as the development of novel molecular probes to elucidate cell signalling mechanisms in response to the stimulation and interaction experiments made possible by optical tweezers. In turn these can be expected to realise advances in emerging application areas such as neuroscience, immunology or synthetic biology. In conclusion, optical tweezers is a unique tool that enables remote, non-contact manipulation and exertion of forces on single cells and sub-cellular structures. As such it is expected to reveal new biology across scales: from molecular pathways to cell behaviour and fate to whole animal behaviour and phenotype.

**Acknowledgements**

LP acknowledges support from Technology Touching Lives grant funded through EPSRC/BBSRC/MRC (Engineering and Physical Sciences Research Council/Biotechnology and Biological Sciences Research Council/Medical Research Council) joint grants EP/R03 5563/1, EP/R035156/1 and EP/R035067/1. PHJ acknowledges financial contribution from the MSCA-ITN-ETN project ActiveMatter sponsored by the European Commission (Horizon 2020, Project Number 812780).

## 35 — Quantifying material properties of living matter

*Kirstine Berg-Sørensen*
Department of Health Technology, Technical University of Denmark

*Younes F. Barooji*, *Lene B. Oddershede*
Niels Bohr Institute, University of Copenhagen, Denmark

**Status**

Optical tweezers are the only nano-tool capable of reaching inside living organisms and soft matter essentially without perturbing the system while making precise measurements of forces, distances, and material properties. Thus, they are a highly valuable tool to quantify physical properties of biological matter that influence processes essential for life. Here, we present their application to the *in vivo* investigation of the material properties of the cell cytoplasm.

A convenient way to characterize material properties of the cellular cytoplasm is to follow the stochastic motion of a tracer particle, which can be endogenous or artificially placed inside the cell, as exemplified in [1] where lipid granules in *S. pombe* cells were tracked. In this case, the optical tweezers is merely used for its ability to precisely track the position, $x(t)$, of a particle with refractive index contrast as a function of time, $t$. Using photodiodes, such tracking measurements can be performed at frequencies up to the MHz range, thus uncovering viscous and (visco)elastic properties in a large frequency range. This is important because the (visco)elastic properties of biological tissue depend strongly on the probing frequency, with passive and active processes taking place at different time scales.

An informative physical parameter which can be extracted from positional time-series is the scaling exponent, $\alpha$. This exponent characterizes the root-mean-squared displacement of the probe particle, $\langle \Delta x^2(t) \rangle \propto t^\alpha$, while also characterizing the power-law behavior of the positional power spectrum, $P(f) \sim f^{-(1+\alpha)}$, which typically is present in the kHz range, relevant for cytoskeletal dynamics. Also, $\alpha$ appears in the complex shear modulus with real and imaginary frequency-dependent parts, $G'(f), G''(f) \sim f^\alpha$, describing the viscous and elastic components. **Figure 51** displays an example of $P(f), G'(f)$, and $G''(f)$ obtained from tracking tracer particles in water and inside living viscoelastic embryonic stem cells (data from [2]). In a viscoelastic medium, the tracer particle is influenced both by the viscous and elastic components of the medium (the cytoplasm) and by the (weak) optical trapping potential, wherefore a simple inverse relation between $\langle x^2 \rangle$ and the trap spring constant does not hold. To address this, one needs to employ appropriate procedures [3,4]. When carefully avoiding the frequency range affected by the trapping potential, an analysis as illustrated in **Figure 51** provides quantification of material properties with sub-celllular precision inside individual cells, tissues or whole organisms. Examples of this are provided in the following section.

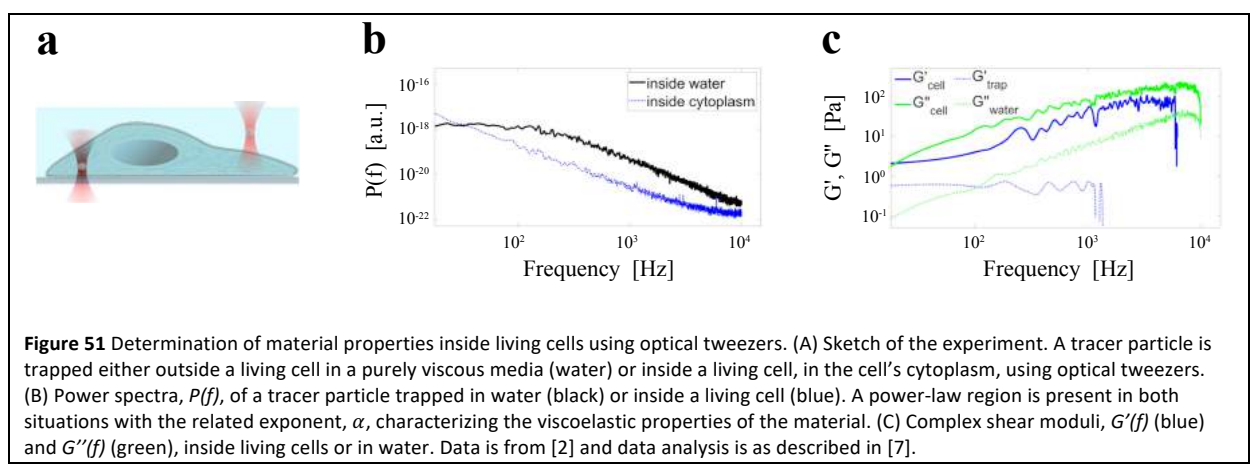

**Figure 51** Determination of material properties inside living cells using optical tweezers. (A) Sketch of the experiment. A tracer particle is trapped either outside a living cell in a purely viscous media (water) or inside a living cell, in the cell's cytoplasm, using optical tweezers. (B) Power spectra, *P(f)*, of a tracer particle trapped in water (black) or inside a living cell (blue). A power-law region is present in both situations with the related exponent, $\alpha$, characterizing the viscoelastic properties of the material. (C) Complex shear moduli, *G'(f)* (blue) and *G''(f)* (green), inside living cells or in water. Data is from [2] and data analysis is as described in [7].

**Current and Future Challenges**

The cell cycle is a life-essential process regulating reproduction and growth for all organisms. Optical tweezers have been applied to study material properties of the fission yeast *S. Pombe* [1], during the cell cycle. By analysing the scaling exponent, $\alpha$, the cell cytoplasm was found to be significantly more viscous and less stiff when the cell is undergoing mitosis (active cell division) in comparison to when it is in the interphase. This was probably caused by higher abundance of cytoskeletal biopolymers, such as actin or microtubules, increasing the stiffness of the cytoplasm during the interphase. A recent study [5] use optical tweezers to address the same important question with Madin-Darby canine kidney (MDCK) cells as model system: phagocytosed 1-µm particles were tracked, and the study concluded that the cytoplasm (however, not the cortical region) became softer and more viscous, during mitosis, in accordance with the observations in *S. pombe* [1]. One way for the cell to control the cytoplasmic viscoelasticity is through protein-mediated crosslinking of actin filaments, as also recently probed by optical tweezers [6].

The differentiation of stem cells is of crucial importance for the development of a functional organism. Stem cell differentiation depend strongly on the mechanical properties of the substrate on which they are cultured, or on the niche in which they grow. Optical tweezers experiments have correlated the development of pancreatic organoids and the viscoelastic properties of the matrix in which they are embedded. For example, these studies have found that the degree of pancreatic branching depends on the stiffness of the matrix [7]. Also, optical tweezers were used to demonstrate that the viscoelastic properties of naive embryonic stem cells depend on culturing conditions, where a culturing condition (in Serum/LIF) pushing the cell towards differentiation causes the cell periphery to become stiffer than if preserved in the naive state [2].

Optical tweezers have also uncovered how the properties and behaviour of cancer cells are linked to the material properties of the cell and its matrix. A tumour is often discovered because the tissue has an increased stiffness, caused by an increased density of collagen fibres in the extracellular matrix. Optical tweezers were used to demonstrate that only invasive cancer cells can adjust their cytoplasmic material properties in response to the material properties of their matrix [8]. Also, it was shown that cancer cells belonging to a spheroid and leading the invasion into collagen matrices are more viscous compared to cells in the spheroid centre.

**Advances in Science and Technology to Meet Challenges**

While optical tweezers are powerful tools to map out material properties in living cells with sub-cellular resolution, one needs to be aware that intracellular material properties change as a function of time. Hence, it is important to control the frequency or timescale of observation. A conclusion drawn at a certain frequency or timescale, might only be relevant for this particular condition. Also, if one uses, e.g., endocytosed or phagocytosed particles, the properties probed are (only) those of the related pathway and not of the entire cytoplasm (unless there has been a specific release from the pathway).

A major challenge is to perform optical tweezers measurements in a nearly non-invasive manner. One concern is how to avoid, or minimize, temperature increase and phototoxicity in live specimens due to absorption of the laser light. Based on absorption in water and of biomolecules like haemoglobin, a laser wavelength in the near-infrared region seems preferable. Heating and phototoxicity may further be reduced by use of a structured or pulsed trapping laser beam rather than a CW Gaussian beam profile [9]. However, even when using CW lasers and typical laser powers (up to some 100s mW) such as in the studies here presented, the observed temperature increase is typically less than 1K. Further, particle types that allow reduction in laser power while still measuring a significant signal magnitude are highly useful. As an example, highly doped upconverted nanoparticles [10] give promise of applying significantly higher forces than thereto proven possible by optical tweezers and with particles which appear to be non-toxic for living cells. This would be an important step towards quantitative force-measurements in living cells and for optical tweezers to approach a force range that until this date has only been achievable by AFMs. While optical tweezers can be operated in an almost non-invasive manner, also deep inside living organisms, AFMs cannot reach inside a living cell, or a living organism, without perturbing the membrane, hence, causing serious physiological damage.

Another interesting perspective is to introduce particle types that may trigger cellular responses or act as intracellular biocompatible sensors, as illustrated in **Figure 52**. One example of such nanoparticles are nanodiamonds with NV defects, a system that has demonstrated capabilities as sensors of temperature or magnetic fields. Plasmonic nanoparticles are also quite interesting, as their temperature increase can be controlled externally by the trapping laser, and their local heating can be used to provoke and control cellular responses.

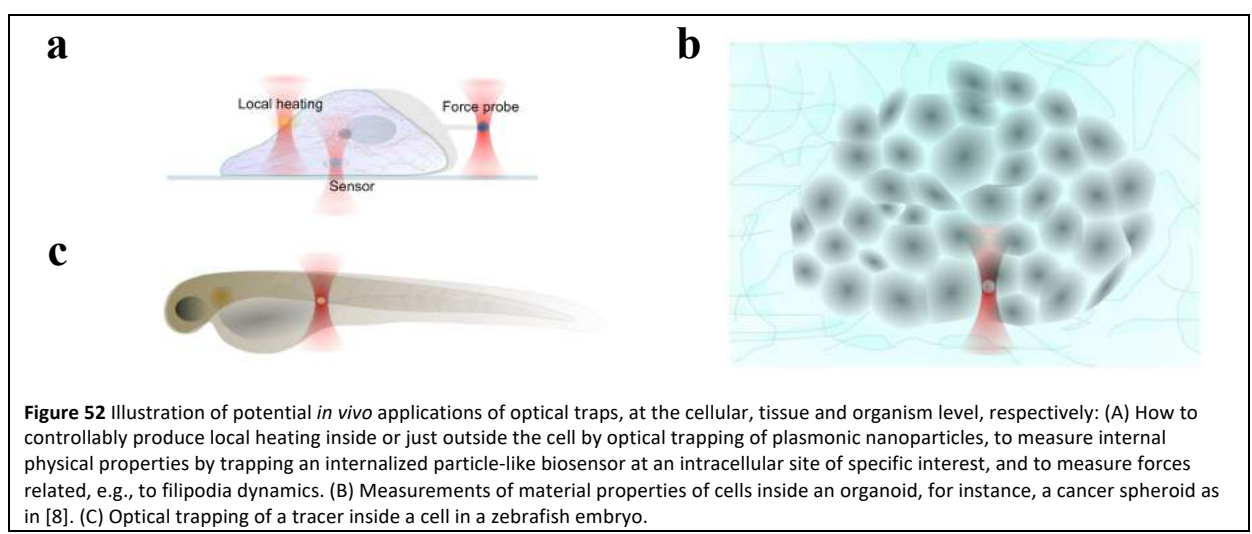

**Figure 52** Illustration of potential *in vivo* applications of optical traps, at the cellular, tissue and organism level, respectively: (A) How to controllably produce local heating inside or just outside the cell by optical trapping of plasmonic nanoparticles, to measure internal physical properties by trapping an internalized particle-like biosensor at an intracellular site of specific interest, and to measure forces related, e.g., to filipodia dynamics. (B) Measurements of material properties of cells inside an organoid, for instance, a cancer spheroid as in [8]. (C) Optical trapping of a tracer inside a cell in a zebrafish embryo.

**Concluding Remarks**

Optical tweezers can provide quantitative information on material properties of living organisms, both at the single molecule, (sub)cellular, tissue, and whole-organism level. Such information is crucial to link material properties to biological function and to obtain a general mechanistic insight into soft and living matter. The raw data from an optical tweezers measurement often consist of a time series carrying information on the physical properties of the probed system. Analyses of such time series are likely to be considerably refined in the years to come, for instance by employing machine-learning-based approaches. As a result, much more information on the probed system may be found. When passive observation of time series of the stochastically moving trapped particle are combined with active relaxation measurements, both trap characteristics and the viscoelastic moduli of the cytoplasm may be extracted [3,4,6], providing promise of precision in quantification of forces also inside live cells. In conclusion, optical tweezers based on near-infrared light are a versatile, almost non-invasive tool, and future refinements in both instrumentation and data analysis as well as a clever choice of probe particles are likely to bring ground-breaking insights into both materials science, biology, biomedicine, and biophysics in the years to come.

**Acknowledgements**

KBS acknowledges support from the Novo Nordisk Foundation (grant no NNF20OC0061673) and Independent Research Fund Denmark (grant no 0135-00142B).

## 36 — Cellular studies using quantitative phase imaging

*Pegah Pouladian, Daryl Preece*
Beckman Laser Institute, University of California Irvine, USA.

**Status**

Of critical importance to optical trapping (OT) is the shape and refractive index of objects under study. It is perhaps natural then that OT should be paired with technologies that can access this information. A variety of different optical techniques have been used to gather information about the phase changes created when light propagates through a transparent microscopic object. These include digital holographic microscopy (DHM), quantitative phase microscopy (QPM), and quantitative phase imaging (QPI). Typically, these techniques utilize interferometry-based imaging technology that retrieves information about an object under study by interrogating the phase structure of the transmitted or reflected light to gain multidimensional information about a sample, such as the three-dimensional morphology of an object, its refractive index and its associated dynamics. This results in a versatile label-free imaging modality that has attracted considerable research interest in quantitative biological cell imaging [1-3]. This method may be used to detect subcellular structures as well as the morphology of entire tissue slices, and because no external labelling is required, it can be used to assess dynamical processes from milliseconds to days [4].

Though many optical trapping experiments utilize dielectric microspheres, most biological experiments at the micron size scale utilize objects which consist of complex shapes with spatially/temporary varying refractive indices. Within this context, DHM/OT enables discrimination of "trappable" objects and apprehension of the effect of optical forces on objects under study. DHM has also been shown to be an effective way to evaluate the amount of photodamage, if any, caused by the trapping beam [5]. This also allows for adaptive trapping of biological specimens and feedback on the effects of optically trapping a biological entity.

Integrating OT with DHM has enabled biologists to quantitatively study cellular forces, morphological changes, and mechanical properties, such as cellular deformation and elasticity. For instance, optical fibers combined with DHM have been used as a portable system for precise dynamic measurements of cells such as red blood cells (RBCs) and lipid cells [6]. Furthermore, this integration has been applied to detect axial changes in RBC shape caused by OT, enabling the determination of the biophysical (both mechanical and material) changes of the cells by measuring the refractive index and elasticity simultaneously [7]. This methodology can also be used to improve DHM imaging systems by introducing mechanical actuation and rotation to objects under study. This in turn facilitates tomographic measurements which can be used to give a truly 3D view of refractive index and morphology.

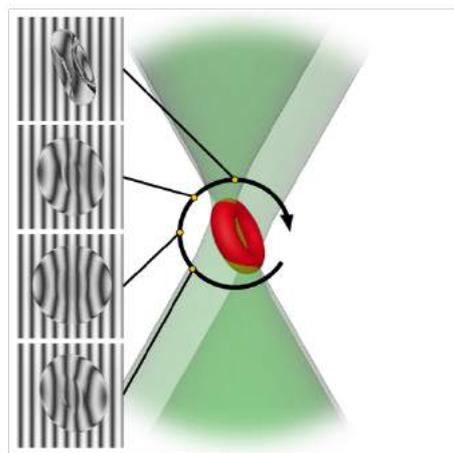

**Figure 53** Tomographic images can be taken of red blood cells using holographic optical tweezers to rotate the cell while information about the phase delay is acquired at different angles. A 3D model of the cell can then be constructed.

**Current and Future Challenges**

As mentioned previously, one of the advantages of this technology is the capability to simultaneously manipulate and interferometrically measure cells of interest. This enables access not only to the cell's refractive index but also to morphological changes taking place in the cell. Since morphology (thickness) and (intracellular RI), are both extracted from the optical path length of light traveling through the cell, calculations typically require one parameter be held constant. Despite this, this functionality has been widely applied to study red blood cells and other simple cell types such as yeast cells, the potential of the technology to conduct experiments with other larger and more physiologically complex cell typeshas yet to be fully utilized. Though, three-dimensional measurement of refractive index and adaptive optical trapping techniques have been shown previously this technology still requires significant development [8]. The increased complexity of less homogenous cells means that combined manipulation and measurement cannot be achieved without substantial finesse.

Another underutilized area of investigation is the trapping of submicron objects. These nanoscale objects are of significant interest in both biological and non-biological contexts and label-free imaging of small particles in the cell such as mitochondria, vesicles, and other organelles are desirable. As with all optical imaging systems DHM, is fundamentally diffraction-limited making a determination of particle properties like refractive index difficult. Despite these nanoparticles can be influenced by optical forces and often exhibit significant refractive index differences from their surrounding environment though so far there has been little attempt to use combined OT/DHM systems to either actuate or measure nanoscale particles.

One challenge for DHM/OT systems which is shared by many biophotonics systems is the physical integration of different technologies into a single optical train. For instance, many DHM systems have traditionally used a reflective coverslip inside a Michelson interferometer setup to achieve high precision phase measurements. When the sample is microscopic, matched objectives are used to eliminate aberration across the arms of the interferometer.

Since this design is not easily compatible with optical trapping, transmissive techniques have been developed. These techniques improve integration between OT and DHM but also can limit resolution and introduce aberration which must be corrected subsequently.

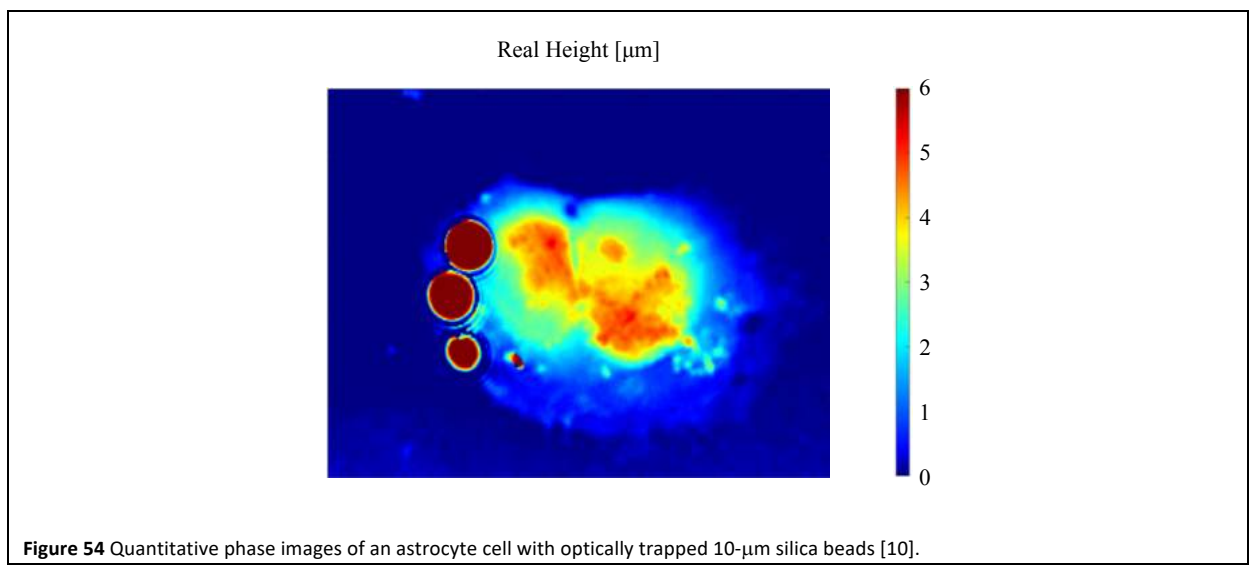
**Figure 54** Quantitative phase images of an astrocyte cell with optically trapped 10-μm silica beads [10].

**Advances in Science and Technology to Meet Challenges**

With improved access to information about the optical properties of objects under study computational processing becomes even more necessary to utilize this information fully. There has recently been an increase in the use of artificial intelligence algorithms for the optimization of optical trapping. This approach may also provide a route to improve the processing of data from DHM/OT systems. Advances in computation and processing could decrease computational overheads while improving trapping performance.

Improvements of digital holographic techniques may provide access to information about submicron particles; several super-resolution phase-contrast methodologies have been proposed that may increase access to submicron particles. However, label-free super-resolution imaging remains challenging. While improving resolution via oblique illumination or by using composite Fourier holography images can lead to improvements in resolution, images are still limited to a few times the native resolution of the objective. In order to achieve the improvement required for nanoparticle or even single-molecule imaging, new phase imaging modalities must be designed. Currently, pump–probe-type experiments have been suggested as possible solutions to imaging small particles. Though this in itself brings with it several problems such as photon flux and phase sensitivity. Ultimately any solution will require extensive study.

Another useful application of optical tweezers is photonic force microscopy (PFM), in which 3D information from a sample can be retrieved by evaluating the interference of scattered light as an optically trapped bead scans the sample. Combining optical tweezers for PFM, to a DHM can overcome the RI/thickness coupling problem, by providing additional information about the morphology of the sample under study.

Though many of the topics discussed so far have involved sophisticated optical setups, it is worth noting that wider adoption of this technology necessitates that it becomes simpler and cheaper. Along these lines, several papers have already shown how consumer products such as DVD pickups can be used to create compact, cost-effective tools [9].

Another area where simplification of the technology may be possible is the integration of spatial light modulator technology which can be used for both holographic optical trapping of particles and for phase shifting and aberration correction in DHM. This has previously been combined effectively to image and trap. Clearly, the simplification of more sophisticated systems poses a challenge in the long term, however, increased usage of computer processing may be able to alleviate some physical constraints [8].

**Concluding Remarks**

Two factors of importance in cellular studies are how cells are observed and how cells are probed or manipulated. While holographic methods allow for advanced label-free observation optical trapping allows for precise non-destructive biomechanical tests. Fundamentally, the technology discussed here represents a way to understand how light propagates and scatters at the cellular level, and in gaining such understanding we improve both aspects. This adds much to biophotonics, but challenges still exist and utilizing the technology to its full potential. Improved digital equipment and processing have enabled phase measurements to be taken faster and more reliably and optical traps to be moved with greater dexterity and ingenuity. One would expect this trend to continue. However, it is unlikely that future systems will be simply refinements of current designs. Investigations into biological phenomena will continue to drive greater integration of different technologies and will also require a diversity of approaches to probing cells. While trapping at the limits of our current proficiency will require the development of new as yet unknown technologies.

**Acknowledgements**

Many thanks to the Beckman Laser Institute and the Airforce office for scientific research (FA9550-20-1-0052) for funding this work and to Toyohiko Yamauchi and Micheal Berns for useful discussions on this topic over the years.

## 37 — Optical and spectral analysis of single cells

*Caroline Beck Adiels*

Department of Physics, University of Gothenburg, 41296 Gothenburg, Sweden

**Status**

All living cells react to external stimuli by a multitude of cellular responses. The response type and amplitude depend on the level of stimuli of the extracellular environment and internal stochastic fluctuations of cell transcription and translation [1]. Cells are rarely clonal, nor their response homogenous. Even in a population of genetically identical cells, the response can be slightly different. However, in traditional biological settings, only averaged response values are considered upon external stimulation of a cell population. As such, the distributions of the reaction kinetics and the response outcomes may be misleading. To overcome these limitations, the research field of single-cell analysis has emerged during the past decades, combining a large set of technologies and analysis tools. Among these are the optical tweezers used in a non-invasive manner to trap and manipulate single cells.

Budding yeast is a commonly used unicellular model organism of eukaryotic cells and they are easily trapped and manipulated using optical tweezers. Combining optical tweezers with advanced imaging or spectroscopic techniques can provide high temporally and spatially resolved information. Such data can complement cellular information attained via traditional methods and reveal details that advance our understanding of underlying biological mechanisms. For example, suspended yeast cells trapped with optical tweezers were analysed based on their biochemical content using Raman spectroscopy, which resulted in probe-free metabolic data acquisition [2-5]. The real-time acquisition modality provided by optical tweezers also permitted monitoring dynamic responses upon extracellular stimuli, such as osmosis [6].

When imaged with fluorescence microscopy, fluorescent reporter proteins expressed in genetically modified yeast provide single-cell information on a sub-cellular level (see, e.g., **Figure 55**). As such, both intensity information (corresponding to the amount of the tagged protein in question) and localisation, migration or diffusion velocities of the fusion protein can be acquired. Incorporating fluorescence imaging and optical tweezers in the same optical setup also allows choosing individual cells to analyse from a cell population based on unique features, such as which reporter proteins they express [7].

Adding microfluidics to the toolbox further augment the options and enables time-lapse [8,9] and perturbations [10,11] studies impossible to perform in bulk without disturbing the equilibrium (see, e.g., **Figure 56**). Advancing the technology to manipulate more sensitive cells (e.g., mammalian cells [12]) in complex environments in junction with probe-free imaging modalities will add to the analysis possibilities. Such data will boost the biological understanding of single-cell and inter-cellular behaviour of importance to clarify the onset and development of human disease.

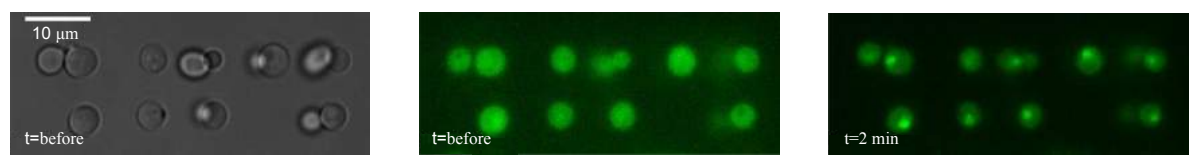

**Figure 55** Example of single yeast cell data acquired from cells trapped with an optical tweezer, attached to the bottom of a microfluidic chamber, and imaged with fluorescence microscopy. The leftmost and middle images display the cell array in the bright field and fluorescence modality before the extracellular environmental change. Initially, the fusion protein (Hog1-GFP) resides in the yeast cytosol

(green). Upon exposing the cells to high osmolarity, the fusion protein migrates and, after 2 minutes, primarily resides in the nuclei, as visualised in the rightmost image.

**Current and Future Challenges**

The use of optical tweezers in combination with other techniques offers a wide range of research opportunities. This combination can significantly increase data quality and experimental efficiency. One area of particular interest regards mammalian cell analysis, ranging from blood and cancerous cell analysis to single-cell and organoid manipulation. However, the technology must be user-friendly and more accessible to reach its full potential and transition from fundamental research fields to life-science-oriented ones. This involves calibration approaches adaptable to the living system in question. Currently, no such standardised protocol is available; instead, specially trained personnel is required.

Further, parallelised, high-throughput setups are preferred to attain trustworthy diagnostics, sensing, or drug testing results. Such endeavours set high standards on the readout and may require novel technologies for imaging, sorting, or sensing in combination with controlling the optical tweezers.

For long-time studies, mammalian cells require a much more controlled environment than, e.g., yeast cells. Mammalian cells need a narrow interval of temperature, humidity and specific gas ratios, and the possibility of a perfused environment is beneficial. Also, biocompatible materials for living cells need to simultaneously be consistent with the wavelength of the optical tweezers and possibly with that of an optical readout. A mismatched scenario might affect the outcome and provide a false readout. Finally, any possible photodamage must be taken into consideration to avoid induced photodamage or thermal stress due to trapping or imaging.

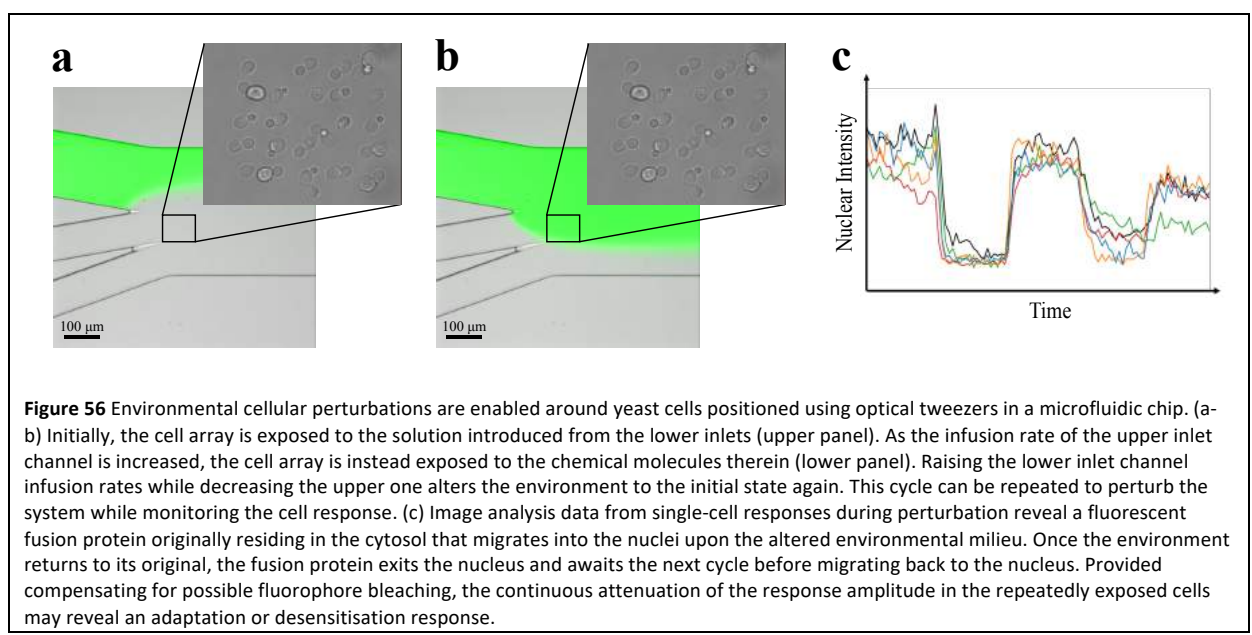

**Figure 56** Environmental cellular perturbations are enabled around yeast cells positioned using optical tweezers in a microfluidic chip. (a-b) Initially, the cell array is exposed to the solution introduced from the lower inlets (upper panel). As the infusion rate of the upper inlet channel is increased, the cell array is instead exposed to the chemical molecules therein (lower panel). Raising the lower inlet channel infusion rates while decreasing the upper one alters the environment to the initial state again. This cycle can be repeated to perturb the system while monitoring the cell response. (c) Image analysis data from single-cell responses during perturbation reveal a fluorescent fusion protein originally residing in the cytosol that migrates into the nuclei upon the altered environmental milieu. Once the environment returns to its original, the fusion protein exits the nucleus and awaits the next cycle before migrating back to the nucleus. Provided compensating for possible fluorophore bleaching, the continuous attenuation of the response amplitude in the repeatedly exposed cells may reveal an adaptation or desensitisation response.

**Advances in Science and Technology to Meet Challenges**

First, portable and affordable setups will make the optical tweezers technology accessible to the conventional biological and medical research community. This involves compact fibre lasers and optical designs that allow simultaneous imaging and optical tweezing.

Second, automatic tweezers alignment and calibration software are necessary to avoid relying on specialised technicians installing the setup. With today's advancements in machine learning, automatised software to initiate cell manipulation or measurement could be applied. Further, such algorithms can also track and trap individual cells in an automatised way for further analysis, e.g., upon cell replication, isolating the daughter cell from the mother cell [13]. Another possible application is automatic detection and manipulation of, e.g., lipid vesicles, augmenting them close a given cell surface. After that, to follow their potential uptake and intracellular fate while monitoring metabolic and morphological cell responses, e.g., via Raman spectroscopy, holographic tomography or fluorescent microscopy.

As a cell sorting tool, one single trap might not meet the required speed for high-throughput compared to what existing available fluorescence-activated cell sorting (FACS)-systems can provide. Holographic optical tweezers would partly overcome this limitation, and such automatic manipulation systems have already been reported [14]. As post-experimental cell analysis is highly valued for different off-chip analyses, combining custom-made microfluidics and machine learning, automatic detection and sorting into compartments for dedicated treatment will be possible. For instance, optical tweezers can facilitate accessing specific cells for downstream intracellular biochemical analysis after an optical cellular analysis.

Further, most adherent cells thrive in a 3D setting surrounded by other cells. Mammalian cells can therefore be cultured as spheres or organoids, useful for cell-cell interactions, tissue engineering and stem cell differentiation studies. Such organoids are established following various protocols, and microfluidics-based approaches have proven highly controlled and consistent. Using optical tweezers to sort by size, cellular content, or other measurable properties opens an avenue of analysis opportunities before not even acknowledged.

**Concluding Remarks**

Moving into the field of mammalian single-cell analysis is highly interdisciplinary, where the requirements of the living system must be made compatible with that of the optical system. Close collaborations across disciplines are necessary to identify suitable and relevant applications and significantly improve the success rate. Physiologically relevant cell environments will ensure reliable data; meanwhile, long-term optical and biochemical cell studies can reveal whether the cells experienced any possible negative impact by the handling and analysis per se. Combining high-frequency imaging modalities of high spatial resolution with sorting capabilities in microfluidic devices, intra- and intercellular behaviours can be monitored, unveiling complex signalling events that can be perturbed in real-time. Such quantitative data can help explain complex mechanisms underlying cell functionality and dysfunctionality to understand disease progression and find possible treatments.

**Acknowledgements**

CBS acknowledges support from the European Commission through the MSCA-ITN "DeLiver" (grant agreement No. 766181) and from the Swedish Foundation for Strategic Research (ITM17-0384).

## 38 — Raman tweezers for single-cell analysis and sorting

*Anna Chiara De Luca*

IEOS-CNR, Institute of Experimental Oncology and Endocrinology "G. Salvatore", Napoli, Italy

**Status**

Understanding biological heterogeneity at single-cell level is a frontier of biomedicine and precision medicine. Indeed, single-cell analysis can shed light on the origin, differentiation and heterogeneity of cancer cells as well as detect the response and resistance of individual cells to drug treatment. Conventional methods based on the use of fluorescence probes or immunochemical staining require as a prerequisite a known cell phenotype or require the use of selective biomarkers or probes for which the expected alterations are known. Labelling can require several extensive and time-consuming sample preparation steps affecting cell viability or perturbing normal cell function. It can suffer from non-specific binding and photobleaching making difficult long-term analysis. Multiplexed detection is limited by the broad spectral profile of the fluorescence and the relative crosstalk. In addition, traditional biological assays such as qRT-PCR and Western blotting are unable to capture differences at the single-cell level resulting in loss of information on cell-to-cell heterogeneity.

Raman spectroscopy is a vibrational spectroscopic approach providing information about the chemical composition, molecular structure and molecular interactions in cells and tissues. Although it is difficult to assign individual bands to single biomolecules because of the cell complexity, a specific fingerprint may be obtained, non-destructively and without needing exogenous labels. The use of chemometric tools of supervised and unsupervised data analysis can then be applied for cell identification and classification [1]. Since 2002 Raman spectroscopy has been applied in combination with optical traps to study single cells, bacteria, spores and other biological systems in their aqueous environment [2]. The Raman tweezers approach allowed real-time and continuous monitoring of the expression of specific proteins of single trapped bacteria under antibiotic stress [3] and for non-destructive sorting living microorganisms [4]. Dual beam Raman tweezers has been developed for stretching single red blood cells and measuring their deformation, allowing both mechanical and chemically properties to be analyzed simultaneously [5] (**Figure 57**).

Although these experiments provide the intrinsic biochemical profile of the cells at a given state, the relatively low sorting throughput greatly hampers their broader applications.

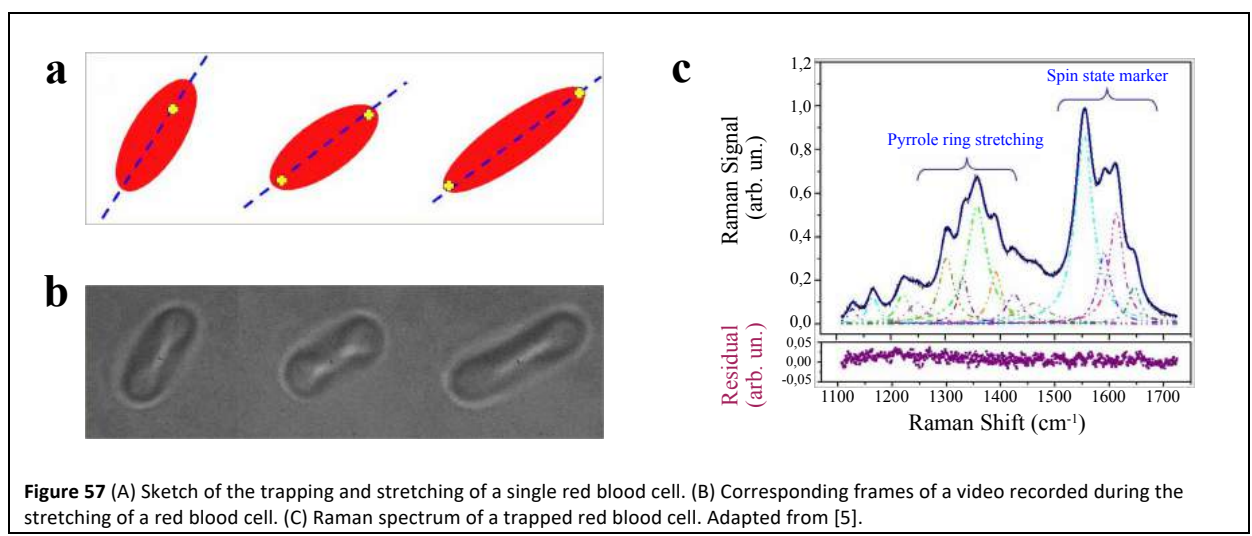

**Figure 57** (A) Sketch of the trapping and stretching of a single red blood cell. (B) Corresponding frames of a video recorded during the stretching of a red blood cell. (C) Raman spectrum of a trapped red blood cell. Adapted from [5].

**Current and Future Challenges**

When the cell trapping and Raman classification is synchronized with multichannel microfluidic devices, automated and fast Raman cell sorting becomes possible. An integrated optofluidic platform for Raman-activated cell sorting (RACS) has already demonstrated to be complementary to the more standard fluorescence-activated cell sorting (FACS) [6]. In the RACS platform, the optical tweezers are employed to trap the single cell from the bulk cell-suspension, the single cell Raman signal provides an intrinsic label-free "fingerprint" of biochemical components in the cell for the purpose of functional classification or identification then the cell of interest is moved and separated from the cell suspension for physical isolation and collection. This system has been used for label-free isolation of yeast and bacterial cells for further amplification and cultivation [7]. However, it still suffers of low throughput due to the intrinsically weak Raman signal and the semi-automated operation, in which cells are identified and manually sorted one by one.

As schematized in **Figure 58**, two challenges should be taken into account to accomplish high-throughput RACS:

1. To improve the rate of single Raman spectra acquisition by employing advanced Raman techniques.

2. To improve the quality and the efficiency of RACS platform by employing microfluidics-based cell manipulation techniques.

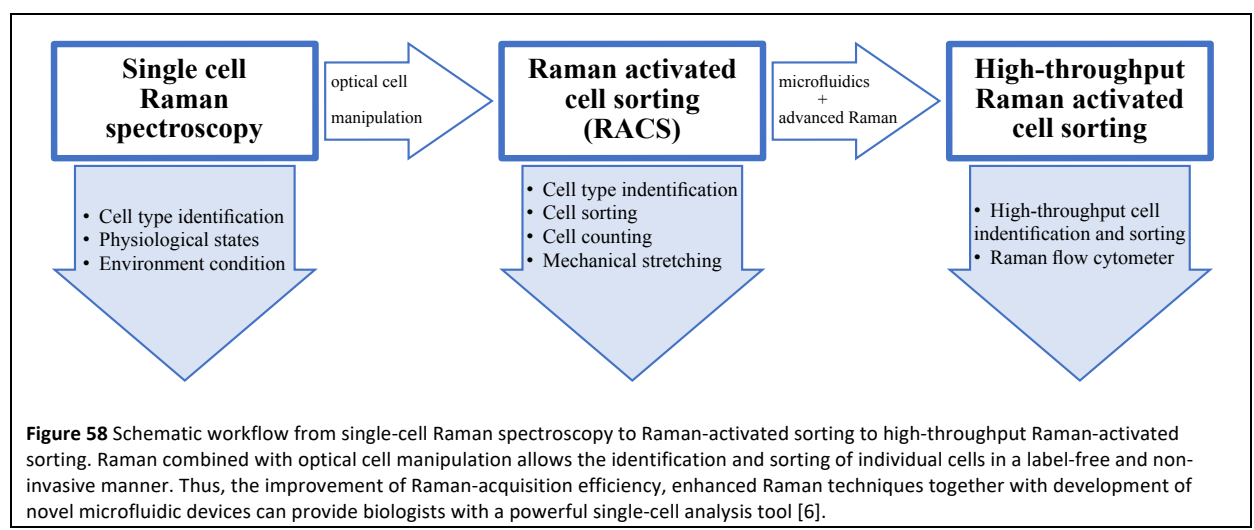

**Figure 58** Schematic workflow from single-cell Raman spectroscopy to Raman-activated sorting to high-throughput Raman-activated sorting. Raman combined with optical cell manipulation allows the identification and sorting of individual cells in a label-free and non-invasive manner. Thus, the improvement of Raman-acquisition efficiency, enhanced Raman techniques together with development of novel microfluidic devices can provide biologists with a powerful single-cell analysis tool [6].

**Advances in Science and Technology to Meet Challenges**

With the current available instruments, a single bacteria Raman signal can be acquired within milliseconds. The sensitivity of Raman detection can be further enhanced by employing novel Raman instruments, including laser sources, optical filters, and low-noise and sensitive EMCCD camera.

A variety of enhanced Raman techniques have been developed and demonstrated, including surface-enhanced Raman spectroscopy (SERS) and tip-enhanced Raman spectroscopy (TERS). The use of metallic nanoparticles or metallic tips can amplify the Raman signal by as much as 10 to 14 orders of magnitude, strongly increasing sensitivity of the signal detection. Indeed, SERS has been used to study the dynamic release of drugs in living cells with resolution of few attograms at millisecond time scales [8].

However, SERS cannot be considered really label-free. Alternatively, Coherent anti-stokes Raman spectroscopy (CARS) and stimulated Raman spectroscopy (SRS) can achieve 3–5 orders of enhancement of a selected Raman spectral band. These techniques can significantly increase the signal intensity, but require a greater complexity for the instrumentation as well as interpretation and reconstruction of the data. On the other hand, the utilization of Raman tags (e.g., alkyne, deuterium, metal–carbonyl complex) can be used to detect small target molecules inside cells. Indeed, they provide specific bands in the silent region of Raman spectra of cells, reducing the background and therefore improving the signal detection.

The alignment in the flow between the single cells and the Raman detection spot is also crucial for high-efficiency signal acquisition. In the cell trap–release strategy, an individual cell can be trapped accurately at the Raman detection spot for signal acquisition before being released by periodically applying a trapping force. Alternatively, an integrated fiber laser for single-cell trapping and single-cell Raman acquisition has been demonstrated for target cell sorting [9]. A microfluidic glass chip incorporates functionalities to separate cells from a reservoir, accommodates laser fibres for the trapping and sorting of single cells according to the Raman-based classification. The developed Raman-activated cell sorting enabled the acquisition of Raman spectra from more than 100 000 individual cells and a fully automated sampling of 1000 single cells in less than 20 minutes [10].

**Concluding Remarks**

The development of advanced optical designs for the Raman spectrometer, the involvement of sophisticated microfluidic devices, and the use of chemometric tools of supervised and unsupervised data analysis has revolutionized the combination of Raman spectroscopy and optical tweezers for single cell analysis. Indeed, the integrated optofluidic platform for Raman-activated cell sorting provides high analytical potential overcoming the problem of the complex sample preparation steps affecting fluorescence-based approaches. It allows measurements of the intrinsic molecular makeup of cells, providing a higher number of parameters in a label-free and non-destructive manner. Because no sample preparation is required, cells can simply be trapped and measured. In summary, the single-cell Raman sorting platform provide biologists with a powerful single-cell analysis tool to explore the scientific questions or applications into many areas of biomedicine.

**Acknowledgements**

ACDL acknowledges the financial support by the Italian Association for Cancer Research (AIRC) IG grant no. 21420 "Correlative optical microscopies for cancer imaging".

## OPTICAL FORCES IN SPACE

## 39 — Space tweezers for cosmic dust characterization

*A. Magazzù, D. Bronte Ciriza, O. M. Maragò, M. A. Iatì*
CNR-IPCF, Messina, Italy

**Status**

Dust is everywhere in the Universe. Up to the mid-20th century, cosmic dust was only considered as an annoying fog obscuring starlight. In about half a century, such view has totally changed: dust plays a vital role in many processes, going from the formation of planets, stars, and galaxies to the ignition of a potentially life-bearing chemistry. The importance that cosmic dust, in the form of interstellar, interplanetary, and planetary dust particles, has gained in recent years is driving the study of its nature, composition, and evolution [1].

Optical tweezers (OT) have shown their remarkable capability to trap, manipulate, and characterize a wide range of microscopic and nanoscopic particles, in liquids, air, and vacuum [2,3] finding applications in a wide range of fields: from biology, soft matter, and ultra-sensitive spectroscopy to atomic physics, nanoscience, photonics, and aerosol science [4]. The coupling of optical tweezers with a Raman spectrometer (Raman tweezers) has opened novel and interesting scenarios, allowing the chemical and physical analysis of trapped particles through their vibrational fingerprints [4]. Despite the significant advances in optical trapping techniques and the many examples of cosmic phenomena driven by radiation pressure, the application of OT to space exploration and dust characterization (space tweezers) has yet to be developed [5,6].

The application of space tweezers to study the solar system opens perspectives for the analysis of cometary particles, dust particles in the Martian atmosphere and/or on the Martian and Lunar surfaces (**Figure 59**a). Such analysis is strategic to characterize the chemical and physical properties of such particles with curation facilities designed for an uncontaminated handling of extraterrestrial samples returned by research expeditions [7]. The direct implementation and application of optical trapping and Raman tweezers to space materials open doors to new information that is currently unreachable without bias. In space missions, different techniques have been successfully used to collect or characterize samples [7]. However, they are limited because of the unavoidable sample modification or contamination due to collecting media. An OT system mounted on stratospheric instruments, onboard cometary probes, or on landers and rovers to Mars or to the Moon, would offer a great opportunity for the *in situ* characterization of extraterrestrial particles, without introducing any contamination.

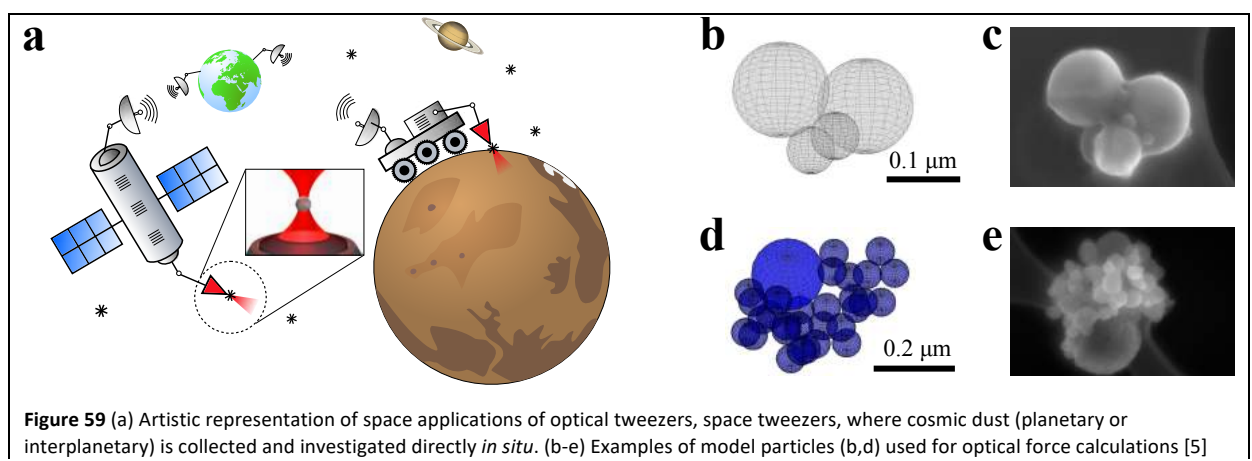

**Figure 59** (a) Artistic representation of space applications of optical tweezers, space tweezers, where cosmic dust (planetary or interplanetary) is collected and investigated directly *in situ*. (b-e) Examples of model particles (b,d) used for optical force calculations [5] corresponding to interplanetary dust grains made of quenched melt silica (b,c) and Ca[O] nanograins on a carbonate melted particle (d,e) collected within the DUSTER project [7]. (b-e) Adapted from Refs. [5,7] with permission of The European Physical Journal (EPJ) and Elsevier.

**Current and Future Challenges**

Challenges for space tweezers combine both theoretical and experimental aspects. From the computational point of view, the diversity of dust particles in an astrophysical context implies a richness of models that need to be developed to calculate the optical forces. For model spherical particles such forces can be calculated by using exact electromagnetic solutions of the scattering problem [4]. However, realistic dust grains are far from exhibiting these simple shapes (**Figure 59**c,e). Aiming to replicate realistic dust scattering properties, different model particles have been proposed that include aggregates (**Figure 59**b,d), stratified, agglomerate debris, or Gaussian random particles [8,9]. While the use of these more sophisticated model geometries better represents physical reality, the increased detail in the morphology comes with the drawback of a higher demand for computational power, limiting the number of configurations that can be numerically explored.

From the experimental point of view, OT have been mostly employed to trap and manipulate micron-sized objects in liquid environment rather than in air or in vacuum. While liquid media minimize the effects of external perturbations on a trapped particle and the effect of inertia on its dynamics [2], trapping in low viscosity medium (air or vacuum) is more challenging. The trapping stability can be affected by external perturbations generated by airflows, acoustic shocks, thermal drift, mechanical vibrations, and also by the intensity noise of the trapping laser [3]. Furthermore, liquid media can suspend the particles for a long time before and after trapping allowing the operator to select and trap a specific single particle within a multitude of particles. In contrast, in air tweezers setups, a large number of particles need to be launched, which eventually approach the laser focus under gravity and are trapped passively [3]. Optical trapping in air or at low pressure poses challenges to select a specific dust grain, position it or store it for subsequent characterization [10]. Once the particle gets out of the trap it is generally lost and it is difficult to retrieve it. In air OT the numerical aperture of the focusing objective can never be greater than unity, giving rise to scattering forces that can overcome the trapping forces. Moreover, long working distance objectives and lenses (having lower NA) are often used in air tweezers setups to prevent lens contamination by the sprayed or launched samples [3]. Because air is less efficient than a liquid in dissipating heat, absorbing particle trapped in air can produce photophoretic forces and convective forces, generated by airflows which often can destabilize the trapped particle dynamics [3].

**Advances in Science and Technology to Meet Challenges**

In recent years much effort has been devoted to the accurate modelling of optical forces acting on complex shaped particles [4], making it possible to calculate such forces for particles in a given position and orientation. However, there is always a compromise between the calculation accuracy and the calculation speed. When studying the particle dynamics of systems where the inertia needs to be considered (for example in air or vacuum), the equations of motion need to be integrated using a very short time step and the optical force needs to be computed in each of these time steps. While the traditional force calculation becomes prohibitively slow to explore these systems, recent advances in machine learning for optical force calculations [11] significantly increase this calculation speed. This is crucial for the accurate modelling of particles of complex shapes. On the other hand, this improvement in the calculation time could allow real-time adjustments of the light intensity and beam shape to enhance the trapping of a given particle in OT experiments.

From the experimental point of view, several experimental configurations have been developed to achieve a stable three-dimensional trapping in air or vacuum. For instance, setups with counterpropagating beams can be used to balance the effect of the radiation pressure on a trapped particle [3] and they can be realized also with optical fibers [12]. Structured light beams, such as Bessel, hollow or vortex beams can be used instead of Gaussian beams either in horizontal or vertical configuration [3,13,14]. In addition, configurations having a single beam, such as the confocal-beam trap, combines the simplicity of a single-beam trap and the robustness of a dual-beam trap. In this configuration, a spherical concave mirror reflects back the incident focused beam, forming a symmetric counter propagating beam [3]. Finally, the stability of a single-beam trap can be increased also by a feedback device that, by controlling the laser power, improves the trapping stability or can even realize laser cooling of the particle motion [2].

**Concluding Remarks**

Space tweezers can be used to trap and characterize extra-terrestrial particulate matter. From the theoretical point of view, it is possible to exploit scattering theory to calculate optical forces on a variety of complex particles of astrophysical interest, simulating atmospheric and planetary environments. On the other side, in recent years much effort has been devoted to build versatile, stable, and more compact setups that can be used to trap micro and nanoscopic particles in controlled laboratory experiments. Space tweezers applications for a non-destructive, non-contact and non-contaminating investigation of extra-terrestrial particles will pave the way to new information on space materials currently unreachable with the instrumentation used in space missions. However, several different challenges still need to be faced, e.g., controlled capture, selection and storage of single grains, improved stability for trapping in extreme environmental conditions, general protocols to optically characterize levitated dust. This means that optical trapping and optical manipulation of particles in space and on planetary body surfaces still need some improvements in the experimental setups. On the other side, applications in curation facilities, designed for the uncontaminated handling and preliminary characterization of extra-terrestrial samples returned by space probes, appear more at hand.

**Acknowledgements**

We acknowledge support from the MSCA-ITN-ETN project ActiveMatter sponsored by the European Commission (Horizon 2020, Project No. 812780) and by the agreement ASI-INAF n.2018-16-HH.0, project "SPACE Tweezers".

## 40 — Diffractive solar sails

*Grover A. Swartzlander, Jr.*

Center for Imaging Science, Rochester Institute of Technology, Rochester (NY), USA

**Status**

Solar sails have recently been tested in space, and more missions are planned in the coming years. Navigating the heavens by means of solar radiation pressure provides multiple advantages not afforded by rocket propulsion: Continuous acceleration, an inexhaustible energy source, and massless propellant, respectively providing non-Keplerian orbits, long durations, and a large delta-V [1]. Modern optical design and fabrication techniques open opportunities to produce more efficient sails compared to the century-old idea of a flat reflective film concept. Just as a jet fighter is designed differently than a crop duster, so too will future solar sails be designed for specific space objectives. Depicted in **Figure 60** are high delta-V missions like solar polar orbiters and long duration missions like sub-L1 halo orbits, both enabling heliophysics science observatories and warning systems for debilitating Earth-bound coronal mass ejections [2].

The current status of solar sailing is akin to the early days of air flight when significant innovations emerged after sustained flight was demonstrated. Over the next decades sailcraft missions may explore difficult-to-reach parts of the solar system, embarking on spiral trajectories toward, away, and over the sun. Other near-term missions may include Earth-orbiting satellites that benefit from the experiences of the ongoing Lightsail-2 mission of the Planetary Society [3]. For longer time spans, one may envision both solar and laser-driven sails that probe interstellar distances on hyperbolic trajectories, perhaps visiting neighboring exoplanets [4]. Both science and the emerging space economy may benefit from relatively low-cost, CubeSat/SmallSat compatible solar sails — especially if optical scientists can enhance the associated figure of merit. For example, diffractive sails were recently proposed to achieve higher momentum transfer efficiencies than a flat reflective sail for spiral trajectory missions [5].

Advancement opportunities include: (1) design and fabrication of sun-facing sails having a high momentum transfer efficiency component perpendicular to the sunline; (2) low mass-to-area ratio sails using metasurface or polarization diffraction grating principles; (3) sail elements that are electro-optically switchable; and (4) robust optical materials that both withstand the harsh space environment (e.g., ionizing radiation and extreme temperature cycles) and are immune to packaging and unfurling damage. The fabrication of diffractive sails may include roll-to-roll imprint technology followed by adhesive stitching to achieve large areas. Before launch the finished product must be folded into a small volume package and mounted to a deployment system comprised of low-mass booms to support the gossamer sail. Once in space, the sail is unfurled and readied for action.

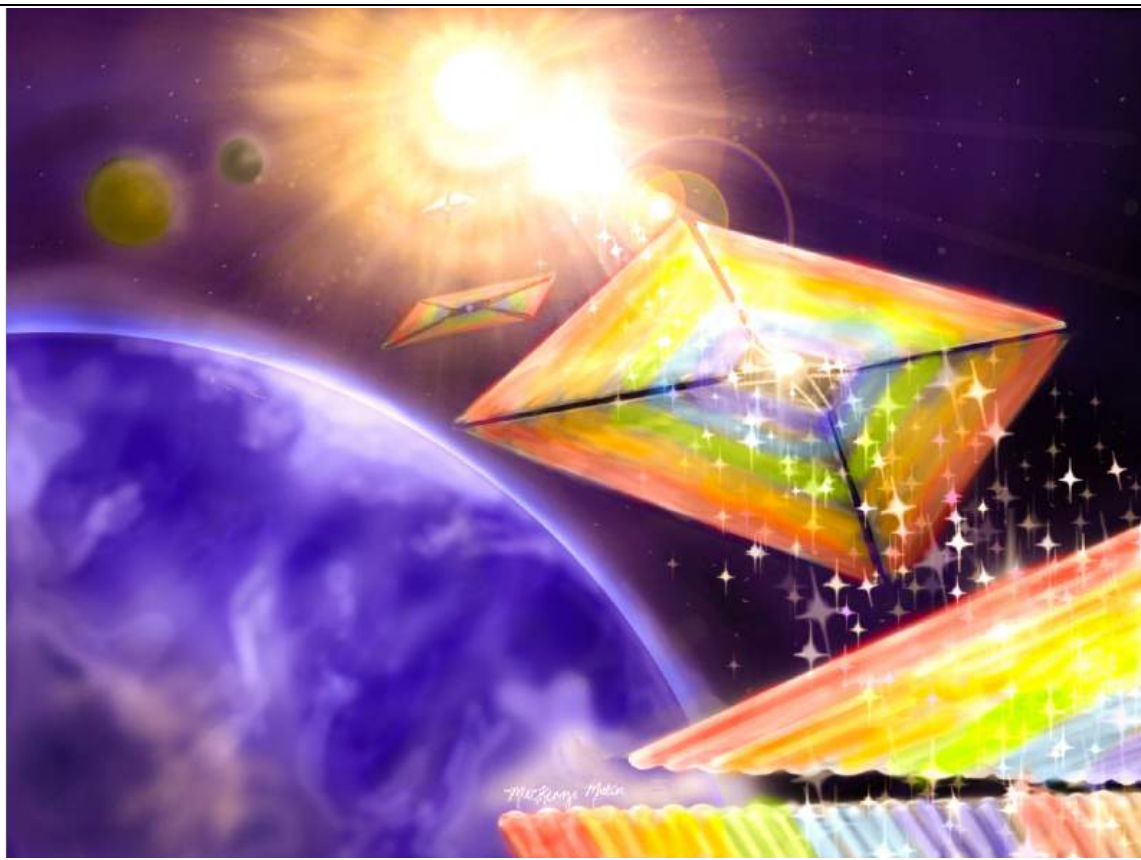

**Figure 60** Artist depiction of a constellation of diffractive light sails leaving Earth en route to the Sun (MacKenzi Martin, RIT, 2013).

**Current and Future Challenges**

Perhaps counter to intuition, the component of radiation pressure force directed away from the sun is less important than the perpendicular component when discussing the inner and outer planets or high solar latitudes [1,5,6]. Whereas the force directed away from the sun elongates orbital ellipticity, the latter enables spiral trajectories that are essential for rendezvous missions. After a rocket positions the spacecraft beyond the influence of Earth gravity, navigation toward (away from) the Sun is achieved by tangentially decelerating (accelerating) from an initial quasi-circular orbit at 1 AU.

A dimensionless figure-of-merit for such missions may be expressed FOM = $(\sigma^*/\sigma)\sin\Theta$, where $\Theta$ is defined as an equivalent scattering angle for a sun-facing sail, $\sin\Theta = \cos\theta_i \sin\theta_s$, where $\theta_i$ is the projection angle of sunlight upon the sail, $\theta_s$ is the net scattering angle measured with respect to the sun line and averaged across the solar spectrum and all scattering angles, $\sigma^* = 1.54$ [g/m$^2$] is a solar constant, and $\sigma = M_{sc}/A$ is the sailcraft areal density where $M_{sc}$ and $A$ are respectively the sailcraft total mass and sail area. Current space technology provides areal densities on the order of 100 $\sigma^*$. For a traditional ideal sail obeying the law of reflection, the optimum incidence angle is $\theta_i = 35.3°$ [1], resulting in a scattering angle $\theta_s = 2\theta_i$ that provides FOM = 0.77 $(\sigma^*/\sigma)$ and $\Theta = 50.3°$.

A challenge for optical scientists is to enhance the FOM. An upper value occurs for the case of a Sun-facing sail $\theta_i = 0°$ along with a uniform grazing scattering angle $\theta_s = 90°$, resulting in $\Theta_{max} = 90°$ and FOM = 1.00 $(\sigma^*/\sigma)$. Further enhancements to FOM may be achieved by reducing the sailcraft mass $M_{sc}$ by using a multi-functional sail that allows for the elimination of mass components, such as those used for attitude control, communication, or thermal management. For example, an electro-optically controlled sail, though higher in mass, may provide net mass savings owing to the elimination of attitude control devices. The first challenge for optical scientists is to achieve an equivalent scattering angle significantly greater than $\Theta = 50.3°$ (preferrable a value approaching 90°).

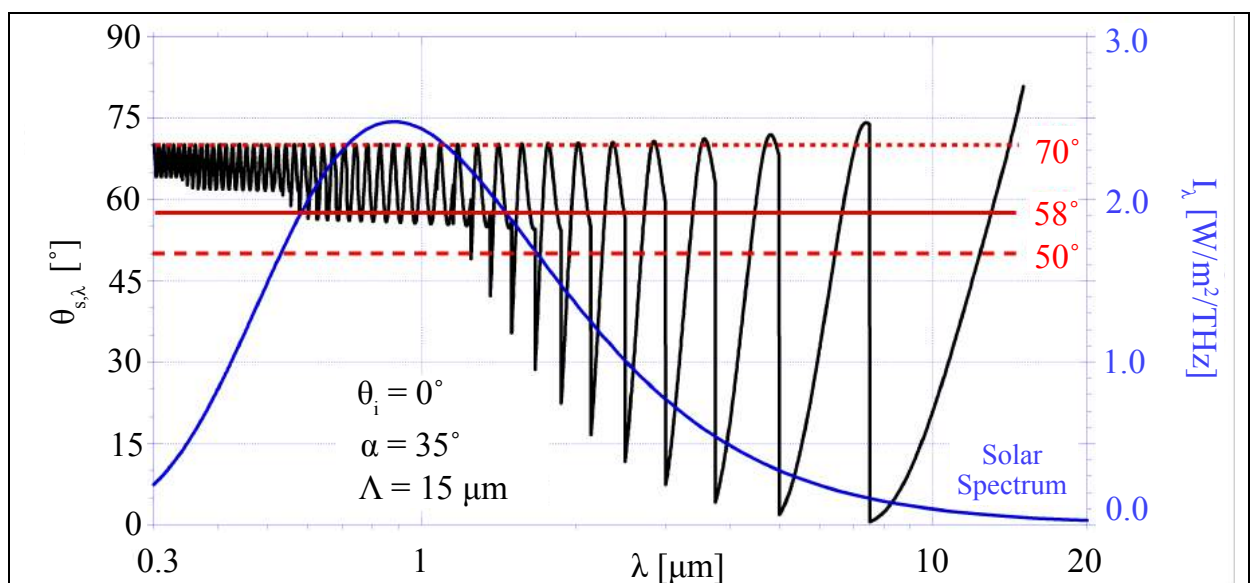

**Figure 61** Semi-log plot of the effective scattering angle for a uniform irradiance distribution and reflective saw-tooth grating with period $\Lambda$, apex angle $\alpha$, and film incidence angle $\theta_i$. Blue: frequency-parameterized blackbody solar spectrum. Red: reflection angle $2\alpha$, wavelength averaged scattering angle weighted for blackbody spectrum $\Theta=58°$, and scattering angle for a flat reflective sail 50°.

**Advances in Science and Technology to Meet Challenges**

A transmission or reflection function resembling $f(x,\lambda) = \exp(2\pi i x \sin(\theta_s)/\lambda)$ is desired, where x is a coordinate that is perpendicular to the sunline, the wavelength $\lambda$ ranges across the solar spectrum, and the scattering angle $\theta_s$ is roughly 90°. Whereas a prism can be made to satisfy these requirements, it would be too massive to function as a solar sail. A periodic sawtooth grating having a surface profile $h(x) = h(x+\Lambda) = h_0 (1 - x/\Lambda)$ may also be used to create a tilted wavefront where $\Lambda$ is the period, $h_0 = \Lambda \tan(\alpha)$ is the height above the surface and $\alpha$ is the apex angle. A short period grating has smaller mass than a long period grating having the same apex angle. On the other hand, a short period grating exhibits pronounced diffraction, producing a smaller equivalent scattering angle $\Theta$ when averaged over the solar spectrum (**Figure 61**). The FOM scales as $\sin\Theta/M_{sc}$. If the mass of the sail is a small fraction of the total sailcraft mass $M_{sc}$, then it is arguably desirable to first optimize the value of $\Theta$.

Advances in optical science in areas like space-variant polarization diffraction gratings [7,8] and metasurfaces [9,10] suggest paths for improving the FOM. Research is needed to design and fabricate reflective or transmissive elements that uniformly scatter light into a single direction or narrow cone with high efficiency. As depicted in **Figure 61**, a simple sawtooth diffraction grating provides opportunities to achieve a larger equivalent scattering angle than a flat reflective sail (58° vs 50°, representing a 10% increase in the FOM), but the $\Theta \sim 90°$ goal is yet beyond reach. At values of $\lambda \sim \Lambda$, the cumulative effect from multiple diffraction orders is to dimmish the effective wavelength-dependent scattering angle, $\theta_{s,\lambda}$ owing to large modulations. The modulation is less pronounced at shorter wavelengths, resulting in a greater momentum transfer efficiency. Research on broadband engineered optical films that enables less wavelength dependent modulation of the scattering angle may provide a path toward greater values of the FOM.

The ability to electro-optically switch the radiation pressure force on advanced diffractive elements (e.g., electro-optically or thermally) may provide significant opportunities to decrease the sailcraft mass, thereby affording greater acceleration. A key target for this is the attitude management system. Spacecraft often suffer a center-of-mass/center-of-pressure offset, resulting in a torque that must be countered to prevent spinning from the desired attitude. Gyroscopes may be used for this purpose; however, they contribute to system mass and they are prone to gimbal lock and failure. Radiation pressure on solar panels have been used in the past to rescue missions from such failures, so it is feasible that switchable diffractive elements could be integrated into a large solar sail to reduce both failure risk and sailcraft mass.

**Concluding Remarks**

Recent coalescing developments including low-cost access to space, CubeSat and SmallSat technology, and the design and fabrication of engineered optical thin film materials, are combining to provide opportunities to significantly enhance the in-space flight performance of solar sails. Besides the benefits afforded to the space science community and to the emerging space economy, the financial impact could amount to trillions of dollars of savings in the form of early warning systems to protect Earth from coronal mass ejections and other space hazards. Replacing a flat reflective solar sail with an advanced diffractive film may provide a ~30% enhancement of the optical momentum transfer needed for spiral trajectory orbits, and what is more, further figure of merit enhancements may be achieved by offsetting sailcraft mass with multi-functional sails. To achieve the required optical bandwidth (e.g., wavelengths ranging from roughly 0.3 to 3 μm) and high scattering efficiency at large angles, innovative ideas from the optical community are sought. Funding for material design, fabrication, and testing, as well as for validation experiments in low Earth orbit will be pivotal. The rapid advancement of aircraft design in the early 1900's serves as a historical guide, pointing toward an exciting future for solar sailing.

**Acknowledgements**

This research was supported by NASA, United States of America Innovative Advanced Concepts Program (NIAC), Grants 80NSSC18K0867 and 80NSSC19K0975. I am grateful to Les Johnson and Andy Heaton (NASA Marshall Space Flight Center) for discussions on solar sailing.